\documentclass[english,aps,prb,twocolumn,nofootinbib,notitlepage]{revtex4-2}
\usepackage{CJK}

\usepackage{amsfonts}
\usepackage{amsmath}
\usepackage{ dsfont }
\usepackage{hyperref}
\usepackage{amssymb}
\usepackage{array,multirow}
\usepackage{longtable}
\usepackage{ragged2e}
\usepackage{xspace}
\usepackage{xcolor}
\usepackage{relsize}
\usepackage[bottom]{footmisc}

\usepackage{tikz}
\usetikzlibrary{calc} 

\newcommand{\be}{\begin{equation}}
\newcommand{\ee}{\end{equation}}
\newcommand{\bea}{\begin{equation} \begin{aligned}}
\newcommand{\eea}{\end{aligned} \end{equation} }
\newcommand{\bi}{\begin{itemize}}
\newcommand{\ei}{\end{itemize}}

\newcolumntype{C}[1]{>{\centering\arraybackslash}p{#1}}

\usepackage{xspace}

\newcommand{\la}{\lambda}
\renewcommand{\be}{\beta}
\newcommand{\eps}{\epsilon}
\renewcommand{\l}{\lambda}

\newcommand{\lp}{\left(}
\newcommand{\rp}{\right)}
\newcommand{\del}{\partial}
\newcommand{\Tr}{\text{Tr} \ }

\newcommand{\mbf}[1]{\mathbf{#1}}
\newcommand{\bpm}{\begin{pmatrix}}
\newcommand{\epm}{\end{pmatrix}}

\renewcommand{\d}{\downarrow}
\DeclareRobustCommand{\Eq}[1]{Eq.~(\ref{#1})}
\renewcommand{\be}{\beta}
\newcommand{\al}{\alpha}
\renewcommand{\l}{\lambda}

\newcommand{\down}{\downarrow}

\usepackage{color}
\usepackage{graphicx}
\usepackage[space]{grffile}
\usepackage{verbatim}
\usepackage{amsmath}
\usepackage{amssymb}
\usepackage{mathrsfs}
\usepackage{wasysym}
\usepackage[caption=false]{subfig}
\usepackage{url}
\usepackage{bbold}
\usepackage{slashed}
\usepackage{epstopdf}
\usepackage{braket}
\usepackage{float}
\usepackage[percent]{overpic}

\DeclareRobustCommand{\App}[1]{App.~#1}
\DeclareRobustCommand{\Apps}[2]{Apps.~#1 and~#2}

\DeclareRobustCommand{\Fig}[1]{Fig.~\ref{#1}}

\DeclareRobustCommand{\Eq}[1]{Eq.~(\ref{#1})}

\DeclareRobustCommand{\Refs}[2]{Refs.~\cite{#1, #2}}

\usepackage{mathtools}
\usepackage{placeins}
\usepackage{booktabs}
\usepackage{setspace}

\allowdisplaybreaks
\graphicspath{{./}{SM/}}

\begin{document}

\title{Exact Fractional Chern Insulators from Hubbard Interactions}

\author{Jonah Herzog-Arbeitman}
\thanks{These authors contributed equally to this work.}
\affiliation{Department of Physics, Massachusetts Institute of Technology, Cambridge, MA 02139, United States}
\author{Rui-Heng Liu}
\thanks{These authors contributed equally to this work.}
\affiliation{Beijing National Laboratory for Condensed Matter Physics and Institute of Physics, Chinese Academy of Sciences, Beijing 100190, China}	
\affiliation{University of Chinese Academy of Sciences, Beijing 100049, China}
\author{Jiangping Hu}
\affiliation{New Cornerstone Science Laboratory, Beijing National Laboratory for Condensed Matter Physics and Institute of Physics, Chinese Academy of Sciences, Beijing 100190, China}
\author{Chen Fang}
\email{cfang@iphy.ac.cn}
\affiliation{New Cornerstone Science Laboratory, Beijing National Laboratory for Condensed Matter Physics and Institute of Physics, Chinese Academy of Sciences, Beijing 100190, China}

\begin{abstract}
Laughlin's wavefunction and its parent interaction, Haldane's pseudopotential, underpin our analytic understanding of fractional quantum Hall effect.
They are an exact pair in the sense that the wavefunction is the unique, zero-energy ground state of its frustration-free parent Hamiltonian.
In this work, we propose exact pairs of many-body wavefunctions for fractional Chern insulators in tight binding models with onsite Hubbard interactions as their parents.
The key to our construction is introduction of a \emph{Landau-dual Chern band}, a multi-orbital Chern band obtained from directly sampling the spin-$s$ Dirac Landau level on a Bravais lattice. 
The exact many-body wavefunctions are given, analytically, in terms of the matrix coordinate $X+iY+J$, where $(X,Y)$ are the lattice coordinates and $J$ is a nilpotent matrix in spin space. The constructed exact pairs cover the fractional Chern states of the Laughlin, the Read-Rezayi, and the Halperin series. Our results prove that FCI ground states exist beyond ideal bands. 
\end{abstract}

\date{\today}
\maketitle

\section{Introduction}

Laughlin's wavefunction \cite{laughlin1983anomalous} showed, in one fell swoop, the existence of a state with fractional Hall conductance in the lowest Landau level (0LL). Although its success was clearly indicated by a large overlap with the Coulomb ground state for a small number of particles \cite{laughlin1983anomalous,haldane1985finite}, Haldane's subsequent discovery of a local ``pseudo-potential" \cite{haldane1983fractional,trugman1985exact} parent Hamiltonian $\nabla^2 \delta(\mbf{r}-\mbf{r}')$ between any two electrons located at $\mathbf{r},\mathbf{r}'$  -- recently proven to be gapped in the thermodynamic limit \cite{openai2026fock,scheer2026hierarchy} -- was decisive evidence that this state describes a physical phase of matter. 

The pair of data, Laughlin's state and Haldane's parent Hamiltonian in the lowest Landau level, are perhaps the most famous example of a frustration-free construction for a correlated electronic ground state.
More broadly, this analytical framework consists of two parts: a model wavefunction $\Psi$ that is a representative of a desired phase of matter, and a positive-semi-definite ``parent" Hamiltonian $\hat{H}$ that annihilates $\Psi$, implying that $\Psi$ is a ground state.  If $H$ is sufficiently well-behaved (for instance, being a sum of short-ranged terms) then such a frustration-free Hamiltonian can provide sufficient conditions for the existence of a phase of matter. Frustration-free Hamiltonians are some of the most versatile and powerful tools in studying strongly interacting physics \cite{affleck1987rigorous,rokhsar1988superconductivity,kitaev2003fault,levin2005string,schroeter2007spin,lian2021twisted,Tovmasyan_2016,herzogarbeitman2026many,han2025exactmodelschiralflatband,herzogarbeitman2026exactstiffnessdynamicalresponses}. 
 
Frustration-free models have provided insight into Fractional Chern Insulators (FCIs) \cite{neupert2011fractional,sheng2011fractional,regnault2011fractional,tang2011high,sun2011nearly,wu2013bloch,behrmann2016model}, defined as gapped states of matter with fractional Hall conductance at zero magnetic field.
Existing constructions \cite{ledwith2020fractional,wang2021exact,wang2023origin,ledwith2023vortexability,estienne2023ideal,guerci2025layer} have produced continuum frustration-free models for fractional Chern insulators by deforming and decorating the Laughlin state while -- crucially -- preserving annihilation by Haldane's pseudo-potential \cite{ledwith2020fractional,wang2021exact,ledwith2023vortexability,wang2023origin,dong2023many}.
These various deformation and decoration operators preserve a property of the kinetic energy null space (the flat band): their Bloch wavefunction is ``ideal," meaning it is holomorphic in $k = k_x + i k_y$ (up to normalization).
This property is equivalent to having equal Berry curvature and the trace of Fubini-Study metric \cite{roy2014band,mera2021kahler,ozawa2021relations,mera2021engineering,wang2021exact,cano2026ideal}. 
This criterion was named by Roy \cite{roy2014band}, who suggested that bands with uniform Berry curvature and ideal geometry were optimal candidates for FCIs \cite{parameswaran2012fractional,wu2012zoology,roy2014band,jackson2015geometric,claassen2015position,jackson2016quantum,lee2017band,varjas2022topological}. 

In lattice models, where FCIs were discovered numerically more than 10 years ago \cite{sheng2011fractional,neupert2011fractional,regnault2011fractional,wang2011fractional,wu2012zoology,bernevig2012emergent,venderbos2012fractional,wang2012nonabelian,liu2012fractional,liu2013beyond,sterdyniak2013series,lauchli2013hierarchy,liu2013nonabelian,kourtis2014fractional}, exact pairs exist in, roughly speaking, two major types.
One type of construction exploits an identification of single-particle orbitals in 0LL and the Chern band (either the Bloch \cite{wu2013bloch,wu2014haldane,luan2018evolution} or the hybrid-Wannier basis \cite{qi2011generic,wu2012gauge,scaffidi2012adiabatic,liu2013from,liu2013bulk}.
The FCI wavefunction is obtained from replacing the Landau level basis vectors in the known FQH wavefunction with the basis vectors of the Chern band; the parent interaction is fine-tuned such that its matrix elements are identical to those of the pseudo-potential in 0LL.
Parent interactions as such generically decay exponentially with distance, and contain non-density-density terms that are difficult to realize in physical settings.
Another type of construction derives an ``ideal'' Chern band from sampling the 0LL orbitals on a non-Bravais lattice \cite{scaffidi2014exact,dong2020exact,shen2026exact,zhao2026ideal,wang2026symmetry,li2026quantum,ji2026generalized}. (In the Kapit-Mueller case, this results in a Hofstadter band with an extended magnetic unit cell.) In \cite{ji2026generalized}, the FCI wavefunction is obtained from displacing the zeros of the FQH wavefunction onto neighbor orbitals in the magnetic unit cell. Its parent Hamiltonian is density-density repulsion of nearby sites, which breaks lattice rotation down to inversion.
On the other hand, many model wavefunctions have been proposed \cite{qi2011generic,wu2012gauge,wu2013bloch} without interactions that annihilate them. Alternatively, analogues of the Haldane pseudo-potentials have been proposed which lack analytically solvable ground states\cite{claassen2015position}.
To the best of our knowledge, a Bravais-lattice fermion FCI wavefunction having a strictly density-density parent interaction is yet to be established.

In this work, we construct exact FCIs that exceed all earlier lattice constructions in simplicity.
We report a general principle for the construction of exact Bravais-lattice FCI wavefunctions with only onsite density interactions as parents, and apply it to the construction of all FCIs in the Laughlin series \cite{laughlin1983anomalous}, the Read-Rezayi series \cite{read1999beyond,moore1991nonabelions,bernevig2008model,bernevig2008properties,bernevig2009anatomy}, and the Halperin series \cite{halperin1983theory,crepel2018matrix,estienne2012spin}.
The method consists of two exact dualities that we call the \emph{scalar-spinor duality} (an extended version of the Burrello-Trembettoni duality \cite{burrello2011ultracold,Yang_2019}) and the \emph{Landau-Chern duality}. The Chern bands we obtain in this way are not ideal and do not have single-signed Berry curvature. Such bands have not been identified by any previous criteria, yet they host exact FCIs. 

The scalar-spinor duality maps wavefunctions from the usual lowest Landau level to the $2s$-th Landau level of the spin-$s$, Dirac (spinor) Hamiltonian, and maps the Haldane pseudo-potential to contact interactions.
First discovered in Ref.~\cite{burrello2011ultracold}, this duality states that scalar-0LL with the pseudo-potential $\hat{V}_1$ is the same as spinor-1LL with contact potential $\hat{V}_0$.
In this work, we extend this duality to higher-spin-higher-Landau levels, and to more types of pseudopotentials.
For instance, we show that the three-body, Moore-Read pseudo-potential $\hat{V}^{(3)}_3$ in scalar-0LL is identical to the three-body contact potential $\delta(\mathbf{r}_i-\mathbf{r}_j)\delta(\mathbf{r}_j-\mathbf{r}_k)\delta_{s_i\neq{s}_j\neq{s}_k}$ in the $S=1$-second-Landau level.

The Landau-Chern duality is a Bravais-lattice version of the Kapit-Müller \cite{kapit2010exact} sampling map.
We sample the magnetic-Bloch wavefunction of the spinor-LL $\phi_\mathbf{k}(\mathbf{r}$ on a square Bravais lattice $\mathbf{R}\in\mathbb{Z}^2$, and identify $\phi_\mathbf{k}(\mathbf{R}=\mathbf{0})$ with the Bloch function $\psi_\mathbf{k}$ after normalization.
Here a sharp contrast appears between the scalar-0LL and the spinor-high-Landau levels:
if $\phi_\mathbf{k}$ is scalar, Premolov's identify enforces $\phi_\mathbf{k}(\mathbf{0})=0$ at some $\mathbf{k}_0$, so the above mapping fails \cite{perelomov1971completeness,greiter2012generalizations}.
This is exactly why all Kapit-Müller-like sampling must involve multiple positions in a magnetic cell. Here we avoid this issue by showing that different components of a spinor $\phi_\mathbf{k}(\mathbf{0})$ vanishes at different crystal momenta, so that the Bloch state is never null and remains well-defined.
Under this Bravais-lattice sampling, the contact potential $\delta(\mathbf{r}_i-\mathbf{r}_j)$ and $\delta(\mathbf{r}_i-\mathbf{r}_j)\delta(\mathbf{r}_j-\mathbf{r}_k)$ become onsite Hubbard interactions.
Unlike the scalar-spinor duality, the Landau-Chern duality does not preserve the matrix elements, but preserves the ground states: if a many-body wavefunction $\Phi(\{\mathbf{r}_i\})$ has no double/triple occupancy at any $\mathbf{r}\in\mathbb{R}^2$, then it certainly has no double/triple occupancy at any integer site $\mathbf{R}\in\mathbb{Z}^2$.

In fact, the simplest Hamiltonian we obtain from this procedure is equivalent, after truncating small exponentially decaying hoppings, to the QWZ model \cite{PhysRevB.74.085308} (half the BHZ model \cite{Bernevig_2006}) with additional next-nearest-neighbor hopping and Hubbard interactions. Hence our construction reveals a simple and physical limit connecting existing, numerically studied models for FCIs to their solvable, frustration-free limit.

The duo of dualities almost enable us to write the exact pair, an analytic FCI wavefunction and its parent Hubbard interaction, for any FCI on a Bravais lattice, once its FQH analogue is known in the scalar-0LL.
However, two questions should be addressed: is the exact wavefunction the \emph{only} ground state of the Hubbard parent? Is there a thermodynamic gap separating the ground state from the excitations?
For the first question, we obtain a rigorous proof for the Laughlin series for number of electrons $N_e>12$, and provide numerical evidence for the uniqueness for all sizes below.
For the Moore-Read series and the Halperin series, we rely on exact-diagonalization to support the claim of uniqueness for $N_e>8$. These calculations support the existence of a thermodynamic gap, which is nonzero and increasing with system sizes currently accessible. 
Finally, we are able to build, for any finite number of electrons, an adiabatic path that smoothly connects the energy spectrum of the Laughlin $1/3$ FCI to its FQH energy spectrum, preserving the dimension of the null space. This continuity suggests that the FCI ground state is also short-range correlated, i.e., gapped.

The rest of the paper is organized as follows. 
In Sec.II, we review the known exact pairs of fractional quantum Hall states in 0LL. 
Sec.III extends the Burrello-Trembettoni duality to higher spins and higher Landau levels, establishing the general scalar-spinor duality. 
We establish the Landau-Chern duality between a spinor-Landau level to a Chern band in Sec.IV. 
Sec.V applies the two dualities to the analytic construction of FCI in the Laughlin series, the Read-Rezayi series, and the Halperin series, each with supporting numerical evidence, such as the energy and the entanglement spectrum, from exact diagonalization.
Finally, we introduce the adiabatic path between FCI and FQH in Sec.VI before concluding in Sec.VII.

\section{Prelude: Fractional Quantum Hall States in the Lowest Landau Level}\label{sec:scalar_ll_parents}

The focus on ideal band geometry as a key predictor of FCI stability \cite{roy2014band,jackson2015geometric,wang2021exact,wang2023origin} relies fundamentally on Haldane's parent Hamiltonian, which realizes an exact Laughlin ground state when the ideal condition is met. We will briefly review the frustration-free construction of FQH model wavefunctions and their parent Hamiltonians. We set $\hbar=e=1$.

Let us start from the first-quantized Haldane parent Hamiltonian
\bea
H_\text{L} &= \omega_c \sum_i a^\dag_i a_i + g^{(2)}_1 \sum_{i<j} \nabla^2 \delta(\mbf{r}_i - \mbf{r}_j)
\eea
with $g_1^{(2)}>0$. We denote the interaction term by $V_1^{(2)}$. Here $a = \sqrt{\frac{2}{B}}(\del_{\bar{z}} + \frac{B z}{4})$ is the LL lowering operator at field $B$ for the $i$th particle, so $a\phi_m =0$ for any basis of states $m$ of the 0LL. We will take the (unnormalized) symmetric gauge wavefunctions $z^n e^{-B|z|^2/4}$ to write expressions on the disk for simplicitly. 
The kinetic term is manifestly positive semi-definite. The interaction term is less obvious, but in fact it is true on fermionic wavefunctions \cite{trugman1985exact}. To see this, expand the wavefunction near $\mbf{r} = \mbf{r}'$ to leading order: $\phi(\mbf{r},\mbf{r}') \sim f(\mbf{r}-\mbf{r}') g(\mbf{r}')$ and $f$ must be odd by fermionic anti-symmetry. Then integration by parts gives
	\bea
	\braket{\phi| V_1^{(2)}|\phi} &= g_1^{(2)}\int d^2r'|g(\mbf{r}')|^2 \, \nabla^2 |f|^2(0) \geq 0 
	\eea
since the curvature of $|f(\mbf{d})|^2$ must be non-negative at $\mbf{d}=0$ since $f(\mbf{d})$ is odd. In contrast, a bosonic wavefunction can have $\phi(\mbf{r},\mbf{r}') \sim 1 - |\mbf{r}-\mbf{r}'|^2$, where $\nabla^2 |\phi(\mbf{r},\mbf{r}')|^2 < 0$ as $\mbf{r} \to \mbf{r}'$. Hence $V_1^{(2)}$ is indefinite for bosons. 
	
From this discussion, we see that $V_1^{(2)}$ is positive semi-definite and has a null model only if $\phi$ has a cubic zero whenever $\mbf{r}=\mbf{r}'$ \cite{laughlin1983anomalous,haldane1983fractional,trugman1985exact}. Although the kinetic and interaction term do not commute, they have a common null state at $1/3$ filling:
	\bea
	\Psi_{\text{L}}(\mbf{r}_i) &= \prod_{i<j} (z_i - z_j)^3 \prod_i e^{- \frac{B}{4}|z_i|^2} \ .
	\eea
To see this, it is convenient to write $\nabla^2 = 4 \del_z \del_{\bar{z}}$ and then integrate the Dirac delta by parts. 
Since each term in $H$ is positive semi-definite, $\Psi_{\text{L}}$ is a ground state. As one can see from this derivation, projection to the 0LL is not required in the Hamiltonian, although the null state is supported only in the 0LL by construction. 
	
While we have focused on the simplest Laughlin $1/3$ state, the full Laughlin series 
\bea
\Psi_{\text{L},m}(\mbf{r}_i) &= \prod_{i<j} (z_i - z_j)^{2m+1} \prod_i e^{- \frac{B}{4}|z_i|^2}
\eea
gives frustration-free ground states of the Trugman-Kivelson interaction $\nabla^{2(2m-1)}\delta(\mbf{r}-\mbf{r}')$ \cite{trugman1985exact}. Counting the factors of $z$ in the usual way, these are ground states at density $1/(2m+1)$. For any finite size, they are the densest ground states at filling $1/(2m+1)$, and at smaller densities there is a zero energy quasi-hole manifold. However, uniqueness at every finite size does not establish the ground state in the thermodynamic limit. It is believed from the plasma analogy that for $2m+1 > 70$, the Laughlin state is not topological and its parent Hamiltonian is gapless \cite{laughlin1983anomalous,Moitra_2026}.

We next consider the Read-Rezayi series \cite{read1999beyond,moore1991nonabelions,bernevig2008model,bernevig2008properties}, in which the simplest non-abelian state is the Moore-Read wavefunction at $\nu=1/2$. The general Read-Rezayi series at filling $k/(k+2)$ requires a $k+1$-body interaction to stabilize zero modes with $k+1$-body clustering. The Laughlin state is obtained for $k=1$, and the Moore-Read state for $k=2$. We will now introduce this state's parent Hamiltonian in detail. For simplicity, we project to the 0LL and consider the interaction 
\bea
H_\text{MR} &= V^{(3)}_1 \mathcal{P}_{0LL}\sum_{i<j<k} \nabla^2 \delta(\mbf{r}_i - \mbf{r}_j)\nabla^4 \delta(\mbf{r}_j - \mbf{r}_k) \mathcal{P}_{0LL} \ .
\eea
An equivalent, although longer, expression that makes symmetry and positivity manifest is given in \App{A}~\cite{SupplementalMaterial}.  Again, it is helpful to write $\nabla^{2m} = 4^m \del^m_z \del^m_{\bar{z}}$ and integrate by parts. A generic wavefunction in the 0LL will have the structure $\Psi \propto z_{12} z_{23} z_{13} F(z_1,z_2,z_3)$ near a three-body zero, where $z_{ij} = z_i - z_j$ and $F$ is a symmetric function of its arguments to ensure anti-symmetry and will generically start with a constant and linear terms. However, requiring $0 = \del_{z_1}\del_{z_3}^2 \Psi|_{z_i = z}$ to implement the null space condition of $H_{\text{MR}}$ requires that these term vanish. The lowest order non-vanishing term is $F(z_1,z_2,z_3) = z_{12}^2+z_{23}^2+z_{13}^2$, leading to a fifth order triple zero. The most dense wavefunction with fifth order zeros for all coinciding triples is the Pfaffian wavefunction \cite{moore1991nonabelions,greiter1992paired}
\bea
\Psi_{\text{MR}}(\mbf{r}_i) &= \text{Pf}\lp \frac{1}{z_i -z_j}\rp \prod_{i<j} (z_i-z_j)^2  \prod_i e^{- \frac{B}{4}|z_i|^2} \ .
\eea
Here we have assumed an even number of particles $N$ so that the Pfaffian is nonzero. On the torus, the number of ground state is 6 for $N$ even and 2 for $N$ odd. Finally, we point out that a Moore-Read-like ground state is found experimentally and numerically in the $n=1$ LL with two-body interactions \cite{willett1987observation,morf1998transition,rezayi2000incompressible} or higher \cite{3by3-jsjw}, but no frustration-free parent Hamiltonian is known in this case.

The final series we will consider is the Halperin family \cite{halperin1983theory}, developed for multi-component systems, e.g. spin-ful electrons with $\sigma = \uparrow,\downarrow$ restricted to the 0LL. We will again take the simplest example of the $33m$ series, whose wavefunction is
\bea
\Psi_{33m}(\mbf{r}_i) &= \prod_{\sigma,i<j} (z^\sigma_i - z^\sigma_j)^3 \prod_{i,j} (z^\uparrow_i - z^\downarrow_j)^m  \prod_{\sigma,i} e^{- \frac{B}{4}|z^\sigma_i|^2}
\eea
for the spin component where the first $N$ particles are all spin $\uparrow$ and the second $N$ particles are all spin $\downarrow$. All other amplitudes are obtained by anti-symmetry. Note that the inter-spin term can be either symmetric or anti-symmetric, since the two spins are distinguishable. By counting powers of $z$, the filling factor of the 33$m$ state is determined to be $\nu = 2/(3+m)$.

It is clear that the parent interaction should contain the term
\bea
\label{eq:scalar_halperin_intra}
H_{\text{intra}} &= W_0\sum_{i<j,\sigma}\nabla^2\delta(\mbf{r}^\sigma_i-\mbf{r}^\sigma_j),
\eea
with $W_0>0$. This term annihilates $\Psi_{33m}$ for any $m$. Next we consider the inter-spin term. The Jastrow factor $(z^\uparrow_i - z^\downarrow_j)^m$ attaches an $m$th-order zero, so the appropriate parent Hamiltonian can be expressed in terms of pseudo-potentials. For $m=0$, $\Psi_{33m}$ is already the unique zero mode of $H_{\text{intra}}$. For $m=1$, we can require the linear zero by adding a contact interaction
\bea
\label{eq:scalar_halperin_inter1}
H_{\text{inter},1} &= W_1\sum_{i,j}\delta(\mbf{r}^\uparrow_i-\mbf{r}^\downarrow_j),
\eea
with $W_1>0$. For $m=2$, adding a term $\nabla^2 \delta(\mbf{r}^\uparrow_i - \mbf{r}^\downarrow_j)$ appears to be natural, but we must take care since the inter-spin wavefunction does not have definite exchange symmetry (it can be odd or even) and hence $\nabla^2 \delta(\mbf{r}^\uparrow_i - \mbf{r}^\downarrow_j)$ is not positive semi-definite. To write down parent Hamiltonians for the rest of the $332$ and $333$ states, we introduce the manifestly positive semi-definite operators
\bea
\mathcal{V}_k &= \sum_{i, j} \del^\dag_{\mu_1} \dots \del^\dag_{\mu_k} \delta(\mbf{r}^\uparrow_i - \mbf{r}^\downarrow_j) \del_{\mu_1} \dots \del_{\mu_k}
\eea
where $\del_\mu = \del_{r^{\uparrow,\mu}_{i}} -\del_{r^{\downarrow,\mu}_j}$
the spatial $\mu=x,y$ indices are implicitly summed over, and $\del^\dag_\mu = - \del_\mu$ in first quantization. The parent Hamiltonians can now written simply:
\bea
H_{\text{inter},m} &= \sum_{k=0}^{m-1} W_k \mathcal{V}_k, \qquad W_k \geq 0 \ .
\eea
We have not yet discussed the symmetry properties of the various ground states. It is known that the $330$ and $331$ wavefunctions do not form irreps of the $\sigma = \uparrow,\downarrow$ SU(2) symmetry \cite{halperin1983theory}. They merely are eigenstates of the $S_z$ symmetry. In contrast, the 332 and 333 wavefunctions are each SU(2) irreps: the 332 state is Halperin's famous spin-singlet and the 333 state is the $S_z = 0$ representative of the maximal spin ``ferromagnetic" SU(2) irrep. This structure is also reflected in their parent Hamiltonians. For 330 and 331, the interaction term breaks SU(2) symmetry, consistent with the SU(2)-breaking frustration-free ground states being unique. For 332 and 333 states, there \emph{is} a choice of coefficients that makes the parent Hamiltonian fully SU(2) symmetric: $\mathcal{W}_0 = W_0,\mathcal{W}_1 = W_0/2$ and $\mathcal{W}_2$ is arbitrary. Since the 332 state is a singlet, it can be gapped in all cases (even if $\mathcal{W}_1 \neq W_0/2$). In contrast, the ferromagnetic 333 state will possess a magnon with quadratic dispersion and hence is gapless. Away from the $\mathcal{W}_1 \neq W_0/2$ point, the model is still frustration-free and will have a quadratic dispersion due to this fine-tuning, despite SU(2) being broken \cite{Nitta_2015,Masaoka_2024}. 

We will only consider $m \leq 3$ since the Halperin $nnm$ states are unstable for $m>n$, and describe phase separation. Hence, in the thermodynamic limit, we expect their parent Hamiltonians to be gapless for $m>3$ and gapped for $m\leq 3$. 

While these classic wavefunctions --- the Laughlin series, Read-Rezayi series, and the Halperin series --- have defined an era of progress in the FQH effect, they overstate the importance of the $V_1$ parent Hamiltonian and the ideality of the $n=0$ Landau level. Our next section remedies this. 
    
\section{Dirac-Laughlin State and the Level-Spin Duality}

All the previous constructions start from the scalar $n=0$ LL and inherit its holomorphicity in some form. Since our goal is to formulate a local parent Hamiltonian on the lattice, we are motivated to consider models with on-site interactions between multiple orbital degrees of freedom --- but these are precisely the models where no holomorphic band wavefunctions can exist for $C=1$ \cite{claassen2015position,jian2013momentum}. In order to both exclude holomorphicity and include additional orbitals from the start, we will look at the simplest such example in the continuum: the $n=1$ Dirac LL. 

Let us consider the Dirac Hamiltonian in $B$ field 
\bea
h &= \Delta \bpm 0 & a \\ a^\dag & 0\epm, \qquad a = \sqrt{\frac{2}{B}}(\del_{\bar{z}} + \frac{B z}{4}) \ .
\eea
Each LL is degenerate with one orbital per flux, generated by the guiding center $b^\dag = \sqrt{\frac{B}{2}}z - a $, obeying $[b, b^\dag]=1$ and $[b,a]=[b,a^\dag]=0$. The $n=0$ LL is sublattice polarized: $\phi_{n=0,m} = (0 , z^m)^T e^{-|z|^2/2}$, leading to the same FQH physics as the scalar 0LL. (Here and throughout, we neglect normalization factors.) However, the $n=1$ orbitals are sublattice-hybridized:
\bea
\phi_{1,m}(\mbf{r}) = \bpm w^m \\ \bar{w} w^m - m w^{m-1}\epm e^{-\frac{|w|^2}{2}}, \quad w = \sqrt{\frac{B}{2}}z
\eea
and are clearly not holomorphic thanks to the lower sublattice. To explore this structure, we first seek a simple expression for the Slater state of a filled $\phi_{1,m}$ band. The essential identity is
\bea
\label{eq:Wattachment}
\phi_{1,m} \propto W^m \phi_{1,0} , \quad W = z - \sqrt{\frac{2}{B}}\sigma^- 
\eea
which replaces the $\del_{\bar{z}}$ in $b^\dag$ with a matrix coordinate $W$. This property is a nontrivial example of Haldane's ``orbital attachment" \cite{PhysRevLett.107.116801} which we write as
\bea
Q (z+J) P &= 0
\eea
where $P$ is the projector onto the $n=1$ LL, $Q = 1-P$, and $J = - \sqrt{2/B} \sigma^-$. Note that $J^2 = 0$. This is a generalization of Roy's $Q z P = 0$ identity in ideal bands \cite{roy2014band} and the deformed $Q z(\mbf{r}) P = 0$ identity in vortexable bands \cite{ledwith2023vortexability}. 

Since $W^m$ is common to each row of the Slater determinant, we have
\bea
\label{eq:PsiC1}
\Psi_1(\{r_i\}) &= \prod_{i<j} (W_i - W_j) \bigotimes_i \bpm 1 \\ \bar{w}_i\epm e^{-|w_i|^2/2} \ .
\eea
We emphasize that $\Psi_1(\{r_i\})$ are the spatial amplitudes in the $2^N$-dimensional sublattice Hilbert space. 

The key observation is that the action of the matrix $W$ preserves the $n=1$ single-particle Hilbert space: for any analytical function $f$, the state $f(W)\phi_{1,0}$ is in the $n=1$ Dirac LL. This is directly analogous to how the coordinate $z$ preserves the (scalar) $n=0$ Hilbert space -- the condition generalized to ideal bands. From here, it is natural to write down a Laughlin state
\bea
\Psi_{1/3}(\{r_i\}) &= \prod_{i<j} (W_i - W_j)^2 \Psi_{1}(\{r_i\})
\eea
which is contained in the $n=1$ Hilbert space.  It remains to show that this state has an interacting parent Hamiltonian. Surprisingly, $V^{(2)}_1$ does not annihilate $\Psi_{1/3}(\{r_i\})$. This can be seen already for $N=2$ where the polynomial factor is simply 
\bea
(W_1 - W_2)^3 \propto w_{12}^3 -  3 w_{12}^2 (\sigma_1^- - \sigma^-_2) - 6 w_{12} \sigma_1^- \sigma^-_2
\eea
since $(\sigma_i^-)^2 = 0$, and $w_{12} = w_1-w_2$. While the first two terms ensure (at least) quadratic zeroes, the final term leaves a nonzero amplitude $\propto \ket{\down \down}$ after integration by parts. This term is not annihilated by $V^{(2)}_1$. However, every term attaches at least one zero to the droplet factor, so the component-independent contact repulsion
\begin{equation}
V_0^{(2)}=\mathcal G_0^{(2)}\sum_{i<j}\delta(\mathbf r_i-\mathbf r_j),\qquad \mathcal G_0^{(2)}>0,
\end{equation}
annihilates the state. Here $\delta$ is the two-dimensional Dirac delta, and $V_0^{(2)}$ acts as the identity on the internal components of each particle. The superscript $(2)$ counts the interacting particles, while $\mathcal G_0^{(2)}$ is the coupling constant:
\begin{equation}
V_0^{(2)}\Psi_{1/3}=0.
\end{equation}
The existence of a fermionic $1/3$ wavefunction with a contact interaction parent Hamiltonian is the key technical property that enables a lattice realization. This is our next step: we will show that sampling this wavefunction on the lattice yields a well-defined parent Hamiltonian in a flat Chern band with onsite Hubbard interaction. 

\subsection{Higher Spin Generalization}

A general form of this duality holds for any number $2s+1, s \in \mathbb{N}/2$ of orbitals with $J^{2s+1} = 0$, which we formulate using a spin-$s$ Dirac Hamiltonian. Taking $S^a$ to be the spin operators normalized such that $[S^a, S^b] = i \eps^{abc} S^c$, the Hamiltonian in magnetic field is
\bea
h_s &= \Delta (a S^+ + a^\dag S^-)
\eea
where $S^{\pm} = S_x \pm i S_y, $ are the raising and lowering operators. From the commutators
\bea
[a^\dag a, h_s] &= \Delta (-a S^+ + a^\dag S^-) \, \\
[S^z, h_s] &= \Delta (a S^+ - a^\dag S^-) \ , \\
\eea
we see that there is a symmetry $[a^\dag a + S_z,h_s] = 0$ in addition to conservation of the guiding centers $b$. This symmetry $\mathcal{N} = a^\dag a + S_z + s \geq 0$ is the generalization of ``LL number" to the Dirac case. By $\mathcal{N}$ conservation, the eigenvectors of $h_s$ must have the LL structure $\phi_n = (c_{s} \ket{n-2s},\dots,c_{-s+1}\ket{n-1},c_{-s}\ket{n})$ where $\ket{j} = 0$ if $j<0$. There is a privileged $\mathcal{N} = 2s$ LL which is the smallest $\mathcal{N}$ sector where all spin components are nonzero and hence $J^{2s+1} = 0$ is possible and nontrivial. 

Our strategy to determine $J$ satisfying $Q(z+J)P=0$ relies on the
guiding-center identity $b^\dag=\sqrt{\frac{B}{2}} z-a$. Since
$b^\dag$ preserves the Landau level, it suffices to construct
a matrix $J$ such that $a\phi_{2s}=-\sqrt{B/2}\,J\phi_{2s}$. The $\mathcal{N}=2s$ block of $h_s$ is tridiagonal with nonnegative entries. Provided its adjacent off-diagonal couplings are nonzero, this block is irreducible, and the Perron--Frobenius theorem ensures that its largest eigenvalue $\eps$ has a normalized eigenvector
\bea
\label{eq:psi}
\phi_{2s} &= (c_s\ket{0},\dots,c_{-s+1}\ket{2s-1},c_{-s}\ket{2s})^T
\eea
with every $c_m$ nonzero. We can therefore define
\bea
a\phi_{2s} &= -\sqrt{\frac{B}{2}}\,J\phi_{2s}, \
J_{m,m'} = -\sqrt{\frac{2}{B}}\sqrt{s-m}\, \frac{c_m}{c_{m+1}}\delta_{m+1,m'} 
\eea
for $ m=-s,\dots,s-1$ with $J_{s,m'}=0$. In the ordered spin basis $m=s,\dots,-s$, this matrix is strictly lower triangular and hence satisfies $J^{2s+1}=0$. Let $P_{2s}$ be the projector onto the Landau level generated by $\phi_{2s}$, including all guiding centers, and set $Q_{2s}=1-P_{2s}$. Since $b^\dag$ changes only the guiding center and commutes with $h_s$, it preserves this Landau level. It follows that
\bea
Q_{2s}(z+J)P_{2s} &= \sqrt{\frac{2}{B}}\,Q_{2s}b^\dag P_{2s} =0 \ .
\eea
This completes the orbital attachment construction of $J$ for
a general $(2s+1)$-orbital Dirac Landau level.

In addition to the simplest $s=1/2$ example, we will extensively investigate the $s=1$ case below. As discussed in \App{B}~\cite{SupplementalMaterial}, $s=1$ is the minimal case where a Moore-Read state can be stabilized by the three-body contact interaction
\begin{equation}
V_0^{(3)}=\mathcal G_0^{(3)}\sum_{i<j<k}\delta(\mathbf r_i-\mathbf r_j)\delta(\mathbf r_j-\mathbf r_k)
\end{equation}
with higher-body contact interactions generalizing the Read-Rezayi series. 

\subsection{Scalar-Spinor Duality to the Lowest Landau Level}

We now exhibit a ``scalar-spinor'' duality between the parent Hamiltonian of the $2s$th Dirac LL and Haldane's parent
Hamiltonian for the Laughlin states in the 0LL. This duality maps the projected Hamiltonians into each other by associating each internal spin component with a different scalar Landau level. In this section, we always assume that wavefunctions are in the 0LL or spin-$s$ $2s$LL and we will omit the projectors to these Hilbert spaces. This result generalizes the $s=1/2$ case discovered in Ref. \cite{burrello2011ultracold}.

Using the normalized eigenvector in \Eq{eq:psi}, we define the single-particle map
\bea
\label{eq:Umap}
\mathcal{U}_i
&=
\left(
c_s,\,
c_{s-1}a_i^\dag,\,
\dots,\,
c_{-s}\frac{a_i^{\dag 2s}}{\sqrt{(2s)!}}
\right)^T .
\eea
In particular, $\mathcal{U}_i\ket{0}=\phi_{2s}$, so $\mathcal{U}_i$ maps the scalar 0LL to the $2s$LL without changing the guiding center. Since $\sum_m c_m^2=1$, this map obeys  $\mathcal{U}_i^\dag\mathcal{U}_i=1$ (assumed to act on the scalar 0LL). This holds because $a^n a^{\dag n} = n^!$ when acting on the 0LL. Next, we see that $\mathcal{U}_i\mathcal{U}_i^\dag = 1$ since \Eq{eq:psi} shows $\mathcal{U}_i^\dag \phi_{2s} = \ket{0}$, reproducing an $0LL$ wavefunction. This means that $\mathcal{U}_i$ is a unitary map between the single-particle spaces of the 0LL and the spin-$s$ $2s$LL. The corresponding many-body map is $\mathcal{U}=\prod_i\mathcal{U}_i$. It maps the many-body scalar 0LL Hilbert space to the many-body spinor LL Hilbert space. In particular,
\bea
\label{eq:entwine}
\Psi_{1/3}(\{\mbf{r}_i\})
&=
\mathcal{U}\Psi_{\mathrm{L}}(\{\mbf{r}_i\}) .
\eea

To express this map compactly, let $e_m$, $m=s,\dots,-s$, denote the unit column vector corresponding to spin component $m$. Our definition of $J$ gives
\bea
J^n e_s &= \left(-\sqrt{\frac{2}{B}}\right)^n
\sqrt{n!}\,\frac{c_{s-n}}{c_s}\,e_{s-n}, \quad n=0,\dots,2s.
\eea
Consequently,
\bea
\mathcal{U}_i &= c_s e^{-\sqrt{B/2}\,J_i a_i^\dag}e_s, \\
\mathcal{U} &= \left[ \prod_i c_s e^{-\sqrt{B/2}\,J_i a_i^\dag}
\right] \bigotimes_i e_s,
\eea
where each exponential terminates because $J^{2s+1}=0$. Using $[z,a^\dag]=\sqrt{2/B}$ and the fact that $J$ acts only in the internal spin space, the Baker--Campbell--Hausdorff formula gives
\bea
e^{-\sqrt{B/2}\,Ja^\dag}\, z\, e^{\sqrt{B/2}\,Ja^\dag} &= z+J \equiv W .
\eea
so $\mathcal{U}$ is an entwiner between the scalar coordinate $z_i$ and the matrix coordinate $z_i+J_i$. Applying this property to move $\mathcal{U}$ through any polynomials $p(z_i)$, we find
\bea
\mathcal{U} p(z_1,\dots,z_N) = p(W_1,\dots,W_N)\mathcal{U}
\eea
and hence \Eq{eq:entwine} is proven. 

Now we consider the mapping of the Hamiltonians. Let
\bea
H_s &= g \sum_{i<j} \delta(\mbf{r}_i - \mbf{r}_j) P^{(k)}_{i} P^{(k)}_{j}
\eea
be an interaction in the spinor Hilbert space, with $P^{(k)}_{i}$ a projector onto the spin components $s,\dots,s-k$ for the $i$th particle. Not only does $\mathcal{U}$ map the FQH ground states between the scalar and spinor Hilbert spaces, it also maps their parent Hamiltonians through the formula
\bea
\label{eq:Hmapping}
\mathcal{U}^\dag H_s \mathcal{U} &= \sum_{n}^{k} \mathcal{G}_n V^{(2)}_n
\eea
where $V^{(2)}_n$ is the $n$-th Haldane pseudo-potential and $\mathcal{G}_k \geq 0$. We will derive \Eq{eq:Hmapping} for the $k=1$ case and leave a more general expression for \Apps{A}{B}~\cite{SupplementalMaterial}. First, note that
\bea
P_i^{(1)}\mathcal{U} &= c_s e_s + c_{s-1} e_{s-1} a^\dag_i \\ 
\eea
and that expanding the interaction yields 
\bea
\mathcal{U}^\dag\delta(\mbf{r}_i-\mbf{r}_j)P_iP_j\mathcal{U} &=p_0^2 \delta(\mbf{r}_i-\mbf{r}_j) +p_1^2a_i a_j \delta(\mbf{r}_i-\mbf{r}_j) a_i^\dag a_j^\dag \\
+p_0 p_1 &  \Big(a_i\delta(\mbf{r}_i-\mbf{r}_j) a_i^\dag +a_j \delta(\mbf{r}_i-\mbf{r}_j) a_j^\dag\Big)\\
\eea
where $p_m = |c_m|^2$. The first line is clearly symmetric in $ij$ and will vanish on the fermionic Hilbert space thanks to the $\delta(\mbf{r}_i - \mbf{r}_j)$. Using anti-symmetry again, we can write $\delta(\mbf{r}_i-\mbf{r}_j) a_i^\dag = \delta(\mbf{r}_i-\mbf{r}_j) (a_i^\dag- a_j^\dag)/2$, leading us to
\bea
\mathcal{U}^\dag H_s \mathcal{U} &= \frac{g}{2} p_0 p_1 \sum_{i<j}  (a_i - a_j) \delta(\mbf{r}_i - \mbf{r}_j) (a^\dag_i - a^\dag_j) \\
 &= \frac{g}{B} p_0 p_1 \sum_{i<j} (\del_{z_i} - \del_{z_j})^\dag \delta(\mbf{r}_i - \mbf{r}_j) (\del_{z_i} - \del_{z_j}) \\
 &= \frac{2g}{B} p_0 p_1 \mathcal{V}_1  \ ,\\
\eea
precisely the $\mathcal{V}_1$ Haldane pseudo-potential, emphasizing again that we have assumed projection to the 0LL. Hence there is a zero energy $1/3$ Laughlin-spinor state $\mathcal{U} \Psi_{\text{L}}$ constructed for each $s$. The many-body spectrum of the projected Hamiltonian is identical to the scalar $\mathcal{V}_1$ spectrum, simply rescaled by the factor of $\frac{2g}{B} p_0 p_1 $. 

The other simplest case is when $k=2s$ and contact interactions are felt between all $2s+1$ spin components. This case maps the contact interaction to positive pseudopotentials up to $n=2s$, so that the highest density frustration-free state is the $\nu = 1/(2s+1)$ Laughlin state. 

To summarize this section: we have shown how to map the Laughlin series --- and more generally any states in the scalar Landau level --- to states in a higher-spin, higher-LL Hilbert space. Crucially, in this Hilbert space with internal indices, the parent Hamiltonians only require contact interactions $\delta(\mbf{r}_i - \mbf{r}_j)$. This key simplification will allow us to overcome the difficulty of defining $\mathcal{V}_{n>1}$ on the lattice, since it is easy to define $\mathcal{V}_{0}$ on the lattice: it is simply the Hubbard interaction. 

\section{Landau-dual Chern bands from lattice sampling}

In the previous section, we established a scalar-spinor duality for general spin $s>0$, relating pseudopotential parents in the scalar 0LL to contact interactions in the corresponding Dirac LLs.
The resulting spinor fractional quantum Hall states are exact zero-energy ground states of these contact interactions. These states, e.g. $\Psi_{1/3}(\{\mbf{r}_i\},\{\al_i\})$ in \Eq{eq:PsiC1}, have contact $\delta(\mathbf r_i-\mathbf r_j)$ interactions as their parents. This means that a lattice wavefunction $\tilde{\Psi}_{1/3}(\{\mbf{R}_i\},\{\al_i\}) = \Psi_{1/3}(\{\mbf{r}_i\},\{\al_i\})|_{\mbf{r}_i = \mbf{R}_i}$ will be annihilated by a (multi-orbital) repulsive Hubbard interaction. However, one must still show that the proposed lattice wavefunction $\tilde{\Psi}_{1/3}$ is within the Hilbert space of a single Chern band with Bloch translations. This is indeed the case, and its proof is the focus of this section. 

We will accomplish this by defining a sampling map $\mathcal{S}$ from the continuum LL Hilbert space to a multi-orbital tight-binding Hilbert space (here taken to be a simple square lattice):
\begin{equation}
    \label{eq:sampling}
    \psi_\al(\mathbf R) =(\mathcal S\phi)_\al(\mathbf R)
    =\left. \phi_\al(x,y) \right|_{(x,y)=(X,Y)}, \quad X,Y \in \mathds{Z}
\end{equation}
where $\phi(\mathbf r)$ is a spinor wavefunction in one of the spin-$s$ Dirac LLs constructed in the previous section.

The sampling map $\mathcal S$ acts by pointwise evaluation, without normalizing the sampled state.
We term a lattice Chern band obtained by such sampling of a continuum LL a \emph{Landau-dual-Chern band}, and call this correspondence the \emph{Landau-Chern duality}.

Sampling preserves the matrix-coordinate $W = z+J$ that implements orbital attachment in the Dirac  LL. This is because $J$ acts locally on the internal components, so pointwise evaluation gives
\begin{equation}
    \label{eq:1LL_raising}
    \mathcal S[(z+J)\phi]=(Z+J)\mathcal S\phi,
    \qquad Z=X+iY  \ .
\end{equation}
Again taking $P$ to be the projector onto the sampled subspace and $Q=1-P$, this gives
\begin{equation}
    \label{eq:C=1_raising}
    Q(Z+J)P=0 \ .
\end{equation}
We must now show that $P$ is a rank-1 projector onto Chern band with Bloch translation symmetry. 

\begin{figure*}[!t]
\centering
\includegraphics[width=\textwidth]{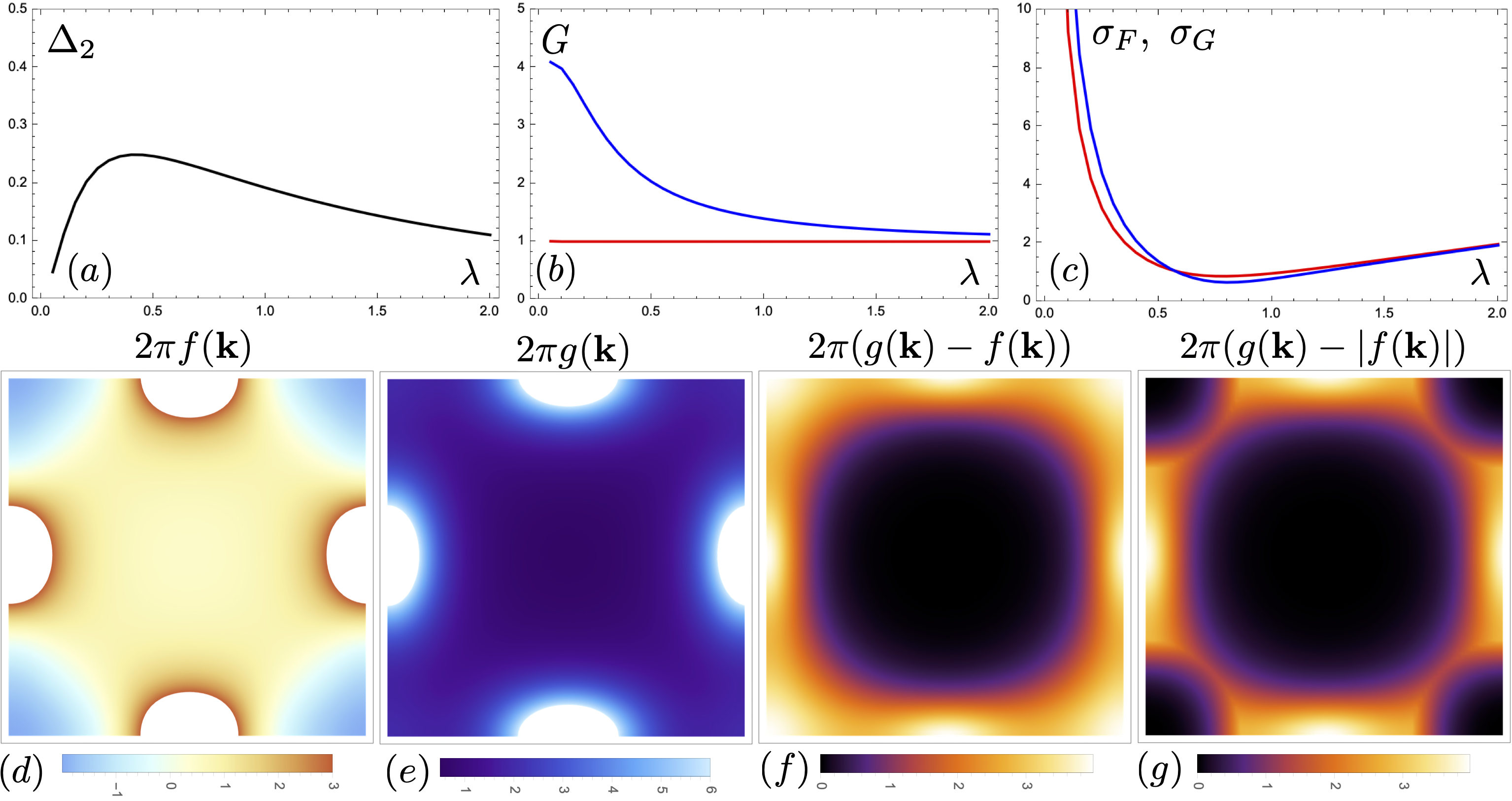}
\caption{Quantum Geometry of the Landau-dual Chern band. $(a)$ shows the average 2-body interaction strength $\Delta_2 = \det \braket{P}_{BZ}$ as a function of $\lambda$. $\Delta_2$ is maximized when the flat band is equally distributed between the 2 orbitals. $(b)$  show the Chern number $C = 2\pi\braket{f}_{BZ} = 1$ and integrated quantum metric $G = 2\pi \braket{g}_{BZ}$ in red and blue respectively. $(c)$ shows their standard deviation, $\sigma_F,\sigma_G$, on the Brillouin zone (BZ). We see that the band is ideal as $\lambda \to \infty$, but that the fluctuation across the BZ is minimized at smaller $\lambda$. $(d)$-$(g)$ show the distribution of the Berry curvature $f$, quantum metric $g$, and their ideality violations $g - f$ and $g-|f|$ as a function of $\mbf{k}$. We see that the Berry curvature is both positive and negative, and that there are strong violations of either $\pm f$ ideality at the $X,Y$ and $M$ points. At $M$, interestingly, the band is anti-ideal and obeys $g = -f$.}
\label{fig:QG}
\end{figure*}

\subsection{Properties of the Sampling Map}

A single lattice band contains $L_xL_y$ states on a periodic  $L_x\times L_y$ lattice (recall we have taken the area of the unit cell to be 1), whereas the selected Dirac LL contains $N_\phi=BL_xL_y/(2\pi)$ states. This suggests that we should choose $B=2\pi$, or one flux quantum per unit cell, for the sampling construction. To show this, it is convenient to work in momentum space. At $B = 2\pi$, magnetic translations by lattice vectors commute and there is a magnetic Brillouin zone \cite{PhysRev.134.A1602}. We define the magnetic Bloch basis of the scalar 0LL $\phi_{\mathbf k}(\mathbf r)$ (see \Refs{Haldane_2018}{PhysRevB.106.085140} for expressions) obeying
\bea
\int d^2r \, \phi^*_{\mathbf k}(\mbf{r}) \phi_{\mathbf k'}(\mbf{r}) =(2\pi)^2\delta(\mathbf k-\mathbf k')
\eea
as a complete and orthonormal basis of the 0LL. \App{C}~\cite{SupplementalMaterial} shows that, in an appropriate Landau gauge, lattice translations give
\begin{equation}
    \label{eq:Landau_translaiton}
    \phi_{\mathbf k}(x+X,y+Y)
    =e^{i2\pi Xy}e^{ik_xX+ik_yY}\phi_{\mathbf k}(x,y).
\end{equation}
At lattice points $(x,y)\in\mathbb Z^2$, the non-Bloch phase factor reduces to unity, so 
\bea
\phi_{\mathbf k}(\mathbf R)=\phi_{\mathbf k}(\mathbf0)e^{i\mathbf k\cdot\mathbf R}
\eea
satisfies the ordinary Bloch translation law with the same crystal momentum $\mathbf k$.  Different crystal momenta are orthogonal because they have distinct eigenvalues under ordinary lattice translations.
We have therefore constructed a set of orthogonal, Bloch wavefunctions. However, we have not shown that it is a complete basis of a band. Indeed, the $\phi_{\mathbf k}(\mathbf R)$ constructed above are \emph{not} a complete basis. To see this, recall that for the scalar 0LL, Perelomov's identity proves that there is one Bloch momentum where a (nonzero) continuum state $\phi_\mbf{k}(\mbf{r})$ vanishes at every lattice site $\mbf{R}$ \cite{perelomov1971completeness,greiter2012generalizations}. Sampling the scalar 0LL at $B=2\pi$ is thus not bijective. 

We will remedy this construction by using the spinor structure introduced above. Since the spinor LL contains e.g. 0LL and 1LL components with nonzero weights, sampling can evade the no-go theorem of Perelomov. 

To show this, we will compute the normalization factors in the sampled spinor LLs and show they are nonzero. 
Let $\phi^{2s}_{\mathbf k}$ be the spinor magnetic Bloch state obtained from $\phi_{\mathbf k}$ by the single-particle map in Eq.~(\ref{eq:Umap}). We define the unnormalized sampled Bloch spinor
\begin{equation}
\label{eq:sampled_bloch_spinor}
v_{\mathbf k}=\phi^{2s}_{\mathbf k}(\mathbf0),\qquad
(\mathcal S\phi^{2s}_{\mathbf k})(\mathbf R)
=v_{\mathbf k}e^{i\mathbf k\cdot\mathbf R}.
\end{equation}
We directly compute 
\bea
\langle\mathcal S\phi_{\mathbf k}|\mathcal S\phi_{\mathbf k'}\rangle=(2\pi)^2\delta(\mathbf k-\mathbf k')s_{\mathbf k}^2, \quad s^2_{\mathbf k} = v^\dag_{\mathbf k} v_\mbf{k} \ .
\eea
We show in \App{C}~\cite{SupplementalMaterial} that $s^2_{\mathbf k} > 0$ for all $\mbf{k}$. Intuitively, this is because the 1LL component is nonzero wherever the 0LL component vanishes. 

The normalized lattice Bloch wavefunction is therefore
\begin{equation}
\label{eq:identification}
\psi_{\mathbf k}(\mathbf R)
=s_{\mathbf k}^{-1}(\mathcal S\phi^{2s}_{\mathbf k})(\mathbf R)
=u_{\mathbf k}e^{i\mathbf k\cdot\mathbf R},
\end{equation}
where $u_{\mathbf k}=s_{\mathbf k}^{-1}v_{\mathbf k}$.
The sampled band inherits $C=1$ because bijective sampling preserves the magnetic Bloch transition phases. The normalized spinor satisfies $u(k_x+2\pi,k_y)=-u(k_x,k_y)$ and $u(k_x,k_y+2\pi)=-e^{ik_x}u(k_x,k_y)$ under the gauge we used here. These boundary conditions give $C=1$ with $\mathcal A_\mu=i u_{\mathbf k}^\dagger\partial_{k_\mu}u_{\mathbf k}$.

Since $s_{\mathbf k}$ is continuous and nonzero on the compact Brillouin zone, it is bounded above and away from zero, allowing us to \emph{define} the inverse map
\bea
\mathcal{S}^{-1}\psi_{\mathbf k}=\phi_{\mathbf k}/s_{\mathbf k}
\eea
Hence $\mathcal{S}$ is a second duality, mapping between the continuum spinor LL and the lattice spinor band. Note that $\mathcal{S}$ is not unitary. $\mathcal{S}$ acts diagonally in momentum space and applies to both single-particle and many-particle wavefunctions. To be explicit, let $\Phi(\{\mathbf k_i\})$ be the coefficients of a continuum many-body state in the magnetic Bloch basis $\{\phi_{\mathbf k}\}$, and let $\Psi(\{\mathbf k_i\})$ be the coefficients of its sampling image in the normalized lattice Bloch basis $\{\psi_{\mathbf k}\}$.
Then
\begin{equation}
    \label{eq:momentum_space_duality}
    \Psi(\{\mathbf k_i\})= \lp \prod_i s_{\mathbf k_i} \rp \Phi(\{\mathbf k_i\}).
\end{equation}
The inverse is obtained by replacing each $s_{\mathbf k_i}$ with $s_{\mathbf k_i}^{-1}$. Note that similar constructions have been used to write down general model wavefunctions in Chern bands \cite{qi2011generic,wu2012gauge,wu2013bloch}. 

\subsection{Bloch Bands and Parent Hamiltonians}

Direct sampling of the $s=1/2$ 1LL and the $s=1$ 2LL gives two- and three-component Bloch spinors which we will include here for concreteness. Write $k=k_x-ik_y$ for the complex momentum, and define the Jacobi theta function
\bea
\vartheta_1(z|e^{i\pi\tau})=\sum_{n=-\infty}^{\infty}e^{\pi i \tau (n+1/2)^2+2i(n+1/2)(z-\pi/2)}
\eea
and covariant derivative
\begin{equation}
D=2\partial_k-\frac{ik_y}{\pi}=\partial_{k_x}+i\partial_{k_y}-\frac{ik_y}{\pi},
\label{eq:D}
\end{equation}
where $\partial_k=\tfrac12(\partial_{k_x}+i\partial_{k_y})$ is the derivative at fixed $\bar k$, so $Dk_y=i$.
The unnormalized Bloch vector in a smooth gauge is
\bea
\label{eq:explicit_dirac_sample}
v^{(1/2)}(\mathbf k)&= e^{-k_y^2/4\pi}
\bpm
\lambda \vartheta_1(k/2|e^{-\pi}) \\
i D \vartheta_1(k/2|e^{-\pi})
\epm  
\eea
for the $s= 1/2$ LL, where $\la = 1/\sqrt{\pi}$. For the $s=1$ LL, we obtain
\bea
v^{(1)}(\mathbf k)&= e^{-k_y^2/4\pi}
\bpm
\lambda^2\,\vartheta_1(k/2|e^{-\pi})\\
\sqrt{3}\lambda\,i D\vartheta_1(k/2|e^{-\pi})\\
(iD)^2\vartheta_1(k/2|e^{-\pi})
\epm \ .
\label{eq:explicit_dirac_sample_one}
\eea
The parameter $\lambda$ here is determined from direct sampling, but there are further hidden tunable parameter preserving the frustration-free structure. For an invertible matrix $M$, we can take $J \to MJM^{-1}$ and $u \to M u$ as a deformation of the wavefunction. Since $M$ only acts on the onsite orbitals, the deformed wavefunction is still in the null space of the Hubbard interaction. The general expression for the spin-$s$ sampling is given in \Apps{B}{C}~\cite{SupplementalMaterial}.

One remarkable feature of these wavefunction is that they are not ideal. This can be most directly seen from the following non-smooth gauge for the 2-orbital spinor
\bea
v^{(1/2)}(\mathbf k) \propto
\bpm \lambda\\ \zeta(k)-\frac{\bar k}{4\pi} \epm
\eea
where the non-holomorphic dependence on $\bar{k}$ is explicit, and the proportionality symbol includes a $\mbf{k}$-dependent normalization and phase. One can see that the wavefunction is periodic on the BZ, for this choice of normalization, thanks to the quasi-periodic identity $\zeta(k +G) = \zeta(k) + \bar{G}/4\pi$ for all complex reciprocal lattice vectors $G \in 2\pi \mathds{Z}+2\pi i\mathds{Z}$. We show the quantum geometric properties of the wavefunction in \Fig{fig:QG}.

Given $v(\mbf{k})$, we can construct a parent Hamiltonian with the kinetic term
\begin{equation}
    \label{eq:flat_band}
    h({\mathbf k}) = v_{\mathbf k}^\dag v_{\mathbf k}\mathbb{1}_{2s+1}-v_{\mathbf k}v_{\mathbf k}^\dag.
\end{equation}
The magnetic Bloch transition phases cancel in the bilinears, making $h_{\mathbf k}$ periodic.
Its lowest band is exactly flat with $C=1$, while the other $2s$ bands have energy $s_{\mathbf k}^2$.
For the two- and three-component families above, \App{D}~\cite{SupplementalMaterial} establishes Gaussian decay of the hopping amplitudes. We can use the normalized spinor $u_\mathbf{k}$ to construct a flat-band Hamiltonian $u^\dag u - u u^\dag$, but with exponential decay.

In second quantization, we write
\bea
H_0 = \sum_{\mbf{k} \al \be} c^\dag_{\mbf{k},\al} h_{\al \be}(\mbf{k}) c_{\mbf{k} \be}
\eea
where $\{c_{\mbf{k} \be},c^\dag_{\mbf{k} \al}\} = \delta_{\al \be}$ are the creation operators at momentum $\mbf{k}$ and spinor-orbital $\al = 1,\dots, 2s+1$.  We will now consider partial filling of this flat band. With the appropriate Hubbard interactions, we will derive FCI wavefunctions for each of the FQH series considered in the previous sections. 

\section{Fractional Chern insulators in Landau-dual Chern bands}

\begin{figure*}[!t]
\centering
\includegraphics[width=\textwidth]{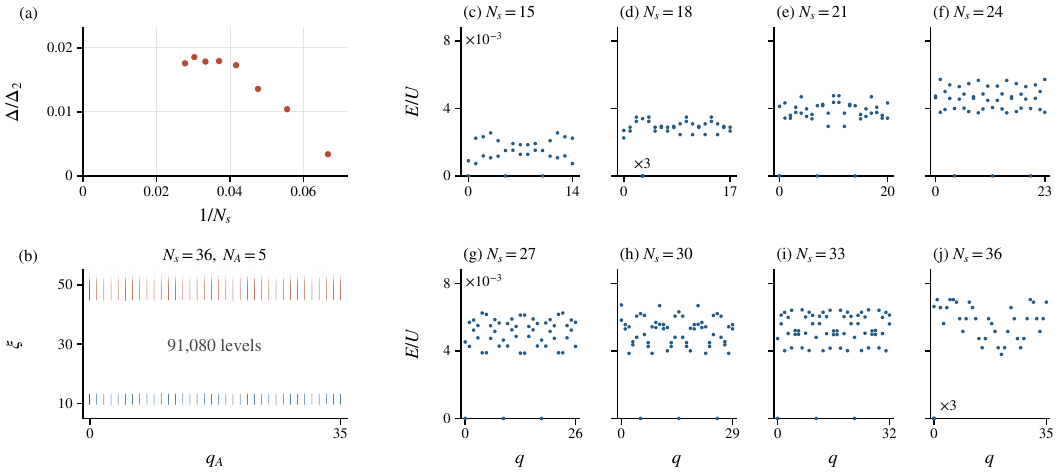}
\caption{Laughlin $1/3$ at $\lambda=0.4$, with three numerical zero modes within $|E/U|<2\times10^{-12}$. (a) Lowest excitation gap $\Delta$ across all sectors, normalized by $\Delta_2=0.21606\,U$. (b) PES of the equal-weight ground-state mixture at $N_s=36$, $N_A=5$; $91{,}080$ levels lie below the entanglement gap. (c--j) Low-energy spectra for $N_s=15,18,\ldots,36$ and $N_e=N_s/3$; $\times n$ marks zero-mode multiplicity.}
\label{fig:laughlin-third-gap-spectra}
\end{figure*}

In the previous section, we established the sampling duality between the selected Dirac LLs and Landau-dual-Chern bands.
In this section, we use this duality to construct exact many-body wavefunctions and parent Hamiltonians for lattice analogues of the Laughlin, Read-Rezayi, and Halperin states. We verify our constructions with exact diagonalization (ED) calculations on the torus. 

Throughout this section, we write $W_i = Z_i + J_i$ as the discrete matrix coordinate in the Landau-dual Chern band, which obeys the orbital-attachment condition $QWP = 0$ which ensures that the action of $W_i$ on any state in the Chern band Hilbert space preserves this Hilbert space. In other words $F(W_i) \Psi$ is a zero mode of the kinetic energy $H_0$ as long as $\Psi$ is a zero-mode of $H_0$. We will construct FCI wavefunctions using this property, and verify that they are zero modes of appropriate contact interactions. For simplicity of presentation, we will write the model wavefunctions on the disk. However, since they are obtained from sampling on the lattice, we can easily obtain wavefunctions on the torus. Some explicit construction is given in \App{D}~\cite{SupplementalMaterial}. 

\subsection{The Laughlin series}

The first lattice FCI wavefunction we consider is the Laughlin $1/3$ wavefunction
\begin{equation}
\label{eq:Laughlin_1/3}
\Psi^{\text{FCI}}_{1/3}=\prod_{i<j}(W_i - W_j)^2 \ket{\text{Slater}_{s=1/2}}
\end{equation}
where $\ket{\text{Slater}_{s=1/2}}$ is the Slater determinant $\prod_{i<j}(W_i - W_j) \Psi_0$ and $\Psi_0$ is the symmetric spinor-gaussian product sampled on the lattice (see \Eq{eq:PsiC1}). $\Psi^{C=1}_{1/3}$ is an exact zero-energy ground state of the Hubbard interaction: its continuum parent has zero double-occupation amplitude at every $\mathbf r$, and sampling preserves this zero at every $\mathbf R$. Hence, the interacting term in its parent Hamiltonian is
\begin{equation}
\label{eq:Hubbard_1/3}
H_{1/3}=U\sum_\mathbf{R}\hat{n}_{\mathbf{R},0}\hat{n}_{\mathbf{R},1} =U\sum_\mathbf{R}(\hat{c}_{\mathbf{R},2}\hat{c}_{\mathbf{R},1})^\dag\hat{c}_{\mathbf{R},2}\hat{c}_{\mathbf{R},1}.
\end{equation}
Note that $\Psi^{C=1}_{1/3}$ is the exact zero mode of the unprojected kinetic and interaction terms, for all $U$. To study this Hamiltonian numerically, we will project the interaction to the flat band, denoted as $\bar{H}_{1/3}$, and diagonalize $\bar{H}_{1/3}$ on the torus. For $5\le N_e\le12$, ED finds exactly three zero modes (to numerical resolution) on the tori studied in Fig.~\ref{fig:laughlin-third-gap-spectra}(c--j), separated from excitations by a finite gap. To compare the gap across different parameters (shown in \App{E}~\cite{SupplementalMaterial}), we normalize the energies by a natural interaction scale set by the $2$-body problem. Since $\bar{H}_{1/3}$ has rank $N$ for a torus of $N$ unit cells, the two body spectrum consists of zero modes with only one nonzero energy state per momentum. This energy is
\bea
E_{2}(\mbf{p}) &= \frac{1}{2N} \sum_\mbf{k}  \Tr P(\mbf{p}+\mbf{k}) \sigma_y P^*(-\mbf{k}) \sigma_y  \ .
\eea
See \App{E}~\cite{SupplementalMaterial} for more details. We denote the average 
\bea
\Delta_2 &=\frac{1}{N} \sum_\mbf{p} E_{2}(\mbf{p}) =  
\det \int \frac{d^2k}{(2\pi)^2} P(\mbf{k})
\eea
where we took the thermodynamic limit. We show the resulting ED specta in \Fig{fig:laughlin-third-gap-spectra}(a), where the 3 zero modes are verified and a gap is visible, approaching $0.02 \Delta_2$ with increasing system size. The particle entanglement spectrum (PES) in Fig.~\ref{fig:laughlin-third-gap-spectra}(b) reproduces the Laughlin $1/3$ quasihole counting in every momentum sector. Further PES results are given in \App{E}~\cite{SupplementalMaterial}. All signatures are fully consistent with a Laughlin $1/3$ FCI \cite{sterdyniak2011extracting,sterdyniak2012particle,regnault2011fractional,bernevig2012emergent,wu2012zoology}. We have confirmed numerically that the single quasi-hole null space has the correct dimension, $N_s = 3N+1$ for $N$ electrons, but we point out that the null space dimension for lower doping does not coincide with that of the Haldane $V^{(2)}_1$ interaction. For instance, the two particle sector has $N_s$ nonzero energies, while the $V^{(2)}_1$ Hamiltonian has $2N_s$ nonzero energies.

\begin{figure*}[!t]
\centering
\includegraphics[width=\textwidth]{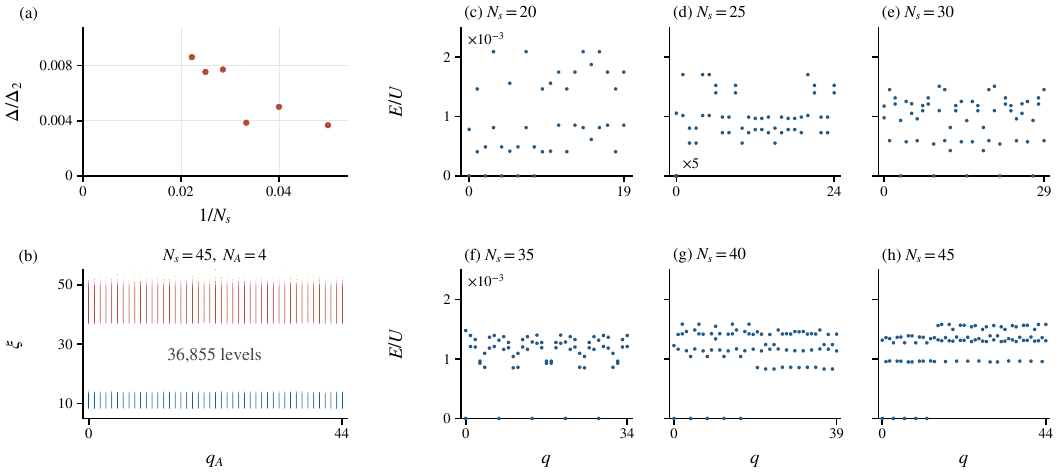}
\caption{Laughlin $1/5$ at $\lambda=0.4$, with five numerical zero modes within $|E/U|<10^{-9}$. (a) Lowest excitation gap $\Delta$ across all sectors, normalized by $\Delta_2=0.11047\,U$. (b) PES of the equal-weight ground-state mixture at $N_s=45$, $N_A=4$; $36{,}855$ levels lie below the entanglement gap. (c--h) Low-energy spectra for $N_s=20,25,\ldots,45$ and $N_e=N_s/5$; $\times n$ marks zero-mode multiplicity.}
\label{fig:laughlin-fifth-gap-spectra}
\end{figure*}

We also study the $1/5$ Laughlin state in the three-orbital $s=1$ band. The model wavefunction is 
\begin{equation}
\label{eq:Wrong_Laughlin_1/(2m+1)}
\Psi^{\text{FCI}}_{1/5}=\prod_{i<j}(W_i - W_j)^{4}\ket{\text{Slater}_{s=1}}
\end{equation}
To test the Hubbard zero-mode condition, set $Z_1=Z_2$. The corresponding pair factor becomes $(J_1-J_2)^{4}$. Since $J_i^3 = 0$, $(J_1-J_2)^4 = 6 J1^2 J2^2$ which is only nonzero on the $\ket{2,2}$ projection of the Slater (indexing the orbitals $\al = 0,1,2$). Since its spin structure is symmetric, its amplitude must be exactly zero in the Slater at $Z_i = Z_j$. Hence the Hubbard interaction
\bea
H_{1/5}=U\sum_{\mathbf R}\bigl(\hat n_{\mathbf R,0}\hat n_{\mathbf R,1}+\hat n_{\mathbf R,1}\hat n_{\mathbf R,2}+\hat n_{\mathbf R,2}\hat n_{\mathbf R,0}\bigr)
\eea
annihilates $\Psi^{\text{FCI}}_{1/5}$. Indeed, Figure~\ref{fig:laughlin-fifth-gap-spectra}(c--h) shows five numerical zero modes separated from excitations by a finite gap. Although the gap is increasing with particle number, it has not showed signs of saturation on accessible systems and a rough extrapolation suggests $\Delta \to .012\Delta_2$. Here, since $H_{1/5}$ is rank-3 per site in the 2-particle Hilbert space, we normalize by $\Delta_2 = \frac{1}{3N} \sum_{\mbf{p},a} E_a(\mbf{p})$ as $N \to \infty$. The PES in Fig.~\ref{fig:laughlin-fifth-gap-spectra}(b) reproduces the Laughlin $1/5$ quasihole counting in every momentum sector, providing strong evidence for the corresponding FCI. Further PES results and parameter scans are given in \App{E}~\cite{SupplementalMaterial}.

\subsection{The Moore-Read state and the Read-Rezayi series}

The Moore-Read state, known for the non-Abelian statistics of its quasiparticles, is a candidate for the first-LL state at filling $5/2$. In the scalar 0LL, it is an exact ground state at filling $1/2$ of the three-body parent introduced in Sec.~\ref{sec:scalar_ll_parents}. (Since the parent Hamiltonian is a local $3$-body interaction, we will consider the $s=1$ spinor since $s= 1/2$ has only two orbitals per site and cannot support this interaction.) To present its exact FCI wavefunction, we introduce the symmetric Moore-Read polynomial
\begin{equation}
\label{eq:Bosonic_Moore_Read}
P_{\text{MR}}(\{z_i\})=\operatorname{Pf}[\frac{1}{z_i-z_j}]\prod_{1\le{i}<j\le{N_e}}(z_i-z_j).
\end{equation}
It vanishes when three particles coincide. Writing their coordinates as $z+\delta z_i$, $i=1,2,3$, gives
\begin{equation}
\label{eq:Moore_Read_expansion}
\begin{aligned}
P(\{z_i\})\propto{}&(\delta z_1-\delta z_2)^2+(\delta z_2-\delta z_3)^2\\
&+(\delta z_3-\delta z_1)^2+O(\delta z_i^3),
\end{aligned}
\end{equation}
vanishing quadratically as the three particles approach coincidence.
After canceling the Pfaffian denominators against the Vandermonde factors in $P$, we substitute $z_i\mapsto z_i+J_i$ in the resulting polynomial to obtain
\begin{equation}
\label{eq:2LL_MooreRead}
\Psi^{\text{FCI}}_{\mathrm{MR}}=P(\{z_i+J_i\})\ket{\text{Slater}_{s=1}}
\end{equation}
Here $J$ is the same three-component matrix as in the $1/5$ construction, with $J^3=0$. The quadratic zero of $P$ and the cubic coincidence zero of the Slater factor give a fifth-order zero of the full scalar Moore-Read state. Mapping from the continuum gives use the parent interaction
\begin{equation}
\label{eq:Chern_MooreRead}
\begin{aligned}
H_{\mathrm{MR}}&=U\sum_{\mathbf R}\hat n_{\mathbf R,0}\hat n_{\mathbf R,1}\hat n_{\mathbf R,2}
\end{aligned}
\end{equation}
which annihilates the state via the following argument: since the Moore-Read wavefunction is at least degree 5 near the three-body zero $Z_1 = Z_2 = Z_3$, it will contain terms with $J^a_1 J^b_2 J^c_3$ where $a+b+c \geq 5$. Since $J^3 = 0$, the allowed terms are $(a,b,c) = (1,2,2),(2,2,2)$ and permutations. However, each such term has at least one repeated exponent, which will lead to a symmetric spin wavefunction. Hence, when $Z_1 = Z_2 = Z_3$ which is required in the null space of $H_{\mathrm{MR}}$, all these amplitudes must be zero. Conversely, if $J^3 \neq 0$, there are allowed term with distinct indices, e.g. $J_1 J_2^2 J_3^3$. 

\begin{figure*}[!t]
\centering
\includegraphics[width=\textwidth]{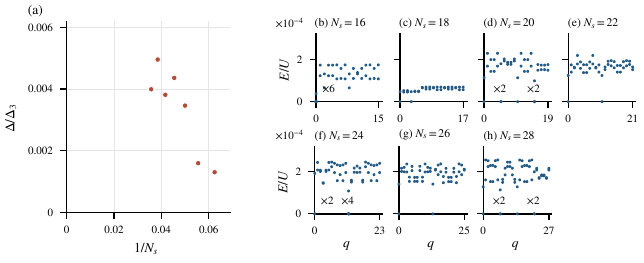}
\caption{Moore-Read state at $\lambda=0.6$, with six numerical zero modes for even $N_e$ and two for odd $N_e$, all satisfying $|E/U|<4\times10^{-10}$. The gap $\Delta$ is the lowest excited energy across all momentum sectors.
(a) Size dependence of $\Delta/\delta_3(\infty)$, with $\delta_3(\infty)=0.02841\,U$.
(b--h) Low-energy spectra for $N_s=16,18,\ldots,28$, with $N_e=N_s/2$.
Each momentum sector $q$ shows all numerical zero modes and up to two lowest nonzero levels; $\times n$ marks zero-mode multiplicity.}
\label{fig:moore-read-gap-spectra}
\end{figure*}

Verifying our construction of the frustration-free parent Hamiltonian, the spectra in Fig.~\ref{fig:moore-read-gap-spectra} reproduce the expected Moore-Read zero-mode counting: 2 zero modes for odd numbers of electrons and 6 zero modes for even numbers of electrons \cite{read2000paired,moore1991nonabelions}. We also observe a finite gap increasing with system size, providing numerical support for the Moore-Read phase in the thermodynamic limit. A rough extrapolation suggests $\Delta > 0.01 \Delta_{3}$, where the normalization is by the average $3$-body energy
$\Delta_3= \frac{1}{N_s} \sum_\mbf{p} E_s(\mbf{p})$ (\App{E}~\cite{SupplementalMaterial}).

Our construction for the Moore-Read state generalizes to the full Read-Rezayi series \cite{read1999beyond,bernevig2008model,bernevig2008properties,bernevig2009anatomy}. In the scalar 0LL, the fermionic $k$-Read-Rezayi state has the form
\begin{equation}
\label{eq:0LL_Read_Rezayi}
\Phi^{\mathrm{0LL}}_{k\mathrm{RR}}=P^{(k)}(\{z_i\})|\mathrm{0LL}\rangle.
\end{equation}
Here $|\mathrm{0LL}\rangle$ is the compact $N_e$-particle scalar Slater droplet.
The symmetric Read-Rezayi factor $P^{(k)}$ vanishes to at least second order when $k+1$ particles coincide. Including the Slater factor, the full fermionic state has clustering order at least $\binom{k+1}{2}+2$.
The filling is $k/(k+2)$, with $k=2$ giving the Moore-Read state.

Under the duality $\mathcal{U}$, the $k$th Read--Rezayi state maps to the spin-$(s=k/2)$ Dirac spinr in the $2s = k$th LL. With matrix coordinate $z+J$ where $J^{k+1}=0$, we obtain the wavefunction
\begin{equation}
\Phi^{k\mathrm{LL}}_{k\mathrm{RR}}=P^{(k)}(\{z_i+J_i\})|k\mathrm{LL}_{s=k/2}\rangle \ .
\end{equation}
Its contact component contains $\binom{k+1}{2}$ LLs, fewer than the scalar clustering order, so this state is an exact zero-energy ground state of the component-independent $(k+1)$-body contact interaction (\App{B}~\cite{SupplementalMaterial})
\begin{equation}
\label{eq:k-body-contact}
V_0^{(k+1)}=\mathcal G_0^{(k+1)}\sum_{i_1<\cdots<i_{k+1}}\prod_{a=1}^{k}\delta(\mathbf r_{i_a}-\mathbf r_{i_{a+1}}),
\end{equation}
with $\mathcal G_0^{(k+1)}>0$. The sampling map in Eq.~(\ref{eq:sampling}) gives the FCI wavefunction and its $(k+1)$-body Hubbard parent
\begin{equation}
\label{eq:RR_WF_Hubbard}
\begin{aligned}
\Psi^{C=1}_{k\mathrm{RR}}&=P^{(k)}(\{Z_i+J_i\})\ket{\text{Slater}_{s=k/2}}\\
H_{k\mathrm{RR}}&=U\sum_{\mathbf R}\prod_{\alpha=0}^{k}\hat n_{\mathbf R,\alpha},\qquad U>0.
\end{aligned}
\end{equation}
We will numerical investigate of these wavefunction to future studies.

\subsection{The Halperin series}
The Halperin $33m$ states describe two-component systems (spins $\uparrow,\downarrow$ or layers $A$ and $B$), with $N_e/2$ electrons in each layer and $m=0,1,2,3$ \cite{halperin1983theory}. Their disk wavefunctions and parent Hamiltonians are defined in Sec.~\ref{sec:scalar_ll_parents}. We denote them by $\hat V^{\mathrm{0LL}}_{33m}$ and their restrictions to the two scalar 0LLs by $\hat H^{\mathrm{0LL}}_{33m}$.  

We now apply the scalar-spinor duality $\mathcal{U}$, replacing $z_i \to W_i$ in the wavefunctions and obtaining the contact-only parent Hamiltonians
\begin{equation}
\label{eq:1LL_33m_parent}
\begin{aligned}
\hat V^{\mathrm{1LL}}_{330}&=\mathcal G\sum_{\substack{i<j\\l_i=l_j}}\delta(\mathbf r_i-\mathbf r_j),\\
\hat V^{\mathrm{1LL}}_{331}&=\hat V^{\mathrm{1LL}}_{330}+\mathcal G\sum_{\substack{i<j\\l_i\ne l_j}}\delta(\mathbf r_i-\mathbf r_j)\delta_{s_i\uparrow}\delta_{s_j\uparrow},\\
\hat V^{\mathrm{1LL}}_{332}&=\hat V^{\mathrm{1LL}}_{331}+\mathcal G\sum_{\substack{i<j\\l_i\ne l_j}}\delta(\mathbf r_i-\mathbf r_j)(1-\delta_{s_i s_j}),\\
\hat V^{\mathrm{1LL}}_{333}
&=\hat V^{\mathrm{1LL}}_{332}
+\mathcal G\sum_{\substack{i<j\\l_i\ne l_j}}
\delta(\mathbf r_i-\mathbf r_j)
\delta_{s_i\downarrow}\delta_{s_j\downarrow}.
\end{aligned}
\end{equation}
with $\mathcal G>0$. After projection to the two Dirac $\mathrm{1LL}$s, these interactions are unitarily equivalent to the corresponding scalar parents with proper couplings (\App{B}~\cite{SupplementalMaterial}). We will specialize to the $s=1$ case and denote the orbitals $A,B$ instead of $0,1$ for notational clarity in this section. 

We then apply the Landau-Chern sampling map $\mathcal{S}$ separately to the two Dirac $\mathrm{1LL}$s, obtaining two identical $C=1$ Landau-dual-Chern bands labeled $A$ and $B$. Substituting $z\to Z$ gives the Halperin-$33m$ FCI wavefunction
\begin{multline}
\label{eq:Halperin_33m_C=1*2}
\Psi^{C=1\otimes C=1}_{33m}=\prod_{\substack{i<j\\l_i=l_j}}(Z_i+J_i-Z_j-J_j)^2\\
{}\times\prod_{\substack{i:l_i=A\\j:l_j=B}}(Z_i+J_i-Z_j-J_j)^m\ket{\text{Slater}_A \otimes \text{Slater}_B} \ .
\end{multline}
Sampling the contact interactions gives onsite lattice parents. With $U>0$, we choose
\begin{equation}
\label{eq:33m_Hamiltonian_1+1}
\begin{aligned}
\hat U_{330}&=U\sum_{\mathbf R} \bigl[\hat n_{\mathbf R A\uparrow}\hat n_{\mathbf R A\downarrow}+\hat n_{\mathbf R B\uparrow}\hat n_{\mathbf R B\downarrow}\bigr] \ ,\\
\hat U_{331}&=\hat U_{330}+U\sum_{\mathbf R}\hat n_{\mathbf R A\uparrow}\hat n_{\mathbf R B\uparrow} \ ,\\
\hat U_{332}&=\hat U_{331}+U\sum_{\mathbf R}\bigl[\hat n_{\mathbf R A\uparrow}\hat n_{\mathbf R B\downarrow}+\hat n_{\mathbf R A\downarrow}\hat n_{\mathbf R B\uparrow}\bigr] \ , \\
\hat U_{333}&=\hat U_{332}+U\sum_{\mathbf R} \hat n_{\mathbf R A\downarrow}\hat n_{\mathbf R B\downarrow} \ .
\end{aligned}
\end{equation}
All three interactions are onsite, positive semidefinite, and invariant under exchanging $A$ and $B$. They annihilate the corresponding sampled Halperin states before band projection (\App{D}~\cite{SupplementalMaterial}).

\begin{figure}[!t]
    \centering
    \includegraphics[width=\columnwidth]{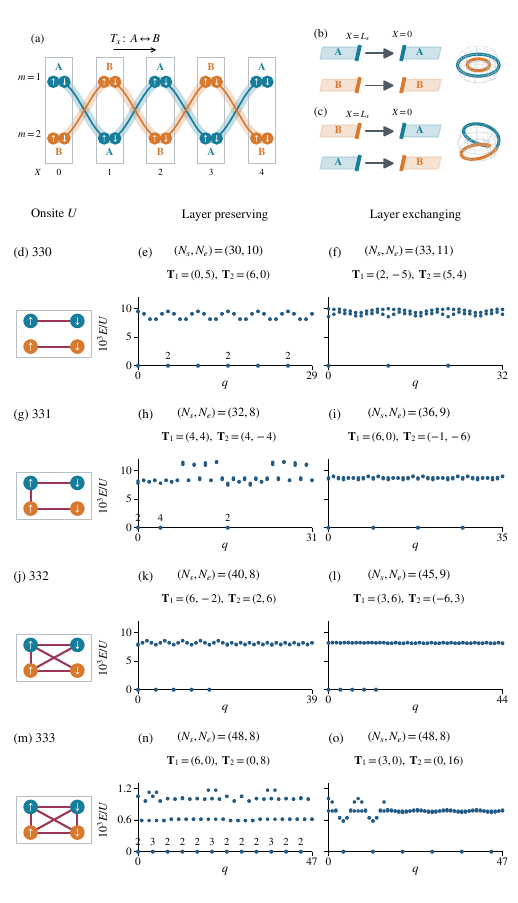}
    \caption{Halperin $33m$ states.
    (a) Layer assignment in the four-component band: gray boxes denote sites, and one-site translation exchanges $A,B$.
    (b,c) Layer-preserving and layer-exchanging boundary identifications.
    Rows below show $330$, $331$, $332$, and $333$; columns show onsite repulsions (red links, strength $U$) and spectra for the two boundary types, respectively.
    The corresponding zero-mode counts are $(9,8,5,27)$ and $(3,4,5,6)$.
    Headers give $(N_s,N_e)$ and torus periods $\mathbf T_1,\mathbf T_2$.
    Each momentum sector $q$ shows the zero modes and up to two lowest available nonzero levels; numbers mark zero-mode multiplicities greater than one.
    Energies are plotted as $10^3E/U$.}
    \label{fig:halperin_layers_channels_ed}
\end{figure}

To construct Halperin states on the lattice, an intrinsic layer structure is required. To make it compatible with the matrix-coordinate construction on the lattice, we introduce a coordinate $Z+J_A$ in layer $A$ and $Z+J_B$ in layer $B$, where $Z=X+iY$ and the nilpotent internal matrices act within their respective layers. We implement this structure through a color-entangled construction in which one-site translation along $x$ exchanges $A$ and $B$ \cite{barkeshli2012topological,wu2013bloch,wu2014haldane,wu2015fractional,sterdyniak2013series}.
We consider the simplest four-band construction, with orbitals $(m,s)=(1\uparrow,1\downarrow,2\uparrow,2\downarrow)$ assigned to layers as follows:
\begin{equation}
\label{eq:alternate_layer}
\begin{aligned}
l(m,X)=A,\quad&
m=1,\ X\text{ even};\quad m=2,\ X\text{ odd},\\
l(m,X)=B,\quad&
m=1,\ X\text{ odd};\quad m=2,\ X\text{ even}.
\end{aligned}
\end{equation}
The real-space layer projectors are
\[
\begin{aligned}
\Pi_A(X)&=\frac{\mathbb{1}_4+(-1)^X\operatorname{diag}(\mathbb{1}_2,-\mathbb{1}_2)}{2},\\
\Pi_B(X)&=\frac{\mathbb{1}_4-(-1)^X\operatorname{diag}(\mathbb{1}_2,-\mathbb{1}_2)}{2}.
\end{aligned}
\]
Let $\hat P_A$ and $\hat P_B$ project onto single bands supported on these layers. We can then decompose the full-band projector as $\hat P=\hat P_A+\hat P_B$.

We first construct the $A$-layer band. Its $2\times1$ primitive cell contains two orbitals at each of two adjacent sites, as illustrated in Fig.~\ref{fig:halperin_layers_channels_ed}, and its Brillouin zone is $k_x\in[0,\pi)$, $k_y\in[0,2\pi)$.

We represent $J_A$ by the same two-component nilpotent matrix $J$ at both sites. Its matrix in the layer cell is therefore $J\oplus J$; embedded in the original four orbitals at each site, it is $\Pi_A(X)(J\oplus J)\Pi_A(X)$. With $\hat Q_A$ the complementary projector of the layer band, we impose
\begin{equation}
\label{eq:halperin_layer_preservation}
\hat Q_A(Z+J\oplus J)\hat P_A=0.
\end{equation}
Here $Z$ uses the integer site positions. Let $u_2^{(1)}(\mathbf k)$ and $u_2^{(2)}(\mathbf k)$ be the even- and odd-site amplitudes in layer $A$, with the site Bloch factor $e^{i\mathbf k\cdot\mathbf R}$ extracted. Due to their one-site relative displacement along $x$, the odd-site block carries $e^{ik_x}$ in the cell spinor:
\begin{equation}
\label{eq:halperin_layer_cell_spinor}
u_A(\mathbf k)=
\begin{pmatrix}
u_2^{(1)}(\mathbf k)\\
e^{ik_x}u_2^{(2)}(\mathbf k)
\end{pmatrix}.
\end{equation}
Since the matrix coordinate acts within each orbital block, the band defined by each $u_2^{(a)}$ must also be preserved. With its projector $\hat P_2^{(a)}$ and complementary projector $\hat Q_2^{(a)}$, this requires
\begin{equation}
\label{eq:halperin_component_preservation}
\hat Q_2^{(a)}(Z+J)\hat P_2^{(a)}=0,
\qquad a=1,2.
\end{equation}
We satisfy this condition by taking each $\hat P_2^{(a)}$ to be a Landau-dual $C=1$ band constructed as above for the $2\times1$ layer cell. Both $u_2^{(1)}$ and $u_2^{(2)}$ are nowhere zero, so projection onto either orbital block identifies the $A$-layer band with the corresponding $C=1$ band, giving $C_A=1$.

We choose the layer cell spinor periodic under $k_x\to k_x+\pi$. The displacement factor in Eq.~(\ref{eq:halperin_layer_cell_spinor}) then makes $u_2^{(1)}$ periodic and $u_2^{(2)}$ antiperiodic under this shift. With $C_A=1$, we choose $u_A(k_x,k_y+2\pi)=e^{-2ik_x}u_A(\mathbf k)$ in the Chern convention of \App{D}~\cite{SupplementalMaterial}. Both blocks inherit this transition, giving
\begin{equation}
\label{eq:Bloch_boundary}
\begin{aligned}
u_2^{(a)}(k_x+\pi,k_y)
&=(-1)^{a-1}u_2^{(a)}(\mathbf k),\\
u_2^{(a)}(k_x,k_y+2\pi)
&=e^{-2ik_x}u_2^{(a)}(\mathbf k),
\end{aligned}
\end{equation}
for $a=1,2$. \App{D}~\cite{SupplementalMaterial} gives explicit amplitudes satisfying Eq.~(\ref{eq:halperin_layer_preservation}) and these boundary relations.

We obtain layer $B$ by one-site translation along $x$, denoted by $\hat T_x$, taking $\hat P_B=\hat T_x\hat P_A\hat T_x^{-1}$ and $J_B=\hat T_xJ_A\hat T_x^{-1}$. Thus $J_B$ has the same local matrix $J$ on the complementary orbital pair. Translation changes $Z$ only by a constant, so $B$ inherits band preservation and $C_B=1$. With consistent Bloch phases, the layer states are
\begin{equation}
\label{eq:subbands}
\psi_{4A,4B}(\mathbf k,\mathbf R)\propto
\begin{pmatrix}
[1\pm(-1)^X]u_2^{(1)}(\mathbf k)\\
[1\mp(-1)^X]u_2^{(2)}(\mathbf k)
\end{pmatrix}
e^{i\mathbf k\cdot\mathbf R}.
\end{equation}
For normalized layer states at $\mathbf k$ in the reduced Brillouin zone, the one-site translation eigenstates are
\begin{align}
\psi_4(\mathbf k)
&=\frac{\psi_{4A}(\mathbf k)+\psi_{4B}(\mathbf k)}{\sqrt2},
\label{eq:unfolded_layer_bands}\\
\psi_4(\mathbf k+(\pi,0))
&=\frac{\psi_{4A}(\mathbf k)-\psi_{4B}(\mathbf k)}{\sqrt2}.
\label{eq:unfolded_layer_bands_shifted}
\end{align}
This unitary transformation unfolds the two layer bands into a single four-component band on the full Brillouin zone $[0,2\pi)^2$ and gives
\begin{equation}
\label{eq:layer_band_projectors}
\begin{aligned}
\sum_{\mathbf q\in\{\mathbf k,\,\mathbf k+(\pi,0)\}}
|\psi_4(\mathbf q)\rangle\langle\psi_4(\mathbf q)|
=
\sum_{l=A,B}|\psi_{4l}(\mathbf k)\rangle
\langle\psi_{4l}(\mathbf k)|.
\end{aligned}
\end{equation}
Summing over the reduced Brillouin zone recovers the full-band projector $\hat P$. The full-band spinor is $u_4^{\mathrm T}\propto (u_2^{(1)},u_2^{(2)})^{\mathrm T}$ with
\begin{equation}
\label{eq:Half_Bloch_boundary}
u_4(k_x+\pi,k_y)=
\operatorname{diag}(\mathbb{1}_2,-\mathbb{1}_2)u_4(\mathbf k).
\end{equation}
Equation~(\ref{eq:Bloch_boundary}) also gives $u_4(k_x,k_y+2\pi)=e^{-2ik_x}u_4(\mathbf k)$, whose phase winds twice over the full Brillouin zone, yielding $C=2$. The resulting filling is $N_e/N_s=1/(3+h)$, while the fractional Chern number is expected to be $2/(3+h)$. 

In the original orbital basis, the two layer actions $J_A$ and $J_B$ together are represented by $J\oplus J$. Since the matrix coordinate preserves both layer bands, their sum obeys
\begin{equation}
\label{eq:ideal_C=2}
\hat Q(Z+J\oplus J)\hat P=0,
\end{equation}
where $\hat Q$ is the complementary projector of the full band. The interactions in Eq.~(\ref{eq:33m_Hamiltonian_1+1}) remain onsite under the layer assignment in Eq.~(\ref{eq:alternate_layer}). Their $A\leftrightarrow B$ symmetry ensures invariance under one-site translation.

Using these properties of $\psi_4$, we obtain the $33m$ wavefunction
\begin{multline}
\label{eq:C=2-33m}
\Psi^{C=2}_{33m}=\prod_{\substack{i<j\\l_i=l_j}}(Z_i+J_i-Z_j-J_j)^2\\
{}\times\prod_{\substack{i:l_i=A\\j:l_j=B}}(Z_i+J_i-Z_j-J_j)^m \ket{\text{Slater}_{C=2}}
\end{multline}
Here $l_i$ follows Eq.~(\ref{eq:alternate_layer}), and $\ket{\text{Slater}_{C=2}}$ is the antisymmetrized product of compact $N_e/2$-particle Slater droplets in the two layer bands. The matrix-coordinate condition keeps the constructed state within the zero-energy band, while its coincidence zeros ensure that it is annihilated by the onsite interaction $\hat U_{33m}$ in Eq.~(\ref{eq:33m_Hamiltonian_1+1}).

On a torus, a period vector with an odd $x$ component exchanges the layers, whereas periods with even $x$ components preserve them. The spectra in Fig.~\ref{fig:halperin_layers_channels_ed} reproduce the corresponding Halperin zero-mode counts: $9,8,5$ for layer-preserving boundaries and $3,4,5$ for layer-exchanging boundaries, for $330,331,332$, respectively \cite{barkeshli2012topological,wu2014haldane} Additional sizes are reported in \App{E}~\cite{SupplementalMaterial}. The finite-size gaps and particle entanglement spectra also provide numerical support for the Halperin phases (\App{E}~\cite{SupplementalMaterial}).

We now discuss the behavior of each state. To do so, it is helpful to first write down the color $SU(2)$ symmetry that acts on the $u_1,u_2$ spinors. On each site $\mathbf R=(X,Y)$ group the four orbitals into two doublets,
$\hat d_{\mathbf R s}=(\hat c_{\mathbf R1s},\,\hat c_{\mathbf R2s})^{\mathsf T}$ with $s=\uparrow,\downarrow$, and let $\sigma^a$ act on the
orbital index $m$. For $N_x$ and $N_y$ even, the color generators are
\begin{equation}
\hat S^a=\frac12\sum_{\mathbf R}\sum_{s=\uparrow,\downarrow}e^{i\mathbf Q_a\cdot\mathbf R}\,
\hat d^\dagger_{\mathbf R s}\,\sigma^a\,\hat d_{\mathbf R s}.
\label{eq:Sa}
\end{equation}
and $\mathbf Q_{x,y,z}=(0,\pi),(\pi,\pi),(\pi,0)$. The algebra $[S^a,S^b]=i\epsilon^{abc}S^c$ holds exactly after projection because the single-particle operators
$s^a=\tfrac12e^{i\mathbf Q_a\cdot\mathbf R}\sigma^a$ commute with the band projector. The doublet invariants $n_{\mathbf R\ell}+n_{\mathbf R,\ell+2}$ and $n_{\mathbf R\ell}n_{\mathbf R,\ell+2}$ commute with all $S^a$,
so $H_{333}$ and $H_{332}$ are fully SU(2) symmetric, although $H_{330}$ and $H_{331}$ only preserve $\hat{S}_z$. Since $H_{330}$ and $H_{331}$ do not posses an $SU(2)$ symmetry and their ground states do not break $\hat{S}_z$, we expect them to be gapped as is consistent with \Fig{fig:halperin_layers_channels_ed}. We see that $H_{332}$ also appears gapped, even though it is $SU(2)$ symmetric. This is because the 332 state is a color singlet, so it is not symmetry broken. In contrast, the $333$ state is the maximal color ferromagnet. It must be gapless by Goldstone's theorem, and will have a type-II quadratic Goldstone mode which is the color magnon. The charge sector is still expected to be gapped \cite{PhysRevLett.117.056806,PhysRevB.47.16419,Yang_1996}. 

Finally, we point out there is a third natural SU(2) symmetric Hamiltonian:
\bea
H_{111}=\sum_{\mathbf R}n_{\mathbf{R}, A \uparrow}n_{\mathbf{R}, B \uparrow} \ .
\eea
This Hamiltonian also has a frustration free ground state at $\nu =1/2$: the $C=1$ ferromagnet. These ground states can be constructed from the maximally polarized Slater determinant
which fills the layer-$A$ band:
\bea
\ket{\Psi_A}&=\prod_{\mathbf k\in [0,\pi)\times[0,2\pi)}\hat a^\dagger_{A\mathbf k} \ket{0} \\
\hat a^\dagger_{A\mathbf k}&=\frac{\hat\gamma^\dagger_{\mathbf k}+\hat\gamma^\dagger_{\mathbf k+(\pi,0)}}{\sqrt2},\qquad
\hat\gamma^\dagger_{\mathbf k}=\sum_{m,s}[u_2^{(m)}(\mathbf k)]_s\,\hat c^\dagger_{\mathbf k ms} \ .
\eea
This state has $N_e=N_s/2$ electrons, $\hat S^z=N_e/2$ and $C=1$.
It is annihilated by $\hat U_{111}$ because $\hat n_{\mathbf RBs}\ket{\Psi_A}=0$. Lowering with $\hat S^-$ generates the $N_e+1$ zero modes of $\hat U_{111}$, ending at $\ket{\Psi_B}=\hat T_x|\Psi_A\rangle$. This state also has type-II magnons because of the broken $SU(2)$.\footnote{The SU(2) can be broken to $U(1) \times \mathds{Z}_2$ by adding a term $=V\sum_{\langle\mathbf R\mathbf R'\rangle}\bigl(\hat n_{\mathbf RA}\hat n_{\mathbf R'B}+\hat n_{\mathbf RB}\hat n_{\mathbf R'A}\bigr)$ while preserving the maximally spin polarized zero modes and gapped the others (at $\nu=1/2$ and the 333 state at $\nu=1/6$. In this Ising symmetry class, we expect a full gap.}

For even $N_y$ the total density is uniform, but the orbital density alternates,
\begin{equation}
\big\langle\hat n_{\mathbf R1}-\hat n_{\mathbf R2}\big\rangle=\frac{(-1)^X}{2},
\end{equation}
a density wave at $(\pi,0)$ that breaks $\hat T_x$. We note that our construct possess an exact SU(2) color symmetry, unlike that of Ref. \cite{dong2023many} 

Having completed the exact constructions above, we now consider three more questions about the properties of these exact zero modes: (i) whether the constructed states exhaust the zero modes, (ii) their fractional Hall conductance, and (iii) whether a thermodynamic gap separates the ground states from excitations. Below we present the corresponding arguments for the Laughlin $1/3$ state. Some of these arguments already extend to the other states while a complete treatment is left to future work.

\section{Formal Properties}
We have introduced the key ideas that allowed us to construct frustration-free FCI wavefunctions and have examined their properties numerically. We now establish several properties of these states which we may have taken for granted, focusing primarily on the simplest case, the Laughlin \(1/3\) state. We prove analytically that, for sufficiently large systems, the sampled wavefunctions span the entire zero-mode space. Further numerical calculations confirm that the degenerate ground-state manifolds have the expected fractional Hall conductances. Finally, we construct an analytic path connecting the lattice Laughlin state to its continuum counterpart. The excitation gap remains open at every point we sample along this path on a finite torus, providing evidence for a nonzero gap in the thermodynamic limit.

\subsection{Completeness of the zero modes}\label{sec:laughlin_completeness}
We have constructed three lattice Hubbard zero modes by sampling the Dirac Laughlin states. We now show that these states span the full zero-energy ground-state space on a rectangular torus with periodic lattice boundary conditions when $N_s=N_\phi=3N_e$ and $N_e\ge13$. We give the proof for the directly sampled band. The result extends to the parameter family introduced in Sec. IV because the onsite orbital transformation and its inverse both preserve the no-double-occupancy condition.

The starting point is the corresponding continuum result. At this filling, the three torus Laughlin states span the zero-mode space of Haldane’s pseudopotential in the scalar 0LL \cite{haldane1985periodic}. Under the scalar–spinor duality of Sec. III, they map to the three exact zero modes of the contact repulsion in the spin-$1/2$ Dirac 1LL. To apply this result to the lattice, consider an arbitrary zero-energy state $\Psi$ of the lattice Hamiltonian. Since the kinetic and Hubbard terms are both positive semidefinite, $\Psi$ lies in the flat band and has no double occupancy. Let $\Phi=\mathcal S^{-1}\Psi$ be its Dirac 1LL preimage. For any pair $i<j$, the onsite constraint gives
\begin{equation}
\label{eq:dirac_onsite_constraint}
\left.\Phi(\mathbf r_1,\ldots,\mathbf r_{N_e})\right|_{\mathbf r_i=\mathbf r_j=\mathbf R}=0
\qquad\text{for every site }\mathbf R.
\end{equation}
Applying the scalar–spinor duality in reverse, we find that the scalar 0LL wavefunction $\mathcal U^{-1}\Phi$ vanishes cubically in the pair separation when the pair is centered at $\mathbf R$. To establish completeness, it therefore remains to show that this cubic zero persists when the pair is centered at an arbitrary point on the continuum torus. Once this is shown, $\mathcal U^{-1}\Phi$ is a zero mode of Haldane’s pseudopotential, and $\Psi$ lies in the span of the three sampled states.

We prove this by zero counting. For a nonzero holomorphic function on the torus, the quasiperiodic boundary conditions fix the total number of zeros, counted with multiplicity. This principle applies directly to the holomorphic part of the scalar 0LL wavefunction. On the other hand, as we have established, the onsite Hubbard constraint requires a cubic pair zero at every lattice site. If this constraint imposes more zeros than the boundary conditions allow, the function must vanish identically. Technically, we first remove the zeros required by fermionic antisymmetry and consider the remaining holomorphic function $G(z_1,\ldots,z_{N_e})$. Applying the zero-counting principle with five particles at a common position $z_1=\cdots=z_5=z$ then shows directly that, for $N_e\ge13$, $G$ must vanish for every $z$ on the torus. If $G$ vanishes whenever $n$ particles coincide, then $n-1$ coincident particles encounter an additional zero wherever they meet another particle. Repeating the same zero count from five particles down to two gives a pair zero of $G$. By symmetry, this zero is at least quadratic, so restoring the fermionic zero gives the required cubic zero and proves completeness. We give the technical details in \App{C}~\cite{SupplementalMaterial}.

\subsection{Fractional Hall conductance}
We determine the fractional Hall response from the total many-body Chern number $C_{\mathrm{tot}}$ of the $d$-dimensional ground-state manifold \cite{niu1985quantized,hatsugai2005characterization,sheng2011fractional}. As usual, we introduce twisted boundary conditions $\pmb\theta=(\theta_x,\theta_y)\in[0,2\pi)^2$ on a torus. At neighboring points on the $\boldsymbol{\theta}$ mesh, we use the ED basis and the band form factors to evaluate the overlaps between the two sets of ground states. Using the standard Wilson loop method and summing the phases of the loop determinants yields $C_{\mathrm{tot}}$, and $C_{\mathrm{tot}}/d$ gives the Hall conductance of the equally weighted manifold in units of $e^2/h$.

Table~\ref{tab:charge-hall} gives the results for the Laughlin and Halperin states. In the numerical convention here, the Laughlin $1/3$ triplet has $C_{\mathrm{tot}}=-1$, giving a Hall conductance of magnitude $e^2/(3h)$. We also check Halperin $(330)$, $(331)$, $(332)$, and $(333)$, the respective conductance are $2/3$, $1/2$, $2/5$, and $1/3$ in units of $e^2/h$, in agreement with the expected charge response $2/(3+m)$ of the $(33m)$ series. The dependence on torus geometry provides a further check. For $(330)$, the $3\times4$ torus has three ground states with $C_{\mathrm{tot}}=-2$, whereas the $6\times3$ torus has nine with $C_{\mathrm{tot}}=-6$. Both give $C_{\mathrm{tot}}/d=-2/3$. The same holds for $(333)$, where the six- and fifteen-state manifolds give $-2/6=-5/15=-1/3$.
\begin{table}[t]
\centering
\caption{Many-body Hall conductance of the Laughlin and Halperin
ground-state manifolds under charge twists. $L_x\times L_y$ denotes
the lattice dimensions, $N_e$ the particle number, and $d$ the
ground-state degeneracy.}
\label{tab:charge-hall}
\begin{tabular}{lc!{\hspace{6pt}}c!{\hspace{6pt}}cccc}\hline\hline
State & $\nu$ & $L_x\times L_y$ & $N_e$ & $d$
      & $C_{\mathrm{tot}}$ & $C_{\mathrm{tot}}/d$ \\
\hline
Laughlin & $1/3$ & $3\times5$            & 5 &  3 & $1$ & $1/3$ \\
$(330)$  & $1/3$ & $3\times4$            & 4 &  3 & $2$ & $2/3$ \\
$(330)$  & $1/3$ & $6\times3$            & 6 &  9 & $6$ & $2/3$ \\
$(331)$  & $1/4$ & $3\times4$            & 3 &  4 & $2$ & $1/2$ \\
$(331)$  & $1/4$ & $4\times4$            & 4 &  8 & $4$ & $1/2$ \\
$(332)$  & $1/5$ & $4\times5$            & 4 &  5 & $2$ & $2/5$ \\
$(333)$  & $1/6$ & $8\times3,\;3\times8$ & 4 &  6 & $2$ & $1/3$ \\
$(333)$  & $1/6$ & $4\times6,\;6\times4$ & 4 & 15 & $5$ & $1/3$ \\
\hline\hline
\end{tabular}
\end{table}

\subsection{Thermodynamic gap}\label{sec:laughlin_gap}
Recently, a proof was given for the Laughlin gap on the sphere for the $V_1$ Hamiltonian \cite{openai2026fock,scheer2026hierarchy}. Using our scalar--spinor duality, this gap result could be tranferred to the corresponding Dirac contact interaction on the sphere. Motivated by this continuum reference point, we construct a path of many-body Hamiltonians on each finite torus connecting the projected Hubbard model to a lattice Hamiltonian with the spectrum of Haldane's $V_1$ Hamiltonian. A gap that stays open along this path at finite size provides evidence for a thermodynamic gap in the lattice model.

Let $\hat U_0$ be the onsite Hubbard repulsion and $\hat P$ the projector onto the $C=1$ flat band. The projected interaction is
\begin{equation}
\hat P\hat U_0\hat P = \frac{U}{N_s^2}\sum_{\mathbf k,\mathbf k',\mathbf q} V^{(0)}_{\mathbf k\mathbf k'}(\mathbf q)\hat\psi_{\mathbf k-\mathbf q}^\dagger\hat\psi_{\mathbf k'+\mathbf q}^\dagger\hat\psi_{\mathbf k'}\hat\psi_{\mathbf k}.
\label{eq:hubbard-endpoint}
\end{equation}
Here $\hat\psi_{\mathbf k}$ annihilates the normalized Bloch orbital $\psi_{\mathbf k}(\mathbf R)/\sqrt{N_s}$, with $\psi_{\mathbf k}(\mathbf R)=e^{i\mathbf k\cdot\mathbf R}u_{\mathbf k}$. The coefficient for the ordered orbital components is
\begin{equation}
V^{(0)}_{\mathbf k\mathbf k'}(\mathbf q) = \sum_{\mathbf R} \psi^*_{\mathbf k-\mathbf q,0}(\mathbf R)\psi^*_{\mathbf k'+\mathbf q,1}(\mathbf R)\psi_{\mathbf k',1}(\mathbf R)\psi_{\mathbf k,0}(\mathbf R).
\end{equation}
Introducing the onsite pair annihilator
\begin{equation}
\hat B_{0,\mathbf R} = \frac{1}{N_s}\sum_{\mathbf k,\mathbf k'} \psi_{\mathbf k,1}(\mathbf R)\psi_{\mathbf k',0}(\mathbf R)\hat\psi_{\mathbf k}\hat\psi_{\mathbf k'},
\label{eq:hubbard-pair}
\end{equation}
we can write the interaction in the compact form
\begin{equation}
\hat P\hat U_0\hat P = U\sum_{\mathbf R} \hat B_{0,\mathbf R}^\dagger\hat B_{0,\mathbf R}.
\end{equation}
At filling $1/3$, direct sampling preserves the contact zero of each Dirac Laughlin state at every lattice site, giving three Hubbard zero modes.

We define the projection of $\hat U_1$ onto the same flat band through the matrix elements of the Dirac 1LL contact repulsion,
\begin{equation}
\hat P\hat U_1\hat P = \frac{\mathcal G_0^{(2)}}{N_s^2}\sum_{\mathbf k,\mathbf k',\mathbf q} V^{(1)}_{\mathbf k\mathbf k'}(\mathbf q)\hat\psi_{\mathbf k-\mathbf q}^\dagger\hat\psi_{\mathbf k'+\mathbf q}^\dagger\hat\psi_{\mathbf k'}\hat\psi_{\mathbf k}.
\label{eq:dirac-endpoint}
\end{equation}
In the normalized Dirac magnetic Bloch orbitals $\phi_{\mathbf k}/\sqrt{N_s}$, the corresponding coefficient is the contact integral over the torus,
\begin{equation}
V^{(1)}_{\mathbf k\mathbf k'}(\mathbf q) = \int d^2r \phi^*_{\mathbf k-\mathbf q,0}(\mathbf r)\phi^*_{\mathbf k'+\mathbf q,1}(\mathbf r)\phi_{\mathbf k',1}(\mathbf r)\phi_{\mathbf k,0}(\mathbf r).
\end{equation}
Introducing the contact pair annihilator
\begin{equation}
\hat A_1(\mathbf r) = \frac{1}{N_s}\sum_{\mathbf k,\mathbf k'} \phi_{\mathbf k,1}(\mathbf r)\phi_{\mathbf k',0}(\mathbf r)\hat\psi_{\mathbf k}\hat\psi_{\mathbf k'},
\label{eq:dirac-pair}
\end{equation}
we obtain the corresponding compact form
\begin{equation}
\hat P\hat U_1\hat P = \mathcal G_0^{(2)}\int d^2r \hat A_1^\dagger(\mathbf r)\hat A_1(\mathbf r).
\end{equation}
Matching these matrix elements in the two Slater bases preserves the entire Dirac contact spectrum at any filling, and hence the scalar $V_1$ spectrum up to the coupling scale. Each Dirac Laughlin state gives a lattice zero mode with the same Slater coefficients.

To interpolate between the two interactions, recall that sampling gives $(\mathcal S\phi_{\mathbf k})(\mathbf R)=s_{\mathbf k}\psi_{\mathbf k}(\mathbf R)$. For $0\le t\le 1$, define
\begin{equation}
\hat D(t) = \prod_{\mathbf k} s_{\mathbf k}^{(1-t)\hat n_{\mathbf k}}, \hat n_{\mathbf k} = \hat\psi_{\mathbf k}^\dagger\hat\psi_{\mathbf k}.
\end{equation}
The commutator $[\hat\psi_{\mathbf k},\hat n_{\mathbf k'}]=\delta_{\mathbf k,\mathbf k'}\hat\psi_{\mathbf k}$ gives
\begin{equation}
\hat\psi_{\mathbf k}\hat D(t) = \hat D(t)s_{\mathbf k}^{1-t}\hat\psi_{\mathbf k}.
\label{eq:s_n_commutation}
\end{equation}

We want the contact pair to annihilate the Laughlin state after the state is transformed by $\hat D(t)$. We therefore define $\hat A_t(\mathbf r)=\hat D(t)\hat A_1(\mathbf r)\hat D(t)^{-1}$. Equation~\eqref{eq:s_n_commutation} gives its momentum expansion,
\begin{equation}
\hat A_t(\mathbf r) = \frac{1}{N_s}\sum_{\mathbf k,\mathbf k'} s_{\mathbf k}^{t-1}s_{\mathbf k'}^{t-1}\phi_{\mathbf k,1}(\mathbf r)\phi_{\mathbf k',0}(\mathbf r)\hat\psi_{\mathbf k}\hat\psi_{\mathbf k'}.
\end{equation}
The onsite pair is the restriction of this contact pair to a lattice site, $\hat B_{t,\mathbf R}=\hat A_t(\mathbf R)$. Sampling contributes one factor $s_{\mathbf k}$ per orbital, giving
\begin{equation}
\hat B_{t,\mathbf R} = \frac{1}{N_s}\sum_{\mathbf k,\mathbf k'} s_{\mathbf k}^{t}s_{\mathbf k'}^{t}u_{\mathbf k,1}u_{\mathbf k',0}e^{i(\mathbf k+\mathbf k')\cdot\mathbf R}\hat\psi_{\mathbf k}\hat\psi_{\mathbf k'}.
\end{equation}
At $t=0$ the sampling factors in the lattice pair equal one, recovering the Hubbard pair of Eq.~\eqref{eq:hubbard-pair}. At $t=1$ the same holds for the continuum pair in Eq.~\eqref{eq:dirac-pair}. We therefore weight the lattice sum by $1-t$ and the continuum integral by $t$ to connect the two interactions,
\begin{equation}
\hat H(t) = (1-t)U\sum_{\mathbf R} \hat B_{t,\mathbf R}^\dagger\hat B_{t,\mathbf R} + t\mathcal G_0^{(2)}\int d^2r \hat A_t^\dagger(\mathbf r)\hat A_t(\mathbf r).
\label{eq:Ut}
\end{equation}
Since $s_{\mathbf k}>0$ on the finite momentum mesh, every matrix element of $\hat H(t)$ is analytic in $t$. Its real-space interactions decay exponentially along the path (\App{C}~\cite{SupplementalMaterial}).

For each Laughlin zero mode $|\Psi(1)\rangle$ of $\hat H(1)$, define
\begin{equation}
|\Psi(t)\rangle = \frac{\hat D(t)|\Psi(1)\rangle}{\|\hat D(t)|\Psi(1)\rangle\|}.
\end{equation}
For any fixed $t\in[0,1]$, the definition of $\hat A_t$ and the Laughlin contact condition give, at every $\mathbf r$,
\begin{equation}
\hat A_t(\mathbf r)\hat D(t)|\Psi(1)\rangle = \hat D(t)\hat A_1(\mathbf r)|\Psi(1)\rangle = 0.
\end{equation}
This identity holds in particular at every lattice site, where $\hat A_t(\mathbf R)=\hat B_{t,\mathbf R}$. Both pair operators thus annihilate the normalized state at the chosen $t$. Substitution in Eq.~\eqref{eq:Ut} gives
\begin{equation}
\hat H(t)|\Psi(t)\rangle = 0.
\end{equation}

\begin{figure}[htbp]
\centering
\includegraphics[width=\columnwidth]{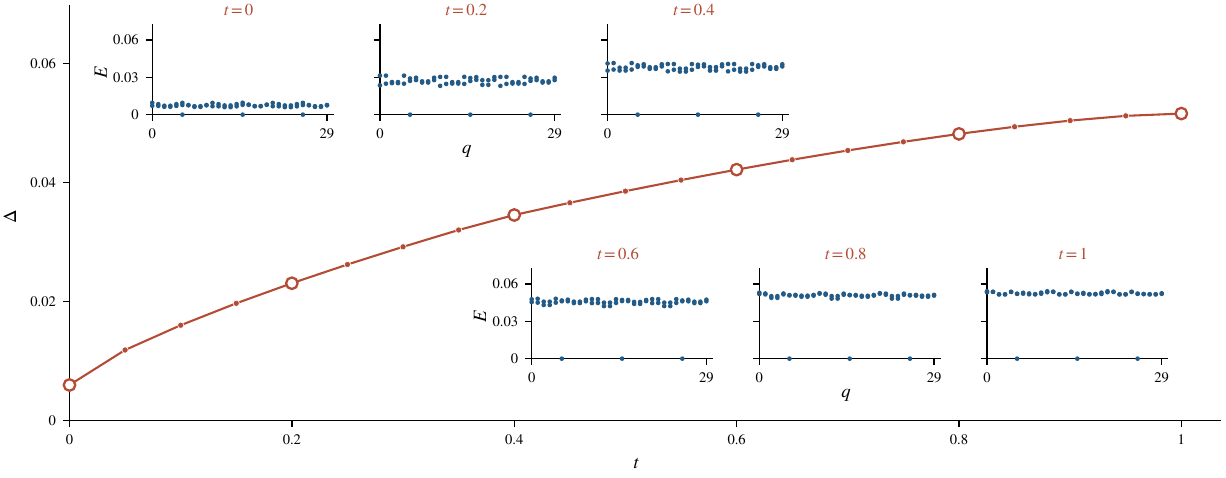}
\caption{Laughlin $1/3$ interpolation on a $6\times5$ torus with $N_e=10$. The lattice band is obtained by directly sampling the Dirac 1LL. We use $U=\mathcal G_0^{(2)}=1$. Open circles mark $t=0,0.2,\ldots,1$. The gap $\Delta$ is the lowest excitation energy across all momentum sectors in these coupling units. The insets show three numerical zero modes and up to two lowest nonzero levels per sector $q$.}
\label{fig:laughlin-third-interpolation}
\end{figure}

We numerically diagonalized $\hat H(t)$ of Eq.~\eqref{eq:Ut} on a $6\times5$ torus with ten electrons. Figure~\ref{fig:laughlin-third-interpolation} shows that the three ground states remain at zero energy and that the excitation gap stays open at every sampled $t$, increasing from the Hubbard endpoint to the $V_1$ endpoint. This interpolation supports an adiabatic connection between the lattice ground states and the continuum Laughlin $1/3$ states, providing evidence for a thermodynamic gap in the lattice model.
\App{C}~\cite{SupplementalMaterial} gives the analogous paths for Laughlin $1/5$ and Moore--Read, using the spin-1 sampling norm in $\hat D(t)$ and, respectively, the three pair channels $01,02,12$ or the three-particle contact channel $012$, as supporting evidence for gap of these states.

\section{Discussion}

Although Roy's suggestion that the quantum geometric properties of a band are key metrics of FCI stability was supported by numerical evidence and has endured through the discovery of frustration-free continuum models for FCIs \cite{wang2021exact,wang2023origin,ledwith2023vortexability,estienne2023ideal,guerci2025layer}, there are simple reasons to believe that these descriptors are not fundamental to the lattice. For one, the quantum metric and Berry curvature are not uniquely defined on the lattice, since they require a choice of ``embedding" which specifies where the locations of the orbitals are within the unit cell \cite{PhysRevB.102.165148,2014arXiv1401.0529H}. The many-body Hamiltonian is independent of this choice, but the integrated metric and the very notion of ideality are dependent. Although there are privileged choices of embedding, for instance via symmetry or minimization \cite{herzogarbeitman2026many,PhysRevB.106.014518,wang2026symmetryquantumgeometrybloch}, fundamentally the relevance of quantum geometry to FCIs must depend on the interaction, whose local real-space structure provides the only relevant notion of embedding. We can therefore understand the ideal band condition as arising from the $V_1$ Haldane pseudo-potential condition in the continuum. 

From the scalar-spinor duality, we can already see that ideality is not crucial for FCIs in the continuum. For instance, the simplest $n=1$ Dirac LL has Berry curvature $f = 1/2\pi$ and quantum metric $g = 2/2\pi$, and therefore is far from ideal. However, it hosts the same Laughlin FCIs as the ideal $n=0$ LL. This is consistent with their differing parent Hamiltonians. Our construction of exact FCIs in non-ideal ands on the lattice reaffirms this picture. We should point out that while the Dirac spinor bands are higher-vortexable \cite{Fujimoto_2025,spasicmlacak2026designinghamiltoniansexactlyflat}, the lattice bands we construct are not. 

Having established that neither uniformity nor ideality are essential for the existence of an FCI \emph{ground state}, we should turn our attention to the FCI gap. It may be possible to generalize the techniques developed recently to prove the Laughlin gap \cite{openai2026fock,scheer2026hierarchy} to our system, and therefore to understand the behavior of both the gap and the quantum geometry as a function of $\lambda$. There is some progress in this direction in the continuum \cite{liu2026evaluate}. 

A second possible use of our results is an investigation of the composite fermion theory \cite{Halperin_2020,PhysRevLett.63.199} on the lattice, where  Hofstadter calculations \cite{Sheffer_2021,Kol2024} can be used to propose model wavefunctions for other ground states and for anyonic excitations \cite{scl58pv6,gattu2026searchfreenonabeliananyons,xu2025dynamicsanyonclustersfractional,Goncalves_2026,yan2026anyon,li2026boundstatesanyonsgeometric}. Our models are also amenable to other powerful numerical techniques for composite fermions\cite{hu2026hyperdetwavefunctionphaseagnosticansatz,shrz-b692,PhysRevB.109.245125}. Finally, we hope that these models may help guide the realization of FCIs in materials, where Hubbard interactions are typically dominant and strong.

\section{Acknowledgment.}
We thank Jiabin Yu for discussing early versions of a manuscript\cite{Zhao2026AnalyticalFCI}, which constructs an analytical tight-binding parent Hamiltonian for the Laughlin state in ideal lattice bands.

J.H.-A. gratefully acknowledges discussions with Eslam Khalaf, Michael Scheer, Zhaoyu Han, Ady Stern, Ahmed Abouelkomsan, and Patrick Ledwith, as well as earlier collaborations with Andrei Bernevig and Nicolas Regnault and support from the MIT Pappalardo Fellowship. C.F. thanks  Kin Fai Mak, Jie Wang, and Hoi Chun Po for helpful discussion and questions.
C.F. and R.H. acknowledge support from the National Natural Science Foundation of China (NSFC) under grant number 12325404, 12547112, 12188101, National Key R\&D Program of China under grant number 2022YFA1403800, 2023YFA1406704 and the 
the Strategic Priority Research Program (B) of the Chinese Academy of Sciences (CAS) (Grant No. XDB1720000) as well as the New Cornerstone Foundation by Tencent.

We used ChatGPT Astra 6.0, Sol 6.1 and Claude Opus 5.5, Fable 5 to search the literature, explore proof strategies, write code, and edit the manuscript.

\bibliography{main_references}

\clearpage
\onecolumngrid
\appendix
\renewcommand{\thesection}{\Alph{section}}
\renewcommand{\thesubsection}{\thesection.\arabic{subsection}}
\numberwithin{equation}{section}
\renewcommand{\theequation}{\thesection.\arabic{equation}}
\renewcommand{\thefigure}{S\arabic{figure}}
\renewcommand{\thetable}{S\arabic{table}}
\setcounter{figure}{0}
\setcounter{table}{0}
\setcounter{equation}{0}
\renewcommand{\d}{\text{d}}
\newcommand{\ii}{\mathrm i}
\newcommand{\one}{\mathbf 1}

\section{Scalar Landau levels and contact parents}
\label{cont:sec:scalar-parents}

In this appendix, we review scalar Landau levels in continuous space and derive contact-interaction parent Hamiltonians for the Laughlin, Moore--Read, and Halperin states.

\subsection{Scalar Landau levels}
\label{cont:sec:lll}

In this subsection, we establish our conventions for scalar Landau levels. We consider fermions of mass $m_{\rm p}$ and charge $q=-e$ ($e>0$) on the infinite plane, with position $\mathbf r=(x,y)$. For a uniform field $\mathbf B=-B\hat{\mathbf z}$, $B>0$, we choose the symmetric gauge $\mathbf A=(By/2,-Bx/2)$. The kinetic Hamiltonian and magnetic scales are
\begin{equation}
H_{\rm kin}=\frac{(-\ii\hbar\nabla+e\mathbf A)^2}{2m_{\rm p}},\qquad \ell_B=\sqrt{\frac{\hbar}{eB}},\qquad \omega_c=\frac{eB}{m_{\rm p}}.
\label{cont:eq:landau-setup}
\end{equation}
Hereafter we set $\hbar=e=1$ and write $\boldsymbol\pi=-\ii\nabla+\mathbf A$. We define $z=x+\ii y$, $\partial_z=(\partial_x-\ii\partial_y)/2$, and $\partial_{\bar z}=(\partial_x+\ii\partial_y)/2$. For convenience, we subtract $\omega_c/2$ to set the lowest Landau level (LLL) energy to zero:
\begin{equation}
a=\sqrt2\ell_B\left(\partial_{\bar z}+\frac{z}{4\ell_B^2}\right),\qquad H=H_{\rm kin}-\omega_c/2=\omega_c a^\dagger a.
\label{cont:eq:lll-parent-single}
\end{equation}
In any gauge with the same uniform field, the kinetic momenta obey $[\pi_x,\pi_y]=\ii B\one$. We write the ladder operators as
\begin{equation}
a=\frac{\ii\ell_B}{\sqrt2}(\pi_x+\ii\pi_y),\qquad
a^\dagger=-\frac{\ii\ell_B}{\sqrt2}(\pi_x-\ii\pi_y),\qquad [a,a^\dagger]=\one.
\label{sample:eq:ladder}
\end{equation}
A single-particle wavefunction $\phi$ in the LLL solves $a\phi=0$, giving
\begin{equation}
\phi(\mathbf r)=f(z)e^{-|z|^2/(4\ell_B^2)},\qquad \partial_{\bar z}f=0.
\label{cont:eq:lll-wavefunction}
\end{equation}
The holomorphic factor labels the LLL states. For a given normalized LLL wavefunction $\phi$, we define the corresponding scalar LL orbitals by
\begin{equation}
\widetilde\phi_n=\frac{(a^\dagger)^n}{\sqrt{n!}}\phi,\qquad n\ge0.
\label{cont:eq:scalar-ll-partners}
\end{equation}
With $P$ the LLL projector, commutation with $z$ gives
\begin{equation}
[a,z]=0,\qquad (\one-P)zP=0.
\label{cont:eq:lll-closure}
\end{equation}
Thus multiplication by a holomorphic polynomial preserves the LLL.

\subsection{The Laughlin parents}
\label{cont:sec:scalar-third}
\label{cont:sec:scalar-fifth}

In this subsection, we derive the scalar Laughlin parent Hamiltonians, beginning with the $1/3$ state. For $N_e$ particles labeled $i=0,\ldots,N_e-1$, the ordinary contact interaction is $g_0^{(2)}\sum_{i<j}\delta(\mathbf r_i-\mathbf r_j), g_0^{(2)}>0$. Here $\delta$ denotes the two-dimensional Dirac delta. Fermionic antisymmetry makes the wavefunction vanish at pair coincidence, so ordinary contact repulsion vanishes for scalar fermions. To penalize pairs with relative angular momentum one in the scalar LLL, we use the Trugman--Kivelson derivative contact interaction~\cite{TrugmanKivelson1985},
\begin{equation}
V_{\rm TK}=g_1^{(2)}\sum_{i<j}\nabla_{\mathbf s}^2\delta(\mathbf s),\qquad \mathbf s=\mathbf r_i-\mathbf r_j,\qquad g_1^{(2)}>0.
\label{cont:eq:tk-interaction}
\end{equation}
To project $V_{\rm TK}$ onto the LLL, we first resolve a general radial pair interaction $V(\mathbf r_0-\mathbf r_1)$ into relative-angular-momentum channels. We define $\mathbf r_{\rm c}=(\mathbf r_0+\mathbf r_1)/2$, $\mathbf s=\mathbf r_0-\mathbf r_1$, and $s=s_x+\ii s_y$. We denote pair wavefunctions by $\chi$. The two-particle LLL states separate into center-of-mass and relative parts:
\begin{equation}
\chi_{M,m}^{\rm LLL}\propto (r_{{\rm c},x}+\ii r_{{\rm c},y})^{M}s^m
e^{-|\mathbf r_{\rm c}|^2/(2\ell_B^2)-|s|^2/(8\ell_B^2)},\qquad M=0,1,\ldots,\quad m=1,3,\ldots{}.
\label{cont:eq:two-body-lll}
\end{equation}
Here $M$ and $m$ label the center-of-mass and relative polynomial degrees. Fermionic antisymmetry requires odd $m$. A radial potential $V(\mathbf s)=V(|\mathbf s|)$ is diagonal in $m$ and independent of the center-of-mass angular momentum $M$. With normalized relative wavefunctions
\begin{equation}
\chi_m(\mathbf s)=\frac{s^m e^{-|s|^2/(8\ell_B^2)}}{\sqrt{\pi(4\ell_B^2)^{m+1}m!}},\qquad m=1,3,\ldots{},
\label{cont:eq:relative-orbitals}
\end{equation}
the two-body energies are
\begin{equation}
v_m=\int\!\d^2s\,V(\mathbf s)|\chi_m(\mathbf s)|^2.
\label{cont:eq:relative-energy}
\end{equation}
Within the LLL, let $\Pi_{ij}^{(m)}$ project the pair $ij$ onto relative angular momentum $m$. With $\mathcal P=\prod_iP_i$ the many-body LLL projector,
\begin{equation}
\mathcal P\left[\sum_{i<j}V(\mathbf r_i-\mathbf r_j)\right]\mathcal P
=\sum_{i<j}\sum_{m\ {\rm odd}}v_m\Pi_{ij}^{(m)}.
\label{cont:eq:lll-pair-decomposition}
\end{equation}
The pair-channel energies $v_m$ are Haldane's pseudopotential coefficients~\cite{Haldane1983}. The ordinary contact interaction gives $v_m=g_0^{(2)}|\chi_m(0)|^2=0$ for every odd $m$, consistent with the coincidence argument above.\\

For $V_{\rm TK}$, integration by parts in Eq.~\eqref{cont:eq:relative-energy} gives
\begin{equation}
v_m=g_1^{(2)}\left.\nabla_{\mathbf s}^2|\chi_m(\mathbf s)|^2\right|_{\mathbf s=0}
=\frac{g_1^{(2)}}{4\pi\ell_B^4}\delta_{m,1}.
\label{cont:eq:tk-pseudopotentials}
\end{equation}
The projected interaction is the $m=1$ Haldane pseudopotential:
\begin{equation}
\mathcal P V_{\rm TK}\mathcal P=\frac{g_1^{(2)}}{4\pi\ell_B^4}\sum_{i<j}\Pi_{ij}^{(1)}.
\label{cont:eq:haldane-v1}
\end{equation}
Excluding $m=1$ forces at least cubic pair zeros. To construct a state with these zeros, we start from the unnormalized Slater determinant $\Phi_1$ of the first $N_e$ LLL orbitals, with one zero per pair,
\begin{equation}
\Phi_1=\prod_{i<j}(z_i-z_j)\exp\left[-\sum_i\frac{|z_i|^2}{4\ell_B^2}\right],
\label{cont:eq:slater}
\end{equation}
and we attach two more zeros per pair to obtain the fermionic Laughlin $1/3$ state~\cite{Laughlin1983}:
\begin{equation}
\Phi_{1/3}=\prod_{i<j}(z_i-z_j)^2\Phi_1
=\prod_{i<j}(z_i-z_j)^3\exp\left[-\sum_i\frac{|z_i|^2}{4\ell_B^2}\right].
\label{cont:eq:laughlin}
\end{equation}
The Jastrow factor preserves the LLL by Eq.~\eqref{cont:eq:lll-closure}. To express the zero-mode condition locally, let $\Phi$ be a scalar fermionic wavefunction smooth near pair coincidence. With pair center $\mathbf r_{\rm c}=(\mathbf r_i+\mathbf r_j)/2$, integration by parts gives
\begin{equation}
\langle\Phi|V_{\rm TK}|\Phi\rangle
=2g_1^{(2)}\sum_{i<j}\int\!\d^2r_{\rm c}\prod_{k\ne i,j}\d^2r_k
\left|\left.\nabla_{\mathbf s}\Phi\right|_{\mathbf s=0}\right|^2.
\label{cont:eq:contact-form}
\end{equation}
Antisymmetry sets $\Phi|_{\mathbf s=0}=0$, so this nonnegative form holds without LLL projection. Within the LLL, its zero modes have at least cubic pair zeros. For $\Phi=\Phi_{1/3}$ in Eq.~\eqref{cont:eq:laughlin}, the cubic pair zeros make it vanish. Together with the LLL condition, this gives a zero-energy ground state of the many-body parent:
\begin{equation}
H_{\mathrm L}=\omega_c\sum_{i=0}^{N_e-1}a_i^\dagger a_i+V_{\rm TK},\qquad H_{\mathrm L}\Phi_{1/3}=0.
\label{cont:eq:many-body-parent}
\end{equation}

The generalization from the Laughlin $1/3$ state to $1/m$ states with odd $m$ is straightforward. Here we briefly discuss the $1/5$ case, whose parent is obtained by adding a sixth-order derivative contact term to the TK interaction:
\begin{equation}
V_{13}^{(2)}=\sum_{i<j}\left[g_1^{(2)}\nabla_{\mathbf s}^2+g_3^{(2)}(\nabla_{\mathbf s}^2)^3\right]\delta(\mathbf s),\qquad \mathbf s=\mathbf r_i-\mathbf r_j.
\label{cont:eq:fifth-derivative-contact}
\end{equation}
Using the relative orbitals of Eq.~\eqref{cont:eq:relative-orbitals} and integrating by parts, we obtain
\begin{equation}
v_m=\left.\left[g_1^{(2)}\nabla_{\mathbf s}^2+g_3^{(2)}(\nabla_{\mathbf s}^2)^3\right]|\chi_m(\mathbf s)|^2\right|_{\mathbf s=0},\qquad m=1,3,\ldots.
\label{cont:eq:fifth-channel-integral}
\end{equation}
The second-order term was evaluated in Eq.~\eqref{cont:eq:tk-pseudopotentials}. For the sixth-order term, we expand the relative Gaussian:
\begin{equation}
|\chi_m(\mathbf s)|^2=\frac1{\pi(4\ell_B^2)^{m+1}m!}\sum_{n=0}^\infty\frac{(-1)^n|s|^{2(m+n)}}{n!(4\ell_B^2)^n},\qquad \nabla_{\mathbf s}^2|s|^{2p}=4p^2|s|^{2p-2}\quad(p\ge1).
\label{cont:eq:fifth-radial-expansion}
\end{equation}
Three Laplacians evaluated at the origin select only $m+n=3$, with derivative factor $4^3(3!)^2$. For odd $m$, this leaves $m=1,3$:
\begin{equation}
\left.(\nabla_{\mathbf s}^2)^3|\chi_m(\mathbf s)|^2\right|_{\mathbf s=0}=\frac{9\delta_{m,1}+3\delta_{m,3}}{2\pi\ell_B^8}.
\label{cont:eq:sixth-contact-channels}
\end{equation}
The projected interaction is
\begin{equation}
\mathcal P V_{13}^{(2)}\mathcal P=\sum_{i<j}\left[v_1\Pi_{ij}^{(1)}+v_3\Pi_{ij}^{(3)}\right],\qquad v_1=\frac{g_1^{(2)}}{4\pi\ell_B^4}+\frac{9g_3^{(2)}}{2\pi\ell_B^8},\qquad v_3=\frac{3g_3^{(2)}}{2\pi\ell_B^8}.
\label{cont:eq:fifth-scalar-parent}
\end{equation}
All higher odd pseudopotential coefficients vanish. We require $v_1,v_3>0$; $g_1^{(2)}$ need not itself be positive. Excluding $m=1,3$ forces at least fifth-order pair zeros. As in Eq.~\eqref{cont:eq:laughlin}, attaching four additional zeros per pair to $\Phi_1$ gives the unnormalized fermionic Laughlin $1/5$ state~\cite{Laughlin1983}:
\begin{equation}
\Phi_{1/5}=\prod_{i<j}(z_i-z_j)^4\Phi_1=\prod_{i<j}(z_i-z_j)^5\exp\left[-\sum_i\frac{|z_i|^2}{4\ell_B^2}\right].
\label{cont:eq:laughlin-fifth}
\end{equation}
The additional fourth-power Jastrow factor preserves fermionic antisymmetry and, being holomorphic, keeps the state in the LLL by Eq.~\eqref{cont:eq:lll-closure}. To obtain the local zero-mode condition as in Eq.~\eqref{cont:eq:contact-form}, let $\Phi$ be a scalar fermionic LLL wavefunction. The normalized relative orbitals in Eq.~\eqref{cont:eq:relative-orbitals} give, with $\partial_s=(\partial_{s_x}-\ii\partial_{s_y})/2$,
\begin{equation}
\langle\Phi|V_{13}^{(2)}|\Phi\rangle=\sum_{i<j}\int\!\d^2r_{\rm c}\prod_{k\ne i,j}\d^2r_k\left[16\pi\ell_B^4v_1\left|\left.\partial_s\Phi\right|_{s=0}\right|^2+\frac{128\pi\ell_B^8}{3}v_3\left|\left.\partial_s^3\Phi\right|_{s=0}\right|^2\right].
\label{cont:eq:fifth-contact-form}
\end{equation}
The first and third derivatives extract the $m=1,3$ relative coefficients. Both vanish for $\Phi_{1/5}$ because of its fifth-order pair zeros, making it a zero-energy ground state of the LLL-projected interaction.\\

The constructions above apply to the plane. On a torus, the wavefunction must also satisfy magnetic boundary conditions. For the Laughlin $1/3$ state, we take a rectangular torus with $N_\phi=L_xL_y/(2\pi\ell_B^2)=3N_e$ and $\tau=\ii L_y/L_x$. In the symmetric gauge, the boundary conditions for each particle are
\begin{equation}
\Phi(x_i+L_x,y_i)=e^{\ii\theta_x+\ii L_xy_i/(2\ell_B^2)}\Phi(x_i,y_i),\qquad \Phi(x_i,y_i+L_y)=e^{\ii\theta_y-\ii L_yx_i/(2\ell_B^2)}\Phi(x_i,y_i).
\label{cont:eq:torus-boundary}
\end{equation}
Here we have included boundary twists $\theta_x$ and $\theta_y$ in addition to the gauge-dependent magnetic phases. The quasiperiodicity of Jacobi theta functions allows us to satisfy these boundary conditions while retaining the cubic pair zeros; their definitions and shift identities are given in Sec.~\ref{special:sec:theta}. For fixed twists, the unnormalized torus Laughlin wavefunctions are~\cite{HaldaneRezayi1985,GreiterSchnellsThomale2016}
\begin{equation}
\Phi_{1/3,\gamma}=e^{-\sum_i|z_i|^2/(4\ell_B^2)}e^{\sum_i z_i^2/(4\ell_B^2)}\vartheta\!\begin{bmatrix}\gamma/3+(N_e-1)/2+\theta_x/(6\pi)\\(N_e-1)/2-\theta_y/(2\pi)\end{bmatrix}\!\left(\frac{3}{L_x}\sum_i z_i\Big|3\tau\right)\prod_{i<j}\vartheta_1\!\left(\frac{\pi(z_i-z_j)}{L_x}\Big|\tau\right)^3.
\label{cont:eq:torus-laughlin}
\end{equation}
Here $\gamma=0,1,2$ labels three independent center-of-mass theta functions, whose argument depends on $\sum_i z_i$. The states are holomorphic after removing the Gaussian. The center-of-mass factor supplies the twists and compensates the quasiperiodic phases of the pair factors, so that the full wavefunctions satisfy Eq.~\eqref{cont:eq:torus-boundary}. With our convention $z=x+\ii y$, the theta shift $z_i\to z_i+\ii L_y$ corresponds to $y_i\to y_i+L_y$. These three states realize the well-known threefold topological ground-state degeneracy of the Laughlin $1/3$ state on the torus and are exact zero modes of the periodized $V_{\rm TK}$. At fixed twists and $N_\phi=3N_e$, every LLL zero mode must contain the cubic pair theta product, leaving only the three-dimensional center-of-mass space; hence these states exhaust the zero-mode space~\cite{HaldaneRezayi1985,GreiterSchnellsThomale2016,LeeThomaleQi2013}.\\

On a torus with $N_\phi=5N_e$, a construction analogous to that in Eq.~\eqref{cont:eq:torus-laughlin} gives five Laughlin $1/5$ states~\cite{HaldaneRezayi1985,GreiterSchnellsThomale2016}. These states realize the well-known fivefold topological ground-state degeneracy and are exact zero modes of the periodized $V_{13}^{(2)}$. Every LLL zero mode must contain the fifth-order pair theta product, leaving only the five-dimensional center-of-mass space; hence these states exhaust the zero-mode space~\cite{HaldaneRezayi1985,GreiterSchnellsThomale2016,LeeThomaleQi2013}.

\subsection{The Moore--Read parent}
\label{cont:sec:scalar-mr}

In this subsection, we derive the scalar Moore--Read parent from a derivative contact interaction that penalizes the lowest allowed three-fermion relative channel:
\begin{equation}
V_3^{(3)}=g_3^{(3)}\sum_{i<j<k}(\nabla_i-\nabla_j)^2(\nabla_j-\nabla_k)^2(\nabla_k-\nabla_i)^2\bigl[\delta(\mathbf r_i-\mathbf r_j)\delta(\mathbf r_j-\mathbf r_k)\bigr],\qquad g_3^{(3)}>0.
\label{cont:eq:mr-derivative-contact}
\end{equation}
The derivatives act on the product of delta functions. Every fermionic LLL polynomial contains the cubic Vandermonde factor for each triplet. Integration by parts supplies three holomorphic derivatives on the wavefunction and three antiholomorphic derivatives on its conjugate. At triple coincidence, these must remove the cubic zero; any additional relative powers leave a residual zero. After LLL projection, the interaction therefore selects only the unique lowest triplet channel, with total relative angular momentum $M=3$~\cite{SimonRezayiCooper2007}. We denote its projector by $\Pi_{ijk}^{(3)}$, with the identity on the center-of-mass and spectator degrees of freedom.\\

To compute its coefficient, we define the relative derivative
\begin{equation}
\mathcal D_{ijk}=(\partial_{z_i}-\partial_{z_j})(\partial_{z_j}-\partial_{z_k})(\partial_{z_k}-\partial_{z_i}),
\label{cont:eq:mr-relative-derivative}
\end{equation}
and choose the normalized Slater determinant of the scalar LLL orbitals with angular momenta $0,1,2$:
\begin{equation}
\chi^{\rm LLL}=\frac1{\sqrt{3!}}\det_{\substack{i=0,1,2\\j=0,1,2}}\left[\frac{z_j^i e^{-|z_j|^2/(4\ell_B^2)}}{\sqrt{2\pi\ell_B^2(2\ell_B^2)^i i!}}\right].
\label{cont:eq:triplet-slater}
\end{equation}
This state has $M=3$ and the lowest center-of-mass orbital. Each relative Laplacian contributes a factor of four, and the polynomial derivative is
\begin{equation}
\mathcal D_{012}\prod_{0\le i<j\le2}(z_j-z_i)=12.
\label{cont:eq:mr-vandermonde-derivative}
\end{equation}
The Slater normalization and the coincidence Gaussian then give
\begin{equation}
\langle\chi^{\rm LLL}|V_3^{(3)}|\chi^{\rm LLL}\rangle=\frac{4^3\,12^2g_3^{(3)}}{768\pi^3\ell_B^{12}}\int\!\d^2r\,e^{-3|z|^2/(2\ell_B^2)}=\frac{8g_3^{(3)}}{\pi^2\ell_B^{10}}.
\label{cont:eq:mr-scalar-coefficient}
\end{equation}
Magnetic translation invariance makes this coefficient independent of the center-of-mass orbital. The LLL projection gives the fermionic Moore--Read (MR) parent, denoted by $H_{\mathrm{MR}}$ in the main text within the scalar LLL:
\begin{equation}
\mathcal P V_3^{(3)}\mathcal P=v_3^{(3)}\sum_{i<j<k}\Pi_{ijk}^{(3)},\qquad v_3^{(3)}=\frac{8g_3^{(3)}}{\pi^2\ell_B^{10}}.
\label{cont:eq:mr-scalar-parent}
\end{equation}

For even $N_e$, its densest zero mode is the MR state at filling $1/2$~\cite{SimonRezayiCooper2007,MooreRead1991}, whose unnormalized wavefunction is
\begin{equation}
\Phi_{\rm MR}=\operatorname{Pf}\!\left(\frac1{z_i-z_j}\right)\prod_{i<j}(z_i-z_j)\Phi_1=\operatorname{Pf}\!\left(\frac1{z_i-z_j}\right)\prod_{i<j}(z_i-z_j)^2e^{-\sum_i|z_i|^2/(4\ell_B^2)}.
\label{cont:eq:mr-wavefunction}
\end{equation}
The Jastrow factors cancel the Pfaffian denominators, leaving an antisymmetric holomorphic polynomial times the Gaussian, hence a scalar LLL state. At triple coincidence, the three squared pair factors supply six zeros, and each Pfaffian term removes at most one of these six zeros, leaving at least a fifth-order zero. For any scalar fermionic LLL wavefunction $\Phi$, integration by parts gives the local form of the contact energy:
\begin{equation}
\langle\Phi|V_3^{(3)}|\Phi\rangle=64g_3^{(3)}\sum_{i<j<k}\int\!\d^2r\prod_{l\ne i,j,k}\d^2r_l\left|\left.\mathcal D_{ijk}\Phi\right|_{\mathbf r_i=\mathbf r_j=\mathbf r_k=\mathbf r}\right|^2.
\label{cont:eq:mr-scalar-contact-form}
\end{equation}
The fifth-order zeros of $\Phi_{\rm MR}$ survive these three derivatives, so it is a zero-energy ground state of the LLL-projected interaction.

\subsection{The Halperin \texorpdfstring{$33h$}{33h} parents}
\label{cont:sec:scalar-halperin}

In this subsection, we extend the Laughlin pair-zero construction to two layers and derive the scalar Halperin $33h$ parents for $h=0,1,2$~\cite{Halperin1983}. We consider two layers $A,B$ in the same magnetic field, with fixed particle numbers $N_A,N_B$ and no tunneling. We write $N_e=N_A+N_B$ and denote the layer of particle $i$ by $l_i\in\{A,B\}$. For $h=0$, we apply the TK interaction separately to each layer:
\begin{equation}
V_{330}=W_0\sum_{\substack{i<j\\l_i=l_j}}\nabla_{\mathbf s}^2\delta(\mathbf r_i-\mathbf r_j),\qquad W_0>0.
\label{cont:eq:halperin-scalar-330}
\end{equation}
The sum counts each same-layer pair once. Equation~\eqref{cont:eq:haldane-v1} gives its projection onto the two scalar LLLs:
\begin{equation}
\mathcal P V_{330}\mathcal P=\frac{W_0}{4\pi\ell_B^4}\sum_{\substack{i<j\\l_i=l_j}}\Pi_{ij}^{(1)}.
\label{cont:eq:halperin-projected-330}
\end{equation}
Excluding $m=1$ forces cubic intralayer zeros, so the product of a Laughlin $1/3$ state in each layer is a zero mode.\\

For $h=1$, we add an interlayer contact interaction:
\begin{equation}
V_{331}=V_{330}+W_1\sum_{\substack{i:l_i=A\\j:l_j=B}}\delta(\mathbf r_i-\mathbf r_j),\qquad W_1>0.
\label{cont:eq:halperin-scalar-331}
\end{equation}
The sum contains all $N_A N_B$ interlayer pairs, each counted once. For an interlayer pair with fixed layer labels, antisymmetry imposes no parity restriction on the relative angular momentum. The pair basis and relative orbitals in Eqs.~\eqref{cont:eq:two-body-lll} and~\eqref{cont:eq:relative-orbitals} therefore extend to all $m\ge0$. Equation~\eqref{cont:eq:lll-pair-decomposition} then applies with the sum over these channels. The contact matrix element is
\begin{equation}
\bra{\chi_m}\delta(\mathbf s)\ket{\chi_m}=|\chi_m(0)|^2=\frac{\delta_{m,0}}{4\pi\ell_B^2}.
\label{cont:eq:halperin-contact-coefficient}
\end{equation}
The projected interaction is
\begin{equation}
\mathcal P V_{331}\mathcal P=\mathcal P V_{330}\mathcal P+\frac{W_1}{4\pi\ell_B^2}\sum_{\substack{i:l_i=A\\j:l_j=B}}\Pi_{ij}^{(0)}.
\label{cont:eq:halperin-projected-331}
\end{equation}
The added positive term excludes interlayer $m=0$, forcing a linear interlayer zero.\\

For $h=2$, we also exclude interlayer $m=1$ by adding
\begin{equation}
V_{332}=V_{331}+W_2\sum_{\substack{i:l_i=A\\j:l_j=B}}(\nabla_{\mathbf s}^2+\ell_B^{-2})\delta(\mathbf r_i-\mathbf r_j),\qquad W_2>0.
\label{cont:eq:halperin-scalar-332}
\end{equation}
For all $m\ge0$, integration by parts gives
\begin{equation}
\bra{\chi_m}\nabla_{\mathbf s}^2\delta(\mathbf s)\ket{\chi_m}=\left.\nabla_{\mathbf s}^2|\chi_m(\mathbf s)|^2\right|_{\mathbf s=0}=\frac{\delta_{m,1}-\delta_{m,0}}{4\pi\ell_B^4}.
\label{cont:eq:halperin-derivative-coefficient}
\end{equation}
By Eq.~\eqref{cont:eq:halperin-contact-coefficient}, the added $\ell_B^{-2}\delta(\mathbf s)$ cancels the negative $m=0$ contribution of the derivative term. The projected interaction is
\begin{equation}
\mathcal P V_{332}\mathcal P=\mathcal P V_{331}\mathcal P+\frac{W_2}{4\pi\ell_B^4}\sum_{\substack{i:l_i=A\\j:l_j=B}}\Pi_{ij}^{(1)}.
\label{cont:eq:halperin-projected-332}
\end{equation}
Excluding interlayer $m=0,1$ forces a quadratic interlayer zero. For fixed layer labels, the corresponding unnormalized Halperin $33h$ wavefunctions are
\begin{equation}
\Phi_{33h}=\prod_{\substack{i<j\\l_i=l_j}}(z_i-z_j)^3\prod_{\substack{i:l_i=A\\j:l_j=B}}(z_i-z_j)^h e^{-\sum_i|z_i|^2/(4\ell_B^2)},\qquad h=0,1,2.
\label{cont:eq:halperin-wavefunctions}
\end{equation}
Antisymmetrizing the spatial component together with the layer kets gives the full fermionic state. Each state lies in the scalar LLLs and has no component in the penalized channels, making it a zero-energy ground state of $\mathcal P V_{33h}\mathcal P$.\\

\section{Dirac Landau levels and contact parents}
\label{cont:sec:dirac}

In this appendix, we study continuum Dirac LLs and their contact interactions in a uniform magnetic field, using the notation and conventions of Sec.~\ref{cont:sec:lll}. Their internal components contain different scalar LLs. Fermions in distinct components can coincide, allowing ordinary contact repulsion to generate nonzero pseudopotentials. We retain one positive-energy Dirac LL per active layer, freeze other flavors, and neglect LL mixing. With velocity $v_F$, Pauli matrices $\sigma_\mu$, and standard spin-1 matrices $S_\mu$, the isotropic massless Hamiltonians are~\cite{ApalkovChakraborty2006,UrbanEtAl2011}
\begin{equation}
H_{\rm D}^{(1/2)}=v_F(\sigma_x\pi_x+\sigma_y\pi_y),\qquad
H_{\rm D}^{(1)}=v_F(S_x\pi_x+S_y\pi_y).
\label{cont:eq:dirac-continuum}
\end{equation}
Here spin specifies the internal Dirac representation; the particles are fermions in both cases. The annihilator in Eq.~\eqref{cont:eq:lll-parent-single} gives
\begin{equation}
\pi_x+\ii\pi_y=-\frac{\ii\sqrt2}{\ell_B}a,\qquad
\pi_x-\ii\pi_y=\frac{\ii\sqrt2}{\ell_B}a^\dagger.
\label{cont:eq:dirac-momenta}
\end{equation}
\subsection{Two-body contact parent of the Laughlin \texorpdfstring{$1/3$}{1/3} state}
\label{cont:sec:dirac-half}

In this subsection, we derive the Laughlin $1/3$ contact parent in the spin-$1/2$ Dirac first LL. We make a unitary change of internal basis to bring Eq.~\eqref{cont:eq:dirac-continuum} to
\begin{equation}
H_{\rm D}^{(1/2)}
=v_F\begin{bmatrix}0&\ii\pi_x-\pi_y\\-\ii\pi_x-\pi_y&0\end{bmatrix}
=\frac{\sqrt2\,v_F}{\ell_B}\begin{bmatrix}0&a\\a^\dagger&0\end{bmatrix}.
\label{cont:eq:dirac-half}
\end{equation}
For a normalized LLL orbital $\phi$ and its scalar 1LL partner $a^\dagger\phi$, the block on $[\phi,0]^{\mathsf T}$ and $[0,a^\dagger\phi]^{\mathsf T}$ is $(\sqrt2\,v_F/\ell_B)\sigma_x$. Its positive-energy eigenstate defines the map from the scalar LLL to this Dirac level~\cite{ApalkovChakraborty2006}:
\begin{equation}
\mathcal U_{1/2}\phi=\frac1{\sqrt2}\begin{bmatrix}\phi\\a^\dagger\phi\end{bmatrix},
\qquad
\mathcal U_{1/2}^\dagger\mathcal U_{1/2}=\one.
\label{cont:eq:half-map}
\end{equation}
The map is unitary from the scalar LLL onto the selected Dirac LL. Applying it to each particle gives $\mathcal U_{1/2}^{\otimes N_e}$, which preserves inner products and fermionic antisymmetry. For arbitrary scalar LLL states $\Phi,\Phi'$, the interaction matrix element between their Dirac images is $\langle\Phi|(\mathcal U_{1/2}^\dagger)^{\otimes N_e}V\mathcal U_{1/2}^{\otimes N_e}|\Phi'\rangle$. Matching means choosing the couplings so that this equals the scalar parent's matrix element for every $\Phi,\Phi'$. The two projected interactions then have identical spectra and corresponding zero-mode spaces.\\

To transport the holomorphic factors used in Appendix~\ref{cont:sec:scalar-parents}, we use the commutator $[a^\dagger,z]=-\sqrt2\ell_B$, which gives
\begin{equation}
\mathcal U_{1/2}(z\phi)=(z\one_2+J_{\rm D}^{(1/2)})\mathcal U_{1/2}\phi,\qquad J_{\rm D}^{(1/2)}=-\sqrt2\ell_B|1\rangle\langle0|.
\label{cont:eq:half-matrix-coordinate}
\end{equation}
Unlike the scalar LLL in Eq.~\eqref{cont:eq:lll-closure}, this Dirac LL is not preserved by multiplication by $z$ alone. The matrix coordinate $z\one_2+J_{\rm D}^{(1/2)}$ restores closure, with $(J_{\rm D}^{(1/2)})^2=0$. More generally, for a holomorphic factor $f$, $\mathcal U_{1/2}[f(z)\phi]=f(z\one_2+J_{\rm D}^{(1/2)})\mathcal U_{1/2}\phi$; the matrix function has a finite Taylor expansion in $J_{\rm D}^{(1/2)}$. On the torus, the full holomorphic factor must have the prescribed quasiperiods in both the scalar and Dirac constructions.\\

For Dirac particles, we write the component-independent contact as $V_0^{(2)}=\mathcal G_0^{(2)}\sum_{i<j}\delta(\mathbf r_i-\mathbf r_j)$, acting as the identity on the internal components, with $\mathcal G_0^{(2)}>0$. For a normalized scalar LLL pair $\chi^{\rm LLL}$ and its image $\chi^{\rm D}=\mathcal U_{1/2}^{\otimes2}\chi^{\rm LLL}$, the contact matrix element is
\begin{equation}
\langle\chi^{\rm D}|\mathcal G_0^{(2)}\delta^{(2)}(\mathbf r_0-\mathbf r_1)|\chi^{\rm D}\rangle=\mathcal G_0^{(2)}\int\!\d^2r_0\,\d^2r_1\,\delta^{(2)}(\mathbf r_0-\mathbf r_1)\sum_{\alpha,\beta=0}^1|\chi^{\rm D}_{\alpha\beta}(\mathbf r_0,\mathbf r_1)|^2
=\mathcal G_0^{(2)}\int\!\d^2r\sum_{\alpha,\beta=0}^1
|\chi^{\rm D}_{\alpha\beta}(\mathbf r,\mathbf r)|^2.
\label{cont:eq:pair-contact-matrix}
\end{equation}
Here $\alpha,\beta$ label the internal components of particles 0 and 1. To identify the relative guiding-center channels contributing to Eq.~\eqref{cont:eq:pair-contact-matrix}, we expand the Dirac pair using Eq.~\eqref{cont:eq:half-map}, with $a_i^\dagger=\sqrt2\ell_B[-\partial_{z_i}+\bar z_i/(4\ell_B^2)]$ acting on particle $i$:
\begin{equation}
[\chi^{\rm D}_{00},\chi^{\rm D}_{01},\chi^{\rm D}_{10},\chi^{\rm D}_{11}]
=\frac12[\chi^{\rm LLL},a_1^\dagger\chi^{\rm LLL},a_0^\dagger\chi^{\rm LLL},a_0^\dagger a_1^\dagger\chi^{\rm LLL}].
\label{cont:eq:half-contact-components}
\end{equation}
An ordered component $pq$ is proportional to $(a_0^\dagger)^p(a_1^\dagger)^q\chi^{\rm LLL}$, also for the higher-spin maps below. At coincidence, antisymmetry leaves only internal states $(|p\rangle_0|q\rangle_1-|q\rangle_0|p\rangle_1)/\sqrt2$, with $p<q$. We call this the $pq$ channel, with amplitude $(\chi^{\rm D}_{pq}-\chi^{\rm D}_{qp})/\sqrt2$.\\

For spin $1/2$, only the $01$ channel survives: $\chi^{\rm D}_{00}=\chi^{\rm D}_{11}=0$ and $\chi^{\rm D}_{10}=-\chi^{\rm D}_{01}$ at coincidence. The two ordered-component probabilities in Eq.~\eqref{cont:eq:pair-contact-matrix} give
\begin{equation}
2\mathcal G_0^{(2)}\int\!\d^2r\,|\chi^{\rm D}_{01}(\mathbf r,\mathbf r)|^2
=\frac{\mathcal G_0^{(2)}}2\int\!\d^2r\,
\left|\left.a_1^\dagger\chi^{\rm LLL}\right|_{\mathbf r_0=\mathbf r_1=\mathbf r}\right|^2.
\label{cont:eq:half-contact-energy}
\end{equation}
The surviving antisymmetric $01$ channel involves $a_1^\dagger-a_0^\dagger$. We therefore separate the cyclotron operators into center-of-mass and relative parts. With $s=z_0-z_1$ as in Sec.~\ref{cont:sec:scalar-third}, define
\begin{equation}
a_{\rm c}^\dagger=\frac{a_0^\dagger+a_1^\dagger}{\sqrt2},\qquad
a_{\rm r}^\dagger=\frac{a_0^\dagger-a_1^\dagger}{\sqrt2}
=2\ell_B\left(-\partial_s+\frac{\bar s}{8\ell_B^2}\right).
\label{cont:eq:pair-cyclotron-ladders}
\end{equation}
Applying $(a_{\rm r}^\dagger)^n$ to the normalized relative LLL orbital $\chi_m(s)$ in Eq.~\eqref{cont:eq:relative-orbitals} and evaluating at $s=0$ gives
\begin{equation}
\left.(a_{\rm r}^\dagger)^n\chi_m(s)\right|_{s=0}
=\frac{(-1)^m\sqrt{m!}}{2\sqrt\pi\ell_B}\delta_{n,m},\qquad n\ge0.
\label{cont:eq:contact-channel-matching}
\end{equation}
Thus $(a_{\rm r}^\dagger)^n$ selects $m=n$ at contact, independently of center-of-mass excitations. The $01$ channel contains one relative raising operator, so only $m=1$ contributes.\\

We next determine the coefficient of this $m=1$ channel by evaluating the same contact matrix element in momentum space~\cite{BurrelloEtAl2015}. We use the Fourier representation
\begin{equation}
\mathcal G_0^{(2)}\delta(\mathbf r_0-\mathbf r_1)
=\mathcal G_0^{(2)}\int\!\frac{\d^2q}{(2\pi)^2}e^{\ii\mathbf q\cdot\mathbf r_0}e^{-\ii\mathbf q\cdot\mathbf r_1}.
\label{cont:eq:pair-fourier}
\end{equation}
Since the map in Eq.~\eqref{cont:eq:half-contact-components} acts on each particle separately, we first transform one density factor. Arbitrary scalar LLL bra and ket orbitals $\phi,\phi'$ give:
\begin{equation}
\langle\mathcal U_{1/2}\phi|e^{\ii\mathbf q\cdot\mathbf r}|\mathcal U_{1/2}\phi'\rangle
=\frac12\left[\langle\phi|e^{\ii\mathbf q\cdot\mathbf r}|\phi'\rangle
+\langle a^\dagger\phi|e^{\ii\mathbf q\cdot\mathbf r}|a^\dagger\phi'\rangle\right].
\label{cont:eq:half-density-matrix}
\end{equation}
To express these matrix elements in the LLL, we set $t=\ell_B^2|\mathbf q|^2$ and use the normalized scalar LL orbitals $\widetilde\phi_n$ in Eq.~\eqref{cont:eq:scalar-ll-partners}, with $\widetilde\phi'_n$ defined likewise from $\phi'$. The relevant commutators are
\begin{equation}
[a,e^{\ii\mathbf q\cdot\mathbf r}]
=\frac{\ii\ell_B}{\sqrt2}(q_x+\ii q_y)e^{\ii\mathbf q\cdot\mathbf r},\qquad
[a^\dagger,e^{\ii\mathbf q\cdot\mathbf r}]
=-\frac{\ii\ell_B}{\sqrt2}(q_x-\ii q_y)e^{\ii\mathbf q\cdot\mathbf r}.
\label{cont:eq:density-ladder-commutators}
\end{equation}
Moving the raising and lowering operators through the plane wave and using $a\phi=a\phi'=0$ gives
\begin{equation}
\langle\widetilde\phi_n|e^{\ii\mathbf q\cdot\mathbf r}|\widetilde\phi'_n\rangle
=L_n(t/2)\langle\phi|e^{\ii\mathbf q\cdot\mathbf r}|\phi'\rangle.
\label{cont:eq:ll-density-matrix-element}
\end{equation}
Here $L_n$ is the ordinary Laguerre polynomial defined in Sec.~\ref{special:sec:laguerre}. Applying Eq.~\eqref{cont:eq:ll-density-matrix-element} to the $n=0,1$ components gives the Dirac density form factor~\cite{ApalkovChakraborty2006}:
\begin{equation}
\langle\mathcal U_{1/2}\phi|e^{\ii\mathbf q\cdot\mathbf r}|\mathcal U_{1/2}\phi'\rangle
=F_{1/2}(t)\langle\phi|e^{\ii\mathbf q\cdot\mathbf r}|\phi'\rangle,\qquad
F_{1/2}(t)=\frac{L_0(t/2)+L_1(t/2)}2=1-\frac t4.
\label{cont:eq:half-form-factor}
\end{equation}
Applying Eq.~\eqref{cont:eq:half-form-factor} separately to the two particles gives
\begin{equation}
\langle\chi^{\rm D}|\mathcal G_0^{(2)}\delta(\mathbf r_0-\mathbf r_1)|\chi^{\rm D}\rangle
=\mathcal G_0^{(2)}\int\!\frac{\d^2q}{(2\pi)^2}F_{1/2}(t)^2
\langle\chi^{\rm LLL}|e^{\ii\mathbf q\cdot(\mathbf r_0-\mathbf r_1)}|\chi^{\rm LLL}\rangle.
\label{cont:eq:projected-pair-matrix}
\end{equation}
The remaining scalar LLL matrix element in Eq.~\eqref{cont:eq:projected-pair-matrix} determines the channel energies. Since $F_{1/2}(t)^2$ is radial, the interaction is diagonal in the relative channels of Eq.~\eqref{cont:eq:lll-pair-decomposition}. Write their energies as $v_m^{(1/2)}$, with the superscript labeling the Dirac representation.\\

For a normalized pair in channel $m$, the plane wave depends only on $\mathbf s=\mathbf r_0-\mathbf r_1$. Integrating out the normalized center-of-mass part gives
\begin{equation}
\begin{aligned}
\langle\chi^{\rm LLL}|e^{\ii\mathbf q\cdot(\mathbf r_0-\mathbf r_1)}|\chi^{\rm LLL}\rangle
&=\int\!\d^2s\,|\chi_m(\mathbf s)|^2e^{\ii\mathbf q\cdot\mathbf s}
=\int\!\frac{\d^2s}{4\pi\ell_B^2}\frac{|\mathbf s|^{2m}}{(4\ell_B^2)^m m!}e^{-|\mathbf s|^2/(4\ell_B^2)+\ii\mathbf q\cdot\mathbf s}\\
&=\frac{(-\nabla_{\mathbf q}^2)^m e^{-t}}{(4\ell_B^2)^m m!}
=\frac1{m!}\frac{\d^m}{\d t^m}(t^m e^{-t})=e^{-t}L_m(t).
\end{aligned}
\label{cont:eq:relative-density}
\end{equation}
The intermediate equalities use the Gaussian Fourier transform and $\nabla_{\mathbf q}^2=4\ell_B^2\partial_t(t\partial_t)$; the last equality uses Rodrigues' formula, Eq.~\eqref{special:eq:laguerre-definition}. Substitution into Eq.~\eqref{cont:eq:projected-pair-matrix} and angular integration, with radial measure $\d t/(4\pi\ell_B^2)$, give~\cite{ApalkovChakraborty2006}
\begin{equation}
v_m^{(1/2)}=\frac{\mathcal G_0^{(2)}}{4\pi\ell_B^2}\int_0^\infty\!\d t\,
e^{-t}F_{1/2}(t)^2L_m(t),\qquad
F_{1/2}(t)^2=\frac58L_0(t)+\frac14L_1(t)+\frac18L_2(t).
\label{cont:eq:dirac-pseudopotentials}
\end{equation}
Laguerre orthogonality leaves only $m=0,1,2$ in Eq.~\eqref{cont:eq:dirac-pseudopotentials}. Since the LL map preserves fermionic antisymmetry, only $m=1$ contributes:
\begin{equation}
v_1^{(1/2)}=\frac{\mathcal G_0^{(2)}}{16\pi\ell_B^2}\overset{B=2\pi}{=}\frac{\mathcal G_0^{(2)}}8,\qquad v_m^{(1/2)}=0\quad(m\ge3\text{ odd}).
\label{cont:eq:half-contact-values}
\end{equation}
The pair decomposition in Eq.~\eqref{cont:eq:lll-pair-decomposition} then gives
\begin{equation}
(\mathcal U_{1/2}^\dagger)^{\otimes N_e} V_0^{(2)}\mathcal U_{1/2}^{\otimes N_e}
=\frac{\mathcal G_0^{(2)}}{16\pi\ell_B^2}\sum_{i<j}\Pi_{ij}^{(1)}.
\label{cont:eq:half-parent}
\end{equation}
The projected interaction has the same matrix elements as the scalar $V_{\rm TK}$ parent in Eq.~\eqref{cont:eq:haldane-v1} when
\begin{equation}
\mathcal G_0^{(2)}=\frac{4g_1^{(2)}}{\ell_B^2}\overset{B=2\pi}{=}8\pi g_1^{(2)}.
\label{cont:eq:half-coupling-match}
\end{equation}
Thus the Laughlin state $\Phi_{1/3}$ in Eq.~\eqref{cont:eq:laughlin} maps to the exact Dirac contact zero mode $\mathcal U_{1/2}^{\otimes N_e}\Phi_{1/3}$: every internal component vanishes whenever two particles coincide. Contact zero modes of mapped Laughlin states in spin-orbit-coupled LLs were established through coincidence conditions~\cite{BurrelloTrombettoni2011} and a $V_{\rm TK}$ pseudopotential calculation~\cite{BurrelloEtAl2015}.\\

The same map applies to the torus states $\Phi_{1/3,\gamma}$ in Eq.~\eqref{cont:eq:torus-laughlin}:
\begin{equation}
\Phi^{\rm D}_{1/3,\gamma}=\mathcal U_{1/2}^{\otimes N_e}\Phi_{1/3,\gamma},\qquad \gamma=0,1,2.
\label{cont:eq:torus-dirac-laughlin}
\end{equation}
\begin{samepage}%
Since $a^\dagger$ commutes with magnetic translations, the map preserves the twists $\theta_x,\theta_y$. At most two derivatives act on a pair, so its cubic zero cannot be removed. All three images are therefore contact zero modes. At fixed twists, the map unitarily relates the periodized, LL-projected parents. The scalar completeness result following Eq.~\eqref{cont:eq:torus-laughlin} therefore makes these the only zero modes at $N_\phi=3N_e$.
\par\end{samepage}%

\subsection{Two-body contact parent of the Laughlin \texorpdfstring{$1/5$}{1/5} state}
\label{cont:sec:dirac-one}

In this subsection, we derive the Laughlin $1/5$ contact parent in the spin-1 Dirac second LL. The spin matrices $S_x,S_y$ couple adjacent internal components with matrix elements of magnitude $1/\sqrt2$. A unitary change of internal basis and Eq.~\eqref{cont:eq:dirac-momenta} give~\cite{UrbanEtAl2011}
\begin{equation}
H_{\rm D}^{(1)}=\frac{v_F}{\ell_B}
\begin{bmatrix}0&a&0\\a^\dagger&0&a\\0&a^\dagger&0\end{bmatrix}.
\label{cont:eq:dirac-one}
\end{equation}
Let $c_n$ be the amplitude of $\widetilde\phi_n$ in component $n=0,1,2$. The relations $a\widetilde\phi_1=\widetilde\phi_0$ and $a\widetilde\phi_2=\sqrt2\widetilde\phi_1$ give the eigenvalue equation
\begin{equation}
\frac{v_F}{\ell_B}
\begin{bmatrix}0&1&0\\1&0&\sqrt2\\0&\sqrt2&0\end{bmatrix}
\begin{bmatrix}c_0\\c_1\\c_2\end{bmatrix}
=E\begin{bmatrix}c_0\\c_1\\c_2\end{bmatrix}.
\label{cont:eq:one-block}
\end{equation}
Its positive-energy eigenstate, with $E=\sqrt3\,v_F/\ell_B$, defines the map from the scalar LLL to this Dirac level:
\begin{equation}
\mathcal U_1\phi=\frac1{\sqrt6}
\begin{bmatrix}\phi\\\sqrt3\,a^\dagger\phi\\(a^\dagger)^2\phi\end{bmatrix},
\qquad \mathcal U_1^\dagger\mathcal U_1=\one.
\label{cont:eq:one-map}
\end{equation}
This isometry maps the scalar LLL onto the $n=2$ Dirac level of Ref.~\cite{UrbanEtAl2011}. It preserves guiding-center labels because $a^\dagger$ acts only on the cyclotron coordinate. For $N_e$ particles, the map is $\mathcal U_1^{\otimes N_e}$.\\

The spin-1 analogue of Eq.~\eqref{cont:eq:half-matrix-coordinate} is
\begin{equation}
\mathcal U_1(z\phi)=(z\one_3+J_{\rm D}^{(1)})\mathcal U_1\phi,\qquad J_{\rm D}^{(1)}=-\sqrt2\ell_B\left(\sqrt3\,|1\rangle\langle0|+\frac2{\sqrt3}|2\rangle\langle1|\right).
\label{cont:eq:one-matrix-coordinate}
\end{equation}
Here $(J_{\rm D}^{(1)})^3=0$, and the same extension to holomorphic factors applies.\\

For a normalized scalar LLL pair $\chi^{\rm LLL}$, the Dirac 2LL image $\chi^{\rm D}=\mathcal U_1^{\otimes2}\chi^{\rm LLL}$ has contact energy
\begin{equation}
\langle\chi^{\rm D}|\mathcal G_0^{(2)}\delta(\mathbf r_0-\mathbf r_1)|\chi^{\rm D}\rangle=\mathcal G_0^{(2)}\int\!\d^2r\sum_{\alpha,\beta=0}^2|\chi^{\rm D}_{\alpha\beta}(\mathbf r,\mathbf r)|^2.
\label{cont:eq:one-contact-matrix}
\end{equation}
\begin{samepage}%
The antisymmetric internal channels are now $01$, $02$, and $12$. Their cyclotron operators acting on $\chi^{\rm LLL}$ are, up to overall component and normalization factors,
\begin{equation}
01:\ a_1^\dagger-a_0^\dagger=-\sqrt2a_{\rm r}^\dagger,\quad
02:\ (a_1^\dagger)^2-(a_0^\dagger)^2=-2a_{\rm c}^\dagger a_{\rm r}^\dagger,\quad
12:\ a_0^\dagger a_1^\dagger(a_1^\dagger-a_0^\dagger)
=\frac{(a_{\rm r}^\dagger)^3-(a_{\rm c}^\dagger)^2a_{\rm r}^\dagger}{\sqrt2}.
\label{cont:eq:one-contact-channels}
\end{equation}
\end{samepage}%
By Eq.~\eqref{cont:eq:contact-channel-matching}, $01$ and $02$ contribute only to $m=1$, whereas $12$ contributes to both $m=1$ and $m=3$.\\

To compute the coefficients, replace $\mathcal U_{1/2}$ by $\mathcal U_1$ in the density-matrix calculation of Eqs.~\eqref{cont:eq:half-density-matrix}--\eqref{cont:eq:dirac-pseudopotentials}:
\begin{equation}
\langle\mathcal U_1\phi|e^{\ii\mathbf q\cdot\mathbf r}|\mathcal U_1\phi'\rangle=F_1(t)\langle\phi|e^{\ii\mathbf q\cdot\mathbf r}|\phi'\rangle,\qquad t=\ell_B^2|\mathbf q|^2.
\label{cont:eq:one-density-matrix}
\end{equation}
The normalized scalar LL components have weights $1/6,1/2,1/3$, so Eq.~\eqref{cont:eq:ll-density-matrix-element} gives
\begin{equation}
F_1(t)=\frac16L_0(t/2)+\frac12L_1(t/2)+\frac13L_2(t/2)
=1-\frac{7t}{12}+\frac{t^2}{24}.
\label{cont:eq:one-form-factor}
\end{equation}
The channel energies $v_m^{(1)}$ follow from Eq.~\eqref{cont:eq:dirac-pseudopotentials} with $F_{1/2}$ replaced by $F_1$. Expanding the square gives
\begin{equation}
F_1(t)^2=\frac{31}{72}L_0(t)+\frac{13}{72}L_1(t)+\frac29L_2(t)+\frac18L_3(t)+\frac1{24}L_4(t).
\label{cont:eq:one-form-factor-square}
\end{equation}
Only $L_1$ and $L_3$ have odd indices, so Laguerre orthogonality gives
\begin{equation}
v_1^{(1)}=\frac{13\mathcal G_0^{(2)}}{288\pi\ell_B^2}\overset{B=2\pi}{=}\frac{13\mathcal G_0^{(2)}}{144},\qquad
v_3^{(1)}=\frac{\mathcal G_0^{(2)}}{32\pi\ell_B^2}\overset{B=2\pi}{=}\frac{\mathcal G_0^{(2)}}{16}.
\label{cont:eq:one-contact-values}
\end{equation}
All higher odd channels vanish, so
\begin{equation}
(\mathcal U_1^\dagger)^{\otimes N_e} V_0^{(2)}\mathcal U_1^{\otimes N_e}
=\frac{\mathcal G_0^{(2)}}{288\pi\ell_B^2}
\sum_{i<j}\left[13\Pi_{ij}^{(1)}+9\Pi_{ij}^{(3)}\right].
\label{cont:eq:one-pair-parent}
\end{equation}
The single Dirac contact strength fixes $v_1:v_3=13:9$. Its projected matrix elements equal those of the scalar $V_{13}^{(2)}$ parent in Eq.~\eqref{cont:eq:fifth-scalar-parent} only when both channel coefficients match, requiring
\begin{equation}
\mathcal G_0^{(2)}=-\frac{36g_1^{(2)}}{7\ell_B^2}=\frac{48g_3^{(2)}}{\ell_B^6}\overset{B=2\pi}{=}-\frac{72\pi g_1^{(2)}}7=384\pi^3g_3^{(2)}.
\label{cont:eq:one-coupling-match}
\end{equation}
Thus $\mathcal U_1^{\otimes N_e}\Phi_{1/5}$, with $\Phi_{1/5}$ from Eq.~\eqref{cont:eq:laughlin-fifth}, is a Dirac contact zero mode. Its fifth-order pair zeros survive the map's at most four derivatives per pair. On a torus with $N_\phi=5N_e$, the map preserves twists and unitarily relates the periodized parents, as in Sec.~\ref{cont:sec:dirac-half}. The scalar completeness result in Sec.~\ref{cont:sec:scalar-fifth} therefore gives exactly five Dirac zero modes.\\

The same construction applies to spin $s=1/2,1,\ldots$. In the internal basis $|n\rangle$, $n=0,\ldots,2s$, the Dirac block built from $\widetilde\phi_n$ is proportional to a real symmetric tridiagonal matrix $K_s$ with zero diagonal. Its highest-energy eigenvector fixes the normalized coefficients $c_n$:
\begin{equation}
(K_s)_{n,n+1}=(K_s)_{n+1,n}=(n+1)\sqrt{2s-n},\qquad 0\le n<2s,\qquad \sum_{n=0}^{2s}c_n^2=1.
\label{cont:eq:general-spin-block}
\end{equation}
\begin{samepage}%
All adjacent couplings are positive, so the highest eigenvector can be chosen with every $c_n>0$. The corresponding isometry is
\begin{equation}
\mathcal U_s\phi=\sum_{n=0}^{2s}c_n|n\rangle\frac{(a^\dagger)^n}{\sqrt{n!}}\phi,\qquad \mathcal U_s^\dagger\mathcal U_s=\one.
\label{cont:eq:general-spin-map}
\end{equation}
\end{samepage}%
Using $[(a^\dagger)^n,z]=-\sqrt2\ell_B n(a^\dagger)^{n-1}$ gives
\begin{equation}
J_{\rm D}^{(s)}=-\sqrt2\ell_B\sum_{n=1}^{2s}\sqrt n\,\frac{c_n}{c_{n-1}}|n\rangle\langle n-1|,\qquad
(J_{\rm D}^{(s)})^{2s+1}=0,\qquad (J_{\rm D}^{(s)})^{2s}\ne0.
\label{cont:eq:general-spin-coordinate}
\end{equation}
For $s=1/2$ and $1$, the coefficient vectors are $(1,1)/\sqrt2$ and $(1,\sqrt3,\sqrt2)/\sqrt6$, respectively, reproducing Eqs.~\eqref{cont:eq:half-matrix-coordinate} and \eqref{cont:eq:one-matrix-coordinate}. For a holomorphic factor $f$,
\begin{equation}
\mathcal U_s[f(z)\phi]=f(z\one_{2s+1}+J_{\rm D}^{(s)})\mathcal U_s\phi,\qquad
f(z\one_{2s+1}+J_{\rm D}^{(s)})=\sum_{r=0}^{2s}\frac{f^{(r)}(z)}{r!}(J_{\rm D}^{(s)})^r.
\label{cont:eq:general-spin-holomorphic}
\end{equation}
Thus the Laughlin state $\Phi_{1/(4s+1)}=\prod_{i<j}(z_i-z_j)^{4s}\Phi_1$ maps to an exact two-body-contact zero mode: its pair zeros of order $4s+1$ survive the map's at most $4s$ derivatives per pair.

\subsection{Three-body contact parent of the Moore--Read state}
\label{cont:sec:dirac-mr}

In this subsection, we retain the spin-1 Dirac second LL and the map $\mathcal U_1$ of Sec.~\ref{cont:sec:dirac-one}, and replace the two-body contact repulsion by
\begin{equation}
V_0^{(3)}=\mathcal G_0^{(3)}\sum_{i<j<k}
\delta(\mathbf r_i-\mathbf r_j)\delta(\mathbf r_i-\mathbf r_k),
\qquad \mathcal G_0^{(3)}>0.
\label{cont:eq:triple-contact}
\end{equation}
The contact acts as the identity on the internal indices $\alpha,\beta,\gamma=0,1,2$. For a normalized scalar LLL triplet $\chi^{\rm LLL}$, its image $\chi^{\rm D}=\mathcal U_1^{\otimes3}\chi^{\rm LLL}$ under Eq.~\eqref{cont:eq:one-map} has contact energy
\begin{equation}
\langle\chi^{\rm D}|V_0^{(3)}|\chi^{\rm D}\rangle
=\mathcal G_0^{(3)}\int\!\d^2r\sum_{\alpha,\beta,\gamma=0}^2
|\chi^{\rm D}_{\alpha\beta\gamma}(\mathbf r,\mathbf r,\mathbf r)|^2.
\label{cont:eq:triple-contact-form}
\end{equation}
Antisymmetry eliminates repeated internal indices at coincidence and relates the six permutations of $012$ by a sign. Thus the component sum is $6|\chi^{\rm D}_{012}|^2$, and Eq.~\eqref{cont:eq:one-map} gives
\begin{equation}
\chi^{\rm D}_{012}(\mathbf r,\mathbf r,\mathbf r)
=\left.\frac{a_1^\dagger(a_2^\dagger)^2}{6\sqrt2}
\chi^{\rm LLL}(\mathbf r_0,\mathbf r_1,\mathbf r_2)\right|_{\mathbf r_0=\mathbf r_1=\mathbf r_2=\mathbf r}.
\label{cont:eq:triplet-component}
\end{equation}
Every fermionic LLL triplet contains a cubic Vandermonde factor. The three derivatives in Eq.~\eqref{cont:eq:triplet-component} can remove this zero only in the unique $M=3$ relative channel; higher relative degrees vanish at coincidence, as in Sec.~\ref{cont:sec:scalar-mr}.\\

\begin{samepage}%
To determine the coefficient, choose the normalized $M=3$ Slater state in Eq.~\eqref{cont:eq:triplet-slater}. At coincidence, only $(-\sqrt2\ell_B)^3\partial_{z_1}\partial_{z_2}^2$ acting on its polynomial survives:
\begin{equation}
\partial_{z_1}\partial_{z_2}^2\prod_{i<j}(z_j-z_i)=2,\qquad
\chi^{\rm D}_{012}
(\mathbf r,\mathbf r,\mathbf r)
=-\frac{e^{-3|z|^2/(4\ell_B^2)}}{6\sqrt6\,(2\pi\ell_B^2)^{3/2}}.
\label{cont:eq:triplet-amplitude}
\end{equation}
\end{samepage}%
Summing the six component probabilities in Eq.~\eqref{cont:eq:triple-contact-form} gives the $M=3$ channel energy
\begin{equation}
\langle\chi^{\rm D}|V_0^{(3)}|\chi^{\rm D}\rangle
=6\mathcal G_0^{(3)}\int\!\d^2r\,
\frac{e^{-3|z|^2/(2\ell_B^2)}}{216(2\pi\ell_B^2)^3}
=\frac{\mathcal G_0^{(3)}}{432\pi^2\ell_B^4}\overset{B=2\pi}{=}\frac{\mathcal G_0^{(3)}}{108}.
\label{cont:eq:triple-coefficient}
\end{equation}
The relative $M=3$ channel is unique, and magnetic translation invariance makes its coefficient independent of the center-of-mass orbital. Summing over triplets gives
\begin{equation}
(\mathcal U_1^\dagger)^{\otimes N_e} V_0^{(3)}\mathcal U_1^{\otimes N_e}
=\frac{\mathcal G_0^{(3)}}{432\pi^2\ell_B^4}\sum_{i<j<k}\Pi_{ijk}^{(3)}.
\label{cont:eq:mr-parent}
\end{equation}
The projected interaction has the same matrix elements as the scalar MR parent in Eq.~\eqref{cont:eq:mr-scalar-parent} when
\begin{equation}
\mathcal G_0^{(3)}=\frac{3456g_3^{(3)}}{\ell_B^6}\overset{B=2\pi}{=}27648\pi^3g_3^{(3)}.
\label{cont:eq:mr-coupling-match}
\end{equation}
For even $N_e$, the scalar MR state at filling $1/2$ in Eq.~\eqref{cont:eq:mr-wavefunction} maps to
\begin{equation}
\Phi_{\rm MR}^{\rm D}=\mathcal U_1^{\otimes N_e}\Phi_{\rm MR}.
\label{cont:eq:mr-dirac-wavefunction}
\end{equation}
The symmetric MR factor supplies two powers at triple coincidence and the Slater factor supplies three. The resulting fifth-order zero survives the three derivatives in Eq.~\eqref{cont:eq:triplet-component}; repeated internal indices already vanish by antisymmetry. Hence $\Phi_{\rm MR}^{\rm D}$ is an exact three-body-contact zero mode.\\

For the $k$-Read--Rezayi state, use $\mathcal U_{k/2}$ and substitute $z\to z\one_{k+1}+J_{\rm D}^{(k/2)}$ in its holomorphic factor, as in Eq.~\eqref{cont:eq:general-spin-holomorphic}. The symmetric RR factor vanishes to at least second order when $k+1$ particles coincide~\cite{ReadRezayi1999}, so the full fermionic wavefunction has clustering order at least $\binom{k+1}{2}+2$. At coincidence, antisymmetry leaves only permutations of the internal indices $0,\ldots,k$, for which the map contains at most $\sum_{n=0}^{k}n=\binom{k+1}{2}$ derivatives on these particles. The contact amplitude therefore vanishes, establishing an exact $(k+1)$-body-contact zero mode.

\subsection{Two-body contact parents of the Halperin \texorpdfstring{$33h$}{33h} states}
\label{cont:sec:dirac-halperin}

In this subsection, we compare contacts in two spin-$1/2$ Dirac first LLs with the scalar Halperin parents of Sec.~\ref{cont:sec:scalar-halperin}. Write the map in Eq.~\eqref{cont:eq:half-map} within layer $a$ as $(\mathcal U_{1/2})_a:\mathrm{LLL}_a\to\mathrm{1LL}^{\rm D}_a$, with $a=A,B$. These maps act separately in the two layers and preserve the layer labels. For $330$, define
\begin{equation}
V_{0,330}^{(2)}=\mathcal G_{0,330}^{(2)}\sum_{\substack{i<j\\l_i=l_j}}\delta(\mathbf r_i-\mathbf r_j),\qquad \mathcal G_{0,330}^{(2)}>0.
\label{cont:eq:halperin-dirac-330}
\end{equation}
Equation~\eqref{cont:eq:half-parent} gives the intralayer $m=1$ coefficient $\mathcal G_{0,330}^{(2)}/(16\pi\ell_B^2)$. These matrix elements equal those of the scalar $330$ parent in Eq.~\eqref{cont:eq:halperin-projected-330} when
\begin{equation}
\mathcal G_{0,330}^{(2)}=\frac{4W_0}{\ell_B^2}\overset{B=2\pi}{=}8\pi W_0.
\label{cont:eq:halperin-330-coupling-match}
\end{equation}

For an interlayer pair with fixed layer labels, every $m\ge0$ is allowed, as in Sec.~\ref{cont:sec:scalar-halperin}. The $331$ parent penalizes $m=0$, while $332$ also penalizes $m=1$. Let $\chi^{\rm LLL}_{AB}(\mathbf r_A,\mathbf r_B)$ be a joint state of one particle in each scalar LLL, with image $\chi^{\rm D}_{AB}=[(\mathcal U_{1/2})_A\otimes(\mathcal U_{1/2})_B]\chi^{\rm LLL}_{AB}$. To select these channels, we separate the mixed components $01,10$ into symmetric and antisymmetric combinations. With $\ket{\alpha\beta}_{ij}=\ket{A,\alpha}_i\ket{B,\beta}_j$ for $l_i=A,l_j=B$ and $\alpha,\beta=0,1$, define
\begin{equation}
p^{00}_{ij}=\ket{00}_{ij}\bra{00}_{ij},\qquad p^\pm_{ij}=\frac12(\ket{01}_{ij}\pm\ket{10}_{ij})(\bra{01}_{ij}\pm\bra{10}_{ij}).
\label{cont:eq:halperin-internal-projectors}
\end{equation}
Here symmetry refers to the internal indices; full particle exchange also swaps the layer labels. With an independent interlayer strength $\mathcal G_{0,331}^{(2)}>0$, define the contact family
\begin{equation}
V_{0,331}^{(2)}(\eta)=V_{0,330}^{(2)}+\mathcal G_{0,331}^{(2)}\sum_{\substack{i:l_i=A\\j:l_j=B}}\delta(\mathbf r_i-\mathbf r_j)\bigl[(1-\eta)p^{00}_{ij}+\eta p^+_{ij}\bigr].
\label{cont:eq:halperin-dirac-331}
\end{equation}
The sum counts every $AB$ pair once, and $0\le\eta\le1$ keeps both contact coefficients nonnegative.\\

For this pair, we write $AB$ in place of $ij$. Set $\mathbf s=\mathbf r_A-\mathbf r_B$ and $\chi^{\rm D}_{AB;\pm}=(\chi^{\rm D}_{AB;01}\pm\chi^{\rm D}_{AB;10})/\sqrt2$. Equations~\eqref{cont:eq:half-contact-components} and~\eqref{cont:eq:pair-cyclotron-ladders}, with particle labels replaced by $A,B$, give
\begin{equation}
\chi^{\rm D}_{AB;00}=\frac12\chi^{\rm LLL}_{AB},\qquad \chi^{\rm D}_{AB;+}=\frac12a_{\rm c}^\dagger\chi^{\rm LLL}_{AB},\qquad \chi^{\rm D}_{AB;-}=-\frac12a_{\rm r}^\dagger\chi^{\rm LLL}_{AB}.
\label{cont:eq:halperin-contact-components}
\end{equation}
The $00,+$ components have relative cyclotron order zero, while $-$ has order one. By Eq.~\eqref{cont:eq:contact-channel-matching}, they select $m=0$ and $m=1$, respectively: the $331$ penalty and the additional channel needed for $332$.\\

For arbitrary bra and ket states and their Dirac images, the $00$ contact matrix element is
\begin{equation}
\bra{\chi^{\rm D}_{AB}}\delta(\mathbf s)p^{00}_{AB}\ket{\chi^{\prime\,\rm D}_{AB}}=\frac14\bra{\chi^{\rm LLL}_{AB}}\delta(\mathbf s)\ket{\chi^{\prime\,\rm LLL}_{AB}}.
\label{cont:eq:halperin-contact-00}
\end{equation}
The symmetric mixed component gives the same result:
\begin{equation}
\bra{\chi^{\rm D}_{AB}}\delta(\mathbf s)p^+_{AB}\ket{\chi^{\prime\,\rm D}_{AB}}=\frac14\bra{\chi^{\rm LLL}_{AB}}a_{\rm c}\delta(\mathbf s)a_{\rm c}^\dagger\ket{\chi^{\prime\,\rm LLL}_{AB}}=\frac14\bra{\chi^{\rm LLL}_{AB}}\delta(\mathbf s)\ket{\chi^{\prime\,\rm LLL}_{AB}}.
\label{cont:eq:halperin-contact-plus}
\end{equation}
The last equality uses $[a_{\rm c},\delta(\mathbf s)]=0$, $[a_{\rm c},a_{\rm c}^\dagger]=\one$, and $a_{\rm c}\chi^{\prime\,\rm LLL}_{AB}=0$. Substituting Eqs.~\eqref{cont:eq:halperin-contact-00} and~\eqref{cont:eq:halperin-contact-plus} into the interlayer term of Eq.~\eqref{cont:eq:halperin-dirac-331} gives
\begin{equation}
\bra{\chi^{\rm D}_{AB}}\mathcal G_{0,331}^{(2)}\delta(\mathbf s)\bigl[(1-\eta)p^{00}_{AB}+\eta p^+_{AB}\bigr]\ket{\chi^{\prime\,\rm D}_{AB}}=\frac{\mathcal G_{0,331}^{(2)}}4[(1-\eta)+\eta]\bra{\chi^{\rm LLL}_{AB}}\delta(\mathbf s)\ket{\chi^{\prime\,\rm LLL}_{AB}}.
\label{cont:eq:halperin-331-matrix}
\end{equation}
These matrix elements equal those of the scalar interlayer contact in Eq.~\eqref{cont:eq:halperin-scalar-331} when
\begin{equation}
\mathcal G_{0,331}^{(2)}=4W_1.
\label{cont:eq:halperin-331-coupling-match}
\end{equation}
Both $00$ and $+$ give one quarter of the scalar contact matrix element for any $B$, so the matching is independent of $\eta$ and $B$. Together with the matched intralayer term, this reproduces the scalar $331$ parent. The choices $\eta=0,1/2,1$ give a pure $00$ contact, equal $00,+$ penalties, and a pure $+$ contact, respectively. The main text uses $\eta=0$ for $331$.\\

For $332$, the quadratic interlayer zero excludes $m=0,1$. Retaining the $331$ penalty on $m=0$, we add the antisymmetric mixed-component contact with independent strength $\mathcal G_{0,332}^{(2)}>0$:
\begin{equation}
V_{0,332}^{(2)}(\eta)=V_{0,331}^{(2)}(\eta)+\mathcal G_{0,332}^{(2)}\sum_{\substack{i:l_i=A\\j:l_j=B}}\delta(\mathbf r_i-\mathbf r_j)p^-_{ij}.
\label{cont:eq:halperin-dirac-332}
\end{equation}
The interlayer $11$ component remains unpenalized. Here $V_{0,331}^{(2)}(\eta)$ denotes the member of the contact family at the same $\eta$ used in $V_{0,332}^{(2)}(\eta)$; this need not be the $\eta=0$ representative chosen separately for the main-text $331$ parent. The added $p^-$ term has pair matrix element
\begin{equation}
\bra{\chi^{\rm D}_{AB}}\mathcal G_{0,332}^{(2)}\delta(\mathbf s)p^-_{AB}\ket{\chi^{\prime\,\rm D}_{AB}}=\frac{\mathcal G_{0,332}^{(2)}}4\bra{\chi^{\rm LLL}_{AB}}a_{\rm r}\delta(\mathbf s)a_{\rm r}^\dagger\ket{\chi^{\prime\,\rm LLL}_{AB}}.
\label{cont:eq:halperin-contact-minus}
\end{equation}
The relative ladder in Eq.~\eqref{cont:eq:pair-cyclotron-ladders} gives
\begin{equation}
[a_{\rm r},\delta(\mathbf s)]=2\ell_B\partial_{\bar s}\delta(\mathbf s),\qquad [[a_{\rm r},\delta(\mathbf s)],a_{\rm r}^\dagger]=\ell_B^2\nabla_{\mathbf s}^2\delta(\mathbf s).
\label{cont:eq:halperin-relative-commutators}
\end{equation}
Commuting the ladders to the bra and ket, where $\bra{\chi^{\rm LLL}_{AB}}a_{\rm r}^\dagger=0$ and $a_{\rm r}\chi^{\prime\,\rm LLL}_{AB}=0$, yields
\begin{equation}
\bra{\chi^{\rm LLL}_{AB}}a_{\rm r}\delta(\mathbf s)a_{\rm r}^\dagger\ket{\chi^{\prime\,\rm LLL}_{AB}}=\bra{\chi^{\rm LLL}_{AB}}\delta(\mathbf s)+[[a_{\rm r},\delta(\mathbf s)],a_{\rm r}^\dagger]\ket{\chi^{\prime\,\rm LLL}_{AB}}=\ell_B^2\bra{\chi^{\rm LLL}_{AB}}(\nabla_{\mathbf s}^2+\ell_B^{-2})\delta(\mathbf s)\ket{\chi^{\prime\,\rm LLL}_{AB}}.
\label{cont:eq:halperin-relative-reduction}
\end{equation}
Substitution into Eq.~\eqref{cont:eq:halperin-contact-minus} gives $(\mathcal G_{0,332}^{(2)}\ell_B^2/4)$ times the scalar matrix element of $(\nabla_{\mathbf s}^2+\ell_B^{-2})\delta(\mathbf s)$. These matrix elements equal those of the added interlayer term in Eq.~\eqref{cont:eq:halperin-scalar-332} when
\begin{equation}
\mathcal G_{0,332}^{(2)}=\frac{4W_2}{\ell_B^2}\overset{B=2\pi}{=}8\pi W_2.
\label{cont:eq:halperin-332-coupling-match}
\end{equation}
Let $v_m^{ab}(33h)$ denote the coefficient of $\Pi_{ij}^{(m)}$ for layers $a,b$. All three parents have the same intralayer coefficients:
\begin{equation}
v_m^{AA}(33h)=v_m^{BB}(33h)=\frac{\mathcal G_{0,330}^{(2)}}{16\pi\ell_B^2}\delta_{m,1}=\frac{W_0}{4\pi\ell_B^4}\delta_{m,1},\qquad m=1,3,\ldots.
\label{cont:eq:halperin-intralayer-pseudopotentials}
\end{equation}
The interlayer coefficients are
\begin{equation}
v_m^{AB}(330)=0,\qquad v_m^{AB}(331)=\frac{\mathcal G_{0,331}^{(2)}}{16\pi\ell_B^2}\delta_{m,0}=\frac{W_1}{4\pi\ell_B^2}\delta_{m,0}.
\label{cont:eq:halperin-331-pseudopotentials}
\end{equation}
\begin{equation}
v_m^{AB}(332)=\frac{\mathcal G_{0,331}^{(2)}\delta_{m,0}+\mathcal G_{0,332}^{(2)}\delta_{m,1}}{16\pi\ell_B^2}=\frac{W_1}{4\pi\ell_B^2}\delta_{m,0}+\frac{W_2}{4\pi\ell_B^4}\delta_{m,1}.
\label{cont:eq:halperin-332-pseudopotentials}
\end{equation}
All other allowed channels have zero coefficient. At one flux quantum per unit square, each of the three nonzero channel coefficients reduces to
\begin{equation}
\frac{\mathcal G_{0,33h}^{(2)}}{16\pi\ell_B^2}\overset{B=2\pi}{=}\frac{\mathcal G_{0,33h}^{(2)}}8,\qquad h=0,1,2.
\label{cont:eq:halperin-channel-values}
\end{equation}

For fixed layer labels, write $\mathcal U_{1/2,AB}=\bigotimes_{i=0}^{N_e-1}(\mathcal U_{1/2})_{l_i}$, with the $i$th factor acting on particle $i$, and set $V_{0,330}^{(2)}(\eta)=V_{0,330}^{(2)}$. With these matching strengths, summing the pair identities and using the preserved spectator inner products gives
\begin{equation}
\mathcal U_{1/2,AB}^\dagger V_{0,33h}^{(2)}(\eta)\mathcal U_{1/2,AB}=\mathcal P V_{33h}\mathcal P,\qquad h=0,1,2.
\label{cont:eq:halperin-parent-equivalence}
\end{equation}
This unitary equivalence gives identical projected spectra and corresponding zero-mode spaces. The scalar states in Eq.~\eqref{cont:eq:halperin-wavefunctions} therefore map to exact contact zero modes
\begin{equation}
\Phi^{\rm D}_{33h}=\mathcal U_{1/2,AB}\Phi_{33h},\qquad h=0,1,2.
\label{cont:eq:halperin-dirac-wavefunctions}
\end{equation}

For $332$, retain the $\eta$ dependence. In the mixed-component basis $(\ket{01}_{ij},\ket{10}_{ij})$, its contact block is
\begin{equation}
\eta\mathcal G_{0,331}^{(2)}p^+_{ij}+\mathcal G_{0,332}^{(2)}p^-_{ij}\ \longleftrightarrow\ \frac12\begin{pmatrix}\eta\mathcal G_{0,331}^{(2)}+\mathcal G_{0,332}^{(2)}&\eta\mathcal G_{0,331}^{(2)}-\mathcal G_{0,332}^{(2)}\\\eta\mathcal G_{0,331}^{(2)}-\mathcal G_{0,332}^{(2)}&\eta\mathcal G_{0,331}^{(2)}+\mathcal G_{0,332}^{(2)}\end{pmatrix}.
\label{cont:eq:halperin-mixed-block}
\end{equation}
The off-diagonal entries exchange internal components. They vanish when
\begin{equation}
\eta\mathcal G_{0,331}^{(2)}=\mathcal G_{0,332}^{(2)},\qquad \eta=\frac{\mathcal G_{0,332}^{(2)}}{\mathcal G_{0,331}^{(2)}}=\frac{W_2}{\ell_B^2W_1}.
\label{cont:eq:halperin-density-condition}
\end{equation}
This choice lies in $0<\eta\le1$ when $0<W_2\le\ell_B^2W_1$; strict inequality makes all three density coefficients positive. With $s_i=0,1$ corresponding to $\uparrow,\downarrow$ in the main text, the identity $p^++p^-=\ket{01}\bra{01}+\ket{10}\bra{10}$ gives
\begin{equation}
V_{0,332}^{(2)}(\eta)=V_{0,330}^{(2)}+\sum_{\substack{i<j\\l_i\ne l_j}}\delta(\mathbf r_i-\mathbf r_j)\bigl[(\mathcal G_{0,331}^{(2)}-\mathcal G_{0,332}^{(2)})\delta_{s_i,0}\delta_{s_j,0}+\mathcal G_{0,332}^{(2)}(1-\delta_{s_i s_j})\bigr].
\label{cont:eq:halperin-expanded-332}
\end{equation}
Equivalently, its $00$ density coefficient is $4(W_1-W_2/\ell_B^2)$ and each mixed density coefficient is $4W_2/\ell_B^2$. The $00$ coefficient therefore differs from that of $V_{0,331}^{(2)}(0)$ at the same $W_1$. Adding only the two mixed density terms to that $331$ representative would also increase the scalar interlayer contact coefficient.\\

To obtain the main-text contacts with a common density strength $\mathcal G>0$, choose
\begin{equation}
\mathcal G_{0,330}^{(2)}=\mathcal G,\qquad (\eta,\mathcal G_{0,331}^{(2)})_{331}=(0,\mathcal G),\qquad (\eta,\mathcal G_{0,331}^{(2)},\mathcal G_{0,332}^{(2)})_{332}=(1/2,2\mathcal G,\mathcal G).
\label{cont:eq:halperin-main-contact-strengths}
\end{equation}
The $331$ interlayer term is then $\mathcal G\delta(\mathbf s)p^{00}$, whereas the $332$ term is $\mathcal G\delta(\mathbf s)(p^{00}+p^++p^-)$. The same projector identity then gives the first-quantized contacts used in the main text:
\begin{equation}
V_{0,330}^{(2)}=\mathcal G\sum_{\substack{i<j\\l_i=l_j}}\delta(\mathbf r_i-\mathbf r_j),
\label{cont:eq:halperin-expanded-330}
\end{equation}
\begin{equation}
V_{0,331}^{(2)}(0)=V_{0,330}^{(2)}+\mathcal G\sum_{\substack{i<j\\l_i\ne l_j}}\delta(\mathbf r_i-\mathbf r_j)\delta_{s_i,0}\delta_{s_j,0},
\label{cont:eq:halperin-expanded-331}
\end{equation}
\begin{equation}
V_{0,332}^{(2)}(1/2)=V_{0,331}^{(2)}(0)+\mathcal G\sum_{\substack{i<j\\l_i\ne l_j}}\delta(\mathbf r_i-\mathbf r_j)(1-\delta_{s_i s_j}).
\label{cont:eq:halperin-density-only}
\end{equation}
Equations~\eqref{cont:eq:halperin-contact-00}, \eqref{cont:eq:halperin-contact-plus}, and~\eqref{cont:eq:halperin-relative-reduction} give the matching scalar strengths. The common intralayer coefficient and the $331$ interlayer coefficient are
\begin{equation}
W_0=\frac{\ell_B^2\mathcal G}{4}\overset{B=2\pi}{=}\frac{\mathcal G}{8\pi},\qquad \left.W_1\right|_{331}=\frac{\mathcal G}{4}.
\label{cont:eq:halperin-main-match-330-331}
\end{equation}
For $332$, both $p^{00}$ and $p^+$ contribute $\mathcal G/4$ to the scalar contact, while $p^-$ supplies the derivative contact, giving
\begin{equation}
\left.W_1\right|_{332}=\frac{\mathcal G}{2},\qquad \left.W_2\right|_{332}=\frac{\ell_B^2\mathcal G}{4}\overset{B=2\pi}{=}\frac{\mathcal G}{8\pi}.
\label{cont:eq:halperin-main-match-332}
\end{equation}
Thus the recursive density form uses different matched scalar $W_1$ values for $331$ and $332$. The general contact family remains valid for independent positive $W_1,W_2$; when $W_2>\ell_B^2W_1$, no $\eta\in[0,1]$ cancels its exchange term.\\

Table~\ref{cont:tab:contact-conversions} summarizes the coupling conversions derived in this appendix.
\begin{table}[H]
\centering
\setlength{\belowcaptionskip}{10pt}
\caption{Coupling conversions at $B=2\pi$. Matched strengths give equal projected interaction matrix elements for states related by the LL map. $\mathcal G$ and $g,W$ denote Dirac and scalar contact strengths; $v_m$ and $v_3^{(3)}$ are pair and triplet pseudopotential energies. Superscript $(n)$ counts interacting particles, $m$ is relative guiding-center angular momentum, and $A,B$ label layers.}
\label{cont:tab:contact-conversions}
\begingroup
\setlength{\tabcolsep}{12pt}
\renewcommand{\arraystretch}{1.9}
\begin{tabular}{@{}ll@{}}
\toprule
State / channel & Matched strengths \\
\midrule
Laughlin $1/3$ & $\mathcal G_0^{(2)}=8\pi g_1^{(2)}=8v_1$ \\
Laughlin $1/5$ & $\mathcal G_0^{(2)}=-\dfrac{72\pi}{7}g_1^{(2)}=384\pi^3g_3^{(2)}=\dfrac{144}{13}v_1=16v_3$ \\
\midrule
Moore--Read & $\mathcal G_0^{(3)}=27648\pi^3g_3^{(3)}=108v_3^{(3)}$ \\
\midrule
Halperin $330$ (intralayer) & $\mathcal G_{0,330}^{(2)}=8\pi W_0=8v_1^{AA}=8v_1^{BB}$ \\
Halperin $331$ (added interlayer $m=0$) & $\mathcal G_{0,331}^{(2)}=4W_1=8v_0^{AB}$ \\
Halperin $332$ (added interlayer $m=1$) & $\mathcal G_{0,332}^{(2)}=8\pi W_2=8v_1^{AB}$ \\
\bottomrule
\end{tabular}
\endgroup
\end{table}

\section{Lattice sampling of Dirac Landau levels}
\label{sec:sampling}

In this appendix, we construct lattice Chern bands by sampling the spin-$1/2$ Dirac 1LL and spin-1 Dirac 2LL of Appendix~\ref{cont:sec:dirac} on the square lattice $\mathbf R=(X,Y)\in\mathbb Z^2$, with all internal components at the same site. At one flux quantum per unit cell, a suitable gauge identifies continuum magnetic Bloch momentum with lattice momentum. Sampling changes the inner product, so we normalize the sampled states before constructing interactions. We construct onsite parents for the sampled Laughlin and Moore--Read states and interpolate their zero modes to those of exponentially decaying interactions with the continuum contact spectra.

\subsection{Magnetic Bloch basis for lattice sampling}
\label{sample:sec:boundary}

In this subsection, we construct a continuum magnetic Bloch basis in a fixed gauge, then show that its lattice samples obey the ordinary Bloch translation law with the same momentum. In a gauge $\mathbf A$, magnetic translation by $\mathbf L=(L_1,L_2)\in\mathbb Z^2$ acts as
\begin{equation}
(M_{\mathbf L}\phi)(\mathbf r)=e^{\ii\theta_{\mathbf L}(\mathbf r)}\phi(\mathbf r-\mathbf L),\qquad
\mathbf A(\mathbf r-\mathbf L)-\nabla\theta_{\mathbf L}(\mathbf r)=\mathbf A(\mathbf r).
\label{sample:eq:magnetic-definition}
\end{equation}
The phase condition ensures that the kinetic momenta $\boldsymbol\pi=-\ii\nabla+\mathbf A$ are invariant:
\begin{equation}
M_{\mathbf L}\boldsymbol\pi M_{\mathbf L}^{-1}
=-\ii\nabla+\mathbf A(\mathbf r-\mathbf L)-\nabla\theta_{\mathbf L}(\mathbf r)
=\boldsymbol\pi.
\label{sample:eq:magnetic-momenta}
\end{equation}
The Landau Hamiltonian is a function of $\boldsymbol\pi$, hence $[M_{\mathbf L},H]=0$: magnetic translations are symmetries and preserve each LL.\\

A single lattice band has one state per unit cell, whereas a selected continuum LL has $B/(2\pi)$ states per unit area. Matching these densities requires $B=2\pi$, or one flux quantum per unit square, with $\ell_B=1/\sqrt{2\pi}$. With our convention $\boldsymbol\pi=-\ii\nabla+\mathbf A$, magnetic translations obey
\begin{equation}
M_{\mathbf L}M_{\mathbf L'}=\underbrace{e^{\ii2\pi(L_1L'_2-L_2L'_1)}}_{=1}\,M_{\mathbf L'}M_{\mathbf L}=M_{\mathbf L'}M_{\mathbf L}.
\label{sample:eq:symmetry-matching}
\end{equation}
Thus the primitive translations $M_x=M_{(1,0)}$ and $M_y=M_{(0,1)}$ commute. Within each LL, the continuum magnetic Bloch states $\phi_{\mathbf k}(\mathbf r)$ are their simultaneous eigenstates:
\begin{equation}
M_x\phi_{\mathbf k}=e^{-\ii k_x}\phi_{\mathbf k},\qquad
M_y\phi_{\mathbf k}=e^{-\ii k_y}\phi_{\mathbf k},\qquad
\mathbf k\in\mathbb R^2/(2\pi\mathbb Z^2).
\label{sample:eq:momentum-definition}
\end{equation}
To match the sampled magnetic Bloch phases to the ordinary lattice Bloch phases, as shown below in Eq.~\eqref{sample:eq:sample-translation}, we change from the symmetric gauge $\mathbf A^{\rm sym}=(\pi y,-\pi x)$ used in Sec.~\ref{cont:sec:lll} to
\begin{equation}
\Lambda_{\rm g}=\pi xy+\pi(x-y),\qquad
\phi=e^{\ii\Lambda_{\rm g}}\phi^{\rm sym},\qquad
\mathbf A=\mathbf A^{\rm sym}-\nabla\Lambda_{\rm g}=(-\pi,-2\pi x+\pi).
\label{sample:eq:gauge}
\end{equation}
This is the standard Landau gauge $(0,-2\pi x)$ plus the pure-gauge shift $-\nabla[\pi(x-y)]=(-\pi,\pi)$. All continuum orbitals and ladder operators below are expressed in this gauge, with magnetic-translation phase
\begin{equation}
\theta_{\mathbf L}(x,y)=2\pi L_1y.
\label{sample:eq:magnetic-translations}
\end{equation}
For integer translations, $M_{\mathbf L}=M_x^{L_1}M_y^{L_2}$. We now construct a full magnetic Bloch basis for the scalar LLL, starting from the orbital centered at the origin and normalized over the full plane:
\begin{equation}
\phi^{\rm LLL}_{\rm loc}(x,y)=e^{\ii\Lambda_{\rm g}(x,y)}e^{-\pi|\mathbf r|^2/2}=e^{\ii\pi xy+\ii\pi(x-y)}e^{-\pi(x^2+y^2)/2}.
\label{sample:eq:localized-orbital}
\end{equation}
Summing magnetic translates of this orbital with coefficients $e^{\ii\mathbf k\cdot\mathbf L}$ gives a state satisfying Eq.~\eqref{sample:eq:momentum-definition}. With $k_\pm=k_x\pm\ii k_y$, it evaluates to
\begin{equation}
\sum_{\mathbf L\in\mathbb Z^2}e^{\ii\mathbf k\cdot\mathbf L}(M_{\mathbf L}\phi^{\rm LLL}_{\rm loc})(\mathbf r)=\sqrt2\,e^{\ii\pi(x-y)+\ii k_y y-\pi x^2+k_yx-k_y^2/(2\pi)}\vartheta_1(k_+/2|\ii)^*\vartheta_1\!\left(\frac{k_+}{2}+\pi y-\ii\pi x\Big|\ii\right).
\label{sample:eq:continuum-seed-sum}
\end{equation}
The spatially independent factor $\vartheta_1(k_+/2|\ii)^*$ makes the sum vanish at $\mathbf k=0$. For $\mathbf k\ne0$, removing this factor preserves the magnetic Bloch state. Continuum normalization then gives
\begin{equation}
\phi^{\rm LLL}_{\mathbf k}(x,y)=2^{1/4}e^{\ii\pi(x-y)+\ii k_y y}e^{-\pi(x-k_y/(2\pi))^2}\vartheta_1\!\left(\frac{k_+}{2}+\pi y-\ii\pi x\Big|\ii\right).
\label{sample:eq:lll}
\end{equation}
It extends continuously to a nonzero state at $\mathbf k=0$, with delta normalization
\begin{equation}
\int_{\mathbb R^2}d^2\mathbf r\,\phi^{\rm LLL}_{\mathbf k}(\mathbf r)^*\phi^{\rm LLL}_{\mathbf k'}(\mathbf r)=(2\pi)^2\delta(\mathbf k-\mathbf k').
\label{sample:eq:lll-inner-product}
\end{equation}
Here momenta lie in one BZ. Combining the magnetic-translation eigenvalues with Eq.~\eqref{sample:eq:magnetic-translations} gives, for $\mathbf r=(x,y)$ and $\mathbf L=(L_1,L_2)\in\mathbb Z^2$,
\begin{equation}
\phi^{\rm LLL}_{\mathbf k}(\mathbf r+\mathbf L)=e^{\ii2\pi L_1y}e^{\ii\mathbf k\cdot\mathbf L}\phi^{\rm LLL}_{\mathbf k}(\mathbf r).
\label{sample:eq:continuum-bloch-law}
\end{equation}
For a continuous wavefunction $\phi$ in a fixed scalar or Dirac Landau level, we define the sampling map $\mathcal S$ by evaluation at lattice sites, componentwise for a spinor:
\begin{equation}
(\mathcal S\phi)(\mathbf R)=\left.\phi(\mathbf r)\right|_{\mathbf r=\mathbf R},\qquad \mathbf R=(X,Y)\in\mathbb Z^2.
\label{sample:eq:sampling-definition}
\end{equation}
In the original symmetric gauge, translation by $(1,0)$ leaves the site-dependent factor $(-1)^Y$, which cannot be absorbed into a fixed Bloch momentum. The gauge in Eq.~\eqref{sample:eq:gauge} instead makes the magnetic phase in Eq.~\eqref{sample:eq:continuum-bloch-law} equal to $e^{\ii2\pi L_1Y}=1$ at every lattice site, giving
\begin{equation}
(\mathcal S\phi^{\rm LLL}_{\mathbf k})(\mathbf R+\mathbf L)=e^{\ii\mathbf k\cdot\mathbf L}(\mathcal S\phi^{\rm LLL}_{\mathbf k})(\mathbf R).
\label{sample:eq:sample-translation}
\end{equation}
This is the usual Bloch condition under ordinary lattice translations $T_{\mathbf L}$: $T_{\mathbf L}\mathcal S\phi^{\rm LLL}_{\mathbf k}=e^{-\ii\mathbf k\cdot\mathbf L}\mathcal S\phi^{\rm LLL}_{\mathbf k}$. The gauge choice thus makes magnetic translations act as ordinary lattice translations on the samples, identifying the continuum magnetic Bloch momentum with the lattice Bloch momentum.

\subsection{Dirac Bloch states and their lattice samples}
\label{sample:sec:dirac-samples}

In this subsection, we apply the Dirac LL maps in Eqs.~\eqref{cont:eq:half-map} and \eqref{cont:eq:one-map} to the scalar magnetic Bloch basis of Eq.~\eqref{sample:eq:lll}, then calculate the samples through Fourier transforms of localized orbitals. For the spin-$1/2$ Dirac 1LL and spin-1 Dirac 2LL, these maps give
\begin{equation}
\phi^{\rm D,1/2}=\mathcal U_{1/2}\phi^{\rm LLL}=\frac1{\sqrt2}
\begin{bmatrix}\phi^{\rm LLL}\\a^\dagger\phi^{\rm LLL}\end{bmatrix},\qquad
\phi^{\rm D,1}=\mathcal U_1\phi^{\rm LLL}=\frac1{\sqrt6}
\begin{bmatrix}\phi^{\rm LLL}\\\sqrt3\,a^\dagger\phi^{\rm LLL}\\(a^\dagger)^2\phi^{\rm LLL}\end{bmatrix}.
\label{sample:eq:dirac}
\end{equation}
These isometries commute with magnetic translations by Eq.~\eqref{sample:eq:magnetic-momenta}, so they give normalized Dirac magnetic Bloch states with the same momentum $\mathbf k$:
\begin{equation}
\phi^{\rm D,s}_{\mathbf k}=\mathcal U_s\phi^{\rm LLL}_{\mathbf k},\qquad \int_{\mathbb R^2}d^2\mathbf r\,\phi^{\rm D,s}_{\mathbf k}(\mathbf r)^\dagger\phi^{\rm D,s}_{\mathbf k'}(\mathbf r)=(2\pi)^2\delta(\mathbf k-\mathbf k').
\label{sample:eq:dirac-bloch}
\end{equation}
The sampling relation in Eq.~\eqref{sample:eq:sample-translation} therefore gives
\begin{equation}
(\mathcal S\phi^{\rm D,s}_{\mathbf k})(\mathbf R)=e^{\ii\mathbf k\cdot\mathbf R}\phi^{\rm D,s}_{\mathbf k}(\mathbf0).
\label{sample:eq:sample-bloch}
\end{equation}
Equation~\eqref{sample:eq:sample-bloch} shows that $\phi^{\rm D,s}_{\mathbf k}(\mathbf0)$ determines the sampled wavefunction at every lattice site. To calculate it, we use $\mathcal U_s\phi^{\rm LLL}_{\rm loc}$, where $\phi^{\rm LLL}_{\rm loc}$ is the normalized LLL orbital centered at the origin in Eq.~\eqref{sample:eq:localized-orbital}. Its Dirac components are coordinate polynomials multiplying this common orbital, so their Fourier transforms follow by differentiating the Fourier transform of its sampled envelope
\begin{equation}
a_{\mathbf R}=\phi^{\rm LLL}_{\rm loc}(X,Y)=e^{\ii\pi XY+\ii\pi(X-Y)}e^{-\pi(X^2+Y^2)/2}=(-1)^{XY+X+Y}e^{-\pi(X^2+Y^2)/2}.
\label{sample:eq:sampled-envelope}
\end{equation}
The Fourier transform of this envelope is
\begin{equation}
\begin{aligned}
F(\mathbf k)=\sum_{\mathbf R}a_{\mathbf R}e^{-\ii\mathbf k\cdot\mathbf R}&=\sum_{X,Y\in\mathbb Z}(-1)^{XY+X+Y}e^{-\pi(X^2+Y^2)/2-\ii k_xX-\ii k_yY}\\
&=\sqrt2\sum_{X,m\in\mathbb Z}(-1)^Xe^{-\pi X^2/2-\ii k_xX}e^{-[k_y-\pi(X+1)-2\pi m]^2/(2\pi)}\\
&=\sqrt2e^{-k_y^2/(2\pi)}\vartheta_1(k_-/2|\ii)\vartheta_1(k_+/2|\ii)=\sqrt2e^{-k_y^2/(2\pi)}|T^{(0)}_{\mathbf k}|^2.
\end{aligned}
\label{sample:eq:scalar-fourier}
\end{equation}
The second line follows by Poisson summation over $Y$; setting $n=m$ and $l=m+X$ then gives the theta series of Eq.~\eqref{bands:eq:theta1-definition}, with $T^{(0)}_{\mathbf k}=\vartheta_1(k_+/2|\ii)$.
Near $\mathbf k=0$, $F(\mathbf k)\propto k_+k_-=k_x^2+k_y^2$, and $F(\mathbf0)=0$. The nonzero continuum state $\phi^{\rm LLL}_{\mathbf k=0}$ in Eq.~\eqref{sample:eq:lll} vanishes at every lattice site, so $\mathcal S\phi^{\rm LLL}_{\mathbf k=0}=0$.\\

The scalar LLL supplies only the first internal component of each Dirac state; we must also evaluate the samples of its higher scalar LL components. Since $(a^\dagger)^j\phi^{\rm LLL}_{\rm loc}=[\sqrt\pi(x-\ii y)]^j\phi^{\rm LLL}_{\rm loc}$ for $j=0,1,2$, the localized lattice orbitals $\psi^{(s)}_0=\mathcal S\mathcal U_s\phi^{\rm LLL}_{\rm loc}$ are
\begin{equation}
\psi^{(1/2)}_0(\mathbf R)=\frac{a_{\mathbf R}}{\sqrt2}
\begin{bmatrix}1\\\sqrt\pi(X-\ii Y)\end{bmatrix},\qquad
\psi^{(1)}_0(\mathbf R)=\frac{a_{\mathbf R}}{\sqrt6}
\begin{bmatrix}1\\\sqrt{3\pi}(X-\ii Y)\\\pi(X-\ii Y)^2\end{bmatrix}.
\label{sample:eq:localized-sample}
\end{equation}
The superscript $s=1/2,1$ labels the Dirac spin, and the subscript $0$ labels the orbital centered at the origin. Since magnetic translations become ordinary translations after sampling, the continuum Bloch sums take the lattice form
\begin{equation}
\sum_{\mathbf L}e^{\ii\mathbf k\cdot\mathbf L}\psi^{(s)}_0(\mathbf R-\mathbf L)=e^{\ii\mathbf k\cdot\mathbf R}\sum_{\mathbf L}e^{-\ii\mathbf k\cdot\mathbf L}\psi^{(s)}_0(\mathbf L).
\label{sample:eq:localized-bloch-sum}
\end{equation}
Commuting $\mathcal U_s$ through the continuum sum in Eq.~\eqref{sample:eq:continuum-seed-sum} and sampling gives
\begin{equation}
\sum_{\mathbf R}e^{-\ii\mathbf k\cdot\mathbf R}\psi^{(s)}_0(\mathbf R)=2^{1/4}e^{-k_y^2/(4\pi)}\vartheta_1(k_+/2|\ii)^*\phi^{\rm D,s}_{\mathbf k}(\mathbf0).
\label{sample:eq:dirac-continuum-sum}
\end{equation}
The polynomial factors in Eq.~\eqref{sample:eq:localized-sample} then give the remaining components by differentiation:
\begin{equation}
\sum_{\mathbf R}(X-\ii Y)^j a_{\mathbf R}e^{-\ii\mathbf k\cdot\mathbf R}=[\ii(\partial_{k_x}-\ii\partial_{k_y})]^jF(\mathbf k),\qquad j=0,1,2.
\label{sample:eq:raised-component}
\end{equation}
We write the component functions as
\begin{equation}
T^{(0)}_{\mathbf k}=\vartheta_1,\qquad T^{(1)}_{\mathbf k}=\frac12\vartheta_1'+\frac{\ii k_y}{2\pi}\vartheta_1,\qquad T^{(2)}_{\mathbf k}=\frac12\vartheta_1''+\frac{\ii k_y}{\pi}\vartheta_1'+\frac{\pi-k_y^2}{2\pi^2}\vartheta_1.
\label{sample:eq:third-theta}
\end{equation}
All theta expressions are evaluated at $(k_+/2|\ii)$ with primes denoting derivatives with respect to the first argument.\\

Applying these derivatives gives the Fourier transforms, with their small-$\mathbf k$ expansions shown up to nonzero constant factors:
\begin{equation}
\sum_{\mathbf R}e^{-\ii\mathbf k\cdot\mathbf R}\psi^{(1/2)}_0(\mathbf R)=e^{-k_y^2/(2\pi)}{T^{(0)}_{\mathbf k}}^*\begin{bmatrix}T^{(0)}_{\mathbf k}\\2\ii\sqrt\pi T^{(1)}_{\mathbf k}\end{bmatrix}\propto\begin{bmatrix}k_+k_-\\2\ii\sqrt\pi\,k_-\end{bmatrix}+O(|\mathbf k|^3).
\label{sample:eq:fourier-half}
\end{equation}
\begin{equation}
\sum_{\mathbf R}e^{-\ii\mathbf k\cdot\mathbf R}\psi^{(1)}_0(\mathbf R)=e^{-k_y^2/(2\pi)}{T^{(0)}_{\mathbf k}}^*\begin{bmatrix}T^{(0)}_{\mathbf k}/\sqrt3\\2\ii\sqrt\pi T^{(1)}_{\mathbf k}\\-2\pi T^{(2)}_{\mathbf k}/\sqrt3\end{bmatrix}\propto\begin{bmatrix}k_+k_-\\2\ii\sqrt{3\pi}\,k_-\\2k_-^2\end{bmatrix}+O(|\mathbf k|^3).
\label{sample:eq:fourier-one}
\end{equation}
Both Fourier vectors vanish at $\mathbf k=0$, but they include the common factor $\vartheta_1(k_+/2|\ii)^*\propto k_-$ in Eq.~\eqref{sample:eq:dirac-continuum-sum}. Dividing out the prefactor for $\mathbf k\ne0$ recovers $\phi^{\rm D,s}_{\mathbf k}(\mathbf0)$: the LLL component still tends to zero as $k_+$, whereas the second, scalar 1LL component tends to a nonzero constant. Taking $\mathbf k\to0$ therefore gives $\mathcal S\phi^{\rm D,s}_{\mathbf k=0}\ne0$ for both $s=1/2$ and $s=1$. The two sampled spinors are explicitly
\begin{equation}
\phi^{\rm D,1/2}_{\mathbf k}(\mathbf0)=2^{-1/4}e^{-k_y^2/(4\pi)}\begin{bmatrix}T^{(0)}_{\mathbf k}\\2\ii\sqrt\pi\,T^{(1)}_{\mathbf k}\end{bmatrix},\qquad
\phi^{\rm D,1}_{\mathbf k}(\mathbf0)=2^{-1/4}e^{-k_y^2/(4\pi)}\begin{bmatrix}T^{(0)}_{\mathbf k}/\sqrt3\\2\ii\sqrt\pi\,T^{(1)}_{\mathbf k}\\-2\pi T^{(2)}_{\mathbf k}/\sqrt3\end{bmatrix}.
\label{sample:eq:normalized-continuum-sample}
\end{equation}
At $\mathbf k=0$, these spinors lie along the second internal direction, giving the projectors
\begin{equation}
P^{(1/2)}(\mathbf0)=\operatorname{diag}[0,1],\qquad P^{(1)}(\mathbf0)=\operatorname{diag}[0,1,0].
\label{sample:eq:origin-projector}
\end{equation}
Away from the origin, $T^{(0)}_{\mathbf k}\ne0$ in the BZ, so the first component is nonzero. The Dirac samples therefore define a nonzero one-dimensional Bloch subspace at every momentum.

\subsection{Normalization and interaction mapping}
\label{sample:sec:norm}

In this subsection, we first treat the spin-$1/2$ Dirac 1LL, then extend the interaction mapping to the spin-1 Dirac 2LL. We denote the unnormalized sampled spinor of the former by $v_{\mathbf k}$. Equations~\eqref{sample:eq:sample-bloch} and \eqref{sample:eq:normalized-continuum-sample} give
\begin{equation}
v_{\mathbf k}=\phi^{\rm D,1/2}_{\mathbf k}(\mathbf0),\qquad
(\mathcal S\phi^{\rm D,1/2}_{\mathbf k})(\mathbf R)=e^{\ii\mathbf k\cdot\mathbf R}v_{\mathbf k}.
\label{sample:eq:sampling-inner-product}
\end{equation}
We first use a finite periodic system with integer lengths $L_x,L_y$, $N_s=L_xL_y$ unit cells, and boundary twists matched as in Sec.~\ref{bands:sec:two-many-body}. In the finite-system operator expressions below, site sums run over these $N_s$ sites and Bloch momentum sums over the corresponding $N_s$-point mesh. With this finite-volume normalization, the unit-normalized continuum orbitals are $\phi^{\rm D,1/2}_{\mathbf k}/\sqrt{N_s}$; their lattice samples $e^{\ii\mathbf k\cdot\mathbf R}v_{\mathbf k}/\sqrt{N_s}$ have squared norm $v_{\mathbf k}^\dagger v_{\mathbf k}$. We therefore normalize the lattice spinors as
\begin{equation}
\psi_{\mathbf k}(\mathbf R)=e^{\ii\mathbf k\cdot\mathbf R}u_{\mathbf k},\qquad
u_{\mathbf k}=\frac{v_{\mathbf k}}{s_{\mathbf k}},\qquad
s_{\mathbf k}=\sqrt{v_{\mathbf k}^\dagger v_{\mathbf k}}.
\label{sample:eq:normalized-lattice-bloch}
\end{equation}
Thus $u_{\mathbf k}^\dagger u_{\mathbf k}=1$ under the lattice unit-cell inner product, and $\psi_{\mathbf k}/\sqrt{N_s}$ is unit normalized on the finite lattice.\\

Equations~\eqref{sample:eq:scalar-fourier} and \eqref{sample:eq:normalized-continuum-sample} give the sampling norm on this mesh:
\begin{equation}
s_{\mathbf k}^{\,2}=F(\mathbf k)+\frac\pi2\nabla_{\mathbf k}^2F(\mathbf k)=\sum_{X,Y\in\mathbb Z}(-1)^{XY+X+Y}\left[1-\frac\pi2(X^2+Y^2)\right]e^{-\pi(X^2+Y^2)/2}e^{-\ii(k_xX+k_yY)}.
\label{sample:eq:sampling-norm}
\end{equation}
Evaluating Eq.~\eqref{sample:eq:sampling-norm} numerically gives $\min s_{\mathbf k}\simeq0.768225$ and $\max s_{\mathbf k}\simeq1.351492$. The nonvanishing of $v_{\mathbf k}$ and continuity on the compact BZ ensure that $s_{\mathbf k}$ is bounded above and away from zero. Sampling therefore gives an invertible map from the continuum Dirac 1LL to the lattice band on every such momentum mesh. It preserves the magnetic Bloch transition phases, so the band inherits $C=1$; its projector $P(\mathbf k)=v_{\mathbf k}v_{\mathbf k}^\dagger/s_{\mathbf k}^2$ is periodic and real analytic.\\

We now define $\hat U_1$ by matching the contact-interaction matrix elements between the normalized Dirac and lattice Bloch bases. Let $\hat\phi_{\mathbf k}^\dagger$ and $\hat\psi_{\mathbf k}^\dagger$ create the unit-normalized orbitals $\phi^{\rm D,1/2}_{\mathbf k}/\sqrt{N_s}$ and $\psi_{\mathbf k}/\sqrt{N_s}$, respectively; both sets obey canonical fermion anticommutation relations. For the contact repulsion $\mathcal G_0^{(2)}\delta(\mathbf r_0-\mathbf r_1)$ in Eq.~\eqref{cont:eq:pair-contact-matrix}, acting as the identity on internal components, this gives
\begin{equation}
\hat U_1=\frac{\mathcal G_0^{(2)}}{N_s^2}\sum_{\mathbf k_1,\ldots,\mathbf k_4}\left[\int\d^2r\,\phi^{\rm D,1/2}_{\mathbf k_1,0}(\mathbf r)^*\phi^{\rm D,1/2}_{\mathbf k_2,1}(\mathbf r)^*\phi^{\rm D,1/2}_{\mathbf k_3,1}(\mathbf r)\phi^{\rm D,1/2}_{\mathbf k_4,0}(\mathbf r)\right]\hat\psi_{\mathbf k_1}^\dagger\hat\psi_{\mathbf k_2}^\dagger\hat\psi_{\mathbf k_3}\hat\psi_{\mathbf k_4}.
\label{sample:eq:explicit-lattice-interaction}
\end{equation}
This is $\hat U_1$ in the main text, with contact strength $\mathcal G_0^{(2)}$ and components labeled $0,1$. On the finite torus of area $L_xL_y$, the Fourier series of the periodic contact interaction gives
\begin{equation}
\langle\psi_{\mathbf k_1},\psi_{\mathbf k_2}|\hat U_1|\psi_{\mathbf k_3},\psi_{\mathbf k_4}\rangle=\frac{\mathcal G_0^{(2)}}{L_xL_y}\sum_{\mathbf q}\left\langle\phi^{\rm D,1/2}_{\mathbf k_1},\phi^{\rm D,1/2}_{\mathbf k_2}\left|e^{\ii\mathbf q\cdot(\mathbf r_0-\mathbf r_1)}\right|\phi^{\rm D,1/2}_{\mathbf k_3},\phi^{\rm D,1/2}_{\mathbf k_4}\right\rangle.
\label{sample:eq:isospectral-contact}
\end{equation}
Here $\mathbf q=2\pi(m_x/L_x,m_y/L_y)$ with $(m_x,m_y)\in\mathbb Z^2$. The sum runs over all reciprocal vectors of the torus and becomes the integral in Eq.~\eqref{cont:eq:pair-fourier} in the infinite-area limit. The pair states are Slater states, $|\psi_{\mathbf k_1},\psi_{\mathbf k_2}\rangle=\hat\psi_{\mathbf k_1}^\dagger\hat\psi_{\mathbf k_2}^\dagger|\mathrm{vac}\rangle$, and likewise in the continuum.\\

To prove spectral equivalence, extend the correspondence between the continuum and lattice Bloch bases to their Slater states:
\begin{equation}
\prod_{i=1}^{N_e}\hat\phi_{\mathbf k_i}^\dagger|\mathrm{vac}\rangle\ \longmapsto\ \prod_{i=1}^{N_e}\hat\psi_{\mathbf k_i}^\dagger|\mathrm{vac}\rangle.
\label{sample:eq:fock-correspondence}
\end{equation}
Because this correspondence maps orthonormal Slater bases onto one another, the matched matrix elements in Eq.~\eqref{sample:eq:isospectral-contact} give, for any states $\Phi,\Phi'$ in the continuum Dirac 1LL and their lattice images $\Psi,\Psi'$,
\begin{equation}
\langle\Psi|\hat U_1|\Psi'\rangle=\langle\Phi|V_0^{(2)}|\Phi'\rangle.
\label{sample:eq:interaction-matrix-identity}
\end{equation}
Here $V_0^{(2)}$ is the Dirac contact interaction with strength $\mathcal G_0^{(2)}$, whose LL projection is given in Eq.~\eqref{cont:eq:half-parent}. Restricted to the target lattice band, $\hat U_1$ therefore has exactly the same energies and degeneracies as the projected continuum interaction at any filling. In particular, let $|\Psi(1)\rangle$ be the normalized lattice image under Eq.~\eqref{sample:eq:fock-correspondence} of a Dirac Laughlin zero mode from Eq.~\eqref{cont:eq:torus-dirac-laughlin}, expressed in the sampling gauge of Eq.~\eqref{sample:eq:gauge}. Then $\hat U_1|\Psi(1)\rangle=0$, so $\hat U_1$ is its parent Hamiltonian. This correspondence preserves coefficients in the normalized Bloch bases.\\

Under the correspondence in Eq.~\eqref{sample:eq:fock-correspondence}, the continuum Dirac contact pair annihilator is represented in the lattice Fock space by $\hat A(\mathbf r)$, with $\mathbf r$ a continuous spatial coordinate:
\begin{equation}
\hat A(\mathbf r)=\frac1{N_s}\sum_{\mathbf k,\mathbf k'}\phi^{\rm D,1/2}_{\mathbf k,1}(\mathbf r)\phi^{\rm D,1/2}_{\mathbf k',0}(\mathbf r)\hat\psi_{\mathbf k}\hat\psi_{\mathbf k'},\qquad
\hat U_1=\mathcal G_0^{(2)}\int\d^2r\,\hat A(\mathbf r)^\dagger\hat A(\mathbf r).
\label{sample:eq:interpolation-pairs}
\end{equation}
Since $|\Psi(1)\rangle$ is a zero mode, positivity and continuity give $\hat A(\mathbf r)|\Psi(1)\rangle=0$ for every $\mathbf r$. At lattice sites, Eq.~\eqref{sample:eq:sample-bloch} gives explicitly
\begin{equation}
\hat A(\mathbf R)|\Psi(1)\rangle
=\frac1{N_s}\sum_{\mathbf k,\mathbf k'}v_{\mathbf k,1}v_{\mathbf k',0}e^{\ii(\mathbf k+\mathbf k')\cdot\mathbf R}\hat\psi_{\mathbf k}\hat\psi_{\mathbf k'}|\Psi(1)\rangle=0.
\label{sample:eq:sampled-contact-pair}
\end{equation}
A continuum contact zero mode vanishes whenever two particles coincide. Its direct lattice sample therefore vanishes whenever two particles occupy the same site, suggesting onsite repulsion as a lattice parent. With $c_{\mathbf R,\alpha}$ annihilating component $\alpha$ at site $\mathbf R$, we define the onsite repulsion and its pair annihilator by
\begin{equation}
\hat B_{\mathbf R}=c_{\mathbf R,1}c_{\mathbf R,0},\qquad H_{1/3}=U\sum_{\mathbf R}\hat B_{\mathbf R}^\dagger\hat B_{\mathbf R}=U\sum_{\mathbf R}n_{\mathbf R,0}n_{\mathbf R,1},\qquad U>0.
\label{sample:eq:onsite-interpolation-endpoint}
\end{equation}
We write $\hat U_0=H_{1/3}$ and work in the target-band Fock space below. Using $u_{\mathbf k}=v_{\mathbf k}/s_{\mathbf k}$, the onsite pair annihilator takes the form
\begin{equation}
\hat B_{\mathbf R}=\frac1{N_s}\sum_{\mathbf k,\mathbf k'}\frac{v_{\mathbf k,1}v_{\mathbf k',0}}{s_{\mathbf k}s_{\mathbf k'}}e^{\ii(\mathbf k+\mathbf k')\cdot\mathbf R}\hat\psi_{\mathbf k}\hat\psi_{\mathbf k'}=\frac1{N_s}\sum_{\mathbf k,\mathbf k'}u_{\mathbf k,1}u_{\mathbf k',0}e^{\ii(\mathbf k+\mathbf k')\cdot\mathbf R}\hat\psi_{\mathbf k}\hat\psi_{\mathbf k'}.
\label{sample:eq:projected-onsite-pair}
\end{equation}
Compared with $\hat A(\mathbf R)$ in Eq.~\eqref{sample:eq:sampled-contact-pair}, each spinor is divided by $s_{\mathbf k}$. Direct sampling multiplies each many-body coefficient of $|\Psi(1)\rangle$ by the product of $s_{\mathbf k}$ over occupied momenta. This defines the filter $\hat D$, with $\hat n_{\mathbf k}=\hat\psi_{\mathbf k}^\dagger\hat\psi_{\mathbf k}$. The number-operator commutator gives its exchange relations for any real $a$:
\begin{equation}
\hat D=\prod_{\mathbf k}s_{\mathbf k}^{\hat n_{\mathbf k}},\qquad
[\hat\psi_{\mathbf k},\hat n_{\mathbf k'}]=\delta_{\mathbf k,\mathbf k'}\hat\psi_{\mathbf k},\quad \hat D^a\hat\psi_{\mathbf k}\hat D^{-a}=s_{\mathbf k}^{-a}\hat\psi_{\mathbf k},\qquad
\hat\psi_{\mathbf k}\hat D^a=\hat D^a s_{\mathbf k}^{a}\hat\psi_{\mathbf k}.
\label{sample:eq:sampling-filter-algebra}
\end{equation}
Using the last identity to cancel the factors $s_{\mathbf k}^{-1}s_{\mathbf k'}^{-1}$ in Eq.~\eqref{sample:eq:projected-onsite-pair} gives
\begin{equation}
\hat B_{\mathbf R}\hat D=\hat D\hat A(\mathbf R),\qquad
\hat B_{\mathbf R}\hat D|\Psi(1)\rangle=\hat D\hat A(\mathbf R)|\Psi(1)\rangle=0.
\label{sample:eq:pair-filter-relation}
\end{equation}
Thus the normalized directly sampled state $|\Psi(0)\rangle\propto\hat D|\Psi(1)\rangle$ is an onsite zero mode. The filter relating the two endpoint states also gives the interpolation
\begin{equation}
|\Psi(t)\rangle=\frac{\hat D^{1-t}|\Psi(1)\rangle}{\sqrt{\langle\Psi(1)|\hat D^{2(1-t)}|\Psi(1)\rangle}},\qquad 0\le t\le1.
\label{sample:eq:interpolating-zero-mode}
\end{equation}
To preserve the contact zero-mode condition along this path, set $\hat A_t(\mathbf r)=\hat D^{1-t}\hat A(\mathbf r)\hat D^{t-1}$. For the onsite pair, set $\hat B_{t,\mathbf R}=\hat D^{-t}\hat B_{\mathbf R}\hat D^t$; Eq.~\eqref{sample:eq:pair-filter-relation} then gives
\begin{equation}
\hat A_t(\mathbf r)=\frac1{N_s}\sum_{\mathbf k,\mathbf k'}s_{\mathbf k}^{t-1}s_{\mathbf k'}^{t-1}\phi^{\rm D,1/2}_{\mathbf k,1}(\mathbf r)\phi^{\rm D,1/2}_{\mathbf k',0}(\mathbf r)\hat\psi_{\mathbf k}\hat\psi_{\mathbf k'},\qquad \hat B_{t,\mathbf R}=\hat A_t(\mathbf R).
\label{sample:eq:discrete-interpolating-pairs}
\end{equation}
The continuous-space pair annihilates the numerator of $|\Psi(t)\rangle$:
\begin{equation}
\hat A_t(\mathbf r)\hat D^{1-t}|\Psi(1)\rangle=\hat D^{1-t}\hat A(\mathbf r)|\Psi(1)\rangle=0.
\label{sample:eq:path-contact-annihilation}
\end{equation}
Since $\hat B_{t,\mathbf R}=\hat A_t(\mathbf R)$, the onsite pair also annihilates $|\Psi(t)\rangle$. These pair operators therefore define a positive two-body interaction
\begin{equation}
\hat U_t=t \mathcal G_0^{(2)}\int\d^2r\,\hat A_t(\mathbf r)^\dagger\hat A_t(\mathbf r)+(1-t)U\sum_{\mathbf R}\hat B_{t,\mathbf R}^\dagger\hat B_{t,\mathbf R},\qquad 0\le t\le1.
\label{sample:eq:interaction-interpolation}
\end{equation}
The endpoints $\hat A_1=\hat A$ and $\hat B_{0,\mathbf R}=\hat B_{\mathbf R}$ give $\hat U_{t=1}=\hat U_1$ and $\hat U_{t=0}=\hat U_0$ within the target band, at any filling. Along the path, positivity makes $|\Psi(t)\rangle$ a zero-energy ground state:
\begin{equation}
\hat U_t|\Psi(t)\rangle=0,\qquad 0\le t\le1.
\label{sample:eq:path-zero-energy}
\end{equation}
On the finite momentum mesh, $s_{\mathbf k}>0$ ensures a nonzero normalization and analytic dependence of $|\Psi(t)\rangle$ on $t$.\\

For the spin-1 Dirac 2LL, we now construct the Laughlin $1/5$ and Moore--Read interpolation paths. Using the three-component sample in Eq.~\eqref{sample:eq:normalized-continuum-sample}, we set
\begin{equation}
v_{\mathbf k}=\phi^{\rm D,1}_{\mathbf k}(\mathbf0),\qquad s_{\mathbf k}=\sqrt{v_{\mathbf k}^\dagger v_{\mathbf k}},\qquad u_{\mathbf k}=\frac{v_{\mathbf k}}{s_{\mathbf k}}.
\label{sample:eq:one-sampling-normalization}
\end{equation}
The spin-1 sample is nonzero throughout the BZ by Sec.~\ref{sample:sec:dirac-samples}. For Laughlin $1/5$, the continuous and projected onsite pair annihilators have three channels, $\alpha\beta=01,02,12$:
\begin{equation}
\hat A_{\alpha\beta}(\mathbf r)=\frac1{N_s}\sum_{\mathbf k,\mathbf k'}\phi^{\rm D,1}_{\mathbf k,\beta}(\mathbf r)\phi^{\rm D,1}_{\mathbf k',\alpha}(\mathbf r)\hat\psi_{\mathbf k}\hat\psi_{\mathbf k'},\qquad \hat B_{\alpha\beta,\mathbf R}=\frac1{N_s}\sum_{\mathbf k,\mathbf k'}e^{\ii(\mathbf k+\mathbf k')\cdot\mathbf R}u_{\mathbf k,\beta}u_{\mathbf k',\alpha}\hat\psi_{\mathbf k}\hat\psi_{\mathbf k'}.
\label{sample:eq:one-contact-pairs}
\end{equation}
The onsite operators $\hat B_{\alpha\beta,\mathbf R}$ give the band projection of
\begin{equation}
U\sum_{\mathbf R,\alpha<\beta}n_{\mathbf R,\alpha}n_{\mathbf R,\beta}.
\label{sample:eq:fifth-onsite-interaction}
\end{equation}
The continuous operators reproduce the Dirac contact parent in Eq.~\eqref{cont:eq:one-pair-parent}. Applying the channelwise filter definitions preceding Eq.~\eqref{sample:eq:discrete-interpolating-pairs} gives the interpolation
\begin{equation}
\hat U_t^{(1/5)}=t\mathcal G_0^{(2)}\sum_{\alpha<\beta}\int\d^2r\,\hat A_{t,\alpha\beta}^\dagger(\mathbf r)\hat A_{t,\alpha\beta}(\mathbf r)+(1-t)U\sum_{\mathbf R,\alpha<\beta}\hat B_{t,\alpha\beta,\mathbf R}^\dagger\hat B_{t,\alpha\beta,\mathbf R}.
\label{sample:eq:fifth-interpolation}
\end{equation}

For Moore--Read, the continuous and projected onsite triplet annihilators are
\begin{equation}
\hat A_{012}(\mathbf r)=\frac1{\sqrt{N_s^3}}\sum_{\mathbf k_1,\mathbf k_2,\mathbf k_3}\prod_{j=1}^3\bigl[\phi^{\rm D,1}_{\mathbf k_j,3-j}(\mathbf r)\hat\psi_{\mathbf k_j}\bigr],\qquad
\hat B_{012,\mathbf R}=\frac1{\sqrt{N_s^3}}\sum_{\mathbf k_1,\mathbf k_2,\mathbf k_3}e^{\ii\sum_j\mathbf k_j\cdot\mathbf R}\prod_{j=1}^3\bigl[u_{\mathbf k_j,3-j}\hat\psi_{\mathbf k_j}\bigr].
\label{sample:eq:one-contact-triplet}
\end{equation}
The products are ordered by increasing $j$. The onsite operator annihilates one fermion of each component, giving the band projection of
\begin{equation}
U\sum_{\mathbf R}n_{\mathbf R,0}n_{\mathbf R,1}n_{\mathbf R,2}.
\label{sample:eq:mr-onsite-interaction}
\end{equation}
The continuous operator reproduces the Dirac three-body contact parent in Eq.~\eqref{cont:eq:mr-parent}. The same filter construction gives
\begin{equation}
\hat U_t^{(\mathrm{MR})}=t\mathcal G_0^{(3)}\int\d^2r\,\hat A_{t,012}^\dagger(\mathbf r)\hat A_{t,012}(\mathbf r)+(1-t)U\sum_{\mathbf R}\hat B_{t,012,\mathbf R}^\dagger\hat B_{t,012,\mathbf R}.
\label{sample:eq:mr-interpolation}
\end{equation}

For both interpolations, $\hat D$ uses the spin-1 norm in Eq.~\eqref{sample:eq:one-sampling-normalization}. The channelwise definitions are
\begin{equation}
\hat A_t(\mathbf r)=\hat D^{1-t}\hat A(\mathbf r)\hat D^{t-1},\qquad \hat B_{t,\mathbf R}=\hat D^{-t}\hat B_{\mathbf R}\hat D^t.
\label{sample:eq:one-channel-filter}
\end{equation}
The filter algebra gives the same relation as Eq.~\eqref{sample:eq:pair-filter-relation}, and hence $\hat B_{t,\mathbf R}=\hat A_t(\mathbf R)$. Taking $|\Psi(1)\rangle$ from the Dirac zero modes of Secs.~\ref{cont:sec:dirac-one} and \ref{cont:sec:dirac-mr} through Eq.~\eqref{sample:eq:fock-correspondence}, the normalized path in Eq.~\eqref{sample:eq:interpolating-zero-mode} satisfies $\hat U_t|\Psi(t)\rangle=0$ by Eq.~\eqref{sample:eq:path-contact-annihilation}. Positivity of the sampling norm again ensures analyticity on every finite mesh.\\

For the spin-$1/2$ and spin-1 interaction constructions above, the infinite-lattice limit keeps $\mathbf R$ discrete and makes the Bloch momentum $\mathbf k$ continuous:
\begin{equation}
\frac1{N_s}\sum_{\mathbf k}\longrightarrow\int_{\rm BZ}\frac{\d^2k}{(2\pi)^2}.
\label{sample:eq:momentum-continuum-limit}
\end{equation}
We now use delta-normalized Bloch orbitals and annihilation operators, with $\{\hat\psi_{\mathbf k},\hat\psi_{\mathbf k'}^\dagger\}=(2\pi)^2\delta(\mathbf k-\mathbf k')$ and likewise for $\hat\phi$. Each discrete canonical operator is replaced by its delta-normalized counterpart divided by $\sqrt{N_s}$. Each momentum sum then carries a factor $1/N_s$, giving two BZ integrals for pair annihilators and four for two-body interactions. For the spin-$1/2$ pair, with $v_{\mathbf k}$ and $s_{\mathbf k}$ as in Eq.~\eqref{sample:eq:normalized-lattice-bloch},
\begin{equation}
\hat A(\mathbf r)=\int_{\rm BZ}\frac{\d^2k\,\d^2k'}{(2\pi)^4}\phi^{\rm D,1/2}_{\mathbf k,1}(\mathbf r)\phi^{\rm D,1/2}_{\mathbf k',0}(\mathbf r)\hat\psi_{\mathbf k}\hat\psi_{\mathbf k'},\quad \hat B_{\mathbf R}=\int_{\rm BZ}\frac{\d^2k\,\d^2k'}{(2\pi)^4}\frac{v_{\mathbf k,1}v_{\mathbf k',0}}{s_{\mathbf k}s_{\mathbf k'}}e^{\ii(\mathbf k+\mathbf k')\cdot\mathbf R}\hat\psi_{\mathbf k}\hat\psi_{\mathbf k'}.
\label{sample:eq:continuous-pair-expressions}
\end{equation}
The same conversion applies to the spin-1 channels, with three BZ integrals for the triplet annihilator. For either Dirac level, using its respective sampling norm gives
\begin{equation}
\hat D=\exp\!\left[\int_{\rm BZ}\frac{\d^2k}{(2\pi)^2}(\ln s_{\mathbf k})\hat\psi_{\mathbf k}^\dagger\hat\psi_{\mathbf k}\right].
\label{sample:eq:continuous-sampling-filter}
\end{equation}
For each model, the normalized Bloch-basis correspondence in Eq.~\eqref{sample:eq:fock-correspondence} makes the $t=1$ interaction $\hat U_1$ unitarily equivalent to the projected continuum contact interaction on finite lattices and in each fixed finite-$N_e$ sector on the infinite lattice. These two interactions are therefore exactly isospectral in these settings.\\

We now examine locality, writing the spin-$1/2$ case explicitly. For spin 1, replacing $\mathcal U_{1/2}$ by $\mathcal U_1$ preserves the kernel decay used below; the triplet interaction involves six kernels rather than four. The inverse unitary map for spin $1/2$ has kernel
\begin{equation}
K_{\alpha\beta}(\mathbf r,\mathbf R)=\int_{\rm BZ}\frac{\d^2k}{(2\pi)^2}\,\phi^{\rm D,1/2}_{\mathbf k,\alpha}(\mathbf r)\psi_{\mathbf k,\beta}(\mathbf R)^*=\int_{\rm BZ}\frac{\d^2k}{(2\pi)^2}\,\frac{\phi^{\rm D,1/2}_{\mathbf k,\alpha}(\mathbf r)v_{\mathbf k,\beta}^*}{s_{\mathbf k}}e^{-\ii\mathbf k\cdot\mathbf R}.
\label{sample:eq:reconstruction-kernel}
\end{equation}
Omitting $s_{\mathbf k}^{-1}$ gives the Dirac-LL projector kernel, obtained by applying the map in Eq.~\eqref{cont:eq:half-map} to both arguments of the scalar LLL projector. Its magnitude is bounded by a Gaussian times a polynomial. Since $s_{\mathbf k}^2$ is positive, periodic, and real analytic by Eq.~\eqref{sample:eq:sampling-norm}, $s_{\mathbf k}^{-1}$ has exponentially decaying Fourier coefficients. Equation~\eqref{sample:eq:sample-bloch} expresses $K$ as their lattice convolution with the projector kernel, giving $|K_{\alpha\beta}(\mathbf r,\mathbf R)|\le C e^{-|\mathbf r-\mathbf R|/\xi}$ for some $C,\xi>0$.\\

Expressing the contact interaction through this kernel, the coefficient of $c_{\mathbf R_1,\alpha_1}^\dagger c_{\mathbf R_2,\alpha_2}^\dagger c_{\mathbf R_3,\alpha_3}c_{\mathbf R_4,\alpha_4}$ is the integral of four such kernels and obeys
\begin{equation}
\left|\mathcal G_0^{(2)}\int_{\mathbb R^2}\d^2r\,K_{0\alpha_1}(\mathbf r,\mathbf R_1)^*K_{1\alpha_2}(\mathbf r,\mathbf R_2)^*K_{1\alpha_3}(\mathbf r,\mathbf R_3)K_{0\alpha_4}(\mathbf r,\mathbf R_4)\right|\le C' e^{-\max_{i,j}|\mathbf R_i-\mathbf R_j|/(2\xi)}.
\label{sample:eq:interaction-locality}
\end{equation}
The bound follows from $\sum_i|\mathbf r-\mathbf R_i|\ge\max_{i,j}|\mathbf R_i-\mathbf R_j|$, leaving an integrable exponential in $\mathbf r$. Thus $\hat U_1$ is exponentially short-ranged. The additional analytic factors $s_{\mathbf k}^{t-1}$ in $\hat A_t$ preserve exponential kernel decay. Since $\hat B_{t,\mathbf R}=\hat A_t(\mathbf R)$, the integral and lattice-sum terms of $\hat U_t$ obey analogous bounds, establishing locality along the interpolation.\\

Finally, restrict the sampling map $\mathcal S$ of Eq.~\eqref{sample:eq:sampling-definition} to the spin-$s$ Dirac level in Eq.~\eqref{cont:eq:general-spin-map}, with $s>0$. At every momentum, at least one sampled component is nonzero: the scalar LLL component away from $\mathbf k=0$, and the scalar 1LL component at $\mathbf k=0$ (Sec.~\ref{sample:sec:dirac-samples}). Their weights are nonzero for all $s>0$, so sampling is bijective between this Dirac LL and its lattice band. On a finite torus, write $\phi_{\mathbf k}$ for its magnetic Bloch spinor, $s_{\mathbf k}=\|\phi_{\mathbf k}(\mathbf0)\|$, and $\psi_{\mathbf k}=\mathcal S\phi_{\mathbf k}/s_{\mathbf k}$. We write $\mathcal S_n$ for sampling an $n$-particle state, with $\mathcal S_0$ denoting the identity operator on the zero-particle space. For unit-normalized Slater determinants,
\begin{equation}
\mathcal S_n\operatorname{Slat}\!\left(\frac{\phi_{\mathbf k_1}}{\sqrt{N_s}},\ldots,\frac{\phi_{\mathbf k_n}}{\sqrt{N_s}}\right)
=\left(\prod_{i=1}^n s_{\mathbf k_i}\right)\operatorname{Slat}\!\left(\frac{\psi_{\mathbf k_1}}{\sqrt{N_s}},\ldots,\frac{\psi_{\mathbf k_n}}{\sqrt{N_s}}\right).
\label{sample:eq:many-body-sampling-bijection}
\end{equation}
Since every $s_{\mathbf k_i}>0$, the Slater bases are in bijection, and every $n$-particle band state has a unique continuum preimage for $0\le n\le N_s$.

\subsection{Completeness of sampled Laughlin zero modes}
\label{sample:sec:laughlin-completeness}

We prove completeness of the sampled Laughlin $1/3$ and $1/5$ zero modes for $n=N_e\ge13$ fermions on a rectangular torus with periodic lattice boundary conditions. We begin with the spin-$1/2$ Dirac 1LL at $N_\phi=N_s=3n$.\\

The three Dirac Laughlin states in Eq.~\eqref{cont:eq:torus-dirac-laughlin} exhaust the continuum contact zero-mode space in the sampling gauge with matching boundary conditions (Sec.~\ref{cont:sec:dirac-half})~\cite{HaldaneRezayi1985,GreiterSchnellsThomale2016,LeeThomaleQi2013,BurbanKlevtsov2025}. Their direct samples remain onsite zero modes, and the sampling bijection in Eq.~\eqref{sample:eq:many-body-sampling-bijection} preserves their independence. It therefore remains to show that the inverse sampling map $\mathcal S_n^{-1}$ sends every onsite zero mode to a continuum contact zero mode. We establish this for $n\ge13$.\\

Let $\hat H$ denote the onsite interaction in Eq.~\eqref{sample:eq:onsite-interpolation-endpoint} projected onto the $n$-particle band space, and take any $\Psi\in\ker\hat H$. Since $U>0$, its zero energy is a sum of nonnegative terms,
\begin{equation}
0=\langle\Psi|\hat H|\Psi\rangle=U\sum_{\mathbf R}\|\hat B_{\mathbf R}\Psi\|^2,
\label{sample:eq:hubbard-local-zero-condition}
\end{equation}
so $\hat B_{\mathbf R}\Psi=0$ at every site. To evaluate the contact amplitudes of its unique Dirac preimage, express it through a scalar LLL state $\Phi$:
\begin{equation}
\Phi^{\rm D}=\mathcal S_n^{-1}\Psi=\mathcal U_{1/2}^{\otimes n}\Phi,\qquad \Phi=\prod_i g(\mathbf r_i)\,f(\boldsymbol z).
\label{sample:eq:zero-mode-continuum-preimage}
\end{equation}
Here $g(\mathbf r)=\phi^{\rm LLL}_{\rm loc}(\mathbf r)=e^{\ii\Lambda_{\rm g}(\mathbf r)-|z|^2/(4\ell_B^2)}$ is the nonvanishing orbital in Eq.~\eqref{sample:eq:localized-orbital}, $\boldsymbol z=(z_1,\ldots,z_n)$, and $f$ is holomorphic (complex analytic) and antisymmetric under exchange of any two particle coordinates.\\

To evaluate $\hat B_{\mathbf R}\Psi$, fix particle arguments $1,2$ at $(\mathbf R,0)$ and $(\mathbf R,1)$. Particles $3,\ldots,n$ are spectators: their coordinates are not fixed by this pair coincidence. The $01$ component of $\mathcal U_{1/2}^{\otimes2}$ supplies $a_2^\dagger/2$, and the two annihilations supply $\sqrt{n(n-1)}$. The remaining particles stay in the target band, so Eq.~\eqref{sample:eq:many-body-sampling-bijection} permits inversion of their sampling:
\begin{equation}
\mathcal S_{n-2}^{-1}\hat B_{\mathbf R}\Psi=-\ell_B\sqrt{\frac{n(n-1)}2}\,g(\mathbf R)^2\mathcal U_{1/2}^{\otimes(n-2)}\left[\prod_{i=3}^n g(\mathbf r_i)\left.\partial_2f\right|_{z_1=z_2=Z}\right],
\label{sample:eq:half-pair-contraction}
\end{equation}
with $\partial_j=\partial/\partial z_j$. The sign change of $f$ under exchange of particles $1,2$ sets $f=0$ at coincidence, so only its derivative survives in $a_2^\dagger\Phi$. Since $\mathcal U_{1/2}$ is injective and $g\ne0$, the zero-mode condition is equivalent to this derivative vanishing. Relabeling the pair gives
\begin{equation}
\hat B_{\mathbf R}\Psi=0 \quad\Longleftrightarrow\quad \left.\partial_j f\right|_{z_i=z_j=Z}=0\quad\text{for all }i<j.
\label{sample:eq:continuous-spectator-constraint}
\end{equation}
Since sampling has been inverted for the spectators, this is an identity in their continuous coordinates. Only the common position $Z$ of the pair remains restricted to a lattice site.\\

To turn the derivative constraint into a condition on function values, we extract the pair zeros of $f$. Its sign change under particle exchange forces zeros at coincidence; the nonvanishing boundary multipliers in Eq.~\eqref{sample:eq:continuum-bloch-law} extend these zeros to coincidence modulo torus periods. Using the Jacobi theta function $\vartheta_1$ of Sec.~\ref{special:sec:theta} and $\tau=\ii L_y/L_x$, set
\begin{equation}
\Delta_n(\boldsymbol z)=\prod_{1\le i<j\le n}E(z_i-z_j),\qquad E(z)=\frac{L_x\vartheta_1(\pi z/L_x\mid\tau)}{\pi\vartheta_1'(0\mid\tau)}.
\label{sample:eq:laughlin-pair-factor}
\end{equation}
The normalization gives $E(z)=z+O(z^3)$, with zeros only at the torus periods $L_x\mathbb Z+L_x\tau\mathbb Z$. Under period shifts,
\begin{equation}
E(z+L_x)=-E(z),\qquad E(z+L_x\tau)=-e^{-\ii\pi(2z/L_x+\tau)}E(z).
\label{sample:eq:laughlin-pair-factor-periods}
\end{equation}
We factor out these pair zeros and a nonvanishing exponential that simplifies the boundary multipliers:
\begin{equation}
f(\boldsymbol z)=e^{\sum_i z_i^2/(4\ell_B^2)}\Delta_n(\boldsymbol z)G(\boldsymbol z).
\label{sample:eq:laughlin-theta-factorization}
\end{equation}
Removing the pair zeros leaves $G$ entire (holomorphic throughout $\mathbb C^n$). The signs of $f$ and $\Delta_n$ under exchange of identical particles cancel, making $G$ invariant under such exchanges. At $z_1=z_2=Z$, only the derivative of $E(z_1-z_2)$ survives in $\partial_2f$. Since $E'(0)=1$ and the remaining pair factors are nonzero for distinct spectators away from $Z$ modulo periods, Eq.~\eqref{sample:eq:continuous-spectator-constraint} gives
\begin{equation}
G(Z,Z,z_3,\ldots,z_n)=0 \quad\text{for every lattice site }Z\text{ and all continuous spectators}.
\label{sample:eq:laughlin-symmetric-site-constraint}
\end{equation}
Analyticity extends this identity to all spectator configurations.\\

We must extend this site condition to $G(z,z,z_3,\ldots,z_n)=0$ at every continuous $z$. To obtain higher-order site zeros, consider clusters of $r$ coincident particles, $2\le r\le n$, with the remaining coordinates fixed:
\begin{equation}
G_r(z;\boldsymbol w)=G(\underbrace{z,\ldots,z}_{r},w_{r+1},\ldots,w_n),\qquad \boldsymbol w=(w_{r+1},\ldots,w_n).
\label{sample:eq:laughlin-cluster-restriction}
\end{equation}
The quotient $G$ is formed before taking coincident coordinates. We will force a larger cluster to vanish by counting its zeros, then descend to the required pair condition $G_2(z;\boldsymbol w)=0$.\\

The common orbital boundary multipliers in Eq.~\eqref{sample:eq:continuum-bloch-law} pass to the scalar many-body preimage. On the periodic momentum mesh, with unchanged arguments suppressed,
\begin{equation}
\Phi(x_i+L_x,y_i)=e^{2\pi\ii L_xy_i}\Phi(x_i,y_i),\qquad \Phi(x_i,y_i+L_y)=\Phi(x_i,y_i).
\label{sample:eq:laughlin-preimage-boundary-x}
\end{equation}
Substituting Eqs.~\eqref{sample:eq:zero-mode-continuum-preimage} and \eqref{sample:eq:laughlin-theta-factorization} into these boundary conditions gives $rN_\phi$ in the exponential coefficient when $r$ coordinates move together. Dividing out $\Delta_n$ subtracts $r(n-r)$ by Eq.~\eqref{sample:eq:laughlin-pair-factor-periods}: the $r(n-r)$ pairs joining moving particles to spectators change, while separations within the cluster stay fixed. Hence
\begin{equation}
G_r(z+L_x)=C_{x,r}G_r(z),\qquad G_r(z+L_x\tau)=C_{\tau,r}e^{-\ii\pi[rN_\phi-r(n-r)](2z/L_x+\tau)}G_r(z).
\label{sample:eq:laughlin-cluster-quasiperiods}
\end{equation}
Here $\boldsymbol w$ is suppressed and $C_{x,r},C_{\tau,r}$ are nonzero constants independent of $z$. The relations extend from distinct to coincident coordinates because $G$ is entire.\\

For a nonzero entire $G_r$, the argument principle~\cite{OlverEtAl2010} counts its zeros in a torus cell $\mathcal P$ whose boundary avoids zeros. By Eq.~\eqref{sample:eq:laughlin-cluster-quasiperiods}, its logarithmic derivative $\partial_zG_r/G_r$ is periodic along $L_x$ and changes by $-2\pi\ii[rN_\phi-r(n-r)]/L_x$ along $L_x\tau$. Thus the number of zeros, counted with multiplicity, is~\cite{Dolgachev2017}
\begin{equation}
d_r=\frac{1}{2\pi\ii}\oint_{\partial\mathcal P}\frac{\partial_zG_r(z;\boldsymbol w)}{G_r(z;\boldsymbol w)}\,\d z=rN_\phi-r(n-r)=r(2n+r).
\label{sample:eq:laughlin-cluster-zero-count}
\end{equation}
More than $d_r$ zeros therefore force $G_r(z;\boldsymbol w)=0$ for every $z$.\\

\begin{samepage}%
At a site $Z$, expand $G(Z+\delta_1,\ldots,Z+\delta_r,\boldsymbol w)$ in independent displacements. Equation~\eqref{sample:eq:laughlin-symmetric-site-constraint} and the invariance of $G$ under particle exchange make it vanish whenever any two displacements are zero. Every nonzero Taylor term therefore contains at least $r-1$ distinct displacements and has degree at least $r-1$. Setting all displacements equal to $\delta$ gives
\begin{equation}
G_r(Z+\delta;\boldsymbol w)=O(\delta^{r-1})\quad\text{at every lattice site }Z.
\label{sample:eq:laughlin-site-multiplicity}
\end{equation}
\end{samepage}%
The sites therefore supply at least $(r-1)N_s$ zeros. Taking $r=5$ and $N_s=3n$ gives
\begin{equation}
4N_s-d_5=2n-25>0\quad(n\ge13)\qquad\Longrightarrow\qquad G_5(z;\boldsymbol w)=0.
\label{sample:eq:laughlin-five-cluster-zero}
\end{equation}
This holds for all $z$ and $\boldsymbol w$. In each descent step below, choose spectators pairwise distinct modulo periods and away from the sites, so their zeros add to the site zeros; analyticity then extends the resulting identity to all spectator configurations. When the four-particle cluster in $G_4$ meets any spectator, five particles coincide. Thus $G_5=0$ supplies $n-4$ additional zeros of $G_4$, and
\begin{equation}
3N_s+(n-4)-d_4=2n-20>0\qquad\Longrightarrow\qquad G_4(z;\boldsymbol w)=0.
\label{sample:eq:laughlin-four-cluster-zero}
\end{equation}
Likewise, $G_4=0$ supplies $n-3$ spectator zeros of $G_3$, giving
\begin{equation}
2N_s+(n-3)-d_3=n-12>0\qquad\Longrightarrow\qquad G_3(z;\boldsymbol w)=0.
\label{sample:eq:laughlin-three-cluster-zero}
\end{equation}
Both inequalities hold for $n\ge13$. To reach $G_2=0$, we need the double zero at each spectator. At $z_1=z_2=z_3=w_3$, exchange of these three particles makes $\partial_1G,\partial_2G,\partial_3G$ equal. Differentiating $G_3=0$ along the common coordinate, and then moving only the first two coordinates, gives
\begin{equation}
0=\left.\partial_uG_3(u;w_4,\ldots,w_n)\right|_{u=w_3}=3\partial_1G,\qquad \left.\partial_zG_2(z;\boldsymbol w)\right|_{z=w_3}=2\partial_1G=0.
\label{sample:eq:laughlin-pair-double-zero}
\end{equation}
Here $\partial_1G$ is evaluated at this triple coincidence. Since $G_2(w_3;\boldsymbol w)=0$ by $G_3=0$, the vanishing derivative makes $w_3$ at least a double zero. Relabeling the spectators supplies $2(n-2)$ additional zeros, so
\begin{equation}
N_s+2(n-2)-d_2=n-8>0\qquad\Longrightarrow\qquad G_2(z;\boldsymbol w)=0.
\label{sample:eq:laughlin-pair-descent-counts}
\end{equation}
Thus $G(z,z,w_3,\ldots,w_n)=0$ for all continuous coordinates when $n\ge13$. For any pair $i,j$, exchange at fixed pair center reverses $z_i-z_j$ and leaves $G$ unchanged. Its Taylor expansion has only even relative powers, so vanishing at coincidence forces at least a quadratic zero. The factor $\Delta_n$ in Eq.~\eqref{sample:eq:laughlin-theta-factorization} supplies one more zero, giving $f$ and hence $\Phi$ at least cubic pair zeros. Each component of $\Phi^{\rm D}=\mathcal U_{1/2}^{\otimes n}\Phi$ contains at most one ladder operator per particle, so at most two derivatives act on a given pair. Every component therefore retains at least a first-order pair zero. Hence $\Phi^{\rm D}$ vanishes at every continuous pair coincidence and is a continuum contact zero mode.\\

\begin{samepage}%
For $n\ge13$, we have therefore shown that the inverse sampling map $\mathcal S_n^{-1}$ sends every onsite zero mode to a continuum contact zero mode. The continuum zero-mode space is already known to be spanned by the three Dirac Laughlin states at matching boundary conditions (Sec.~\ref{cont:sec:dirac-half}). Since sampling is bijective and their samples are onsite zero modes, these three sampled states exhaust the lattice zero-mode space.\\
\par\end{samepage}%

For Laughlin $1/5$, take $N_\phi=N_s=5n$ and write the Dirac preimage of an onsite zero mode as $\mathcal U_1^{\otimes n}\Phi$, with scalar holomorphic factor $f$ as in Eq.~\eqref{sample:eq:zero-mode-continuum-preimage}. The three onsite pair annihilators $01,02,12$ must each annihilate the state. Inverting the sampling of the spectators as above leaves their positions $\boldsymbol w=(z_3,\ldots,z_n)$ arbitrary. By Eq.~\eqref{cont:eq:one-map}, the three channels act on the pair through $a_2^\dagger$, $(a_2^\dagger)^2$, and $a_1^\dagger(a_2^\dagger)^2$, respectively, up to nonzero constants. At coincident coordinates each ladder operator differentiates the holomorphic factor once in its highest-order term.\\

Whenever $f$ has a zero of order at least $p$ at every pair coincidence, for $p=1$ or $3$ write
\begin{equation}
f=e^{\sum_i z_i^2/(4\ell_B^2)}\Delta_n^p H^{(p)},\qquad
H^{(p)}_2(z)=H^{(p)}(z,z,\boldsymbol w).
\label{sample:eq:fifth-pair-quotient}
\end{equation}
The quotient is entire and symmetric. Moving both particles around a period gives the same count as in Eq.~\eqref{sample:eq:laughlin-cluster-quasiperiods}, now with $2(n-2)$ moving-particle--spectator factors each raised to power $p$. Thus a nonzero $H^{(p)}_2$ has exactly $d_2^{(p)}=2N_\phi-2p(n-2)$ zeros per torus cell, counted with multiplicity. Choose distinct spectators away from the sites modulo periods.\\

Antisymmetry supplies the first pair zero ($p=1$). At each site $Z$, the $01$ constraint sets $\partial_2 f(Z,Z,\boldsymbol w)=0$. The $02$ constraint then gives $\partial_2^2 f(Z,Z,\boldsymbol w)=0$, since both lower-order terms vanish there. Set $b(z)=\partial_2 f(z,z,\boldsymbol w)$. Antisymmetry gives $\partial_1\partial_2 f(z,z,\boldsymbol w)=0$, so $b'(Z)=(\partial_1\partial_2 f+\partial_2^2 f)|_{Z,Z}=0$. Equation~\eqref{sample:eq:fifth-pair-quotient} and $E'(0)=1$ express $b(z)$ as $H^{(1)}_2(z)$ times a holomorphic factor nonzero at each site. Hence every site contributes at least a double zero. Since $2N_s=10n>8n+4=d_2^{(1)}$ for $n>2$, $H^{(1)}_2$ vanishes identically. The symmetric quotient is even in the relative coordinate at fixed pair center, so it contributes two more pair zeros. Relabeling the pair and continuing in the spectator coordinates gives the factor $\Delta_n^3$ globally.\\

With $p=3$, all pair derivatives of $f$ of total order below three vanish at coincidence. The $12$ constraint therefore reduces to $\partial_1\partial_2^2 f(Z,Z,\boldsymbol w)=0$. Since $\partial_1\partial_2^2(z_1-z_2)^3=6$, this sets $H^{(3)}_2(Z)=0$ at all $N_s$ sites. Now $N_s=5n>4n+12=d_2^{(3)}$ for $n\ge13$, so $H^{(3)}_2$ vanishes identically. The same exchange symmetry and analytic continuation give a fifth-order zero for every pair. At most four derivatives act on that pair under $\mathcal U_1^{\otimes n}$, so the Dirac preimage vanishes at every continuum pair coincidence. The continuum parent has exactly five zero modes (Sec.~\ref{cont:sec:dirac-one}). Their samples remain independent by Eq.~\eqref{sample:eq:many-body-sampling-bijection} and thus exhaust the onsite zero-mode space. The conclusion also holds under the invertible onsite transformation in Eq.~\eqref{bands:eq:three-prototype-deformation}, since it and its inverse preserve pair-coincidence zeros.

\section{Lattice Chern bands and parent interactions}
\label{sec:lattice-models}

In this appendix, we extend the Dirac-sampled bands and onsite zero modes of Appendix~\ref{sec:sampling} to two- and three-component $C=1$ parameter families. We retain the sampled theta functions and vary their relative weights and mixing through constant invertible component transformations, with the Dirac weights defining a specific point in each family. The zero-mode interpolations of Appendix~\ref{sec:sampling} also extend to these families. Matrix coordinates give explicit many-body wavefunctions and verify their onsite parents. We then combine two $C=1$ layers into a four-component $C=2$ model supporting Halperin states.

\subsection{Two-component \texorpdfstring{$C=1$}{C=1} model}
\label{bands:sec:two-component}
\subsubsection{Single-particle construction}

In this subsection, we express the two-component sample of Eq.~\eqref{sample:eq:normalized-continuum-sample} through its component ratio and then introduce the variable weights. For a two-component Bloch spinor $[a,b]^{\mathsf T}$ with $a\ne0$, the ratio $r=b/a$ determines its band. Although $r$ can diverge at $a=0$, the band projector can remain smooth. We denote ratio representatives by $\widetilde v$ and regular theta vectors by $v$. The local ratio representation is
\begin{equation}
\widetilde v(\mathbf k)=\bigl[1,r(\mathbf k)\bigr]^{\mathsf T},\qquad P(\mathbf k)=\dfrac{\widetilde v(\mathbf k)\widetilde v^\dagger(\mathbf k)}{\widetilde v^\dagger(\mathbf k)\widetilde v(\mathbf k)}.
\label{bands:eq:two-component-ray}
\end{equation}
For a normalized local frame $u$, we define $\mathcal A_\mu=-\ii u^\dagger\partial_{k_\mu}u$ and $\Omega=\partial_{k_x}\mathcal A_y-\partial_{k_y}\mathcal A_x$. We take the positive BZ boundary orientation to be counterclockwise in the $(k_x,k_y)$ plane. The Chern number follows from the phase winding around poles of $r$~\cite{kohmoto1985}. The normalized local frames $u_{\mathrm{reg}}$ and $u_{\mathrm{pole}}$ obtained from $[1,r]^{\mathsf T}$ and $[1/r,1]^{\mathsf T}$ obey $u_{\mathrm{pole}}=e^{-\ii \arg r}u_{\mathrm{reg}}$. The Berry connection transforms as $\boldsymbol{\mathcal A}_{\mathrm{pole}}=\boldsymbol{\mathcal A}_{\mathrm{reg}}-\nabla_{\mathbf k}\arg r$. Stokes' theorem gives
\begin{equation}
C=\dfrac{1}{2\pi}\int_{\mathrm{BZ}}(\partial_{k_x}\mathcal A_y-\partial_{k_y}\mathcal A_x)\,\d^2k=-\dfrac{1}{2\pi}\sum_a\oint_{\gamma_a}\d\arg r.
\label{bands:eq:chern-pole-count}
\end{equation}
Here $\gamma_a$ encircles each pole counterclockwise, and the second frame is assumed smooth across it.\\

With $k_\pm=k_x\pm\ii k_y$ and $\partial_+=\partial_{k_x}+\ii\partial_{k_y}$, a periodic ratio satisfying $\partial_+r=0$ away from its poles is elliptic in $k_+$. Each order-$m$ pole contributes $m$ to $C$, so $C=1$ requires one simple pole per cell. As noted in Ref.~\cite{jian2013}, this is impossible for an elliptic function, whose residues sum to zero. A nonconstant elliptic ratio therefore has $C\ge2$, with $r=\wp_\Lambda(k_+)$ realizing $C=2$. Here $\Lambda=2\pi(\mathbb Z+\ii\mathbb Z)$; the relevant Weierstrass identities are collected in Sec.~\ref{special:sec:weierstrass}.\\

The sampled Dirac 1LL in Eq.~\eqref{sample:eq:normalized-continuum-sample} shows that a minimal relaxation of the elliptic-function condition already permits $C=1$. After a constant relative rescaling of its components, its periodic ratio $r=T^{(1)}_{\mathbf k}/T^{(0)}_{\mathbf k}$ satisfies
\begin{equation}
\partial_+r(\mathbf k)=-\dfrac{1}{2\pi}.
\label{bands:eq:constant-derivative-ansatz}
\end{equation}
Thus strict holomorphicity is replaced by a constant nonzero $\partial_+r$, whose coefficient corresponds to one residue-one pole. To make the nonholomorphic term explicit, we write this ratio away from the pole as
\begin{equation}
r(\mathbf k)=f(k_+)-\dfrac{k_-}{4\pi}.
\label{bands:eq:affine-antiholomorphic-solution}
\end{equation}
Here $f$ is meromorphic, and periodicity of $r$ requires
\begin{equation}
f(k_++2\pi)=f(k_+)+\dfrac12,\qquad f(k_++2\pi\ii)=f(k_+)-\dfrac{\ii}{2}.
\label{bands:eq:f-required-quasiperiods}
\end{equation}
Since $f$ need only be quasiperiodic, the elliptic residue constraint no longer excludes a single simple pole. The required quasiperiods are supplied by $f=\zeta_\Lambda$: its shifts under the two periods cancel those of $k_-/(4\pi)$, making $r(\mathbf k)=\zeta_\Lambda(k_+)-k_-/(4\pi)$ periodic. Near $\mathbf k=0$,
\begin{equation}
r(\mathbf k)=\dfrac1{k_+}-\dfrac{k_-}{4\pi}+O(k_+^3),\qquad \dfrac1{r(\mathbf k)}=k_++O(|\mathbf k|^3).
\label{bands:eq:r-pole-patch}
\end{equation}
Thus $[1/r,1]^{\mathsf T}$ extends smoothly to $[0,1]^{\mathsf T}$, making $P$ regular, and Eq.~\eqref{bands:eq:chern-pole-count} gives $C=1$.\\

To remove the coordinate pole from the Bloch vector, we use the theta representation from Eq.~\eqref{bands:eq:zeta-theta},
\begin{equation}
r(\mathbf k)=\dfrac12\dfrac{\vartheta_1'(k_+/2|\ii)}{\vartheta_1(k_+/2|\ii)}+\dfrac{\ii k_y}{2\pi}.
\label{bands:eq:r-theta}
\end{equation}
We cancel the denominator by multiplying $[1,r]^{\mathsf T}$ by $\vartheta_1(k_+/2|\ii)$:
\begin{equation}
v(\mathbf k)=\left[\vartheta_1(k_+/2|\ii),\ \dfrac12\vartheta_1'(k_+/2|\ii)+\dfrac{\ii k_y}{2\pi}\vartheta_1(k_+/2|\ii)\right]^{\mathsf T}.
\label{bands:eq:r-theta-spinor}
\end{equation}
Since $\vartheta_1$ has simple zeros, $v$ never vanishes.\\

We now vary the relative weight and mixing of the sampled components while preserving the constant-derivative structure of Eq.~\eqref{bands:eq:constant-derivative-ansatz}, introducing $\lambda>0$ and $\mu\in\mathbb R$:
\begin{equation}
\widetilde v^{(1/2)}_{\lambda,\mu}(\mathbf k)=\bigl[\lambda,r(\mathbf k)+\mu\bigr]^{\mathsf T}=\left[\lambda,\zeta_\Lambda(k_+)-\dfrac{k_-}{4\pi}+\mu\right]^{\mathsf T}.
\label{bands:eq:lambda-mu-weierstrass}
\end{equation}
Here $r$ remains the function in Eq.~\eqref{bands:eq:r-theta}, while the component ratio of the family is $(r+\mu)/\lambda$. Multiplying by $T^{(0)}_{\mathbf k}$ gives the regular vector
\begin{equation}
v^{(1/2)}_{\lambda,\mu}(\mathbf k)=\left[\lambda\vartheta_1(k_+/2|\ii),\ \dfrac12\vartheta_1'(k_+/2|\ii)+\left(\mu+\dfrac{\ii k_y}{2\pi}\right)\vartheta_1(k_+/2|\ii)\right]^{\mathsf T}.
\label{bands:eq:lambda-mu-theta}
\end{equation}
Let $v^{(1/2)}_{\rm D}(\mathbf k)=\phi^{\rm D,1/2}_{\mathbf k}(\mathbf0)$ be the direct Dirac sample in Eq.~\eqref{sample:eq:normalized-continuum-sample}. The Dirac component weights correspond to
\begin{equation}
\lambda_{\rm D}^{(1/2)}=\frac1{2\sqrt\pi}\simeq0.282095,\qquad \mu_{\rm D}^{(1/2)}=0.
\label{sample:eq:parameters}
\end{equation}
The component transformation relating the family to this sample is
\begin{equation}
v^{(1/2)}_{\lambda,\mu}(\mathbf k)\propto M_{1/2}v^{(1/2)}_{\rm D}(\mathbf k),\qquad M_{1/2}=\begin{bmatrix}\lambda/\lambda_{\rm D}^{(1/2)}&0\\\mu/\lambda_{\rm D}^{(1/2)}&-\ii\end{bmatrix}.
\label{bands:eq:two-prototype-deformation}
\end{equation}
At the Dirac parameters, $M_{1/2}=\operatorname{diag}(1,-\ii)$, so the band is precisely the directly sampled band of Appendix~\ref{sec:sampling} after a fixed internal phase change. Throughout the stated parameter range, $M_{1/2}$ remains invertible and preserves $C=1$. The proportionality absorbs a nowhere-zero, momentum-dependent scalar factor: the regular theta vector specifies the band, while $v^{(1/2)}_{\rm D}$ retains the normalization inherited from the continuum. Following Eq.~\eqref{sample:eq:normalized-lattice-bloch}, the transformed sample is normalized as $M_{1/2}v^{(1/2)}_{\rm D}/\|M_{1/2}v^{(1/2)}_{\rm D}\|$. Its norm remains bounded above and away from zero because the original sampling norm has these bounds and $M_{1/2}$ is constant and invertible.\\

For this family and the three- and four-component families below, a regular vector $v(\mathbf k)$ with $n_{\rm c}$ components and Chern number $C$ defines a flat-band Hamiltonian with Gaussian-decaying hopping:
\begin{equation}
h(\mathbf k)=w_C(\mathbf k)\bigl[(v^\dagger v)\one_{n_{\rm c}}-vv^\dagger\bigr],\qquad w_C(\mathbf k)=\sqrt{2C}\,e^{-Ck_y^2/(2\pi)}.
\label{bands:eq:gaussian-construction}
\end{equation}
The weight $w_C$ cancels the quasiperiodic amplitude of $vv^\dagger$, making $h$ periodic, with eigenvalue zero along $v$ and $w_Cv^\dagger v>0$ in its orthogonal complement. For two components, $w_1|T^{(0)}_{\mathbf k}|^2=F(\mathbf k)$ has Gaussian Fourier coefficients $a_{\mathbf R}$ by Eq.~\eqref{sample:eq:scalar-fourier}. Weighted bilinears involving $T^{(1)}_{\mathbf k}$ are finite linear combinations of $F$ and its momentum derivatives, whose Fourier coefficients are $a_{\mathbf R}$ times polynomials in $X,Y$. Varying $\lambda,\mu$ changes only their coefficients. Higher theta derivatives and shifted arguments in the models below likewise yield polynomial, phase, and parity factors multiplying $e^{-\pi(X^2+Y^2)/(2C)}$. Thus all these families have Gaussian-decaying, rather than finite-range, hopping.\\

As an explicit example, choose $(\lambda,\mu)=(1,0)$, so $v=v^{(1/2)}_{1,0}$ and the band energies are $E_0=0$ and $E_1=F(1+|r|^2)$. Equation~\eqref{bands:eq:gaussian-construction} then Fourier transforms to
\begin{equation}
h_{\mathbf R}=\int_{\mathrm{BZ}}\dfrac{\d^2k}{(2\pi)^2}\,e^{\ii \mathbf k\cdot\mathbf R}h(\mathbf k)=a_{\mathbf R}\begin{bmatrix}\dfrac14\bigl(\pi^{-1}-X^2-Y^2\bigr)&\dfrac{\ii}{2}(X+\ii Y)\\\dfrac{\ii}{2}(X-\ii Y)&1\end{bmatrix}.
\label{bands:eq:gaussian-hopping}
\end{equation}
Here $a_{\mathbf R}$ is defined in Eq.~\eqref{sample:eq:sampled-envelope}, and $h_{-\mathbf R}=h_{\mathbf R}^\dagger$. The upper band remains positive at the zero of $F$, with $E_1(0)=[\vartheta_1'(0|\ii)]^2/(2\sqrt2)$.

\subsubsection{Many-body states and onsite interactions}
\label{bands:sec:two-many-body}

Because $M_{1/2}$ acts only on internal components, it commutes with lattice sampling and preserves vanishing pair-coincidence amplitudes. Applying it to each particle therefore carries the directly sampled onsite zero modes into the $(\lambda,\mu)$ family.\\

The same transformation conjugates the Dirac matrix coordinate in Eq.~\eqref{cont:eq:half-matrix-coordinate}, while sampling replaces $z$ by $Z=X+\ii Y$. Writing $v=v^{(1/2)}_{\lambda,\mu}$, the resulting lattice coordinate is
\begin{equation}
W=Z\one_2+J,\qquad J=\frac{\ii}{2\pi\lambda}|1\rangle\langle0|,\qquad J^2=0.
\label{bands:eq:two-matrix-coordinate}
\end{equation}
Here $J=M_{1/2}J_{\rm D}^{(1/2)}M_{1/2}^{-1}$. To verify band preservation, let $P$ project onto the band and $Q=\one-P$. With our Fourier convention, multiplication by $Z$ acts as $\ii\partial_+$ in momentum space. The relation $\partial_+r=-1/(2\pi)$ ensures that the regular vector in Eq.~\eqref{bands:eq:lambda-mu-theta} satisfies $(\ii\partial_++J)v=0$. Any state in the band has the form $f(\mathbf k)v(\mathbf k)$, so
\begin{equation}
QW(fv)=\ii(\partial_+f)Qv=0\qquad\Longrightarrow\qquad QWP=0.
\label{bands:eq:two-band-preservation}
\end{equation}
Thus $W$ preserves the band, just as $z$ preserves the LLL. The holomorphic-factor identity of Sec.~\ref{cont:sec:dirac-half} is preserved by the same conjugation, so sampling gives the substitution $z_i\to W_i$ in many-body holomorphic factors.\\

In the sampling formulas below, the continuum states and maps $\mathcal U_s$ are expressed in the gauge of Eq.~\eqref{sample:eq:gauge}. Let $\Psi_1$ be the unnormalized image of $\Phi_1$ in Eq.~\eqref{cont:eq:slater} under $(M_{1/2}\mathcal U_{1/2})^{\otimes N_e}$ followed by lattice sampling. Applying this substitution to the Jastrow factors in Eq.~\eqref{cont:eq:laughlin} gives
\begin{equation}
\Psi_{1/3}=\left.(M_{1/2}\mathcal U_{1/2})^{\otimes N_e}\Phi_{1/3}\right|_{\mathbf r_i=\mathbf R_i}=\prod_{i<j}(W_i-W_j)^2\Psi_1.
\label{bands:eq:two-laughlin}
\end{equation}
Here $W_i$ acts on particle $i$; the squared factors preserve fermionic antisymmetry and keep the state in the zero-energy band.\\

To verify the onsite parent of Eq.~\eqref{sample:eq:onsite-interpolation-endpoint} throughout this family, we examine the internal channels at coincidence. Since all pair factors commute, we can apply the factor for a chosen pair $i,j$ last. At $\mathbf R_i=\mathbf R_j$, its scalar coordinates cancel:
\begin{equation}
\left.(W_i-W_j)^2\right|_{\mathbf R_i=\mathbf R_j} =(J_i-J_j)^2=-2J_iJ_j =\frac{1}{2\pi^2\lambda^2}|11\rangle\langle00|.
\label{bands:eq:two-contact-zero}
\end{equation}
Here $|\alpha\beta\rangle=|\alpha\rangle_i\otimes|\beta\rangle_j$ labels the two particles' internal components. The output has no $01$ or $10$ component, so $\Psi_{1/3}$ has no opposite-component pair at any site. These are precisely the channels penalized by the onsite repulsion, whose first-quantized form is
\begin{equation}
H_{1/3}=U\sum_{i<j}\delta_{\mathbf R_i,\mathbf R_j} \bigl(|01\rangle\langle01|+|10\rangle\langle10|\bigr)_{ij},\qquad U>0.
\label{bands:eq:two-contact-interaction}
\end{equation}
The absence of opposite-component pairs gives $c_{\mathbf R,1}c_{\mathbf R,0}\Psi_{1/3}=0$ at every site. In the notation of Sec.~\ref{sample:sec:norm}, the same onsite interaction therefore satisfies
\begin{equation}
H_{1/3}=U\sum_{\mathbf R}n_{\mathbf R,0}n_{\mathbf R,1},\qquad H_{1/3}\Psi_{1/3}=0.
\label{bands:eq:two-many-body-parent}
\end{equation}
Positivity makes this local pair-annihilation condition necessary and sufficient for zero Hubbard energy. Since fermionic antisymmetry also excludes same-component coincidences, Hubbard zero modes vanish whenever two particles occupy the same site. The constructed states are exact zero-energy ground states of the full nonnegative Hamiltonian for any $U>0$, since both its single-particle and interaction parts annihilate them.\\

On a rectangular torus, the constant invertible $M_{1/2}$ preserves boundary twists and, together with the injective sampling of Sec.~\ref{sample:sec:norm}, linear independence. The three continuum states in Eq.~\eqref{cont:eq:torus-laughlin} therefore yield three independent lattice Hubbard zero modes. For integer $L_x,L_y$ with $L_xL_y=3N_e$, the sampling gauge of Eq.~\eqref{sample:eq:gauge} gives $\theta_x^{\rm lat}=\theta_x+\pi L_x$ and $\theta_y^{\rm lat}=\theta_y-\pi L_y$ modulo $2\pi$. Choosing $\theta_x=-\pi L_x$ and $\theta_y=\pi L_y$ gives the periodic lattice states
\begin{equation}
\Psi_{1/3,\gamma}=\left.(M_{1/2}\mathcal U_{1/2})^{\otimes N_e}\Phi_{1/3,\gamma}\right|_{\mathbf r_i=\mathbf R_i},\qquad \gamma=0,1,2.
\label{sample:eq:torus-laughlin-samples}
\end{equation}
We apply the same coordinate substitution to the full holomorphic torus Laughlin factor, including its center-of-mass theta function. The cubic pair factor includes the zero already contained in $\Psi_1$ in Eq.~\eqref{bands:eq:two-laughlin}. With $\tau=\ii L_y/L_x$ and $\psi^{(1/2)}_0$ from Eq.~\eqref{sample:eq:localized-sample}, the explicit states are
{\small
\begin{equation}
\Psi_{1/3,\gamma}=e^{\sum_iW_i^2/(4\ell_B^2)}\vartheta\!\begin{bmatrix}\gamma/3+(N_e-1)/2-L_x/6\\(N_e-1)/2-L_y/2\end{bmatrix}\!\left(\frac3{L_x}\sum_iW_i\Big|3\tau\right)\prod_{i<j}\vartheta_1\!\left(\frac{\pi(W_i-W_j)}{L_x}\Big|\tau\right)^3\bigotimes_i\left[M_{1/2}\psi^{(1/2)}_0(\mathbf R_i)\right].
\label{sample:eq:torus-laughlin-explicit}
\end{equation}
}\par\noindent
Each $W_i$ acts on particle $i$, with the identity on the others; its powers are matrix products, and distinct $W_i$ commute. The exponential and theta functions have finite Taylor expansions in the nilpotent $J_i$. An expression analogous to Eq.~\eqref{sample:eq:torus-laughlin-explicit} can be written for an arbitrary torus.\\

These states satisfy
\begin{equation}
H_{1/3}\Psi_{1/3,\gamma}=0,\qquad \gamma=0,1,2.
\label{sample:eq:torus-onsite-zero-modes}
\end{equation}
Since the states lie in the zero-energy band, they are three independent zero-energy ground states of the model. At the Dirac parameters in Eq.~\eqref{sample:eq:parameters}, these reduce to the directly sampled Dirac Laughlin states up to fixed internal phases. The inverse transformation $(M_{1/2}^{-1})^{\otimes N_e}$ also preserves vanishing at pair coincidence and maps every band state back to the directly sampled band. It therefore gives a bijection between the two onsite zero-mode spaces. The completeness result of Sec.~\ref{sample:sec:laughlin-completeness} consequently holds throughout this parameter family on rectangular periodic tori with $N_s=3N_e$ and $N_e\ge13$.

\subsection{Three-component \texorpdfstring{$C=1$}{C=1} model}
\label{bands:sec:three-component}
\subsubsection{Single-particle construction}

In this subsection, we extend the same differential structure to the sampled spin-1 Dirac 2LL and vary its component weights. For the two-component model, Eq.~\eqref{bands:eq:constant-derivative-ansatz} closes differentiation within the two components:
\begin{equation}
\partial_+\left[1,r\right]^{\mathsf T}=-\dfrac1{2\pi}\begin{bmatrix}0&0\\1&0\end{bmatrix}\left[1,r\right]^{\mathsf T}.
\label{bands:eq:two-component-differential-chain}
\end{equation}
The additional sampled component, $t(\mathbf k)=T^{(2)}_{\mathbf k}/T^{(0)}_{\mathbf k}$ with $T^{(2)}_{\mathbf k}$ from Eq.~\eqref{sample:eq:third-theta}, extends this chain to three components:
\begin{equation}
\partial_+\left[1,r,t\right]^{\mathsf T}=-\dfrac1{2\pi}\begin{bmatrix}0&0&0\\1&0&0\\0&4&0\end{bmatrix}\left[1,r,t\right]^{\mathsf T}.
\label{bands:eq:three-component-differential-chain}
\end{equation}
The Weierstrass representation of $t$ is
\begin{equation}
t(\mathbf k)=2\bigl[r(\mathbf k)^2-\wp_\Lambda(k_+)\bigr].
\label{bands:eq:three-component-chain}
\end{equation}
The $\wp_\Lambda$ term cancels the double pole of $r^2$ without changing its $\partial_+$ derivative. Varying the component weights preserves this closure under differentiation and gives the family
\begin{equation}
v^{(1)}_{\lambda,\mu}(\mathbf k)=\left[\lambda^2T^{(0)}_{\mathbf k},\ -2\mu T^{(1)}_{\mathbf k},\ T^{(2)}_{\mathbf k}\right]^{\mathsf T},\qquad \lambda>0,\quad\mu\in\mathbb R\setminus\{0\}.
\label{eq:three-component-regular}
\end{equation}
Let $v^{(1)}_{\rm D}(\mathbf k)=\phi^{\rm D,1}_{\mathbf k}(\mathbf0)$ be the direct sample in Eq.~\eqref{sample:eq:normalized-continuum-sample}. The Dirac component weights correspond to
\begin{equation}
\lambda_{\rm D}^{(1)}=\frac1{\sqrt{2\pi}}\simeq0.398942,\qquad \mu_{\rm D}^{(1)}=\frac{\sqrt3}{2\sqrt\pi}\simeq0.488603.
\label{bands:eq:three-sampling-parameters}
\end{equation}
The corresponding component transformation is
\begin{equation}
v^{(1)}_{\lambda,\mu}(\mathbf k)\propto M_1v^{(1)}_{\rm D}(\mathbf k),\qquad M_1=\operatorname{diag}\!\left[\frac{\lambda^2}{\bigl(\lambda_{\rm D}^{(1)}\bigr)^2},\frac{\ii\mu}{\mu_{\rm D}^{(1)}},-1\right].
\label{bands:eq:three-prototype-deformation}
\end{equation}
At these Dirac parameters, $M_1=\operatorname{diag}(1,\ii,-1)$, recovering the three-component sample of Appendix~\ref{sec:sampling} up to fixed internal phases. Throughout the stated range, $M_1$ is invertible and commutes with sampling, so the same construction preserves $C=1$ and the pair- and triple-coincidence zeros. The normalized transformed sample is $M_1v^{(1)}_{\rm D}/\|M_1v^{(1)}_{\rm D}\|$, following Eq.~\eqref{sample:eq:one-sampling-normalization}. The regular theta vector is smooth and nowhere zero: its first component is nonzero away from the theta zeros, and its second is nonzero at them. The Gaussian Hamiltonian follows from Eq.~\eqref{bands:eq:gaussian-construction}:
\begin{equation}
h^{(1)}_{\lambda,\mu}(\mathbf k)=w_1(\mathbf k)\bigl[\|v^{(1)}_{\lambda,\mu}(\mathbf k)\|^2\one_3-v^{(1)}_{\lambda,\mu}(\mathbf k)v^{(1)\dagger}_{\lambda,\mu}(\mathbf k)\bigr].
\label{bands:eq:three-gaussian-parent}
\end{equation}
Its hopping entries are $a_{\mathbf R}$ times polynomials of degree at most four, since each component contains at most two theta derivatives. The zero-energy band is isolated from the two positive-energy bands.

\subsubsection{Many-body states and onsite interactions}
\label{bands:sec:three-many-body}

The differential relations in Eqs.~\eqref{bands:eq:two-component-differential-chain} and \eqref{bands:eq:three-component-differential-chain} determine the matrix coordinate for the three-component family as in Sec.~\ref{bands:sec:two-many-body}. Write $v=v^{(1)}_{\lambda,\mu}$, let $P$ be its band projector, and set $Q=\one-P$. The lattice coordinate is
\begin{equation}
W=Z\one_3+J,\qquad J=-\frac{\ii\mu}{\pi\lambda^2}|1\rangle\langle0| -\frac{\ii}{\pi\mu}|2\rangle\langle1|,\qquad J^3=0.
\label{bands:eq:three-matrix-coordinate}
\end{equation}
Here $J=M_1J_{\rm D}^{(1)}M_1^{-1}$, with $J_{\rm D}^{(1)}$ given in Eq.~\eqref{cont:eq:one-matrix-coordinate}. The regular vector satisfies $(\ii\partial_++J)v=0$, and Eq.~\eqref{bands:eq:two-band-preservation} gives $QWP=0$.\\

Using the same sampling gauge, let $\Psi_1$ be the unnormalized image of $\Phi_1$ under $(M_1\mathcal U_1)^{\otimes N_e}$ followed by lattice sampling. Applying the same substitution to the scalar Laughlin state gives
\begin{equation}
\Psi_{1/5}=\left.(M_1\mathcal U_1)^{\otimes N_e}\Phi_{1/5}\right|_{\mathbf r_i=\mathbf R_i}=\prod_{i<j}(W_i-W_j)^4\Psi_1.
\label{bands:eq:three-laughlin}
\end{equation}
The fourth powers preserve fermionic antisymmetry and, together with the Slater factor, give filling $1/5$.\\

To verify the onsite parent of Eq.~\eqref{sample:eq:fifth-onsite-interaction} throughout this family, apply the factor for a chosen pair last and set $\mathbf R_i=\mathbf R_j$. Since $J^3=0$,
\begin{equation}
\left.(W_i-W_j)^4\right|_{\mathbf R_i=\mathbf R_j} =(J_i-J_j)^4=6J_i^2J_j^2 =\frac{6}{\pi^4\lambda^4}|22\rangle\langle00|.
\label{bands:eq:three-contact-zero}
\end{equation}
The output has no pair in the $01$, $02$, or $12$ channels, so the same three-channel onsite repulsion remains a parent:
\begin{equation}
H_{1/5}=U\sum_{\mathbf R} \bigl(n_{\mathbf R,0}n_{\mathbf R,1}+n_{\mathbf R,0}n_{\mathbf R,2}+n_{\mathbf R,1}n_{\mathbf R,2}\bigr),\qquad U>0,\qquad H_{1/5}\Psi_{1/5}=0.
\label{bands:eq:three-pair-parent}
\end{equation}
Thus $\Psi_{1/5}$ is a zero-energy ground state of the full nonnegative Hamiltonian.\\

The same substitution in the scalar Moore--Read state of Eq.~\eqref{cont:eq:mr-wavefunction} gives, for even $N_e$,
\begin{equation}
\Psi_{\mathrm{MR}}= \left[\operatorname{Pf}\!\left(\frac{1}{z_i-z_j}\right) \prod_{i<j}(z_i-z_j)\right]_{z_i\rightarrow W_i}\Psi_1.
\label{bands:eq:three-mr-state}
\end{equation}
The Pfaffian denominators cancel against the pair factors before $z_i\to W_i$, leaving a symmetric polynomial whose substitution preserves the zero-energy band. The triple-coincidence zeros of Sec.~\ref{cont:sec:dirac-mr} survive $M_1$ and sampling, so the onsite parent in Eq.~\eqref{sample:eq:mr-onsite-interaction} also applies throughout this family:
\begin{equation}
H_{\mathrm{MR}}=U\sum_{\mathbf R}n_{\mathbf R,0}n_{\mathbf R,1}n_{\mathbf R,2},\qquad U>0,\qquad H_{\mathrm{MR}}\Psi_{\mathrm{MR}}=0.
\label{bands:eq:three-mr-parent}
\end{equation}
Thus $\Psi_{\mathrm{MR}}$ is a zero-energy ground state of the full nonnegative Hamiltonian. Zero-mode completeness on the torus is supported by the numerical results in Sec.~\ref{num:sec:ed-moore-read}; an analytic proof remains open.\\

The Laughlin $1/3$, Laughlin $1/5$, and Moore--Read onsite zero modes above also admit the interpolations of Appendix~\ref{sec:sampling} at each fixed parameter choice. The constant invertible component transformations commute with sampling and preserve many-body coincidence zeros. Restoring continuum normalization only rescales each transformed Bloch orbital, leaving its normalized lattice sample unchanged. The resulting continuum subspaces therefore sample to the same lattice families and supply contact zero modes, so the interpolation construction applies with their corresponding sampling norms. Its $t=1$ endpoint has the spectrum of contact repulsion projected into the corresponding transformed continuum subspace.

\subsection{Four-component \texorpdfstring{$C=2$}{C=2} model}
\label{bands:sec:four-component}
\subsubsection{Single-particle construction}

We first construct a $C=1$ layer $A$ with a $2\times1$ primitive cell labeled by $(X',Y')\in\mathbb Z^2$. Its four components are two orbitals at each original site $(X,Y)=(2X',Y')$ and $(2X'+1,Y')$, in that order. For the two orbital blocks, define
\begin{equation}
T^{(0)}_{\mathbf k,j}=\vartheta_{3-j}(k_+|2\ii),\qquad T^{(1)}_{\mathbf k,j}=\vartheta_{3-j}'(k_+|2\ii)+\frac{\ii k_y}{\pi}T^{(0)}_{\mathbf k,j},\qquad j=0,1.
\label{bands:eq:four-theta-components}
\end{equation}
With cell Bloch factor $e^{\ii (2k_xX'+k_yY')}$ and $\lambda>0$, we take on $[0,\pi)\times[0,2\pi)$
\begin{equation}
v_A(\mathbf k)=\left[\lambda T^{(0)}_{\mathbf k,0},\ T^{(1)}_{\mathbf k,0},\ e^{\ii k_x}\lambda T^{(0)}_{\mathbf k,1},\ e^{\ii k_x}T^{(1)}_{\mathbf k,1}\right]^{\mathsf T}.
\label{bands:eq:c2-layer-spinor}
\end{equation}
The factor $e^{\ii k_x}$ accounts for the second site's displacement from the cell origin. The theta functions have no common zero, so $v_A$ never vanishes. Their translation identities give the normalized vector $u_A$ the boundary conditions
\begin{equation}
u_A(k_x+\pi,k_y)=u_A(k_x,k_y),\qquad u_A(k_x,k_y+2\pi)=e^{-2\ii k_x}u_A(k_x,k_y).
\label{bands:eq:c2-layer-sewing}
\end{equation}
The boundary phase winds once clockwise over $0\leq k_x<\pi$. For the normalized layer frame $u_A$,
\begin{equation}
C_A=\dfrac1{2\pi}\int_0^\pi\bigl[\mathcal A_{A,x}(k_x,0)-\mathcal A_{A,x}(k_x,2\pi)\bigr]\,\d k_x=1.
\label{bands:eq:c2-layer-chern}
\end{equation}
We add layer $B$ as an orthogonal copy of $A$ translated by one site along $x$, so $C_B=1$. Translation by one site exchanges the layers, reducing the combined system's $x$ period from two to one and giving a $1\times1$ primitive cell~\cite{wu2013}. Each site now has four orbitals: $A$ occupies components $0,1$ at even $X$ and $2,3$ at odd $X$, while $B$ occupies the complementary pair. We accordingly redefine the Fourier transform using the one-site cell, with Brillouin zone $[0,2\pi)^2$:
\begin{equation}
\psi_\alpha(\mathbf k)=\sum_{X,Y}e^{-\ii (k_xX+k_yY)}\psi_\alpha(X,Y),\qquad \mathbf k\in[0,2\pi)^2.
\label{bands:eq:c2-site-fourier}
\end{equation}
The eigenstates of one-site translation combine the two layer states at each reduced-zone momentum and carry momenta $\mathbf k$ and $\mathbf k+(\pi,0)$ in the enlarged Brillouin zone, forming a single band. In this convention, the position factor $e^{\ii k_x}$ in Eq.~\eqref{bands:eq:c2-layer-spinor} is included in the site Bloch phase $e^{\ii(k_xX+k_yY)}$. Factoring out this phase gives the regular theta vector
\begin{equation}
v^{(4)}_\lambda(\mathbf k)=\left[\lambda T^{(0)}_{\mathbf k,0},\ T^{(1)}_{\mathbf k,0},\ \lambda T^{(0)}_{\mathbf k,1},\ T^{(1)}_{\mathbf k,1}\right]^{\mathsf T}.
\label{bands:eq:c2-spinor}
\end{equation}
We write $u=u^{(4)}_\lambda$ for its normalized frame and $P$ for its projector. The theta translation identities in the one-site basis involve the matrices
\begin{equation}
D=\operatorname{diag}[1,1,-1,-1],\qquad S=\begin{bmatrix}0&0&1&0\\0&0&0&1\\1&0&0&0\\0&1&0&0\end{bmatrix}.
\label{bands:eq:c2-orbital-matrices}
\end{equation}
Then
\begin{equation}
u(k_x+\pi,k_y)=D u(k_x,k_y),\qquad u(k_x,k_y+\pi)=e^{-\ii k_x}S u(k_x,k_y).
\label{bands:eq:c2-half-sewing}
\end{equation}
Since $D$ and $S$ are constant unitary matrices, $\Omega(k_x+\pi,k_y)=\Omega(k_x,k_y+\pi)=\Omega(\mathbf k)$.\\

Fourier transforming $P(k_x+\pi,k_y)=D P(k_x,k_y)D$ gives, for original sites $\mathbf R_a=(X_a,Y_a)$,
\begin{equation}
(P)_{\mathbf R_1\alpha,\mathbf R_2\beta}=0\quad\text{unless}\quad(-1)^{X_1}D_{\alpha\alpha}=(-1)^{X_2}D_{\beta\beta},\qquad \alpha,\beta=0,1,2,3.
\label{bands:eq:c2-layer-selection}
\end{equation}
The conserved sign labels layer $A$ $(+1)$ or $B$ $(-1)$, so $P$ preserves each layer. To retain this layer assignment across torus boundaries, each period vector must have an even $x$ component.\\

To evaluate the Chern number in the one-site cell, we use the full-period relations
\begin{equation}
u(k_x+2\pi,k_y)=u(k_x,k_y),\qquad u(k_x,k_y+2\pi)=e^{-2\ii k_x}u(k_x,k_y).
\label{bands:eq:c2-full-sewing}
\end{equation}
The boundary phase winds twice clockwise, giving
\begin{equation}
C=\dfrac1{2\pi}\int_0^{2\pi}\bigl[\mathcal A_x(k_x,0)-\mathcal A_x(k_x,2\pi)\bigr]\,\d k_x=2.
\label{bands:eq:c2-chern}
\end{equation}
Each fixed component pair $(0,1)$ or $(2,3)$, normalized separately, also has $C=2$ over this Brillouin zone; the $C=1$ layers follow the site-dependent assignment in Eq.~\eqref{bands:eq:c2-layer-selection}.\\

Equation~\eqref{bands:eq:gaussian-construction} with $C=2$ gives
\begin{equation}
h^{(4)}_\lambda(\mathbf k)=w_2(\mathbf k)\bigl[\|v^{(4)}_\lambda(\mathbf k)\|^2\one_4-v^{(4)}_\lambda(\mathbf k)v^{(4)\dagger}_\lambda(\mathbf k)\bigr].
\label{bands:eq:four-gaussian-parent}
\end{equation}
Its hopping entries have the envelope $e^{-\pi(X^2+Y^2)/4}$ multiplied by polynomials of degree at most two and orbital-dependent lattice phases and parity factors. The Hamiltonian preserves the layer assignment in Eq.~\eqref{bands:eq:c2-layer-selection} and has one zero-energy band separated from three positive-energy bands.

\subsubsection{Many-body states and onsite interactions}
\label{bands:sec:four-many-body}

We label a local state by $|a,s\rangle$, with layer $a=A,B$ and internal component $s=0,1$. The correspondence with orbital numbers at even $X$ is
\begin{equation}
(|A,0\rangle,|A,1\rangle,|B,0\rangle,|B,1\rangle) =(|0\rangle,|1\rangle,|2\rangle,|3\rangle).
\label{bands:eq:four-layer-labels}
\end{equation}
At odd $X$, the $A$ and $B$ assignments are exchanged. We use layer labels in the wavefunctions and orbital numbers in the onsite interactions below.\\

Write $v=v^{(4)}_\lambda$ and set $Q=\one-P$. The theta derivatives in Eq.~\eqref{bands:eq:four-theta-components} lead us to define
\begin{equation}
W=Z\one_4+J,\qquad J=\frac{\ii}{\pi\lambda}\bigl(|1\rangle\langle0|+|3\rangle\langle2|\bigr),\qquad J^2=0.
\label{bands:eq:four-matrix-coordinate}
\end{equation}
Here $(\ii\partial_++J)v=0$ again ensures $QWP=0$. The layer projectors $\Pi_A(\mathbf R)=[\one_4+(-1)^X D]/2$ and $\Pi_B=\one_4-\Pi_A$ commute with both $P$ and $W$. Thus $W$ also preserves each layer separately, with restrictions
\begin{equation}
W_a=Z\one_2+J_a,\qquad J_a=\frac{\ii}{\pi\lambda}|a,1\rangle\langle a,0|,\qquad J_a^2=0,\qquad a=A,B.
\label{bands:eq:four-layer-coordinate}
\end{equation}

Since $W_a$ preserves each layer, we first construct a Laughlin state within each layer and then add interlayer factors. For the droplet construction, choose the unnormalized localized orbital
\begin{equation}
\psi_0^A(\mathbf R)=e^{-\pi(X^2+Y^2)/4+\ii\pi XY/2}\left[\lambda|A,0\rangle-\frac{\ii}{2}(X-\ii Y)|A,1\rangle\right],\qquad \psi_0^B=T_x\psi_0^A.
\label{bands:eq:four-layer-seed}
\end{equation}
The layer kets follow the site-dependent assignment above, and $T_x$ translates by one site in the $x$ direction. The cell Fourier transform of $\psi_0^A$ is proportional to $v_A(\mathbf k)$ in Eq.~\eqref{bands:eq:c2-layer-spinor}, so each seed lies in its layer band. For even $N_e$, we take $N_e/2$ particles in each layer. Let $\Psi_1^a$ be the Slater determinant of the orbitals $\psi_m^a=W_a^m\psi_0^a$, $m=0,\ldots,N_e/2-1$, in layer $a=A,B$. Their component-$0$ amplitudes contain $Z^m$, supplying one Vandermonde factor in the all-$0$ component. As in Eq.~\eqref{bands:eq:two-laughlin}, two additional factors per intralayer pair give
\begin{equation}
\Psi_{1/3}^a=\prod_{i<j\in a}(W_{a,i}-W_{a,j})^2\Psi_1^a,\qquad a=A,B.
\label{bands:eq:four-layer-laughlin}
\end{equation}
Multiplying by interlayer factors gives the balanced Halperin $33h$ states~\cite{Halperin1983,BarkeshliQi2012,WuJainSun2015}:
\begin{equation}
\Psi_{33h}=\mathcal A_{\mathrm F}\left[\prod_{i\in A,\,j\in B}(W_{A,i}-W_{B,j})^h\,\Psi_{1/3}^A\Psi_{1/3}^B\right],\qquad h=0,1,2.
\label{bands:eq:four-halperin}
\end{equation}
Here $\mathcal A_{\mathrm F}$ antisymmetrizes over all particles, including exchanges between layers. For $h=0$, this is the antisymmetrized product of the two layer Laughlin states. The intralayer and interlayer exponents $3$ and $h$ give the bilayer filling $2/(3+h)$, counting the total particle number per flux quantum in either layer. The full $C=2$ band contains twice as many single-particle states as one layer, so its filling is $1/(3+h)$. Since the matrix coordinates preserve the zero-energy band, the single-particle energy vanishes. The onsite parents follow from the coincidence conditions below.\\

On a torus, a period with an odd $x$ component exchanges the layers, joining the two layer subspaces through the boundary identification. The corresponding boundary conditions and zero-mode counts are discussed in Sec.~\ref{num:sec:ed-halperin}.\\

All three states share the same intralayer squared factor. For two particles in the same layer, it reduces to $(J_{a,i}-J_{a,j})^2=-2J_{a,i}J_{a,j}$ at coincidence. As in Sec.~\ref{bands:sec:two-many-body}, its output has both particles in internal component $1$ and no mixed-component pair. Thus the physical intralayer double occupancies $01$ and $23$ are absent in all three states.\\

The distinction between the three states comes from the interlayer contact factor $\Delta J^h$, where $\Delta J=J_{A,i}-J_{B,j}$. We write $|A,s_i;B,s_j\rangle=|A,s_i\rangle_i|B,s_j\rangle_j$. Then
\begin{equation}
\Delta J|A,0;B,0\rangle=\frac{\ii}{\pi\lambda}\bigl(|A,1;B,0\rangle-|A,0;B,1\rangle\bigr),\qquad (\Delta J)^2=\frac{2|A,1;B,1\rangle\langle A,0;B,0|}{\pi^2\lambda^2}.
\label{bands:eq:four-contact-action}
\end{equation}
The first power also maps $|A,0;B,1\rangle$ and $|A,1;B,0\rangle$ to multiples of $|A,1;B,1\rangle$ and annihilates the latter. Thus $h=0$ allows all interlayer pairs, $h=1$ retains their antisymmetric mixed-component combination and the $11$ pair, and $h=2$ retains only the $11$ pair.\\

For the onsite interactions, use the Fock states $|\alpha\beta\rangle=c_{\mathbf R,\alpha}^\dagger c_{\mathbf R,\beta}^\dagger|\mathrm{vac}\rangle$, with $\alpha<\beta$. The contact amplitudes are confined to the following onsite pair subspaces
\begin{equation}
\begin{array}{c|l} \text{state}&\text{allowed onsite pair space}\\\hline \Psi_{330}&\operatorname{span}\{|02\rangle,|03\rangle,|12\rangle,|13\rangle\}\\ \Psi_{331}&\operatorname{span}\{|12\rangle-|03\rangle,|13\rangle\}\\ \Psi_{332}&\operatorname{span}\{|13\rangle\} \end{array}.
\label{bands:eq:four-contact-spaces}
\end{equation}
These spaces are unchanged when the layers exchange at odd $X$, so the same constraints hold at every site and after full antisymmetrization.\\

We now choose nonnegative onsite interactions that annihilate these contact amplitudes. The intralayer density penalties are common to all three states. For $331$, Eq.~\eqref{bands:eq:four-contact-spaces} gives two independent interlayer zero conditions: the $02$ amplitude vanishes, and the $03$ and $12$ amplitudes sum to zero. The antisymmetric mixed-component combination remains allowed. The first condition alone gives
\begin{equation}
c_{\mathbf R,2}c_{\mathbf R,0}\Psi_{331}=0.
\label{bands:eq:four-331-density-zero}
\end{equation}
Thus a positive $02$ density penalty already supplies an onsite parent. For $332$, only $13$ survives, so all three interlayer density penalties $02$, $03$, and $12$ annihilate the state.\\

These are precisely the density channels selected by the continuum contacts in Eqs.~\eqref{cont:eq:halperin-expanded-330}--\eqref{cont:eq:halperin-density-only}. Sampling preserves their vanishing coincidence amplitudes. For two distinct local states, the lattice pair penalty is
\begin{equation}
(c_{\mathbf R,b,t}c_{\mathbf R,a,s})^\dagger(c_{\mathbf R,b,t}c_{\mathbf R,a,s})=n_{\mathbf R,a,s}n_{\mathbf R,b,t},\qquad (a,s)\ne(b,t).
\label{bands:eq:four-sampled-density}
\end{equation}
Here $c_{\mathbf R,a,s}$ annihilates the layer component $|a,s\rangle$, with $0,1$ corresponding to $\uparrow,\downarrow$ in the main text. The intralayer penalties become $n_{\mathbf R,0}n_{\mathbf R,1}+n_{\mathbf R,2}n_{\mathbf R,3}$. For the interlayer terms, the site-dependent layer assignment gives, at both even and odd $X$,
\begin{equation}
n_{\mathbf R,A,0}n_{\mathbf R,B,0}=n_{\mathbf R,0}n_{\mathbf R,2},\qquad n_{\mathbf R,A,0}n_{\mathbf R,B,1}+n_{\mathbf R,A,1}n_{\mathbf R,B,0}=n_{\mathbf R,0}n_{\mathbf R,3}+n_{\mathbf R,1}n_{\mathbf R,2}.
\label{bands:eq:four-layer-density-identities}
\end{equation}
With an independent lattice strength $U>0$, the main-text layer interactions become
\begin{subequations}
\label{bands:eq:four-onsite-parents}
\begin{equation}
H_{330}=U\sum_{\mathbf R}\bigl[n_{\mathbf R,0}n_{\mathbf R,1}+n_{\mathbf R,2}n_{\mathbf R,3}\bigr],
\end{equation}
\begin{equation}
H_{331}=H_{330}+U\sum_{\mathbf R}n_{\mathbf R,0}n_{\mathbf R,2},
\end{equation}
\begin{equation}
H_{332}=H_{331}+U\sum_{\mathbf R}\bigl[n_{\mathbf R,0}n_{\mathbf R,3}+n_{\mathbf R,1}n_{\mathbf R,2}\bigr].
\end{equation}
\end{subequations}
Each density term annihilates the corresponding state, so
\begin{equation}
H_{33h}\Psi_{33h}=0,\qquad h=0,1,2.
\label{bands:eq:four-many-body-zero}
\end{equation}
Together with the vanishing single-particle energy, this makes each constructed state a zero-energy ground state of the nonnegative single-particle and onsite interaction terms. This zero-mode proof does not require a separate penalty for every absent contact component. The boundary-dependent zero-mode counts are tested numerically in Sec.~\ref{num:sec:ed-halperin}; an analytic proof of completeness remains open.\\

For $331$, the second zero condition also permits an optional positive penalty for the symmetric mixed-component combination. Its annihilator is
\begin{equation}
Q_{\mathbf R}=\frac{c_{\mathbf R,3}c_{\mathbf R,0}+c_{\mathbf R,2}c_{\mathbf R,1}}{\sqrt2},\qquad Q_{\mathbf R}\Psi_{331}=0.
\label{bands:eq:four-q-channel}
\end{equation}
Hence $H_{331}+\kappa\sum_{\mathbf R}Q_{\mathbf R}^\dagger Q_{\mathbf R}$, with $\kappa\ge0$, is also a parent; the pure-density choice above has $\kappa=0$. The optional term expands as
\begin{equation}
Q_{\mathbf R}^\dagger Q_{\mathbf R}=\frac12\bigl[n_{\mathbf R,0}n_{\mathbf R,3}+n_{\mathbf R,1}n_{\mathbf R,2}+c_{\mathbf R,0}^\dagger c_{\mathbf R,3}^\dagger c_{\mathbf R,2}c_{\mathbf R,1}+c_{\mathbf R,1}^\dagger c_{\mathbf R,2}^\dagger c_{\mathbf R,3}c_{\mathbf R,0}\bigr].
\label{bands:eq:four-q-expansion}
\end{equation}
Separate $03$ and $12$ density penalties would also remove the allowed antisymmetric combination and therefore cannot replace this optional term for $331$.

\section{Numerical methods and results}
\label{sec:numerics}

We test the parents constructed in Appendix~\ref{sec:lattice-models} on finite tori using exact diagonalization (ED) for zero-mode counts and low-energy spectra, and the particle entanglement spectrum (PES) for Laughlin and Halperin quasihole counting. The torus calculations use the projected Hamiltonians with periodic or twisted boundary conditions.

\FloatBarrier
\subsection{Exact diagonalization}
\label{num:sec:numerical-methods}
\FloatBarrier
\subsubsection{General implementation}
\label{num:sec:ed-general}

We consider $n_{\mathrm c}$ internal components per site, labeled by $\alpha=0,\ldots,n_{\mathrm c}-1$. The operator $c_{\mathbf R,\alpha}$ annihilates a fermion in component $\alpha$ at lattice site $\mathbf R$, and $n_{\mathbf R,\alpha}=c_{\mathbf R,\alpha}^\dagger c_{\mathbf R,\alpha}$. We specify the torus by two independent integer periods $\mathbf T_i=(T_{ix},T_{iy})$, $i=1,2$. Writing $D=\det(\mathbf T_1,\mathbf T_2)$, it contains $N_s=|D|$ sites; for $N_e$ fermions, the band filling is $\nu=N_e/N_s$. Unless specified otherwise, we impose periodic boundary conditions
\begin{equation}
c_{\mathbf R+\mathbf T_i,\alpha}=c_{\mathbf R,\alpha},\qquad i=1,2.
\label{num:eq:ed-torus-boundary}
\end{equation}
The boundary conditions require $\mathbf k\cdot\mathbf T_i=2\pi m_i$ with integers $m_i$. Solving gives the momentum mesh, with $d=\gcd(|T_{2x}|,|T_{2y}|)$:
\begin{equation}
\mathcal M=\left\{\frac{2\pi}{D}\left(m_1T_{2y}-m_2T_{1y},\,m_2T_{1x}-m_1T_{2x}\right)\bmod2\pi: \ 0\le m_1<\frac{N_s}{d},\ 0\le m_2<d\right\}.
\label{num:eq:ed-torus-momenta}
\end{equation}
Equation~\eqref{num:eq:ed-torus-momenta} gives $N_s$ distinct momenta for any integer torus periods. Modulo $2\pi$ reduces each component $k_x$ and $k_y$ to $[0,2\pi)$, including in sums and differences. Real-space sums cover the $N_s$ torus sites. We use the Fourier convention
\begin{equation}
c_{\mathbf k,\alpha}=\frac{1}{\sqrt{N_s}}\sum_{\mathbf R}e^{-\ii \mathbf k\cdot\mathbf R}c_{\mathbf R,\alpha},\qquad c_{\mathbf R,\alpha}=\frac{1}{\sqrt{N_s}}\sum_{\mathbf k\in\mathcal M}e^{\ii \mathbf k\cdot\mathbf R}c_{\mathbf k,\alpha}.
\label{num:eq:ed-fourier}
\end{equation}
This transform preserves the canonical anticommutators. The total momentum is
\begin{equation}
\mathbf K=\sum_{j=0}^{N_e-1}\mathbf k_j\pmod{2\pi\mathbb Z^2},\qquad \mathbf K\in\mathcal M.
\label{num:eq:ed-many-body-momentum}
\end{equation}

For the Bloch vector $v(\mathbf k)$ of Appendix~\ref{sec:lattice-models}, set $u(\mathbf k)=v(\mathbf k)/\sqrt{v^\dagger(\mathbf k)v(\mathbf k)}$ and $P(\mathbf k)=u(\mathbf k)u^\dagger(\mathbf k)$. For $U$ much smaller than the band gap, the low-energy problem is described to leading order by the Hubbard interaction projected to the flat band. We project the interaction into the spectrally flattened band~\cite{Regnault2011,Bergholtz2013}. The analytic band is already isolated and flat at zero energy, so the single-particle Hamiltonian vanishes in the projected subspace before and after flattening.\\

The band operators defined by $u_\alpha(\mathbf k)$ are
\begin{equation}
\hat\psi^\dagger_{\mathbf k}=\sum_{\alpha=0}^{n_{\mathrm c}-1}u_\alpha(\mathbf k)c^\dagger_{\mathbf k,\alpha},\qquad \hat\psi_{\mathbf k}=\sum_{\alpha=0}^{n_{\mathrm c}-1}u_\alpha^*(\mathbf k)c_{\mathbf k,\alpha}.
\label{num:eq:ed-band-annihilator}
\end{equation}
Normalization ensures $\{\hat\psi_{\mathbf k},\hat\psi^\dagger_{\mathbf k'}\}=\delta_{\mathbf k,\mathbf k'}$. Projection with $P=uu^\dagger$ gives $c_{\mathbf k,\alpha}\to\sum_\beta P_{\alpha\beta}(\mathbf k)c_{\mathbf k,\beta}=u_\alpha(\mathbf k)\hat\psi_{\mathbf k}$. Inverse Fourier transformation then gives
\begin{equation}
c_{\mathbf R,\alpha}\longrightarrow\widetilde c_{\mathbf R,\alpha} =\frac{1}{\sqrt{N_s}}\sum_{\mathbf k\in\mathcal M} e^{\ii \mathbf k\cdot\mathbf R}u_\alpha(\mathbf k)\hat\psi_{\mathbf k}.
\label{num:eq:ed-band-projection}
\end{equation}
Hermitian conjugation gives the projected creation operators; substituting into the normal-ordered interaction defines $\widehat H_{\mathrm{proj}}$.\\

For onsite two-body interactions, define the form factor at momentum transfer $\mathbf q\in\mathcal M$
\begin{equation}
\mathcal F_{\alpha\beta}[u](\mathbf k,\mathbf k',\mathbf q) =u_\alpha^*(\mathbf k+\mathbf q)u_\beta^*(\mathbf k'-\mathbf q) u_\beta(\mathbf k')u_\alpha(\mathbf k).
\label{num:eq:ed-form-factors}
\end{equation}
For $\alpha<\beta$, normal ordering gives $n_{\mathbf R,\alpha}n_{\mathbf R,\beta}=c_{\mathbf R,\alpha}^\dagger c_{\mathbf R,\beta}^\dagger c_{\mathbf R,\beta}c_{\mathbf R,\alpha}$. Substituting Eq.~\eqref{num:eq:ed-band-projection} and summing over sites enforces momentum conservation, giving
\begin{equation}
\sum_{\mathbf R}n_{\mathbf R,\alpha}n_{\mathbf R,\beta} \longrightarrow\frac{1}{N_s}\sum_{\mathbf k,\mathbf k',\mathbf q\in\mathcal M} \mathcal F_{\alpha\beta}[u](\mathbf k,\mathbf k',\mathbf q) \hat\psi^\dagger_{\mathbf k+\mathbf q}\hat\psi^\dagger_{\mathbf k'-\mathbf q} \hat\psi_{\mathbf k'}\hat\psi_{\mathbf k}.
\label{num:eq:ed-projected-density-pair}
\end{equation}
Summing these density pairs with real couplings $U_{\alpha\beta}$ gives the projected two-body interaction
\begin{equation}
\widehat H_{\mathrm{proj},U} =\sum_{\mathbf k,\mathbf k',\mathbf q\in\mathcal M} \mathcal V(\mathbf k,\mathbf k',\mathbf q) \hat\psi^\dagger_{\mathbf k+\mathbf q}\hat\psi^\dagger_{\mathbf k'-\mathbf q} \hat\psi_{\mathbf k'}\hat\psi_{\mathbf k},
\label{num:eq:ed-projected-interaction}
\end{equation}
with
\begin{equation}
\mathcal V(\mathbf k,\mathbf k',\mathbf q)=\frac{1}{N_s} \sum_{\alpha<\beta}U_{\alpha\beta} \mathcal F_{\alpha\beta}[u](\mathbf k,\mathbf k',\mathbf q).
\label{num:eq:ed-general-vertex}
\end{equation}
Each unordered component pair appears once.\\

\begin{samepage}%
We diagonalize fixed-$(N_e,\mathbf K)$ sectors directly when small and use fully reorthogonalized block Lanczos otherwise, exploiting $C_4$ symmetry when shared by the Hamiltonian and torus. Plots label momentum sectors by $M_yj_x+j_y$, with $(j_x,j_y)\in\mathbb Z_{M_x}\times\mathbb Z_{M_y}$ and $M_xM_y=N_s$; the mesh figures give their physical momenta. At fixed $N_e$, $E_1\le E_2\le\cdots$ orders energies across all sectors, including degeneracies. Throughout this appendix, we count levels with $|E/U|<10^{-9}$ as numerical zero modes.

\end{samepage}%
\FloatBarrier
\subsubsection{Two-component model: Laughlin $1/3$ state}
\label{num:sec:ed-laughlin-third}

\begin{figure}[!htbp]
\centering
\includegraphics[width=\textwidth]{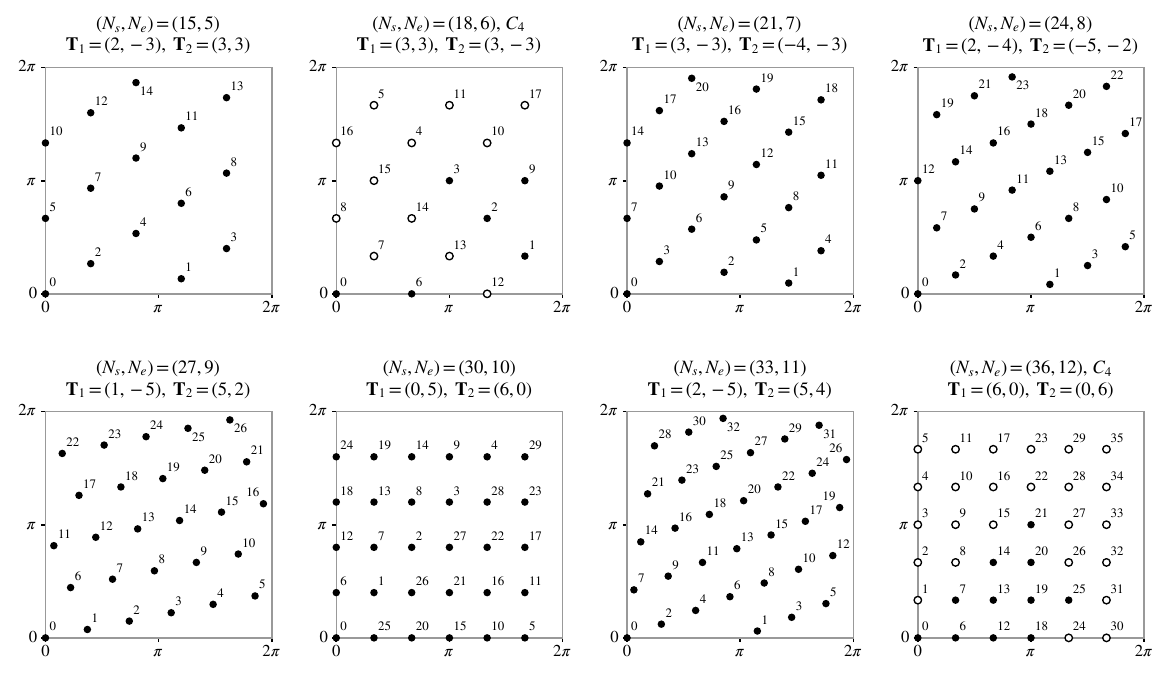}
\caption{Momentum meshes for the Laughlin $1/3$ systems. For $C_4$-symmetric meshes, filled points are orbit representatives and hollow points are generated by their $C_4$ rotations.}
\label{num:fig:mesh-atlas-1}
\end{figure}

For the Laughlin $1/3$ calculations at $N_s=15,18,\ldots,36$, we choose tori as close to $C_4$ symmetry as possible at each size. Figure~\ref{num:fig:mesh-atlas-1} shows their momentum meshes $\mathcal M$ defined by Eq.~\eqref{num:eq:ed-torus-momenta}.\\

\begin{figure}[!htbp]
\centering
\includegraphics[width=0.56\textwidth]{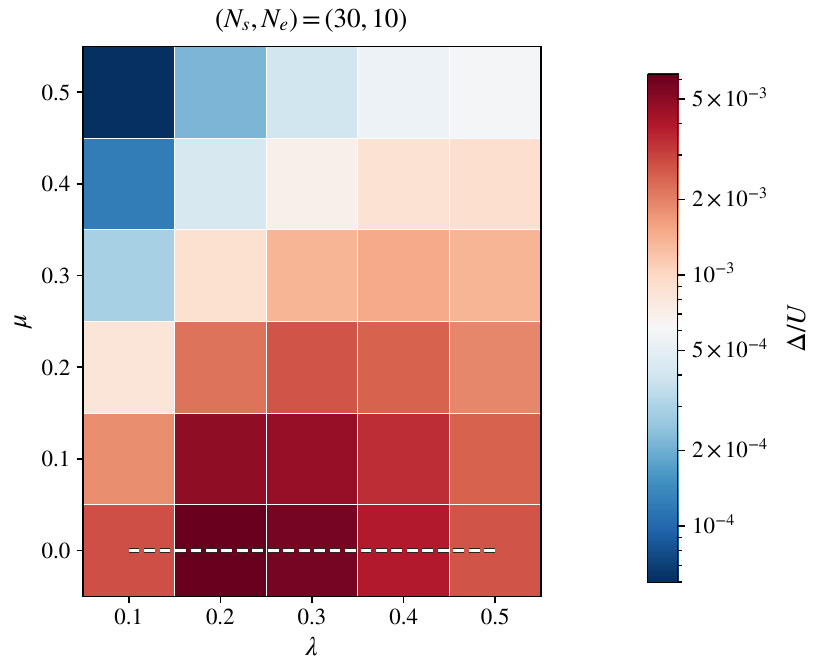}
\caption{Many-body gap $\Delta/U$ of the Laughlin $1/3$ model at $(N_s,N_e)=(30,10)$. The scan uses $\lambda=0.1,0.2,\ldots,0.5$ and $\mu=0,0.1,\ldots,0.5$. Colors show the gap on a logarithmic scale; each cell corresponds to one sampled parameter pair. The dashed line marks the cut $\mu=0$ used in the following spectra.}
\label{num:fig:gap-scan-laughlin-third}
\end{figure}

\begin{samepage}%
We project the onsite parent $H_{1/3}$ of Eq.~\eqref{bands:eq:two-many-body-parent} into the two-component band defined by $v^{(1/2)}_{\lambda,\mu}$ in Eq.~\eqref{bands:eq:lambda-mu-theta}. Let $u^{\mathrm{L}_{1/3}}$ be its normalized vector, $\mathcal F^{\mathrm{L}_{1/3}}_{\alpha\beta}=\mathcal F_{\alpha\beta}[u^{\mathrm{L}_{1/3}}]$, and $H_{1/3,\mathrm{proj}}$ the projected parent. Its coefficient in Eq.~\eqref{num:eq:ed-projected-interaction} is
\par\end{samepage}%
\begin{equation}
\mathcal V^{\mathrm{L}_{1/3}}(\mathbf k,\mathbf k',\mathbf q) =\frac{U}{N_s}\mathcal F^{\mathrm{L}_{1/3}}_{01}(\mathbf k,\mathbf k',\mathbf q).
\label{num:eq:ed-two-vertex}
\end{equation}
At $\nu=1/3$, we define the gap above the three ground states as $\Delta=E_4-E_3$. Figure~\ref{num:fig:gap-scan-laughlin-third} shows the two-parameter scan at $(N_s,N_e)=(30,10)$.\\

For every sampled $\lambda$, the gap decreases monotonically with $\mu$ over the scanned range. We therefore fix $\mu=0$ and vary $\lambda$.\\

Figures~\ref{num:fig:ed-laughlin-third-1} and~\ref{num:fig:ed-laughlin-third-2} compare $N_s=15,18,\ldots,36$ at $\nu=1/3$, with $\mu=0$ and $U=1$. Each spectrum contains three numerical zero modes, matching the torus ground-state degeneracy of the Laughlin $1/3$ state.\\

\begin{figure}[!htbp]
\centering
\includegraphics[width=\textwidth]{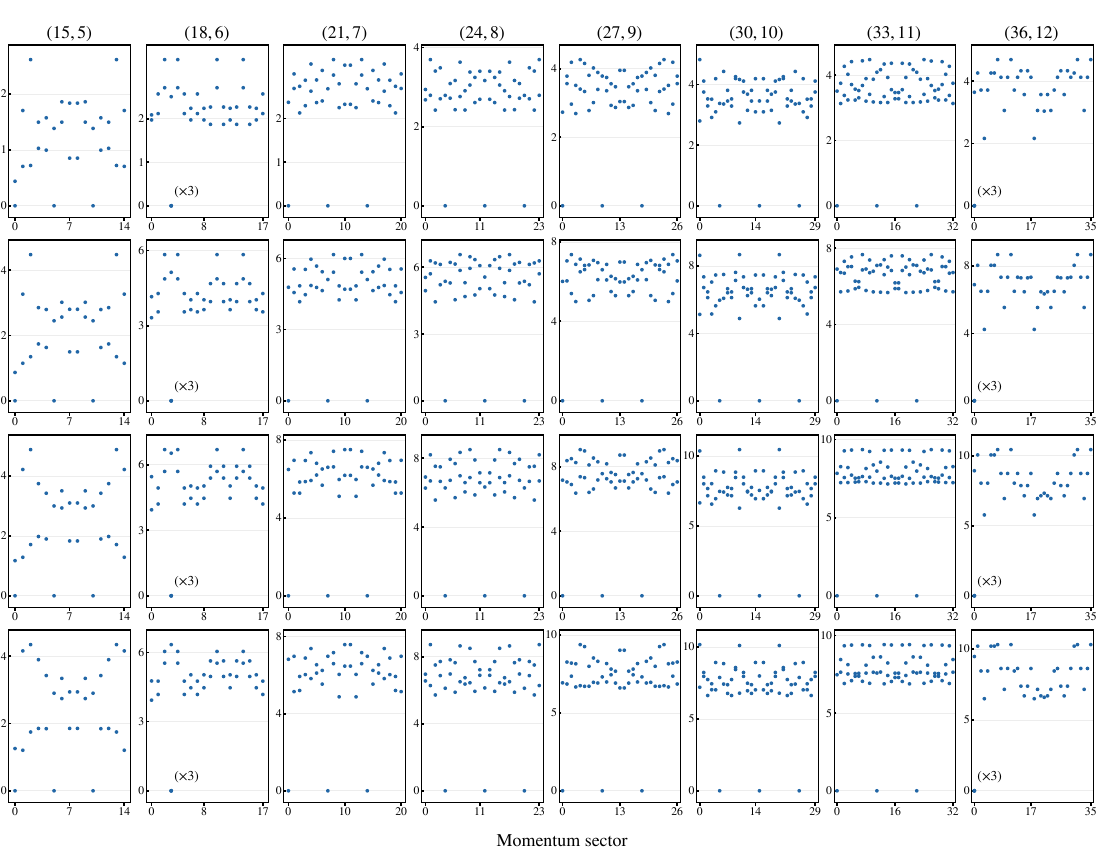}
\caption{ED spectra at $\nu=1/3$ and $U=1$. Column headings give $(N_s,N_e)$. Rows from top to bottom show $\lambda=0.1,0.15,0.2,0.25$ at $\mu=0$. Energies are plotted as $10^3E/U$. Zero-mode sectors show all zero modes and the lowest excitation; other sectors show the two lowest levels (one for $N_s=36$). A label $(\times n)$ gives the zero-mode multiplicity in that sector.}
\label{num:fig:ed-laughlin-third-1}
\end{figure}

\begin{figure}[!htbp]
\centering
\includegraphics[width=\textwidth]{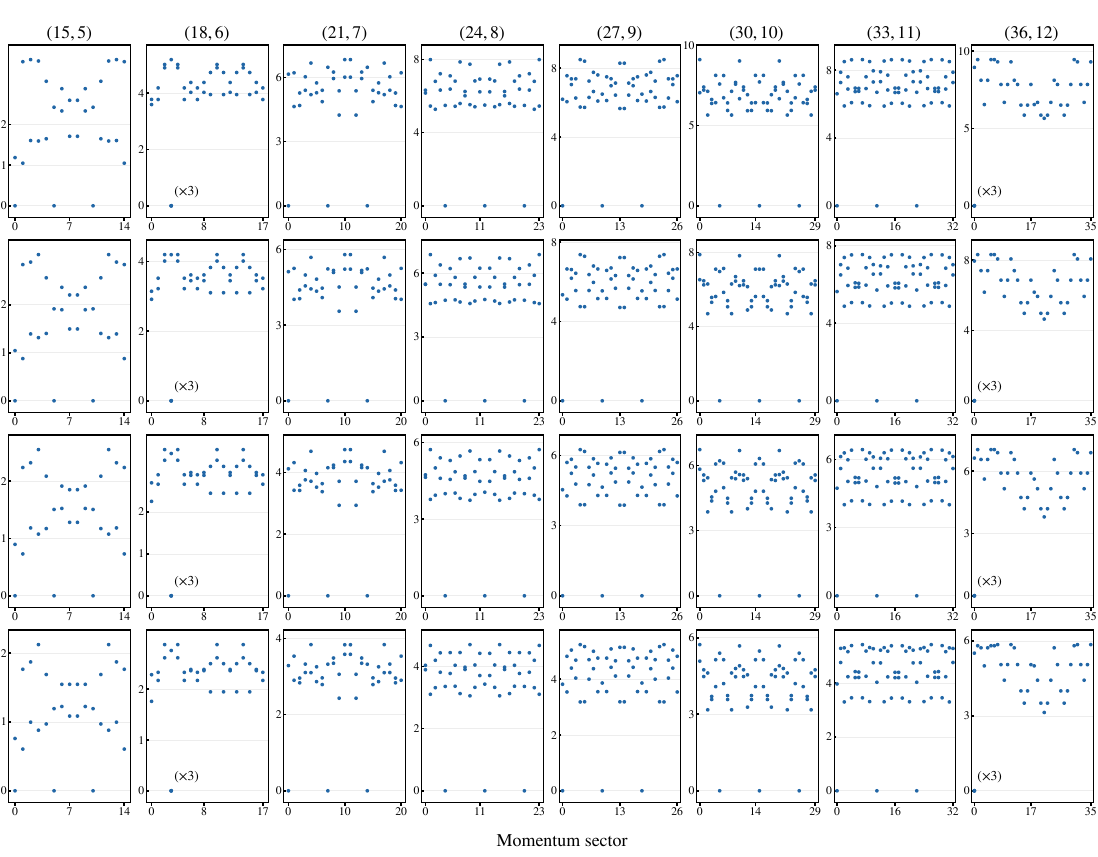}
\caption{Continuation of Fig.~\ref{num:fig:ed-laughlin-third-1}. Rows from top to bottom show $\lambda=0.3,0.35,0.4,0.45$.}
\label{num:fig:ed-laughlin-third-2}
\end{figure}


We normalize all system sizes by the thermodynamic local pair-interaction scale $\Delta_2$ at the same band parameters. Throughout this appendix, $\Delta_j(N_s)$ denotes the finite-size interaction scale and $\Delta_j=\lim_{N_s\to\infty}\Delta_j(N_s)$ its thermodynamic limit ($j=2,3$). Set $u=u^{\mathrm{L}_{1/3}}$ and $P(\mathbf k)=u(\mathbf k)u^\dagger(\mathbf k)$, and use the band-restricted pair annihilator $\hat B_{\mathbf R}$ of Eq.~\eqref{sample:eq:projected-onsite-pair} with this $u$. In the two-particle space, $U \hat B_{\mathbf R}^\dagger \hat B_{\mathbf R}$ has rank at most one. The onsite anticommutators from Eq.~\eqref{num:eq:ed-band-projection} are
\begin{equation}
\{\widetilde c_{\mathbf R,\alpha},\widetilde c_{\mathbf R,\beta}^\dagger\}=(\overline P_{N_s})_{\alpha\beta},\qquad \overline P_{N_s}=\frac{1}{N_s}\sum_{\mathbf k\in\mathcal M}u(\mathbf k)u^\dagger(\mathbf k).
\label{num:eq:charge2-projector-average}
\end{equation}
Let $\ket{\mathrm{vac}}$ be the empty Fock state and $\ket F=\prod_{\mathbf k\in\mathcal M}\hat\psi_{\mathbf k}^\dagger\ket{\mathrm{vac}}$ the filled-band Slater state. Throughout this appendix, $\operatorname{Tr}_n$ denotes the trace over the $n$-particle Fock space of the target band. The trace per site equals $U$ times the squared norm of $\hat B_{\mathbf R}^\dagger\ket{\mathrm{vac}}$, which Wick contraction in $\ket F$ also gives:
\begin{equation}
\Delta_2(N_s)=\frac{1}{N_s}\operatorname{Tr}_2H_{1/3,\mathrm{proj}}=U\bra{\mathrm{vac}} \hat B_{\mathbf R}\hat B_{\mathbf R}^\dagger\ket{\mathrm{vac}}=U\bra F n_{\mathbf R,0}n_{\mathbf R,1}\ket F.
\label{num:eq:charge2-two-particle-trace}
\end{equation}
The contraction gives
\begin{equation}
\Delta_2(N_s)=U\det\overline P_{N_s}.
\label{num:eq:charge2-determinant}
\end{equation}
\begin{samepage}%
Thus $\Delta_2(N_s)$ is the filled-band interaction energy per site. The common normalization uses the Brillouin-zone average:
\begin{equation}
\Delta_2=\lim_{N_s\to\infty}\frac{\operatorname{Tr}_2H_{1/3,\mathrm{proj}}}{N_s}=U\det\overline P_\infty,\qquad \overline P_\infty=\int_{\mathrm{BZ}}\frac{\d^2k}{(2\pi)^2}P(\mathbf k).
\label{num:eq:charge2-continuum}
\end{equation}
\end{samepage}%
Table~\ref{num:tab:charge2-scales} lists $\Delta_2/U$ at $\mu=0$.\\

\begin{table}[!htbp]
\centering
\begingroup
\small
\setlength{\tabcolsep}{3pt}
\renewcommand{\arraystretch}{1.05}
\begin{tabular*}{\textwidth}{@{\extracolsep{\fill}}c*{8}{r}@{}}
\toprule
$\lambda$ & $0.10$ & $0.15$ & $0.20$ & $0.25$ & $0.30$ & $0.35$ & $0.40$ & $0.45$\\
\midrule
$\Delta_2/U$ & 0.20309 & 0.23983 & 0.24980 & 0.24725 & 0.23887 & 0.22793 & 0.21606 & 0.20412\\
\bottomrule
\end{tabular*}
\endgroup
\caption{Thermodynamic local two-particle scale $\Delta_2/U$ for the Laughlin $1/3$ model at $\mu=0$, from Eq.~\eqref{num:eq:charge2-continuum}. Values are rounded to five decimal places.}
\label{num:tab:charge2-scales}
\end{table}

\begin{figure}[!htbp]
\centering
\includegraphics[width=0.78\textwidth]{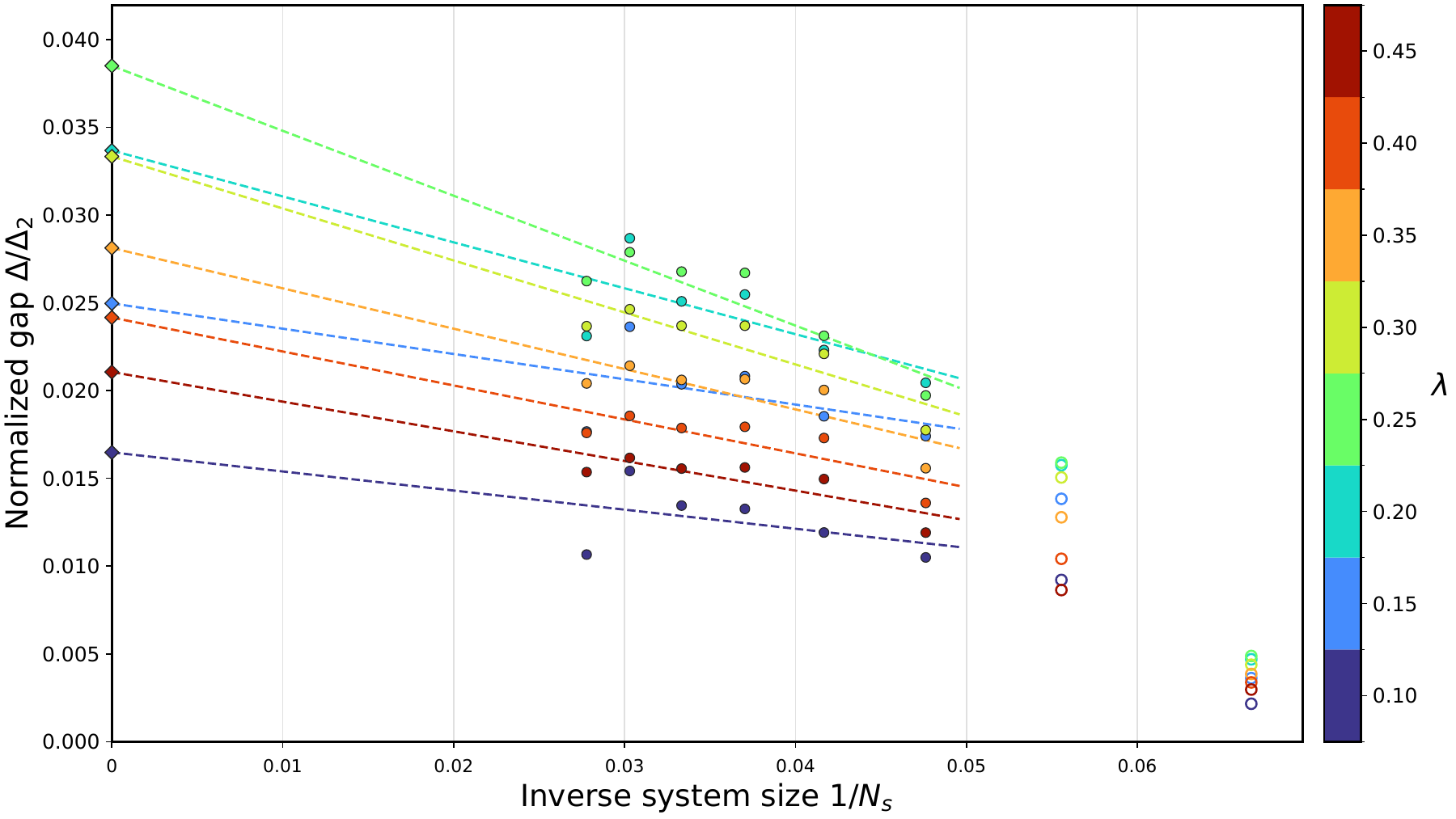}
\caption{Finite-size scaling of the normalized Laughlin $1/3$ gap $\Delta(N_s)/\Delta_2$ at $\mu=0$ for $\lambda=0.10,0.15,\ldots,0.45$. Colors identify $\lambda$. Filled circles mark $N_s=21,24,\ldots,36$, the six sizes in each unweighted linear fit. Open circles show $N_s=15,18$. Each dashed line extends from the intercept $g_\infty$ at $1/N_s=0$ (diamond) to just beyond $1/21$.}
\label{num:fig:gap-scaling-normalized-third}
\end{figure}

\begin{samepage}%
Figure~\ref{num:fig:gap-scaling-normalized-third} shows the gaps at $\mu=0$ for $N_s=15,18,\ldots,36$, normalized by the thermodynamic scales in Table~\ref{num:tab:charge2-scales}. At each fixed $\lambda$, we use an unweighted linear fit over $N_s\ge21$:
\par\end{samepage}%
\begin{equation}
g(N_s)=\frac{\Delta(N_s)}{\Delta_2}\simeq g_\infty+\frac{a}{N_s}.
\label{num:eq:normalized-gap-scaling}
\end{equation}
The intercept $g_\infty$ estimates the thermodynamic normalized gap, and $a$ is the slope; both depend on $\lambda$.\\

\FloatBarrier
\subsubsection{Three-component model: Laughlin $1/5$ state}
\label{num:sec:ed-three-component}
\label{num:sec:ed-laughlin-fifth}

\begin{figure}[!htbp]
\centering
\includegraphics[width=0.90\textwidth]{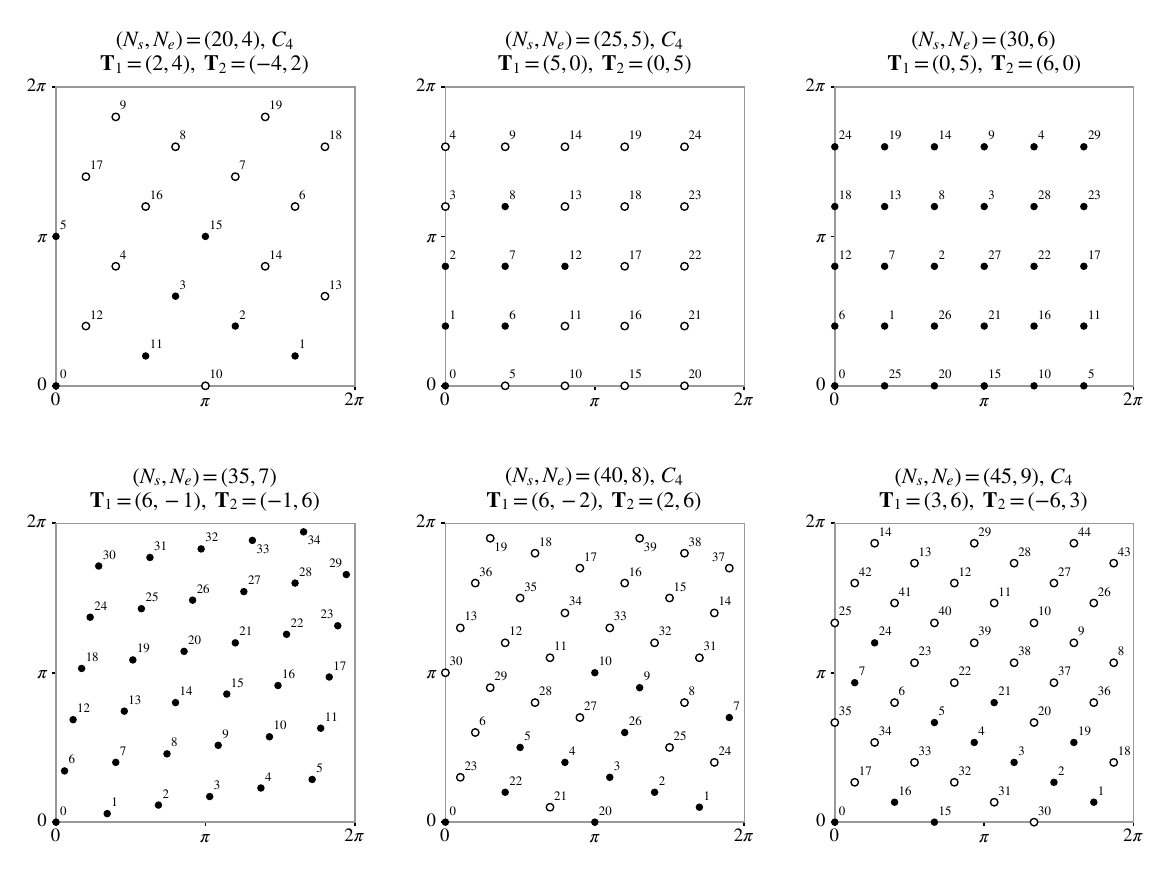}
\caption{Momentum meshes for the Laughlin $1/5$ systems. For $C_4$-symmetric meshes, filled points are orbit representatives and hollow points are generated by their $C_4$ rotations.}
\label{num:fig:mesh-atlas-2}
\end{figure}

For the Laughlin $1/5$ calculations at $N_s=20,25,\ldots,45$, we choose tori as close to $C_4$ symmetry as possible at each size. Figure~\ref{num:fig:mesh-atlas-2} shows their momentum meshes $\mathcal M$ defined by Eq.~\eqref{num:eq:ed-torus-momenta}.\\

\begin{figure}[!htbp]
\centering
\includegraphics[width=0.56\textwidth]{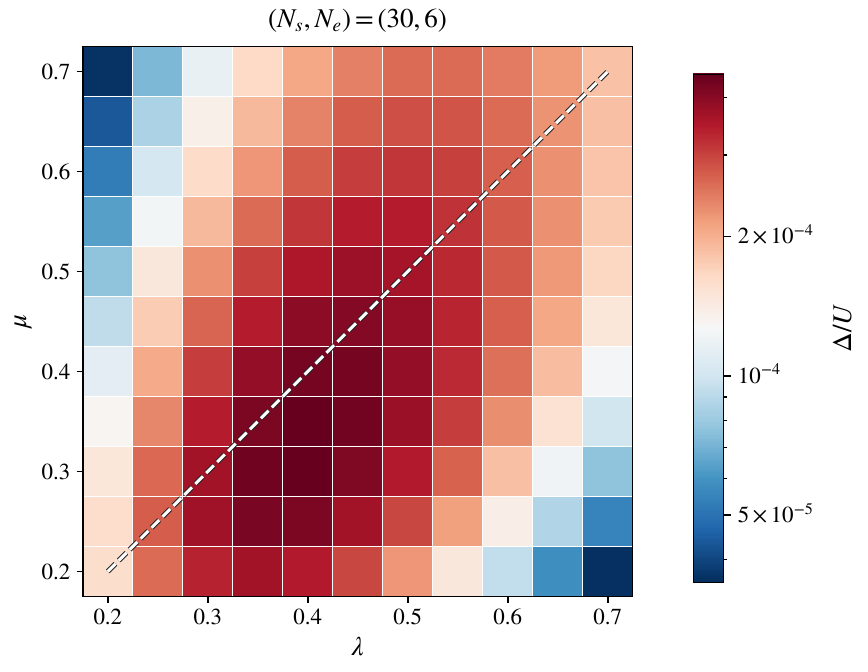}
\caption{Many-body gap $\Delta/U$ of the Laughlin $1/5$ model at $(N_s,N_e)=(30,6)$. Both parameters run from $0.2$ to $0.7$ in steps of $0.05$. Colors show the gap on a logarithmic scale; each cell corresponds to one sampled parameter pair. The dashed line marks the cut $\lambda=\mu$ used in the following spectra.}
\label{num:fig:gap-scan-laughlin-fifth}
\end{figure}

\begin{samepage}%
We project the onsite parent $H_{1/5}$ of Eq.~\eqref{bands:eq:three-pair-parent} into the three-component band defined by $v^{(1)}_{\lambda,\mu}$ in Eq.~\eqref{eq:three-component-regular}. Let $u^{\mathrm{L}_{1/5}}$ be its normalized vector, $\mathcal F^{\mathrm{L}_{1/5}}_{\alpha\beta}=\mathcal F_{\alpha\beta}[u^{\mathrm{L}_{1/5}}]$, and $H_{1/5,\mathrm{proj}}$ the projected parent. Its coefficient in Eq.~\eqref{num:eq:ed-projected-interaction} is
\par\end{samepage}%
\begin{equation}
\mathcal V^{\mathrm{L}_{1/5}}(\mathbf k,\mathbf k',\mathbf q)= \frac{U}{N_s}\left[\mathcal F^{\mathrm{L}_{1/5}}_{01}(\mathbf k,\mathbf k',\mathbf q)+\mathcal F^{\mathrm{L}_{1/5}}_{02}(\mathbf k,\mathbf k',\mathbf q)+\mathcal F^{\mathrm{L}_{1/5}}_{12}(\mathbf k,\mathbf k',\mathbf q)\right].
\label{num:eq:ed-three-vertex}
\end{equation}
At $\nu=1/5$, we define the gap above the five ground states as $\Delta=E_6-E_5$. Figure~\ref{num:fig:gap-scan-laughlin-fifth} shows the two-parameter scan at $(N_s,N_e)=(30,6)$.\\

The scan shows large gaps near $\lambda=\mu$, so we fix $\mu=\lambda$ and vary $\lambda$.\\

\begin{figure}[!htbp]
\centering
\includegraphics[width=\textwidth]{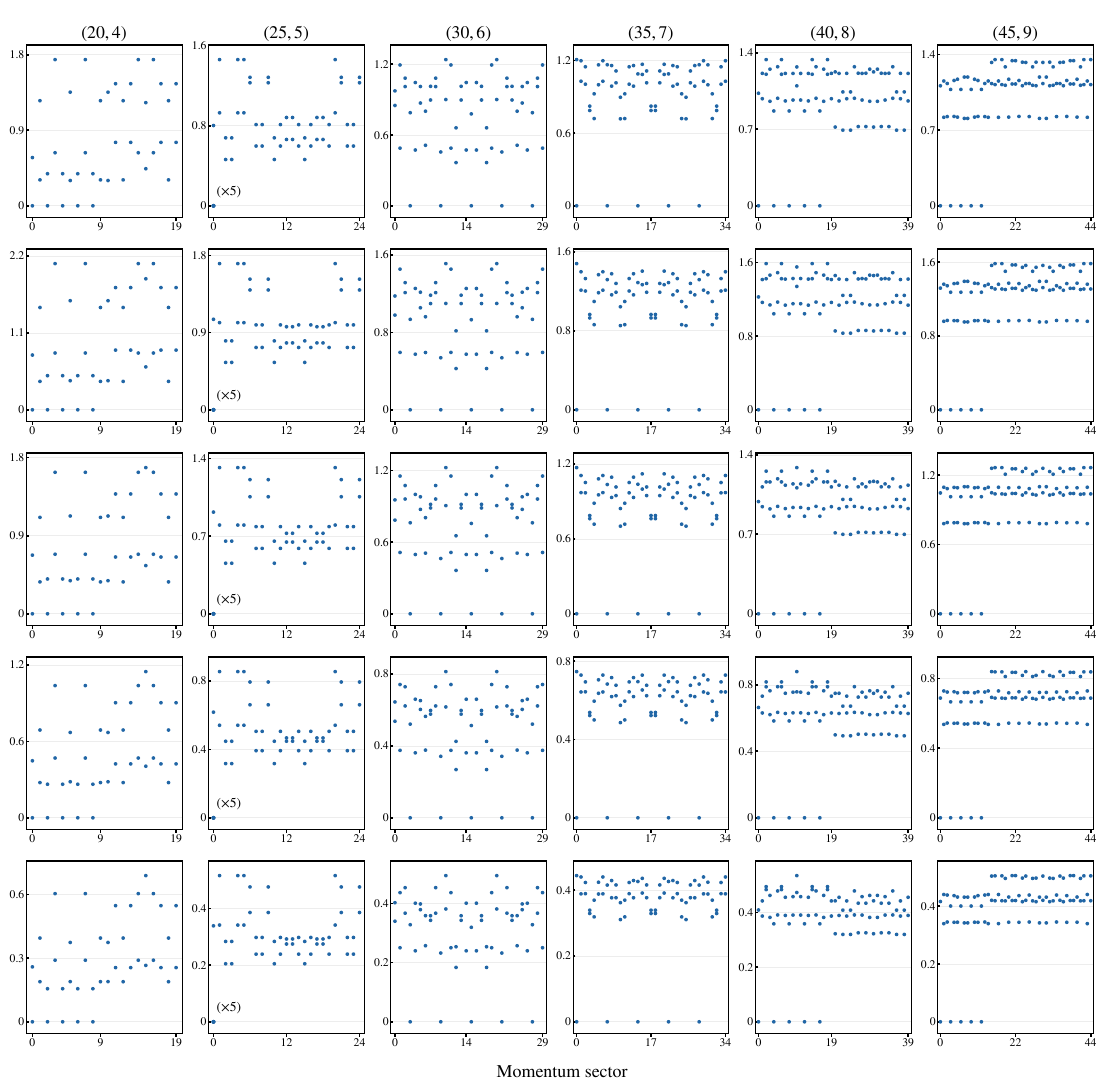}
\caption{ED spectra at $\nu=1/5$ and $U=1$. Column headings give $(N_s,N_e)$. Rows from top to bottom show $\lambda=\mu=0.3,0.4,0.5,0.6,0.7$. Energies are plotted as $10^3E/U$. Zero-mode sectors show all zero modes and the lowest excitation; other sectors show the two lowest levels. A label $(\times n)$ gives the zero-mode multiplicity in that sector.}
\label{num:fig:ed-laughlin-fifth-1}
\end{figure}

Figure~\ref{num:fig:ed-laughlin-fifth-1} compares $N_s=20,25,\ldots,45$ at $\nu=1/5$, with $\mu=\lambda$ and $U=1$. Each spectrum contains five numerical zero modes, matching the torus ground-state degeneracy of the Laughlin $1/5$ state.\\

We use $u=u^{\mathrm{L}_{1/5}}$ in Eq.~\eqref{num:eq:charge2-projector-average} and use the pair annihilators $\hat B_{\alpha\beta,\mathbf R}$ of Eq.~\eqref{sample:eq:one-contact-pairs} with this $u$. Averaging over the three pair channels as in Eq.~\eqref{num:eq:charge2-two-particle-trace} gives
\begin{equation}
\Delta_2(N_s)=\frac{1}{3N_s}\operatorname{Tr}_2H_{1/5,\mathrm{proj}}=\frac{U}{3}\sum_{\alpha<\beta}\bra{\mathrm{vac}}\hat B_{\alpha\beta,\mathbf R}\hat B_{\alpha\beta,\mathbf R}^\dagger\ket{\mathrm{vac}}=\frac{U}{3}\sum_{\alpha<\beta}\bra F n_{\mathbf R,\alpha}n_{\mathbf R,\beta}\ket F.
\end{equation}
Summing the principal $2\times2$ minors and using $\Tr\overline P_{N_s}=1$ gives
\begin{equation}
\Delta_2(N_s)=\frac{U}{3}\sum_{\alpha<\beta}\left[(\overline P_{N_s})_{\alpha\alpha}(\overline P_{N_s})_{\beta\beta}-\left|(\overline P_{N_s})_{\alpha\beta}\right|^2\right]=\frac{U}{6}\left[1-\Tr\!\left(\overline P_{N_s}^{\,2}\right)\right].
\label{num:eq:charge2-fifth-mean}
\end{equation}
Thus $\Delta_2(N_s)$ is the filled-band interaction energy per site and pair channel. We normalize all sizes by its thermodynamic limit
\begin{equation}
\Delta_2=\lim_{N_s\to\infty}\frac{\operatorname{Tr}_2H_{1/5,\mathrm{proj}}}{3N_s}=\frac{U}{6}\left[1-\Tr\!\left(\overline P_\infty^{\,2}\right)\right].
\label{num:eq:charge2-fifth-continuum}
\end{equation}
Table~\ref{num:tab:charge2-fifth-scales} lists $\Delta_2/U$ at $\mu=\lambda$.\\

\begin{table}[!htbp]
\centering
\begingroup
\small
\setlength{\tabcolsep}{3pt}
\renewcommand{\arraystretch}{1.05}
\begin{tabular*}{\textwidth}{@{\extracolsep{\fill}}c*{5}{r}@{}}
\toprule
$\lambda$ & $0.30$ & $0.40$ & $0.50$ & $0.60$ & $0.70$\\
\midrule
$\Delta_2/U$ & 0.10095 & 0.11047 & 0.10940 & 0.10263 & 0.09403\\
\bottomrule
\end{tabular*}
\endgroup
\caption{Thermodynamic mean local two-particle scale $\Delta_2/U$ for the Laughlin $1/5$ model at $\mu=\lambda$, from Eq.~\eqref{num:eq:charge2-fifth-continuum}. Values are rounded to five decimal places.}
\label{num:tab:charge2-fifth-scales}
\end{table}

\begin{figure}[!htbp]
\centering
\includegraphics[width=0.72\textwidth]{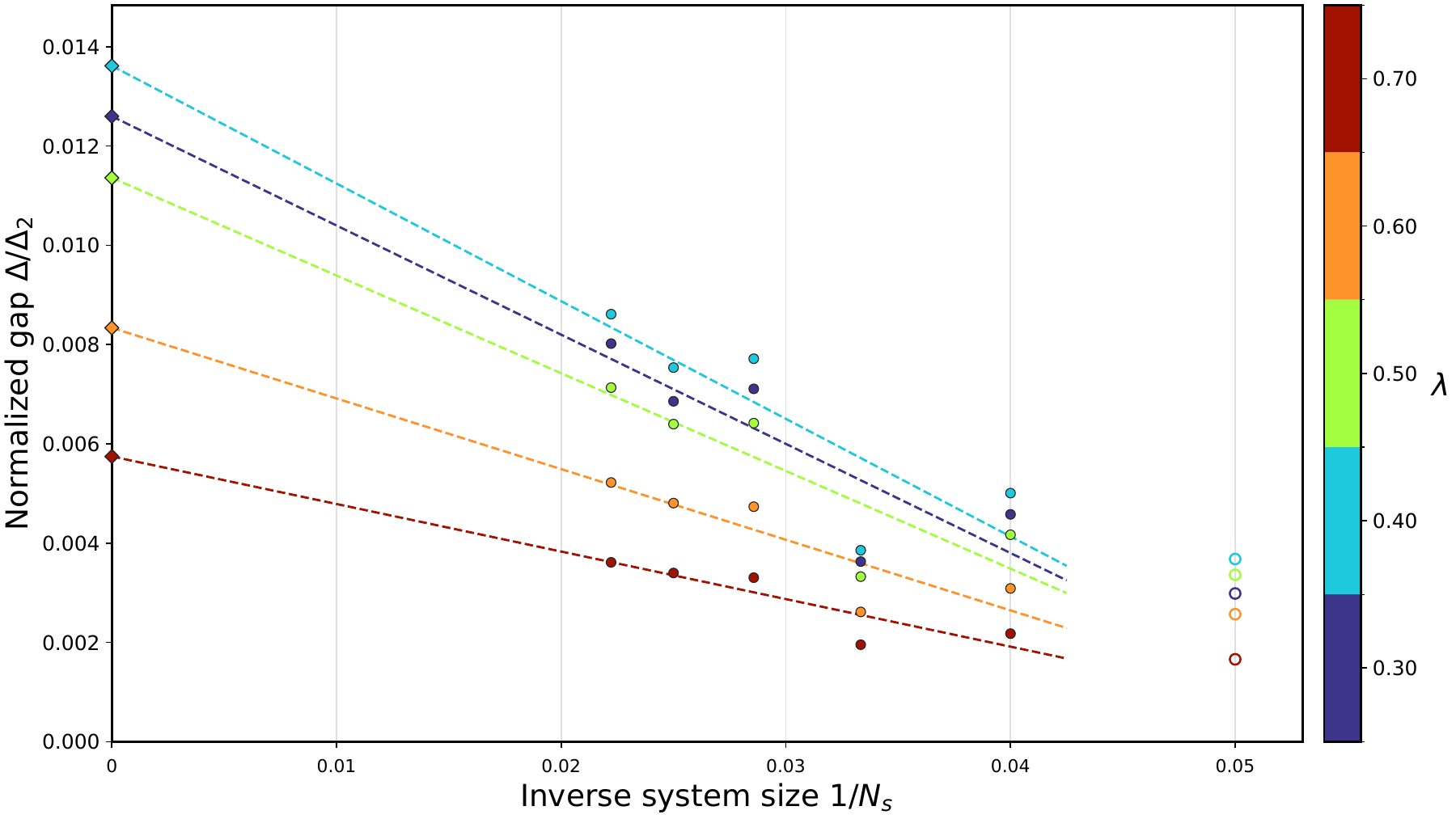}
\caption{Finite-size scaling of the normalized Laughlin $1/5$ gap $\Delta(N_s)/\Delta_2$ at $\mu=\lambda$ for $\lambda=0.30,0.40,\ldots,0.70$. Colors identify $\lambda$. Filled circles mark $N_s=25,30,\ldots,45$, the five sizes in each unweighted linear fit. Open circles show $N_s=20$. Each dashed line extends from the intercept $g_\infty$ at $1/N_s=0$ (diamond) to just beyond $1/25$.}
\label{num:fig:gap-scaling-normalized-fifth}
\end{figure}

\begin{samepage}%
Figure~\ref{num:fig:gap-scaling-normalized-fifth} shows the gaps at $\mu=\lambda$ for $N_s=20,25,\ldots,45$, normalized by the thermodynamic scales in Table~\ref{num:tab:charge2-fifth-scales}. At each fixed $\lambda$, we use an unweighted linear fit over $N_s\ge25$ to Eq.~\eqref{num:eq:normalized-gap-scaling}.\\
\end{samepage}%

\FloatBarrier
\subsubsection{Three-component model: Moore--Read state}
\label{num:sec:ed-moore-read}

\begin{figure}[!htbp]
\centering
\includegraphics[width=\textwidth]{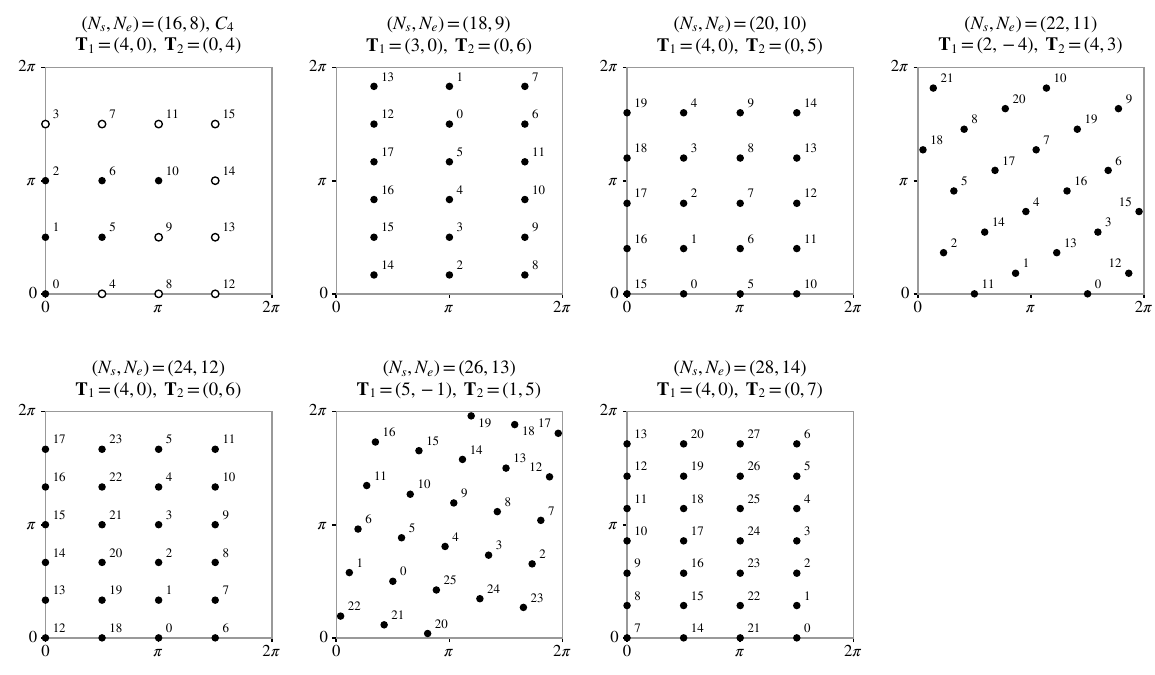}
\caption{Many-body momentum meshes for the Moore--Read systems. Points at $(K_x,K_y)$ are labeled by $M_yj_x+j_y$; $(M_x,M_y)$ are the rectangular dimensions or $(1,N_s)$ for tilted tori. For the $C_4$-symmetric mesh, filled points are orbit representatives and hollow points are generated by their $C_4$ rotations.}
\label{num:fig:mesh-mr}
\end{figure}

For the Moore--Read calculations at $N_s=16,18,\ldots,28$, Fig.~\ref{num:fig:mesh-mr} shows the many-body momentum meshes. We impose boundary twists $\theta_i$ along the periods $\mathbf T_i$:
\begin{equation}
c_{\mathbf R+\mathbf T_i,\alpha}=e^{\ii\theta_i}c_{\mathbf R,\alpha},\qquad
\mathbf k\cdot\mathbf T_i=\theta_i\pmod{2\pi},\qquad
\mathbf K\cdot\mathbf T_i=N_e\theta_i\pmod{2\pi}.
\label{num:eq:mr-boundary-twists}
\end{equation}
We use $(\theta_1,\theta_2)=(\pi,\pi)$ for rectangular tori, $(\pi,0)$ for $N_s=22$, and $(0,\pi)$ for $N_s=26$. The single-particle mesh $\mathcal M$ carries these twists. Momentum transfers lie on the periodic mesh of Eq.~\eqref{num:eq:ed-torus-momenta}.\\

We project the onsite three-body parent $H_{\mathrm{MR}}$ of Eq.~\eqref{bands:eq:three-mr-parent}, with $U>0$, into the three-component band of Eq.~\eqref{bands:eq:three-gaussian-parent}. Let $u^{\mathrm{MR}}$ be the normalized form of $v^{(1)}_{\lambda,\mu}$ in Eq.~\eqref{eq:three-component-regular}.\\

We use the projected triplet annihilator $\hat B_{012,\mathbf R}$ of Eq.~\eqref{sample:eq:one-contact-triplet}, with $u=u^{\mathrm{MR}}$.
For each ordered triple $I=(\mathbf k_1,\mathbf k_2,\mathbf k_3)$ with $\mathbf k_1\prec\mathbf k_2\prec\mathbf k_3$ in a fixed ordering of $\mathcal M$, define $\Gamma_I=\hat\psi_{\mathbf k_3}\hat\psi_{\mathbf k_2}\hat\psi_{\mathbf k_1}$. Antisymmetrizing over the six permutations $p\in S_3$, with fermionic sign $\operatorname{sgn}(p)$, gives
\begin{equation}
\mathcal G(I)=\sum_{p\in S_3}\operatorname{sgn}(p)\prod_{j=1}^{3}u_{j-1}^{\mathrm{MR}}(\mathbf k_{p(j)})=\det\!\left[u^{\mathrm{MR}}_\alpha(\mathbf k_j)\right]_{\substack{\alpha=0,1,2\\j=1,2,3}}.
\label{num:eq:ed-three-body-form-factor}
\end{equation}
With $\mathbf K(I)=\mathbf k_1+\mathbf k_2+\mathbf k_3$ modulo $2\pi\mathbb Z^2$, the annihilator becomes
\begin{equation}
\hat B_{012,\mathbf R}=\frac{1}{(\sqrt{N_s})^3}\sum_I e^{\ii\mathbf K(I)\cdot\mathbf R}\mathcal G(I)\Gamma_I.
\label{num:eq:ed-ordered-triple-annihilator}
\end{equation}
For ordered triples $I,J$, the site sum in $\hat B_{012,\mathbf R}^{\dagger}\hat B_{012,\mathbf R}$ gives $N_s\delta_{\mathbf K(J),\mathbf K(I)}$, with momenta compared modulo $2\pi\mathbb Z^2$. Hence
\begin{equation}
H_{\mathrm{MR},\mathrm{proj}}=U\sum_{\mathbf R}\hat B_{012,\mathbf R}^{\dagger}\hat B_{012,\mathbf R}=\sum_{J,I}\mathcal V^{\mathrm{MR}}_{JI}\Gamma_J^{\dagger}\Gamma_I.
\label{num:eq:ed-three-body-projected-hamiltonian}
\end{equation}
Each ordered triple occurs once. In the normalized basis $\ket I=\Gamma_I^\dagger\ket{\mathrm{vac}}$, the matrix element is
\begin{equation}
\mathcal V^{\mathrm{MR}}_{JI}=\bra J H_{\mathrm{MR},\mathrm{proj}}\ket I=\frac{U}{N_s^2}\mathcal G(J)^*\mathcal G(I)\delta_{\mathbf K(J),\mathbf K(I)}.
\label{num:eq:ed-three-body-vertex}
\end{equation}

At $\nu=1/2$, we define the gap above the ground states as $\Delta=E_7-E_6$ for even $N_e$ and $\Delta=E_3-E_2$ for odd $N_e$.\\

\begin{figure}[!htbp]
\centering
\includegraphics[width=\textwidth]{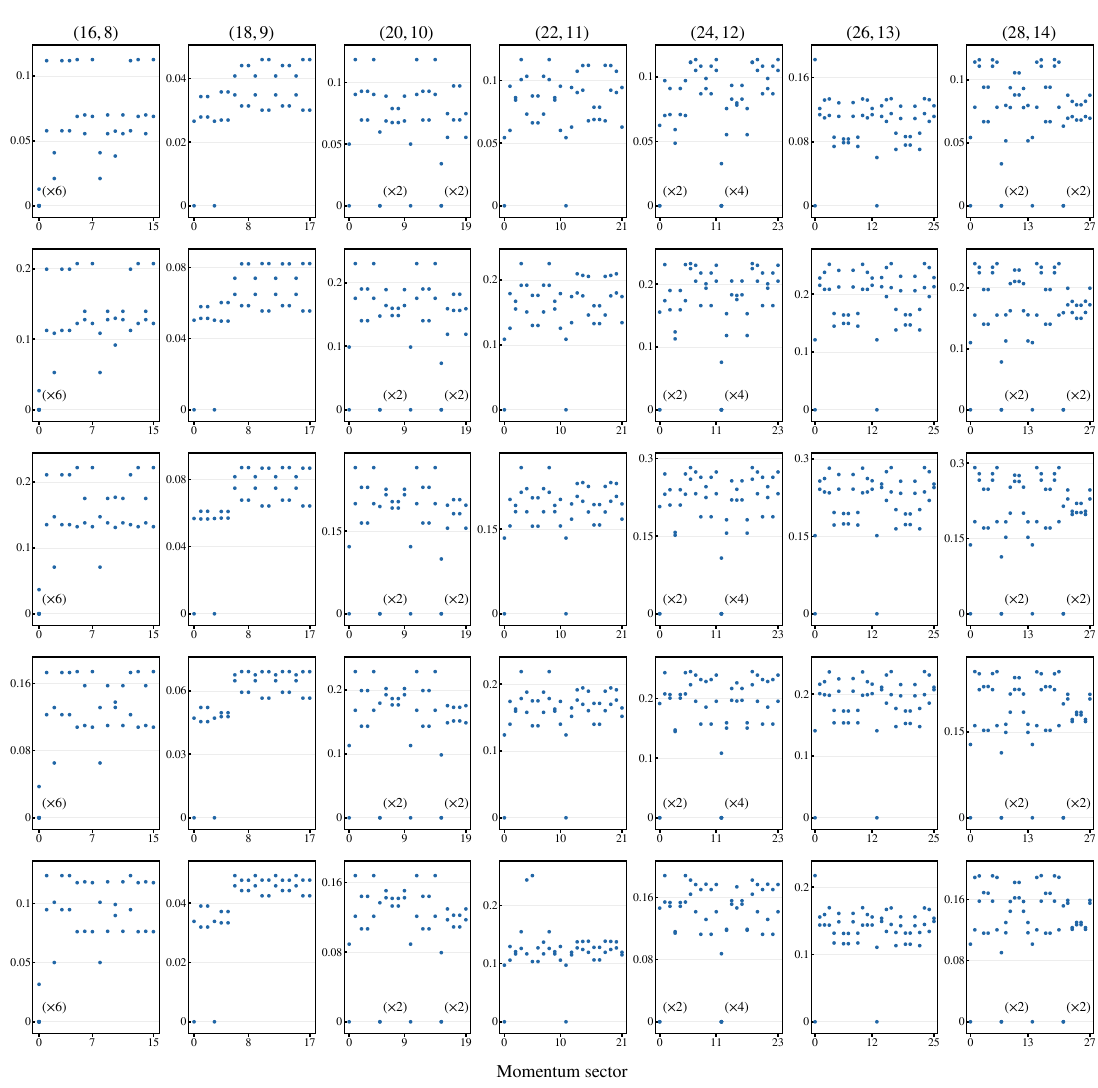}
\caption{ED spectra at $\nu=1/2$ and $U=1$. Column headings give $(N_s,N_e)$. Rows from top to bottom show $\lambda=\mu=0.3,0.4,0.5,0.6,0.7$. Energies are plotted as $10^3E/U$. Zero-mode sectors show all zero modes and the lowest excitation; other sectors show the two lowest levels. A label $(\times n)$ gives the zero-mode multiplicity in that sector.}
\label{num:fig:ed-moore-read}
\end{figure}

Figure~\ref{num:fig:ed-moore-read} compares $N_s=16,18,\ldots,28$ at $\mu=\lambda$ and $U=1$. Each spectrum contains six numerical zero modes for even $N_e$ and two for odd $N_e$, matching the torus ground-state degeneracy of the Moore--Read state~\cite{SeidelYang2011}.\\

With $u=u^{\mathrm{MR}}$ in Eq.~\eqref{num:eq:charge2-projector-average}, the local operator $U\hat B_{012,\mathbf R}^\dagger\hat B_{012,\mathbf R}$ has rank at most one in the three-particle space. The trace per site gives
\begin{equation}
\Delta_3(N_s)=\frac{1}{N_s}\operatorname{Tr}_3H_{\mathrm{MR},\mathrm{proj}}=U\bra{\mathrm{vac}}\hat B_{012,\mathbf R}\hat B_{012,\mathbf R}^\dagger\ket{\mathrm{vac}}=U\bra F n_{\mathbf R,0}n_{\mathbf R,1}n_{\mathbf R,2}\ket F.
\end{equation}
Wick contraction gives
\begin{equation}
\Delta_3(N_s)=U\det\overline P_{N_s}.
\label{num:eq:charge3-determinant}
\end{equation}
Thus $\Delta_3(N_s)$ is the filled-band interaction energy per site. Its thermodynamic limit is
\begin{equation}
\Delta_3=\lim_{N_s\to\infty}\frac{\operatorname{Tr}_3H_{\mathrm{MR},\mathrm{proj}}}{N_s}=U\det\overline P_\infty.
\label{num:eq:charge3-continuum}
\end{equation}
Table~\ref{num:tab:charge3-scales} lists $\Delta_3/U$ at $\mu=\lambda$.\\

\begin{table}[!htbp]
\centering
\begingroup
\small
\setlength{\tabcolsep}{3pt}
\renewcommand{\arraystretch}{1.05}
\begin{tabular*}{\textwidth}{@{\extracolsep{\fill}}c*{5}{r}@{}}
\toprule
$\lambda$ & $0.30$ & $0.40$ & $0.50$ & $0.60$ & $0.70$\\
\midrule
$\Delta_3/U$ & 0.02765 & 0.03637 & 0.03524 & 0.02841 & 0.02071\\
\bottomrule
\end{tabular*}
\endgroup
\caption{Thermodynamic local three-particle scale $\Delta_3/U$ for the Moore--Read model at $\mu=\lambda$, from Eq.~\eqref{num:eq:charge3-continuum}. Values are rounded to five decimal places.}
\label{num:tab:charge3-scales}
\end{table}

\begin{figure}[!htbp]
\centering
\includegraphics[width=0.78\textwidth]{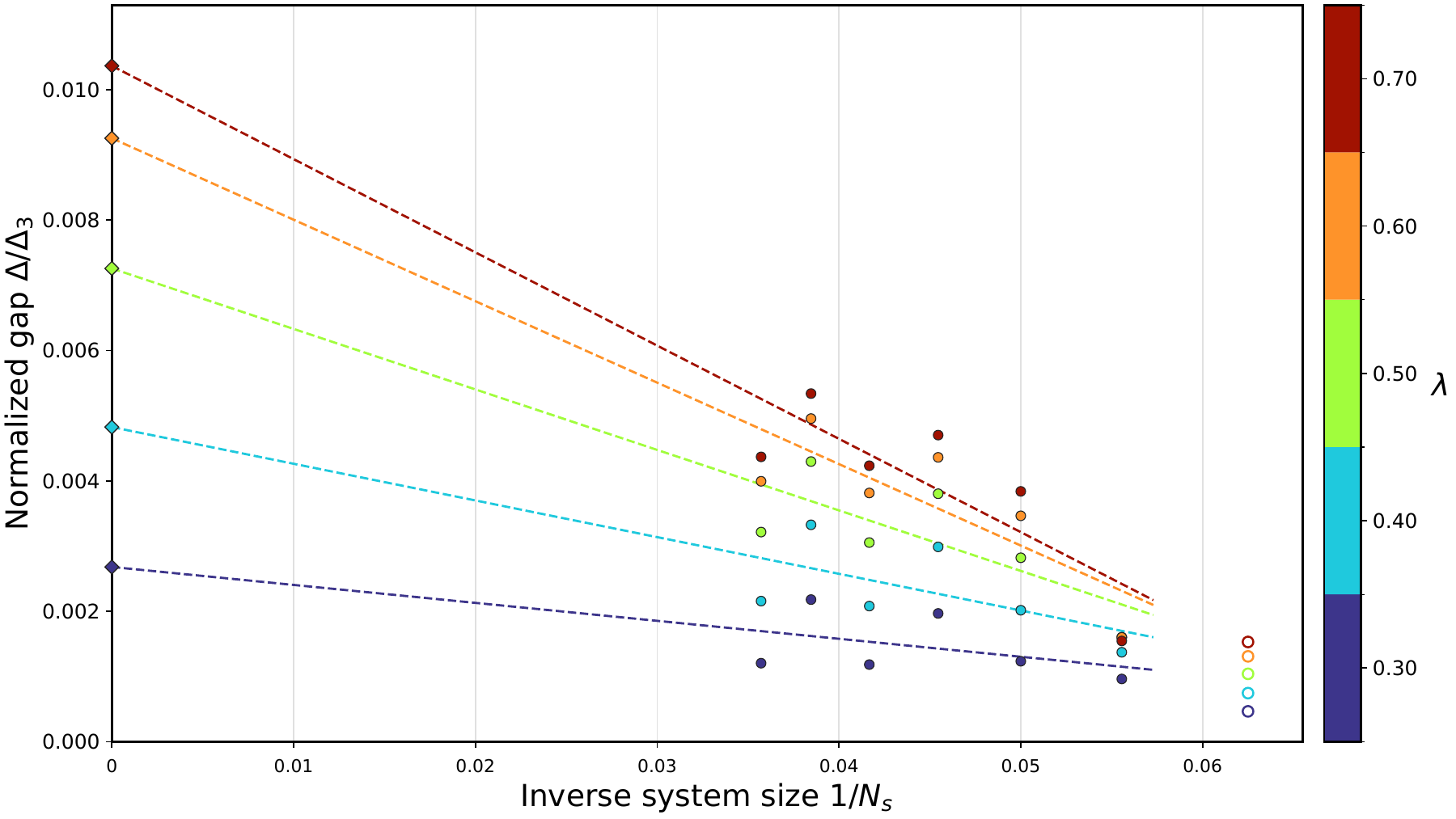}
\caption{Finite-size scaling of the normalized Moore--Read gap $\Delta(N_s)/\Delta_3$ at $\mu=\lambda$ for $\lambda=0.30,0.40,\ldots,0.70$. Colors identify $\lambda$. Filled circles mark $N_s=18,20,\ldots,28$, the six sizes in each unweighted linear fit. Open circles show $N_s=16$. Each dashed line extends from the intercept $g_\infty$ at $1/N_s=0$ (diamond) to just beyond $1/18$.}
\label{num:fig:gap-scaling-normalized-mr}
\end{figure}

Figure~\ref{num:fig:gap-scaling-normalized-mr} shows the gaps at $\mu=\lambda$ for $N_s=16,18,\ldots,28$, normalized by the thermodynamic scales in Table~\ref{num:tab:charge3-scales}. At each fixed $\lambda$, we use an unweighted linear fit over $N_s\ge18$ to Eq.~\eqref{num:eq:normalized-gap-scaling}, with $\Delta_2$ replaced by $\Delta_3$.\\

\FloatBarrier
\subsubsection{Four-component model: Halperin states}
\label{num:sec:ed-halperin}

For the Halperin $33h$ states, $h=0,1,2$, we use $n_{\mathrm c}=4$ and write $u^{33h}$ for the normalized form of $v^{(4)}_\lambda$ in Eq.~\eqref{bands:eq:c2-spinor}. We project the three onsite parents in Eq.~\eqref{bands:eq:four-onsite-parents} into the zero-energy band of Eq.~\eqref{bands:eq:four-gaussian-parent}, at $\lambda=1$ and $U>0$. The fillings are $\nu=1/3,1/4,1/5$ for $330,331,332$, respectively. The density terms follow Eq.~\eqref{num:eq:ed-projected-interaction} with $u=u^{33h}$ and the parent couplings.
The Halperin $33h$ state is described by the topological $K$ matrix~\cite{WenZee1992}
\begin{equation}
K=\begin{bmatrix}3&h\\h&3\end{bmatrix}.
\label{num:eq:ed-halperin-k-matrix}
\end{equation}
Translation by one site along $x$ exchanges the effective $C=1$ layers; a cycle along $\mathbf T_i$ therefore exchanges them exactly when $T_{ix}$ is odd. The torus degeneracy is $|\det K|=(3+h)(3-h)$ when both cycles preserve the layers. A layer-exchanging cycle makes the relative-layer sectors gauge equivalent, leaving $3+h$ states~\cite{BarkeshliQi2012}. For layer-preserving boundaries, equal layer flux requires equal particle populations in the Halperin zero modes, hence even $N_e$. The predicted zero-mode count is
\begin{equation}
N_0^{\mathrm{pred}}= \begin{cases} 9-h^2,&T_{1x},T_{2x}\ \text{both even},\quad N_e\ \text{even}\\ 0,&T_{1x},T_{2x}\ \text{both even},\quad N_e\ \text{odd}\\ 3+h,&\text{at least one of }T_{1x},T_{2x}\ \text{odd} \end{cases}.
\label{num:eq:ed-halperin-counting}
\end{equation}

\begin{figure}[p]
\setstretch{1.0}\small
\centering
\includegraphics[width=0.92\textwidth]{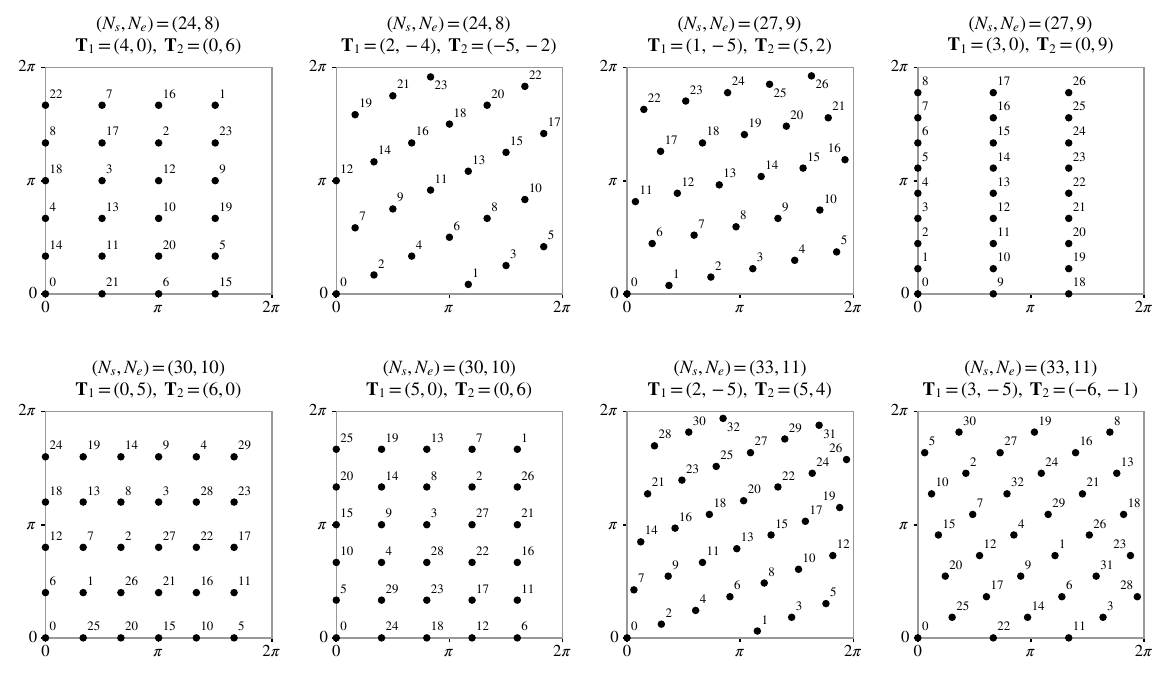}
\caption{Momentum meshes for the Halperin 330 calculations. For $C_4$-symmetric meshes, filled points are orbit representatives and hollow points are generated by their $C_4$ rotations.}
\label{num:fig:halperin-mesh-330}
\medskip
\centering
\includegraphics[width=0.85\textwidth]{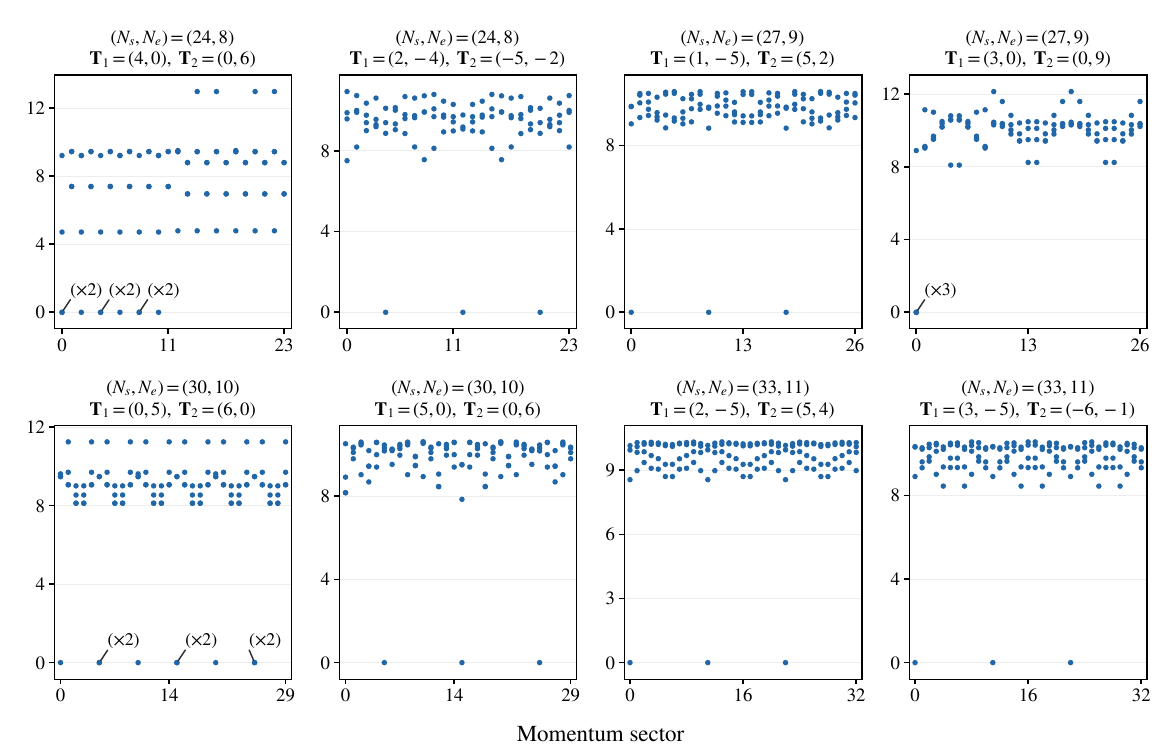}
\caption{ED spectra of the Halperin 330 model at $\nu=1/3$, $\lambda=1$, $U=1$ and zero boundary twists. Panel order matches Fig.~\ref{num:fig:halperin-mesh-330}; headers give $(N_s,N_e)$ and the real-space periods. Energies are plotted as $10^3E/U$. Each sector shows its four lowest levels. Numerical zero-mode totals in panel order are $9,3,3,3,9,3,3,3$. A label $(\times n)$ gives the zero-mode multiplicity in that sector.}
\label{num:fig:ed-halperin-330}
\end{figure}

\begin{figure}[p]
\setstretch{1.0}\small
\centering
\includegraphics[width=0.92\textwidth]{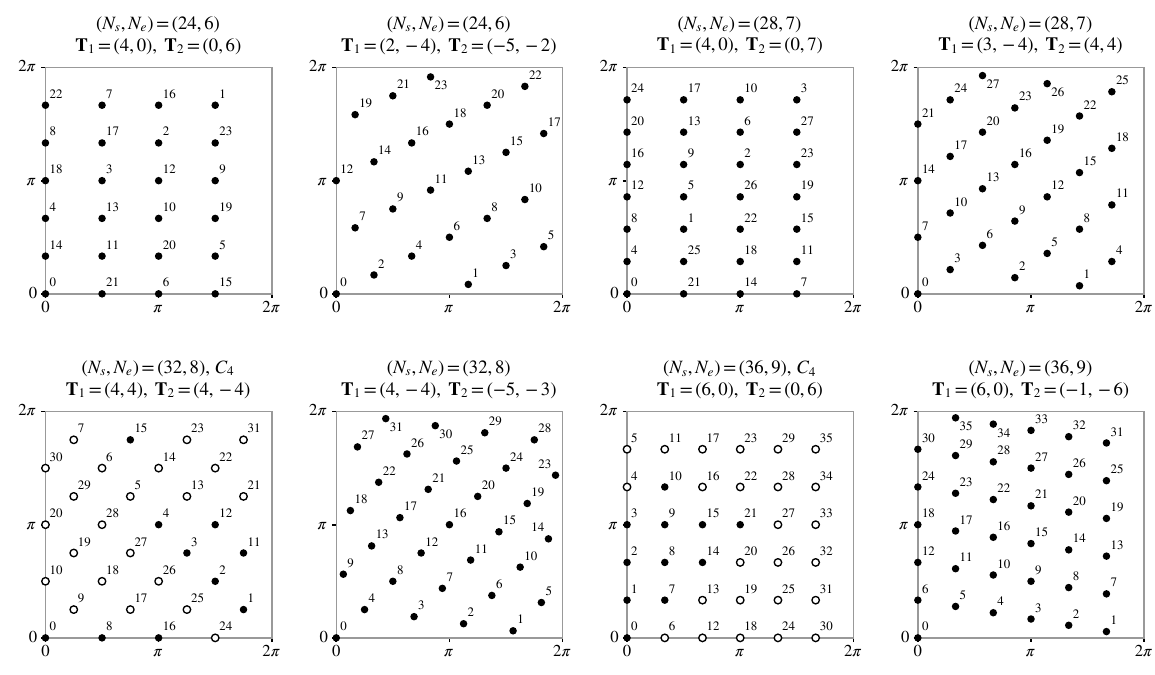}
\caption{Momentum meshes for the Halperin 331 calculations. For $C_4$-symmetric meshes, filled points are orbit representatives and hollow points are generated by their $C_4$ rotations.}
\label{num:fig:halperin-mesh-331}
\medskip
\centering
\includegraphics[width=0.85\textwidth]{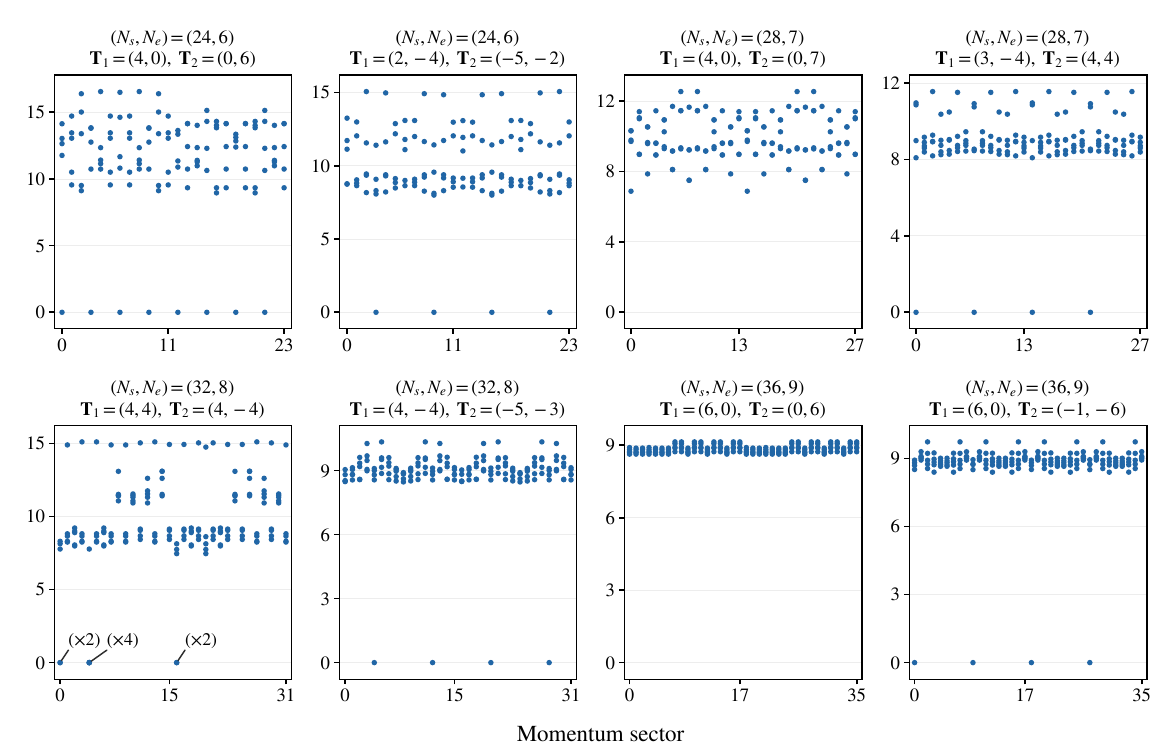}
\caption{ED spectra of the Halperin 331 model at $\nu=1/4$, $\lambda=1$, $U=1$ and zero boundary twists. Panel order matches Fig.~\ref{num:fig:halperin-mesh-331}; headers give $(N_s,N_e)$ and the real-space periods. Energies are plotted as $10^3E/U$. Each sector shows its five lowest levels. Numerical zero-mode totals in panel order are $8,4,0,4,8,4,0,4$. A label $(\times n)$ gives the zero-mode multiplicity in that sector.}
\label{num:fig:ed-halperin-331}
\end{figure}

\begin{figure}[p]
\setstretch{1.0}\small
\centering
\includegraphics[width=0.92\textwidth]{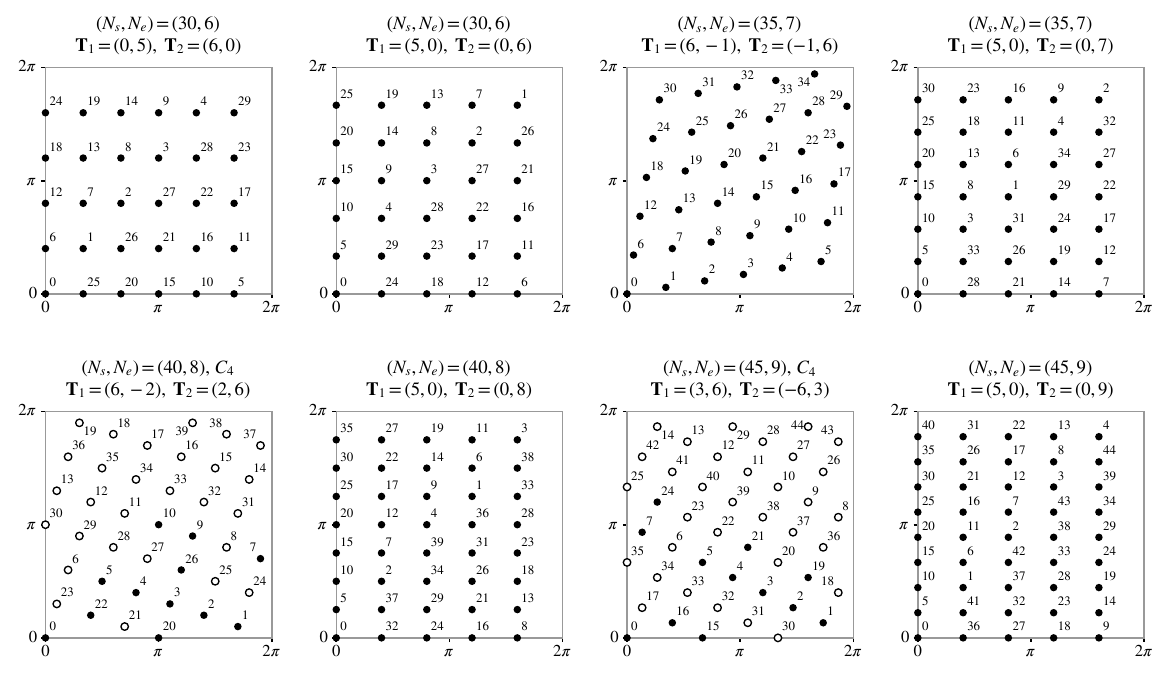}
\caption{Momentum meshes for the Halperin 332 calculations. For $C_4$-symmetric meshes, filled points are orbit representatives and hollow points are generated by their $C_4$ rotations.}
\label{num:fig:halperin-mesh-332}
\medskip
\centering
\includegraphics[width=0.85\textwidth]{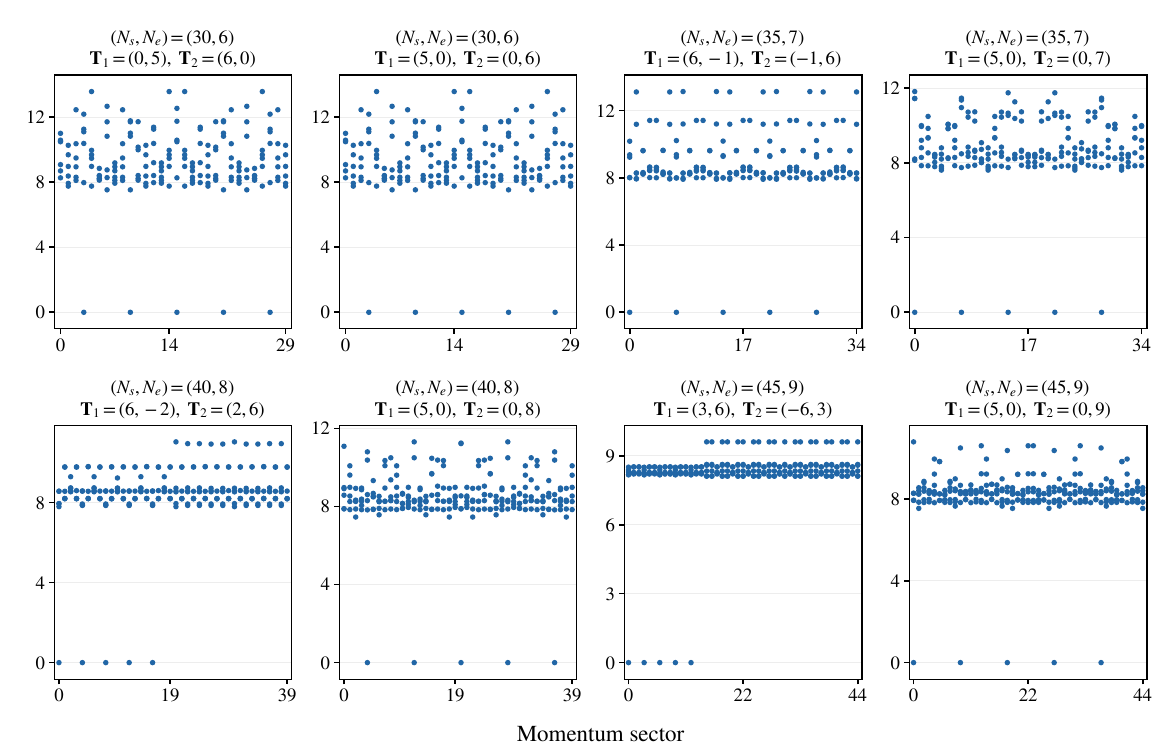}
\caption{ED spectra of the Halperin 332 model at $\nu=1/5$, $\lambda=1$, $U=1$ and zero boundary twists. Panel order matches Fig.~\ref{num:fig:halperin-mesh-332}; headers give $(N_s,N_e)$ and the real-space periods. Energies are plotted as $10^3E/U$. Each sector shows its six lowest levels. Numerical zero-mode totals in panel order are $5,5,5,5,5,5,5,5$. A label $(\times n)$ gives the zero-mode multiplicity in that sector.}
\label{num:fig:ed-halperin-332}
\end{figure}

Figures~\ref{num:fig:ed-halperin-330}, \ref{num:fig:ed-halperin-331}, and \ref{num:fig:ed-halperin-332} show the $330$, $331$, and $332$ spectra in the panel order of Figs.~\ref{num:fig:halperin-mesh-330}, \ref{num:fig:halperin-mesh-331}, and \ref{num:fig:halperin-mesh-332}. All numerical zero-mode counts agree with Eq.~\eqref{num:eq:ed-halperin-counting}.

\clearpage

\clearpage

\FloatBarrier
\subsection{Particle entanglement spectrum}
\label{num:sec:pes}
\FloatBarrier
\subsubsection{General implementation}
\label{num:sec:pes-general}

We test fractional quantum Hall (FQH) correlations in the Laughlin and Halperin ground states by comparing the number of PES levels below the entanglement gap with the expected quasihole counts~\cite{Regnault2011,Sterdyniak2011}. For FQH model states, the particle reduced density matrix is supported in the quasihole space~\cite{Sterdyniak2011,Zozulya2007}; the quasihole count therefore bounds its rank but need not be saturated.\\

For $N_e$ fermions in $N_s$ band orbitals, we retain $N_\alpha$ particles in subsystem $\alpha$ and trace out the remaining $N_\beta=N_e-N_\alpha$ in $\beta$, with $1\le N_\alpha\le N_\beta$~\cite{Zozulya2007}. The retained particles occupy the same $N_s$ orbitals, so their Hilbert-space dimension is
\begin{equation}
D_{\alpha}=\binom{N_s}{N_{\alpha}}.
\label{num:eq:pes-full-dimension}
\end{equation}
Let $d_{\mathrm G}$ be the dimension of the ground-state manifold and $\ket{\Psi_a}$, $a=1,\ldots,d_{\mathrm G}$, an orthonormal basis for it. We use the equal-weight density matrix and its particle partial trace~\cite{Regnault2011}
\begin{equation}
\rho_{\mathrm G}=\frac1{d_{\mathrm G}}\sum_{a=1}^{d_{\mathrm G}}\ket{\Psi_a}\bra{\Psi_a},\qquad \rho_{\alpha}=\operatorname{Tr}_{\beta}\rho_{\mathrm G}.
\label{num:eq:pes-ground-mixture}
\end{equation}
Using orthonormal configurations labeled by occupied-orbital sets $I_{\alpha}$ and $I_{\beta}$ for retained and traced particles, we expand
\begin{equation}
\ket{\Psi_a}=\sum_{I_{\alpha},I_{\beta}}C_a(I_{\alpha},I_{\beta})\ket{I_{\alpha}}\ket{I_{\beta}}.
\label{num:eq:pes-entanglement-matrix}
\end{equation}
For disjoint $I_{\alpha}$ and $I_{\beta}$, $C_a(I_{\alpha},I_{\beta})$ is the Fock coefficient of $I_{\alpha}\cup I_{\beta}$, multiplied by the fermionic partition sign and divided by $\sqrt{\binom{N_e}{N_{\alpha}}}$. It vanishes when the two sets overlap. Tracing over the orthonormal $\beta$ configurations gives
\begin{equation}
(\rho_{\alpha})_{I_{\alpha}I_{\alpha}'}=\frac1{d_{\mathrm G}}\sum_{a=1}^{d_{\mathrm G}}\sum_{I_{\beta}}C_a(I_{\alpha},I_{\beta})C_a(I_{\alpha}',I_{\beta})^*=\frac1{d_{\mathrm G}}\sum_{a=1}^{d_{\mathrm G}}(C_aC_a^\dagger)_{I_{\alpha}I_{\alpha}'}.
\label{num:eq:pes-reduced-matrix}
\end{equation}
Normalization gives $\operatorname{Tr}\rho_{\alpha}=1$. Its eigenvalues $p_j$ give the entanglement levels $\xi_j=-\ln p_j$~\cite{LiHaldane2008}.\\

Translation symmetry separates the PES into retained-particle momentum sectors $\mathbf K_{\alpha}$~\cite{Regnault2011,Bernevig2012}. Here $\mathbf K_{\alpha}$ sums the momenta of the $N_{\alpha}$ retained particles and takes the $N_s$ values in $\mathcal M\subset[0,2\pi)^2$ from Eq.~\eqref{num:eq:ed-torus-momenta}. Writing $\mathbf K_{\beta}$ for the total momentum of the traced particles, a ground state in sector $\mathbf K$ satisfies
\begin{equation}
\mathbf K_{\alpha}+\mathbf K_{\beta}=\mathbf K\pmod{2\pi\mathbb Z^2},\qquad \mathbf K_{\alpha},\mathbf K_{\beta}\in\mathcal M.
\label{num:eq:pes-subsystem-momenta}
\end{equation}
For each ground state, the traced $\beta$ configurations have $\mathbf K_{\beta}=\mathbf K-\mathbf K_{\alpha}$ with its own total $\mathbf K$. Contracting identical $\beta$ configurations connects only $\alpha$ configurations of equal $\mathbf K_{\alpha}$. Equation~\eqref{num:eq:pes-reduced-matrix} then gives blocks $\rho_{\alpha}(\mathbf K_{\alpha})$, whose spectra are $\xi_{j,\mathbf K_{\alpha}}=-\ln p_{j,\mathbf K_{\alpha}}$, with global normalization $\sum_{\mathbf K_{\alpha},j}p_{j,\mathbf K_{\alpha}}=1$.\\

A common probability threshold $p^*$ separates the low-lying PES levels from the numerical null-space tail. The sector and total counts are
\begin{equation}
\mathcal N_{\mathrm{PES}}(\mathbf K_{\alpha})=\#\{j:p_{j,\mathbf K_{\alpha}}>p^*\},\qquad \mathcal N_{\mathrm{PES}}=\sum_{\mathbf K_{\alpha}}\mathcal N_{\mathrm{PES}}(\mathbf K_{\alpha}).
\label{num:eq:pes-observed-counts}
\end{equation}
We use $p^*=10^{-10}$, equivalently $\xi^*=-\ln p^*=10\ln10\simeq23.03$, for all Laughlin and Halperin spectra. All PES figures show $\xi=-\ln p$ for $p>0$, with blue points for $p>p^*$ and red points for $0<p\le p^*$. We compare both total and momentum-sector counts with theory for the Laughlin states; for the Halperin states, we compare only the total counts.

\FloatBarrier
\subsubsection{Laughlin 1/3 state}
\label{num:sec:pes-laughlin-third}

For a fermionic Laughlin $1/m$ state, with odd $m\ge3$, let $N_\phi$ denote the number of Landau-level orbitals on the torus. We obtain the quasihole count $\mathcal N_{\mathrm{qh}}^{(1,m)}(N_\phi,N_\alpha)$ by counting $(1,m)$-admissible $N_\alpha$-particle configurations: those with at most one particle in any $m$ consecutive orbitals, including intervals crossing the periodic boundary~\cite{Bernevig2012,BernevigHaldane2008}. This gives
\begin{equation}
\mathcal N_{\mathrm{qh}}^{(1,m)}(N_\phi,N_\alpha)=\frac{N_\phi}{N_\alpha}\binom{N_\phi-(m-1)N_\alpha-1}{N_\alpha-1},\qquad N_\phi\ge mN_\alpha.
\label{num:eq:pes-laughlin-count}
\end{equation}
For a $C=1$ fractional Chern insulator (FCI), we set $N_\phi=N_s$ to obtain the total quasihole count for the band~\cite{Regnault2011}.\\

The predicted momentum-sector counts are the dimensions of the simultaneous translation eigenspaces in the quasihole space~\cite{Bernevig2012}. They depend on the torus geometry and sum to Eq.~\eqref{num:eq:pes-laughlin-count}.\\

For the two-component band in Eq.~\eqref{bands:eq:lambda-mu-theta} with the onsite parent in Eq.~\eqref{bands:eq:two-many-body-parent}, we compute the PES at $(N_s,N_e)=(36,12)$ on the torus with periods $\mathbf T_1=(6,0)$ and $\mathbf T_2=(0,6)$, using $\mu=0$ and $\lambda=0.1,0.2,0.3,0.4$. We label subsystem momenta by $(j_x,j_y)\in\mathbb Z_6\times\mathbb Z_6$, with
\begin{equation}
\frac{\mathbf K_\alpha}{2\pi}=\left(\frac{j_x}{6},\frac{j_y}{6}\right)\pmod{\mathbb Z^2}.
\label{num:eq:pes-third-momentum-labels}
\end{equation}

Table~\ref{num:tab:pes-counting-third} gives the total counts from Eq.~\eqref{num:eq:pes-laughlin-count} at $m=3$ and $N_\phi=N_s=36$, and the sector counts on the $6\times6$ mesh.\\

\begin{table}[!htbp]
\centering
\begin{tabular}{rrl}
\toprule
$N_{\alpha}$ & Predicted total & Predicted per sector\\
\midrule
2 & 558 & $14\ (9),\ 16\ (27)$\\
3 & 4872 & $135\ (32),\ 138\ (4)$\\
4 & 26325 & $728\ (27),\ 741\ (9)$\\
5 & 91080 & $2530\ (36)$\\
\bottomrule
\end{tabular}
\caption{Theoretical PES counts for the Laughlin $1/3$ state at $(N_s,N_e)=(36,12)$. Sector counts are specific to the $6\times6$ momentum mesh in Eq.~\eqref{num:eq:pes-third-momentum-labels}. The notation $l\ (s)$ denotes $l$ levels in each of $s$ sectors.}
\label{num:tab:pes-counting-third}
\end{table}

\begin{figure}[!htbp]
\centering
\includegraphics[width=0.86\textwidth]{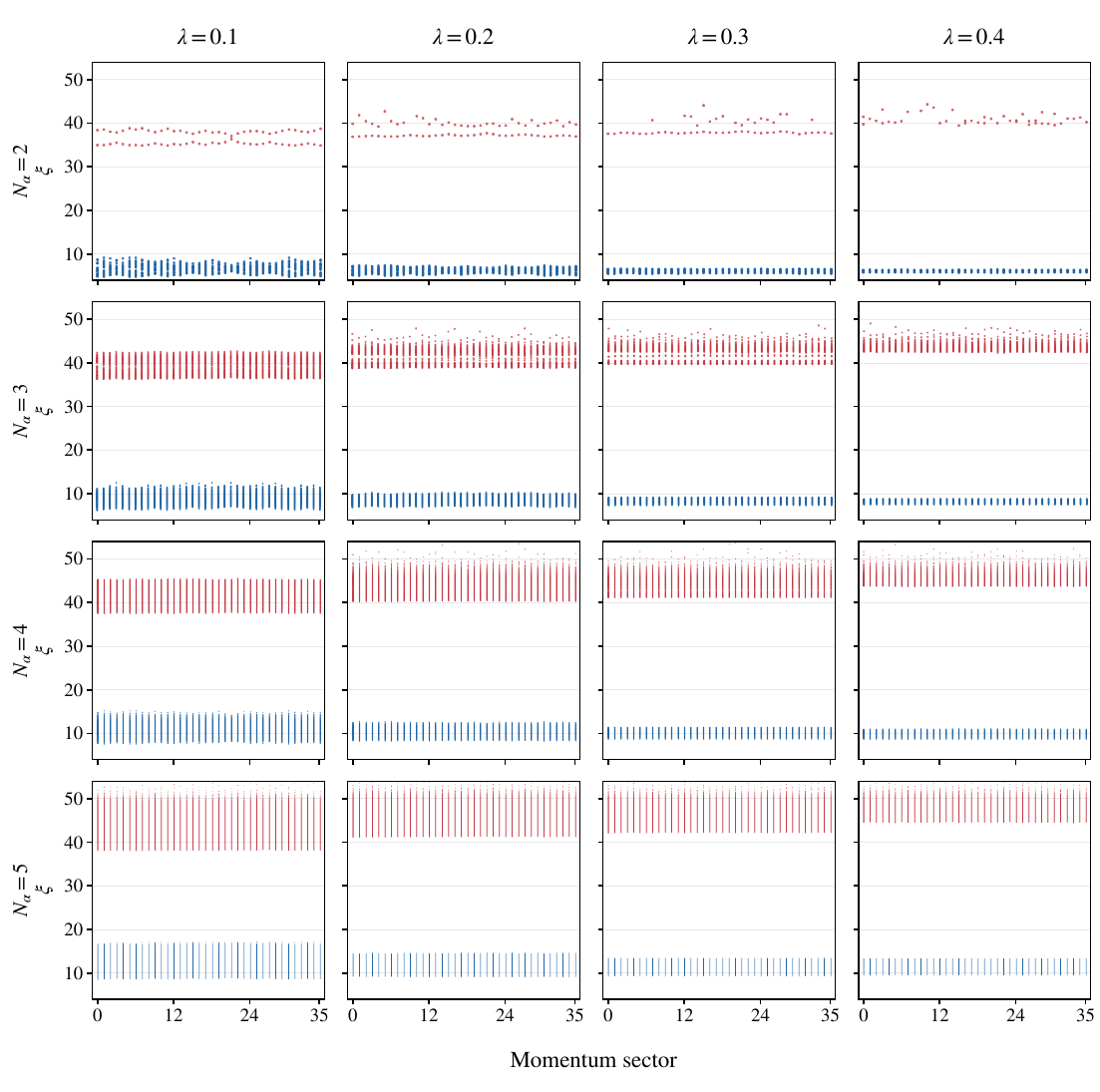}
\caption{PES of the Laughlin $1/3$ model at $(N_s,N_e)=(36,12)$ and $\mu=0$. Columns show $\lambda=0.1,0.2,0.3,0.4$ from left to right; rows show $N_\alpha=2,3,4,5$ from top to bottom. The torus periods are $(\mathbf T_1,\mathbf T_2)=((6,0),(0,6))$. Momentum labels $(j_x,j_y)$ range over $\mathbb Z_6\times\mathbb Z_6$, and the horizontal axis orders sectors by $6j_x+j_y$. Blue points have $p>10^{-10}$; red points have $0<p\le10^{-10}$.}
\label{num:fig:pes-laughlin-third}
\end{figure}

In all 16 spectra in Fig.~\ref{num:fig:pes-laughlin-third}, the counts above $p^*$ agree with Table~\ref{num:tab:pes-counting-third}, both in total and in each of the 36 sectors.\\

\FloatBarrier
\subsubsection{Laughlin 1/5 state}
\label{num:sec:pes-laughlin-fifth}

For the three-component band in Eq.~\eqref{eq:three-component-regular} with the onsite parent in Eq.~\eqref{bands:eq:three-pair-parent}, we compute the PES at $(N_s,N_e)=(45,9)$ on the torus with periods $\mathbf T_1=(3,6)$ and $\mathbf T_2=(-6,3)$, using $\lambda=\mu=0.2,0.3,0.4,0.5$. We label subsystem momenta by $(j_x,j_y)\in\mathbb Z_3\times\mathbb Z_{15}$, with
\begin{equation}
\frac{\mathbf K_\alpha}{2\pi}=\left(\frac{j_x}{3}-\frac{2j_y}{15},\frac{j_y}{15}\right)\pmod{\mathbb Z^2}.
\label{num:eq:pes-fifth-momentum-labels}
\end{equation}

The $(1,5)$-admissible counts from Eq.~\eqref{num:eq:pes-laughlin-count} at $m=5$ and $N_\phi=N_s=45$ are listed in Table~\ref{num:tab:pes-counting-fifth}, together with sector counts on the $3\times15$ mesh.\\

\begin{table}[!htbp]
\centering
\begin{tabular}{rrl}
\toprule
$N_{\alpha}$ & Predicted total & Predicted per sector\\
\midrule
2 & 810 & $18\ (45)$\\
3 & 7440 & $165\ (40),\ 168\ (5)$\\
4 & 36855 & $819\ (45)$\\
\bottomrule
\end{tabular}
\caption{Theoretical PES counts for the Laughlin $1/5$ state at $(N_s,N_e)=(45,9)$. Sector counts are specific to the $3\times15$ momentum mesh in Eq.~\eqref{num:eq:pes-fifth-momentum-labels}. The notation $l\ (s)$ denotes $l$ levels in each of $s$ sectors.}
\label{num:tab:pes-counting-fifth}
\end{table}

\begin{figure}[!htbp]
\centering
\includegraphics[width=0.86\textwidth]{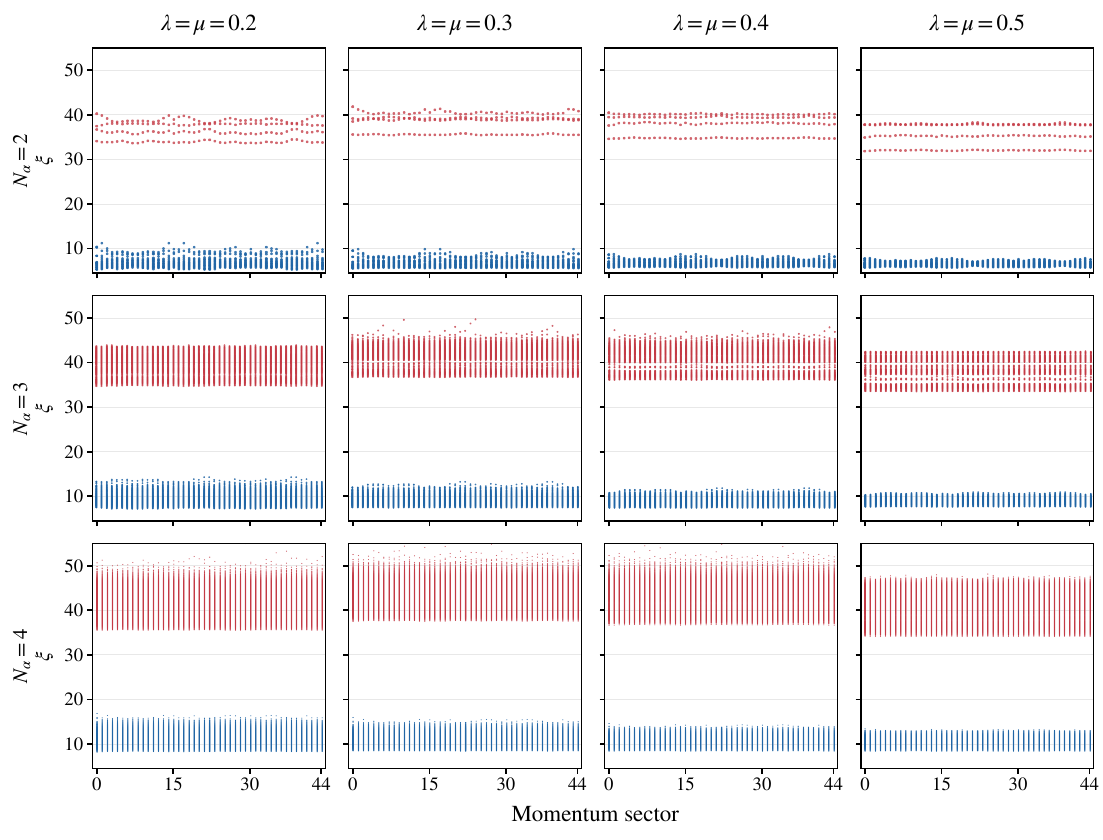}
\caption{PES of the Laughlin $1/5$ model at $(N_s,N_e)=(45,9)$ and $\mu=\lambda$. Columns show $\lambda=0.2,0.3,0.4,0.5$ from left to right; rows show $N_\alpha=2,3,4$ from top to bottom. The torus periods are $(\mathbf T_1,\mathbf T_2)=((3,6),(-6,3))$. Momentum labels $(j_x,j_y)$ range over $\mathbb Z_3\times\mathbb Z_{15}$, and the horizontal axis orders sectors by $15j_x+j_y$. Blue points have $p>10^{-10}$; red points have $0<p\le10^{-10}$.}
\label{num:fig:pes-laughlin-fifth}
\end{figure}

In all 12 spectra in Fig.~\ref{num:fig:pes-laughlin-fifth}, the counts above $p^*$ agree with Table~\ref{num:tab:pes-counting-fifth}, both in total and in each of the 45 sectors.\\

\FloatBarrier
\subsubsection{Halperin 330 state}
\label{num:sec:pes-halperin-330}

The Halperin 330 state consists of two independent Laughlin-1/3 layers~\cite{Halperin1983}. We use superscripts $\mathrm{unex}$ and $\mathrm{ex}$ for boundaries without and with layer exchange, respectively. Without layer exchange, $N_\phi^\uparrow=N_\phi^\downarrow=N_s/2$, and summing Eq.~\eqref{num:eq:pes-laughlin-count} over the retained layer populations gives
\begin{equation}
\mathcal N_{\mathrm{qh},330}^{\mathrm{unex}}(N_s,N_\alpha)=\sum_{N_\alpha^\uparrow=0}^{N_\alpha}\mathcal N_{\mathrm{qh}}^{(1,3)}(N_\phi^\uparrow,N_\alpha^\uparrow)\mathcal N_{\mathrm{qh}}^{(1,3)}(N_\phi^\downarrow,N_\alpha-N_\alpha^\uparrow).
\label{num:eq:pes-halperin-330-no}
\end{equation}
Here $\mathcal N_{\mathrm{qh}}^{(1,3)}(\ell,0)=1$, and the count vanishes for $\ell<3a$ when $a>0$.\\

With layer exchange, the gluing construction in Ref.~\cite{BarkeshliQi2012} joins the layers into a single Laughlin-1/3 torus with $N_s$ orbitals, so
\begin{equation}
\mathcal N_{\mathrm{qh},330}^{\mathrm{ex}}(N_s,N_\alpha)=\mathcal N_{\mathrm{qh}}^{(1,3)}(N_s,N_\alpha).
\label{num:eq:pes-halperin-330-ex}
\end{equation}

For even $N_e$, $N_s=3N_e$, and $1\le N_\alpha\le N_e/2$, substituting Eq.~\eqref{num:eq:pes-laughlin-count} and expanding the binomial sum gives
\begin{equation}
\mathcal N_{\mathrm{qh},330}^{\mathrm{unex}}(N_s,N_\alpha)=\mathcal N_{\mathrm{qh},330}^{\mathrm{ex}}(N_s,N_\alpha).
\label{num:eq:pes-halperin-330-equality}
\end{equation}

The unexchanged 330 PES is constrained by each layer's environment separately. Even when $N_\alpha\le N_e/2$, a layer retaining $a>N_e/4$ of its $N_e/2$ particles has a smaller environment, and the other layer cannot supply its Schmidt partners. In the single-layer Laughlin cuts, the environment contains all $N_e-N_\alpha$ traced particles.\\

Applying Eq.~\eqref{num:eq:pes-entanglement-matrix} to a pure ground state of one layer gives the Schmidt decomposition
\begin{equation}
\ket{\Psi_g}=\sum_{j=1}^{r_g}\sqrt{p_j^{(g)}}\ket{u_j^{(g)}}_\alpha\ket{v_j^{(g)}}_\beta,\qquad p_j^{(g)}>0.
\label{num:eq:pes-halperin-schmidt}
\end{equation}
The $r_g$ nonzero eigenvalues require $r_g$ orthonormal states on each side. Both sides have quasihole support~\cite{Sterdyniak2011}, so $r_g$ is bounded by $\mathcal N_{\mathrm{qh}}^{(1,3)}(N_\phi^\uparrow,N_e/2-a)$. For the equal mixture of three layer ground states, let $R_a$ denote the rank. The reduced matrices add as in Eq.~\eqref{num:eq:pes-reduced-matrix}; the rank of their sum is bounded by both the sum of individual ranks and the retained quasihole dimension:
\begin{equation}
R_a\le\min\left\{\mathcal N_{\mathrm{qh}}^{(1,3)}(N_\phi^\uparrow,a),\;3\mathcal N_{\mathrm{qh}}^{(1,3)}(N_\phi^\uparrow,N_e/2-a)\right\}.
\label{num:eq:pes-halperin-330-rankbound}
\end{equation}

The nine-state 330 mixture factorizes into two layer mixtures, so each retained-population block $(a,N_\alpha-a)$ has tensor-product rank $R_aR_{N_\alpha-a}$. Summing over orthogonal blocks gives
\begin{equation}
\operatorname{rank}\rho_\alpha^{\mathrm{unex}}=\sum_{a=0}^{N_\alpha}R_aR_{N_\alpha-a},
\label{num:eq:pes-halperin-330-totalrank}
\end{equation}
with $R_a=0$ outside $0\le a\le N_e/2$. \\

For the four-component 330 parent studied in Sec.~\ref{num:sec:ed-halperin}, we compute the PES at $(N_s,N_e)=(30,10)$ for $N_\alpha=2,3,4,5$ under both layer-boundary conditions, using $\lambda=1$ and $U=1$.\\

Without layer exchange, each layer has five particles in 15 orbitals. Retaining four leaves a one-particle environment of dimension 15, so $R_4\le3\times15=45$ rather than 75 quasihole states, a deficit of 30. At $N_\alpha=4$, blocks $(4,0)$ and $(0,4)$ each contain an empty retained layer of rank one, reducing the total bound by $2\times30\times1=60$. At $N_\alpha=5$, blocks $(4,1)$ and $(1,4)$ contain a one-particle retained layer of rank 15, giving a reduction of $2\times30\times15=900$. Assuming saturation gives Table~\ref{num:tab:pes-halperin-330-pes}.\\

\begin{table}[!htbp]
\centering
\begin{tabular}{rrr}
\toprule
$N_\alpha$ & Predicted total (unex) & Predicted total (ex)\\
\midrule
2 & 375 & 375\\
3 & 2530 & 2530\\
4 & 9915 & 9975\\
5 & 22356 & 23256\\
\bottomrule
\end{tabular}
\caption{Predicted total PES counts for Halperin 330 at $(N_s,N_e)=(30,10)$. The unex prediction assumes saturation of the layer environment rank bounds; the ex prediction is the complete quasihole count.}
\label{num:tab:pes-halperin-330-pes}
\end{table}

\begin{figure}[!htbp]
\centering
\includegraphics[width=0.86\textwidth]{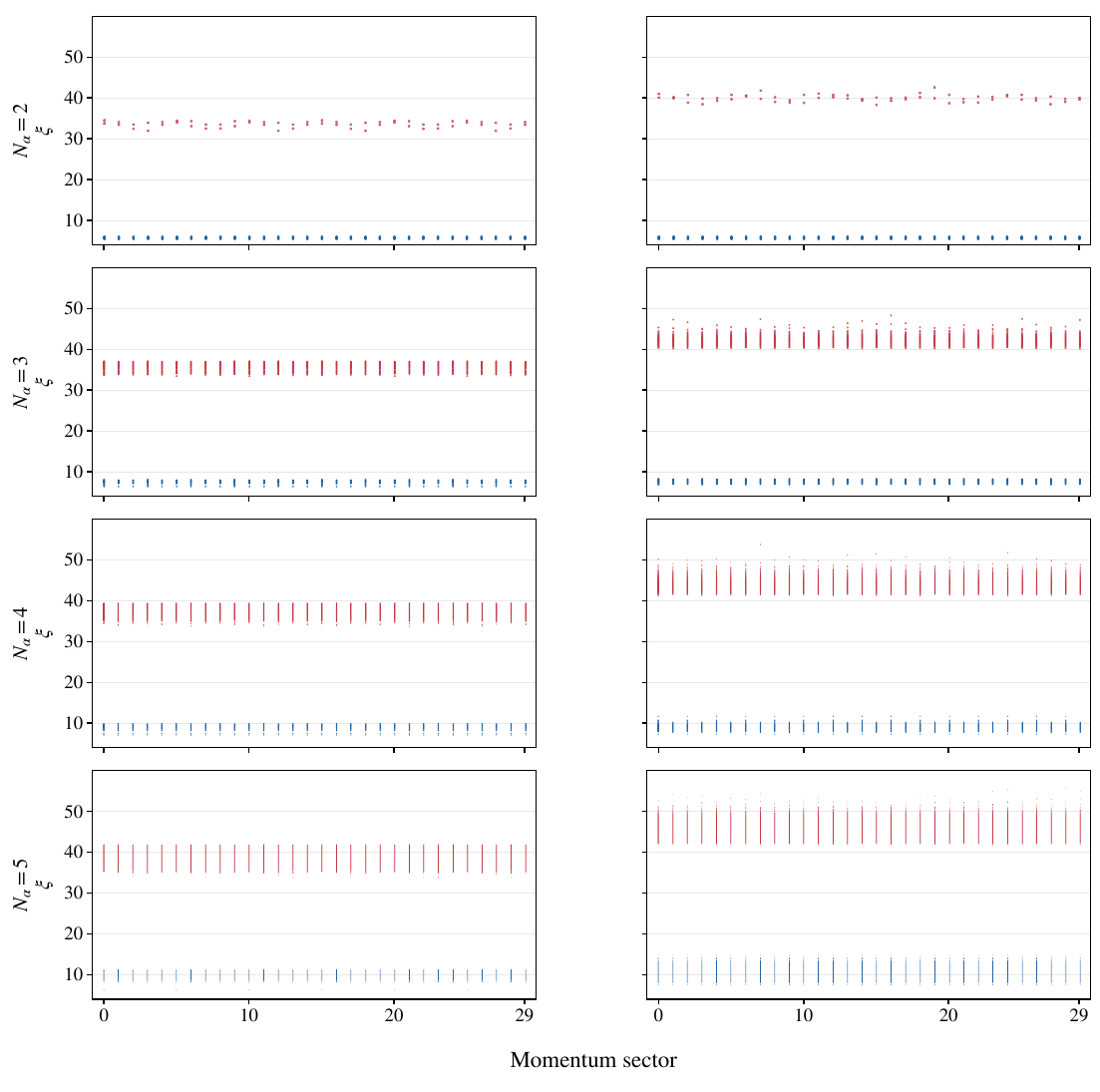}
\caption{PES of the Halperin 330 model at $(N_s,N_e)=(30,10)$, $\lambda=1$, and $U=1$. Columns show boundaries without (left, unex) and with (right, ex) layer exchange; rows show $N_\alpha=2,3,4,5$ from top to bottom. The torus periods $(\mathbf T_1,\mathbf T_2)$ are $((0,5),(6,0))$ on the left and $((5,0),(0,6))$ on the right. Momentum labels $(j_x,j_y)$ range over $\mathbb Z_1\times\mathbb Z_{30}$ in both columns, and the horizontal axis orders sectors by $j_y$. Blue points have $p>10^{-10}$; red points have $0<p\le10^{-10}$.}
\label{num:fig:pes-halperin-330}
\end{figure}

In all eight spectra in Fig.~\ref{num:fig:pes-halperin-330}, the total counts above $p^*$ agree with Table~\ref{num:tab:pes-halperin-330-pes}.\\

\FloatBarrier
\subsubsection{Halperin 331 state}
\label{num:sec:pes-halperin-331}

For the Halperin 331 state without layer exchange, $N_\phi^\uparrow=N_\phi^\downarrow=N_s/2$. Reference~\cite{SeidelYang2011} obtains equal torus quasihole counts for the 331 and Haldane--Rezayi states by counting thin-torus domain-wall arrangements and spin configurations. Simplifying the resulting combinatorial sum, we identify the $(1,4)$-admissible count
\begin{equation}
\mathcal N_{\mathrm{qh},331}^{\mathrm{unex}}(N_s,N_\alpha)=\frac{N_s}{N_\alpha}\binom{N_s-3N_\alpha-1}{N_\alpha-1}=\mathcal N_{\mathrm{qh}}^{(1,4)}(N_s,N_\alpha),\qquad N_s>4N_\alpha.
\label{num:eq:pes-halperin-331-simplified}
\end{equation}

For layer-exchanging boundaries, we conjecture the same total quasihole count, $\mathcal N_{\mathrm{qh},331}^{\mathrm{ex}}=\mathcal N_{\mathrm{qh},331}^{\mathrm{unex}}$, in this regime, but have not derived it.\\

For the four-component 331 parent studied in Sec.~\ref{num:sec:ed-halperin}, we compute the PES at $(N_s,N_e)=(32,8)$ for $N_\alpha=2,3,4$ under both layer-boundary conditions, using $\lambda=1$ and $U=1$. The unexchanged parent has four particles per layer, so these cuts exclude no retained layer-population sector. Table~\ref{num:tab:pes-halperin-331-counts} lists the predicted total counts.\\

\begin{table}[!htbp]
\centering
\begin{tabular}{rrr}
\toprule
$N_\alpha$ & Predicted total (unex) & Predicted total (ex)\\
\midrule
2 & 400 & 400\\
3 & 2464 & 2464\\
4 & 7752 & 7752\\
\bottomrule
\end{tabular}
\caption{Predicted total PES counts for Halperin 331 at $(N_s,N_e)=(32,8)$. The unex prediction follows from Eq.~\eqref{num:eq:pes-halperin-331-simplified}; the same ex count is conjectured.}
\label{num:tab:pes-halperin-331-counts}
\end{table}

\begin{figure}[!htbp]
\centering
\includegraphics[width=0.86\textwidth]{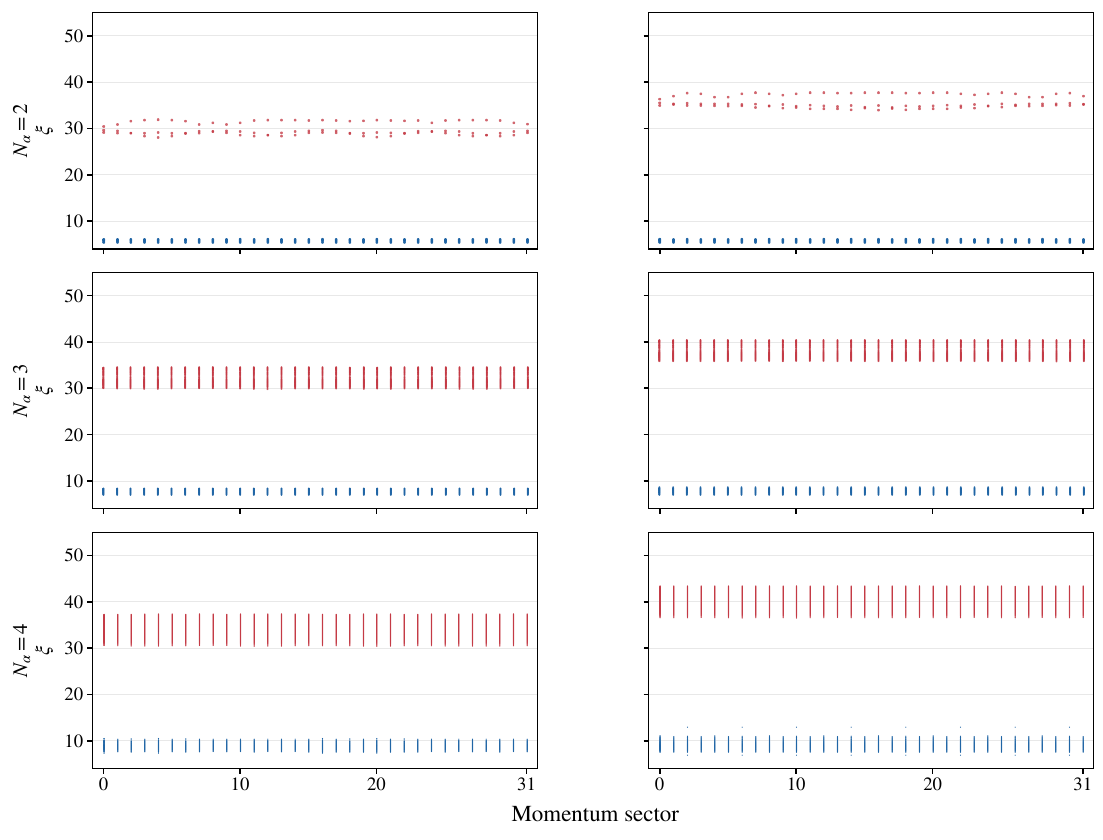}
\caption{PES of the Halperin 331 model at $(N_s,N_e)=(32,8)$, $\lambda=1$, and $U=1$. Columns show boundaries without (left, unex) and with (right, ex) layer exchange; rows show $N_\alpha=2,3,4$ from top to bottom. The torus periods $(\mathbf T_1,\mathbf T_2)$ are $((4,4),(4,-4))$ on the left and $((4,-4),(-5,-3))$ on the right. Momentum labels $(j_x,j_y)$ range over $\mathbb Z_4\times\mathbb Z_8$ on the left and $\mathbb Z_1\times\mathbb Z_{32}$ on the right; the horizontal axis orders sectors by $8j_x+j_y$ and $j_y$, respectively. Blue points have $p>10^{-10}$; red points have $0<p\le10^{-10}$.}
\label{num:fig:pes-halperin-331}
\end{figure}

In all six spectra in Fig.~\ref{num:fig:pes-halperin-331}, the total counts above $p^*$ agree with Table~\ref{num:tab:pes-halperin-331-counts}, supporting the conjecture for layer-exchanging boundaries.\\

\FloatBarrier
\subsubsection{Halperin 332 state}
\label{num:sec:pes-halperin-332}

Numerical studies of fermionic $C=2$ FCIs report total quasihole and PES counts consistent with the $(1,5)$-admissible rule~\cite{LiuEtAl2012,WangLiu2022}. Based on these results and the analytic construction of Halperin 332 states in ideal $C=2$ bands~\cite{DongEtAl2023}, we expect the total quasihole count for both layer-boundary conditions to follow Eq.~\eqref{num:eq:pes-laughlin-count} with $m=5$:
\begin{equation}
\mathcal N_{\mathrm{qh},332}^{\mathrm{ex}}(N_s,N_\alpha)=\mathcal N_{\mathrm{qh},332}^{\mathrm{unex}}(N_s,N_\alpha)=\frac{N_s}{N_\alpha}\binom{N_s-4N_\alpha-1}{N_\alpha-1},\qquad N_s\ge5N_\alpha.
\label{num:eq:pes-halperin-332-no}
\end{equation}

The unexchanged case requires even $N_s$, with $N_\phi^\uparrow=N_\phi^\downarrow=N_s/2$; the exchanged case also allows odd $N_s$. For $(N_s,N_e)=(40,8)$ and $N_\alpha=2,3,4$, Table~\ref{num:tab:pes-halperin-332-counts} lists the predicted total PES counts under both layer-boundary conditions.\\

\begin{table}[!htbp]
\centering
\begin{tabular}{rrr}
\toprule
$N_\alpha$ & Predicted total (unex) & Predicted total (ex)\\
\midrule
2 & 620 & 620\\
3 & 4680 & 4680\\
4 & 17710 & 17710\\
\bottomrule
\end{tabular}
\caption{Predicted total PES counts for Halperin 332 at $(N_s,N_e)=(40,8)$, following Eq.~\eqref{num:eq:pes-halperin-332-no} for both layer-boundary conditions.}
\label{num:tab:pes-halperin-332-counts}
\end{table}

\begin{figure}[!htbp]
\centering
\includegraphics[width=0.86\textwidth]{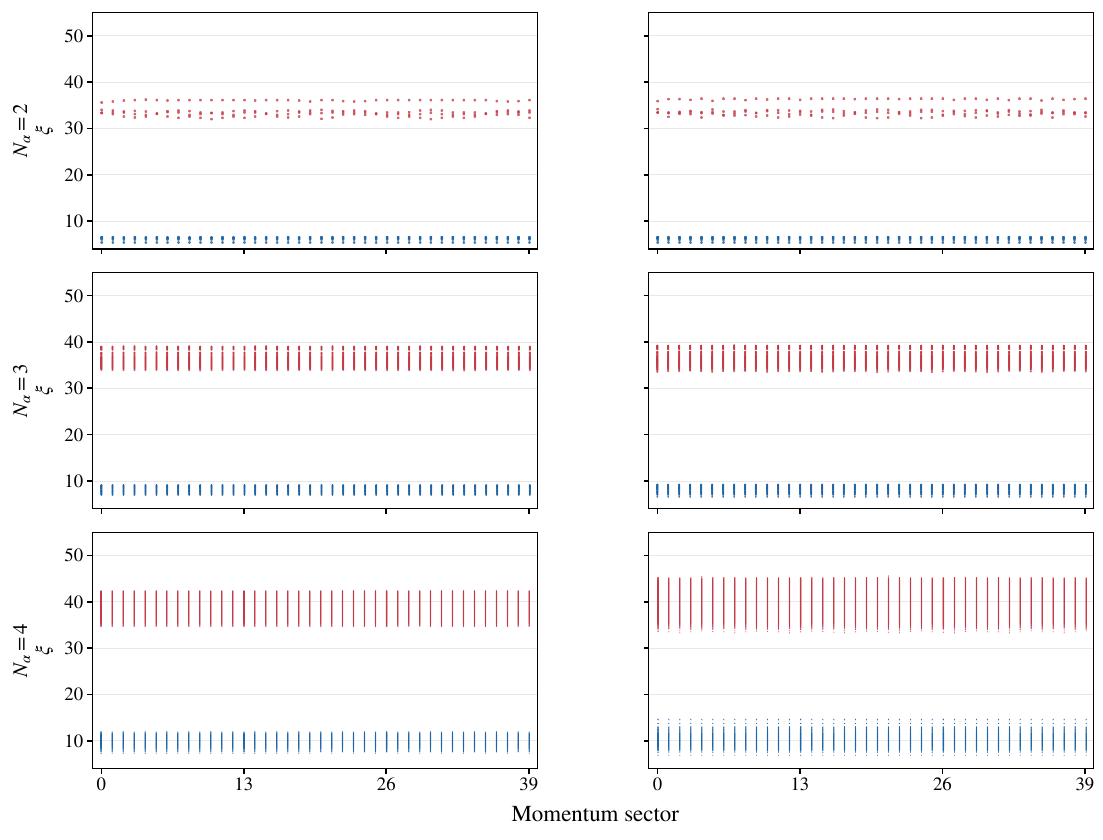}
\caption{PES of the Halperin 332 model at $(N_s,N_e)=(40,8)$, $\lambda=1$, and $U=1$. Columns show boundaries without (left, unex) and with (right, ex) layer exchange; rows show $N_\alpha=2,3,4$ from top to bottom. The torus periods $(\mathbf T_1,\mathbf T_2)$ are $((6,-2),(2,6))$ on the left and $((5,0),(0,8))$ on the right. Momentum labels $(j_x,j_y)$ range over $\mathbb Z_2\times\mathbb Z_{20}$ on the left and $\mathbb Z_1\times\mathbb Z_{40}$ on the right; the horizontal axis orders sectors by $20j_x+j_y$ and $j_y$, respectively. Blue points have $p>10^{-10}$; red points have $0<p\le10^{-10}$.}
\label{num:fig:pes-halperin-332}
\end{figure}

Figure~\ref{num:fig:pes-halperin-332} compares the two boundary conditions at $\lambda=1$ and $U=1$, using an equal-weight mixture of the five ground states on each torus. The total counts above $p^*$ match Table~\ref{num:tab:pes-halperin-332-counts} for all six spectra.\\

\FloatBarrier

\section{Special-function identities}
\label{bands:sec:special-functions}

In this appendix, we present useful special-function definitions and properties.

\subsection{Jacobi theta functions}
\label{special:sec:theta}

We follow the DLMF conventions for theta functions. For $w\in\mathbb C$, $\operatorname{Im}\tau>0$, and real characteristics $\alpha,\beta$, define~\cite{OlverEtAl2010}
\begin{equation}
\vartheta\!\begin{bmatrix}\alpha\\\beta\end{bmatrix}(w|\tau)=\sum_{n\in\mathbb Z}e^{\ii\pi\tau(n+\alpha)^2+2\pi\ii(n+\alpha)(w+\beta)}.
\label{special:eq:theta-characteristics}
\end{equation}
Shifting the argument and relabeling the summation index give
\begin{equation}
\vartheta\!\begin{bmatrix}\alpha\\\beta\end{bmatrix}(w+1|\tau)=e^{2\pi\ii\alpha}\vartheta\!\begin{bmatrix}\alpha\\\beta\end{bmatrix}(w|\tau),
\label{special:eq:theta-characteristic-real-shift}
\end{equation}
\begin{equation}
\vartheta\!\begin{bmatrix}\alpha\\\beta\end{bmatrix}(w+\tau|\tau)=e^{-\ii\pi\tau-2\pi\ii(w+\beta)}\vartheta\!\begin{bmatrix}\alpha\\\beta\end{bmatrix}(w|\tau).
\label{special:eq:theta-characteristic-tau-shift}
\end{equation}
Integer shifts of the characteristics satisfy
\begin{equation}
\vartheta\!\begin{bmatrix}\alpha+m\\\beta+n\end{bmatrix}(w|\tau)=e^{2\pi\ii\alpha n}\vartheta\!\begin{bmatrix}\alpha\\\beta\end{bmatrix}(w|\tau),\qquad m,n\in\mathbb Z.
\label{special:eq:theta-characteristic-integer-shift}
\end{equation}
Theta functions with characteristics use argument shifts $1,\tau$, while the classical Jacobi functions use $\pi,\pi\tau$. They are related by~\cite{OlverEtAl2010}
\begin{equation}
\vartheta_1(w|\tau)=-\vartheta\!\begin{bmatrix}1/2\\1/2\end{bmatrix}(w/\pi|\tau)=-\ii \sum_{n\in\mathbb Z}(-1)^n e^{\ii \pi\tau(n+1/2)^2+\ii (2n+1)w}.
\label{bands:eq:theta1-definition}
\end{equation}
\begin{equation}
\vartheta_2(w|\tau)=\vartheta\!\begin{bmatrix}1/2\\0\end{bmatrix}(w/\pi|\tau)=\sum_{n\in\mathbb Z} e^{\ii \pi\tau(n+1/2)^2+\ii (2n+1)w}.
\end{equation}
\begin{equation}
\vartheta_3(w|\tau)=\vartheta\!\begin{bmatrix}0\\0\end{bmatrix}(w/\pi|\tau)=\sum_{n\in\mathbb Z} e^{\ii \pi\tau n^2+2\ii nw}.
\end{equation}
\begin{equation}
\vartheta_4(w|\tau)=\vartheta\!\begin{bmatrix}0\\1/2\end{bmatrix}(w/\pi|\tau)=\sum_{n\in\mathbb Z}(-1)^n e^{\ii \pi\tau n^2+2\ii nw}.
\end{equation}
For purely imaginary $\tau$, $\vartheta_j(\bar w|\tau)=\overline{\vartheta_j(w|\tau)}$. Table~\ref{bands:tab:theta-properties} collects the parity, zeros, and translation identities of the four special cases~\cite{OlverEtAl2010}.\\

\begin{table}[H]
\centering
\renewcommand{\arraystretch}{1.35}
\begin{tabular}{lcccc}
\toprule
& $\vartheta_1$ & $\vartheta_2$ & $\vartheta_3$ & $\vartheta_4$\\
\midrule
Parity & odd & even & even & even\\
Zero & $0$ & $\pi/2$ & $(\pi+\pi\tau)/2$ & $\pi\tau/2$\\
\midrule
$\vartheta_j(w+\pi|\tau)$ & $-\vartheta_1$ & $-\vartheta_2$ & $\vartheta_3$ & $\vartheta_4$\\
$e^{\ii \pi\tau+2\ii w}\vartheta_j(w+\pi\tau|\tau)$ & $-\vartheta_1$ & $\vartheta_2$ & $\vartheta_3$ & $-\vartheta_4$\\
\midrule
$\vartheta_j(w+\pi/2|\tau)$ & $\vartheta_2$ & $-\vartheta_1$ & $\vartheta_4$ & $\vartheta_3$\\
$e^{\ii \pi\tau/4+\ii w}\vartheta_j(w+\pi\tau/2|\tau)$ & $\ii \vartheta_4$ & $\vartheta_3$ & $\vartheta_2$ & $\ii \vartheta_1$\\
\bottomrule
\end{tabular}
\caption{Parity, zeros, and positive full- and half-period shifts along $\pi$ and $\pi\tau$ (real and imaginary for purely imaginary $\tau$). Left-hand prefactors remove the quasiperiodic exponentials. Unshifted arguments are $(w|\tau)$; zeros are simple and given modulo $\pi\mathbb Z+\pi\tau\mathbb Z$.}
\label{bands:tab:theta-properties}
\end{table}

\subsection{Weierstrass functions}
\label{special:sec:weierstrass}

For $w\in\mathbb C$, let $\Lambda=2\pi(\mathbb Z+\ii \mathbb Z)$. We use the Weierstrass functions~\cite{OlverEtAl2010}:
\begin{equation}
\sigma_\Lambda(w)=w\prod_{\omega\in\Lambda\setminus\{0\}} \left(1-\dfrac{w}{\omega}\right) \exp\!\left(\dfrac{w}{\omega}+\dfrac{w^2}{2\omega^2}\right).
\label{bands:eq:sigma-product}
\end{equation}
The odd entire function $\sigma_\Lambda$ has $\sigma_\Lambda'(0)=1$ and simple zeros on $\Lambda$. Define
\begin{equation}
\zeta_\Lambda(w)=\dfrac{\d}{\d w}\ln\sigma_\Lambda(w)=\dfrac{\sigma_\Lambda'(w)}{\sigma_\Lambda(w)} =\dfrac{1}{w}+\sum_{\omega\in\Lambda\setminus\{0\}} \left(\dfrac{1}{w-\omega}+\dfrac{1}{\omega}+\dfrac{w}{\omega^2}\right).
\label{bands:eq:zeta-definition}
\end{equation}
\begin{equation}
\wp_\Lambda(w)=-\zeta_\Lambda'(w) =\dfrac{1}{w^2}+\sum_{\omega\in\Lambda\setminus\{0\}} \left(\dfrac{1}{(w-\omega)^2}-\dfrac{1}{\omega^2}\right).
\label{bands:eq:wp-definition}
\end{equation}
The odd function $\zeta_\Lambda$ has simple poles of residue one on $\Lambda$, while the even elliptic function $\wp_\Lambda$ has double poles of zero residue there.\\

The local expansions of these functions are determined by the lattice sums
\begin{equation}
\mathfrak g_2=60\sum_{\omega\in\Lambda\setminus\{0\}}\omega^{-4}, \qquad \mathfrak g_3=140\sum_{\omega\in\Lambda\setminus\{0\}}\omega^{-6}.
\label{bands:eq:weierstrass-invariants}
\end{equation}
For the specified square lattice~\cite{OlverEtAl2010},
\begin{equation}
\mathfrak g_2=\dfrac{256\,\Gamma(5/4)^8}{\pi^6},\qquad \mathfrak g_3=0.
\label{bands:eq:square-invariants}
\end{equation}
Here $\Gamma$ is the Euler gamma function. Near $w=0$~\cite{OlverEtAl2010},
\begin{equation}
\sigma_\Lambda(w)=w-\dfrac{\mathfrak g_2}{240}w^5+O(w^9),\quad \zeta_\Lambda(w)=\dfrac1w-\dfrac{\mathfrak g_2}{60}w^3+O(w^7),\quad \wp_\Lambda(w)=\dfrac1{w^2}+\dfrac{\mathfrak g_2}{20}w^2+O(w^6).
\label{bands:eq:square-expansions}
\end{equation}
The zeta quasiperiods are
\begin{equation}
\zeta_\Lambda(w+2\pi)=\zeta_\Lambda(w)+\dfrac12,\qquad \zeta_\Lambda(w+2\pi \ii)=\zeta_\Lambda(w)-\dfrac{\ii}{2}.
\label{bands:eq:zeta-square-shifts}
\end{equation}

For theta functions, primes denote derivatives with respect to the first argument. The theta representation is~\cite{OlverEtAl2010}
\begin{equation}
\sigma_\Lambda(w)= 2e^{w^2/(8\pi)}\dfrac{\vartheta_1(w/2|\ii)}{\vartheta_1'(0|\ii)}.
\label{bands:eq:sigma-theta}
\end{equation}
By logarithmic differentiation, we obtain
\begin{equation}
\zeta_\Lambda(w)=\dfrac{w}{4\pi} +\dfrac12\dfrac{\vartheta_1'(w/2|\ii)}{\vartheta_1(w/2|\ii)},
\label{bands:eq:zeta-theta}
\end{equation}
and
\begin{equation}
\wp_\Lambda(w)=-\dfrac{1}{4\pi} -\dfrac14\left[ \dfrac{\vartheta_1''(w/2|\ii)}{\vartheta_1(w/2|\ii)} -\left(\dfrac{\vartheta_1'(w/2|\ii)}{\vartheta_1(w/2|\ii)}\right)^2 \right].
\label{bands:eq:wp-theta}
\end{equation}

\subsection{Laguerre polynomials}
\label{special:sec:laguerre}

We use ordinary Laguerre polynomials $L_n(t)$ with $n=0,1,\ldots$ and parameter $\alpha=0$~\cite{OlverEtAl2010}. Their finite series and Rodrigues formula are
\begin{equation}
L_n(t)=\sum_{k=0}^n\binom nk\frac{(-t)^k}{k!},\qquad
L_n(t)=\frac{e^t}{n!}\frac{\d^n}{\d t^n}\left(t^n e^{-t}\right).
\label{special:eq:laguerre-definition}
\end{equation}
In particular, $L_0(t)=1$, $L_1(t)=1-t$, and $L_2(t)=1-2t+t^2/2$. They are orthonormal on $[0,\infty)$ with weight $e^{-t}$:
\begin{equation}
\int_0^\infty\!\d t\,e^{-t}L_j(t)L_m(t)=\delta_{jm},\qquad j,m=0,1,\ldots{}.
\label{cont:eq:laguerre-orthogonality}
\end{equation}
A function $f$ square-integrable with this weight has the mean-square expansion
\begin{equation}
f(t)=\sum_{m=0}^\infty b_m L_m(t),\qquad b_m=\int_0^\infty\!\d t\,e^{-t}f(t)L_m(t).
\label{special:eq:laguerre-expansion}
\end{equation}

\clearpage
\begingroup
\setstretch{1.15}
\makeatletter\let\auto@bib@innerbib\@empty\let\auto@bib\@empty\makeatother

\endgroup

\end{document}